\documentclass[epj,nopacs]{svjour}
\usepackage{graphics}
\usepackage[latin1]{inputenc}
\usepackage[dvips]{graphicx,epsfig,color}
\usepackage{wrapfig,rotating}
\usepackage{amssymb,amsmath,array}
\usepackage{color}
\newcommand{\up}{\raisebox{1.7ex}[-1.7ex]}

\newcommand{\cosths}{\ensuremath{\cos\theta^{*}}}
\newcommand{\qcosths}{\ensuremath{q_{\ell}\cdot\cosths}}
\newcommand{\rpv}{\mbox{$R_p \!\!\!\!\!\! / \;\;\;$}}
\newcommand{\neu}{\tilde{\chi}_1^0}
\newcommand{\grav}{\tilde{G}}

\def\gsim{\,\lower.25ex\hbox{$\scriptstyle\sim$}\kern-1.30ex%
\raise 0.55ex\hbox{$\scriptstyle >$}\,}
\def\lsim{\,\lower.25ex\hbox{$\scriptstyle\sim$}\kern-1.30ex%
\raise 0.55ex\hbox{$\scriptstyle <$}\,}

\begin{document}
\title{Review of Searches for Rare Processes and Physics Beyond the Standard Model at HERA}
\author{David M. South\inst{1} \and Monica Turcato\inst{2}\footnote{now at European X-ray Free-Electron Laser facility GmbH, Hamburg, Germany}% etc
}                     % Do not remove
\authorrunning{D. M. South, M. Turcato}
\titlerunning{Review of Searches for Rare Processes and BSM Physics at HERA}
%
%\offprints{}  % Insert a name or remove this line
%
%\institute{Deutsches Elektronen Synchrotron, Notkestrasse 85, 22607 Hamburg, Germany \and
%Institut f\"ur Experimentalphysik, Luruper Chaussee 149, 22761 Hamburg, Germany}
\institute{Deutsches Elektronen Synchrotron, Hamburg, Germany \and
Hamburg University, Institute of Experimental Physics, Hamburg, Germany}
\date{Received: date / Revised version: date}
\abstract{
The electron-proton collisions collected by the H1 and ZEUS experiments
at HERA comprise a unique particle physics data set, and a comprehensive range
of measurements has been performed to provide new insight into the structure of
the proton.
The high centre of mass energy at HERA has also allowed rare processes to be
studied, including the production of $W$ and $Z^{0}$ bosons and events with multiple
leptons in the final state.
The data have also opened up a new domain to searches for physics beyond
the Standard Model including contact interactions, leptoquarks, excited fermions
and a number of supersymmetric models.
This review presents a summary of such results, where the analyses reported 
correspond to an integrated luminosity of up to $1$~fb$^{-1}$, representing the
complete data set recorded by the H1 and ZEUS experiments.
\PACS{
      {PACS-key}{discribing text of that key}   \and
      {PACS-key}{discribing text of that key}
     } % end of PACS codes
} %end of abstract
\maketitle

\section{Introduction}
\label{sec:intro}

The Standard Model (SM) of particle
physics~\cite{Glashow:1961tr,Weinberg:1967tq,Salam:1968}, which
describes the fundamental building blocks of nature and their
interactions, is one of the success stories of the last 50 years in science.
The theoretical development and subsequent experimental confirmation
of the SM, which describes the elementary particles and their weak,
electromagnetic and strong interactions, has been made possible
by a variety of particle accelerators and their associated
experimental programmes operated during this time.
The successful exploration of the electroweak sector, from the
discovery of weak neutral currents in the bubble chamber experiments of
the early 1970's to the subsequent observation of the weak bosons
at the SPS at CERN in the early 1980's, greatly influenced the direction
of research and detector design, as well as the type and energy of
machines developed in the following decades.
Different types of programmes have been formed more recently,
based on making precision measurements such as the LEP experiments at
CERN, BaBar at SLAC and Belle at KEK, to all out discovery machines such as the
Tevatron at Fermilab or the LHC at CERN.
In the last few years, the LHC has produced one of the major physics
results of this exciting half century of particle physics, with the
observation~\cite{higgsATLAS,higgsCMS} of a new narrow resonance
consistent with the long sought Higgs
boson~\cite{Englert:1964et,Higgs:1964pj,Guralnik:1964eu}.

%%%

%The HERA accelerator at DESY is a unique achievement, in that it is
%the world's only electron\footnote{The term ``electron'' is used
%  generically to refer to both electrons and positrons, unless
%  otherwise stated.}-proton collider to be constructed, thus providing
%an unrivalled physics programme to the high energy physics community.
%
% Hack to get the author footnote in
%
The HERA accelerator at DESY is a unique achievement, in that it is
the world's only electron\footnote{The term ``electron'' is used
  generically to refer to both electrons and positrons, unless
  otherwise stated.\\ ~~~$^{\rm a}$ Now at European X-ray Free-Electron
  Laser facility GmbH, Hamburg, Germany.}-proton
collider to be constructed, thus providing
an unrivalled physics programme to the high energy physics community.
%%%
%%%
%
Bringing into collision point-like leptons with finite sized hadrons,
HERA may be thought of as a powerful electron microscope, a tool to
make precise measurements of the structure of the proton and to
investigate the nature of strong force binding it together.
The experimental data from the $ep$ collisions at HERA also allow rare
processes to be studied, typically those involving the electroweak
gauge bosons, and to search for physics beyond the Standard Model (BSM),
in analyses complementary to those at other colliders and sometimes
unique to $ep$ physics.

%%%

This review presents a summary of the measurements of rare processes
and BSM searches based on experimental data taken by the H1 and ZEUS
experiments at HERA.
The majority of the presented analyses utilise the complete data sets
of the experiments, which in combination amounts to an integrated
luminosity of $1$~fb$^{-1}$.
A brief introduction to the kinematics of $ep$ scattering is followed
by a description of HERA and the DESY accelerator facility and the H1
and ZEUS experiments.
An outline of particle identification, event reconstruction and
simulation is then given, describing the key methods
employed in the data analysis performed by each experiment.

%%%

To effectively perform searches for rare processes and new physics
in the data, it is important to have a good understanding and level
of confidence in the description of SM physics.
Therefore, a summary of the main experimental results on deep
inelastic scattering (DIS) at large momentum transfer is given in
section~\ref{sec:sm}, which are performed in a similar kinematic
region to the majority of the results presented in this review.
This is particularly important for the first category of searches,
where deviations from SM expectation of DIS events may reveal
new physics, namely searches for contact interactions and
leptoquarks as described in section~\ref{sec:ci}.
Dedicated searches for first, second and third generation
leptoquarks are then presented in section~\ref{sec:lq}.

%%%

Rare processes with high transverse momentum leptons in the final
state are investigated in sections~\ref{sec:mlep} and
\ref{sec:isolep}, featuring measurements of lepton pair and $W$
production.
Potential new signals are also investigated in the context of these
analyses in sections~\ref{sec:higgs} and~\ref{sec:singletop}.
The production of $Z^{0}$ bosons is examined by the ZEUS experiment
using the hadronic decay channel in section~\ref{sec:z0}.
A model independent ``general'' search for new physics is performed
by H1 as described in section~\ref{sec:gs}, where all final states are
examined containing high transverse momentum entities.
Searches for new physics in the context of specific models are then
presented, for excited fermions and supersymmetry in
sections~\ref{sec:fstar} and~\ref{sec:susy}, respectively. 
Finally a novel search for magnetic monopoles performed by the H1
experiment is presented in section~\ref{sec:monopoles}, before a
summary of the presented results is given together with an outlook in
section~\ref{sec:summary}.

\section{The kinematics of electron-proton scattering}
\label{sec:kinematics}

\begin{figure}[h]
\centerline{\includegraphics[width=0.75\columnwidth]{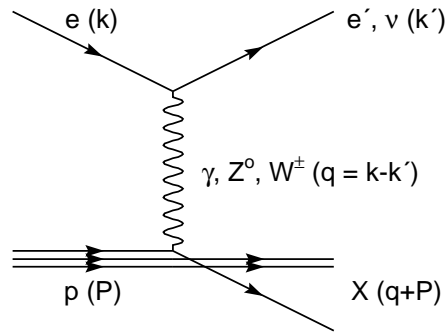}}
\caption{The scattering of electrons and protons at HERA. The four-momenta of the particles
  are indicated in the parentheses. The exchanged gauge boson is a photon ($\gamma$) or
  $Z^{0}$ boson in NC interactions and a $W$ boson in CC interactions.}
\label{fig:epscattering}
\end{figure}

Figure~\ref{fig:epscattering} shows a diagram of the interaction
$ep \rightarrow \ell X$, where a virtual photon ($\gamma$), or heavy
vector boson ($Z^{0}$ or $W$) is exchanged between the incoming
electron ($e$) and proton ($p$).
As the mediator boson between the electron and the proton can be
either a photon or a heavy vector boson, due to the high centre of
mass energy at HERA, QED and weak interactions may be studied
simultaneously, testing the electroweak theory.
The outgoing particles are made up of the scattered lepton ($\ell$)
and those contained in the hadronic final state ($X$). 
In the case of neutral current (NC) interactions, the exchange is
mediated by a $\gamma$ or $Z^{0}$, so that an electron ($e'$) is
present in the final state.
In charged current (CC) interactions, the weak exchange of the $W$ boson
results in a final state neutrino ($\nu$).
The four momenta of the initial state electron and proton are denoted $k$ and
$P$, so that the centre of mass energy $\sqrt{s}$ is given by:
\begin{equation}
  s  = (k + P)^2 .
  \label{eq:s}
\end{equation}
Neglecting the masses of the incoming particles, this can be approximated by
$\sqrt{s} \approx 4 E_{e}^{0} E_{p}^{0}$, where $E_{e}^{0}$ ($E_{p}^{0}$) is the energy of the initial
state electron (proton).
The square of the four-momentum transfer (which is equal to the mass squared of the virtual boson),
$q^{2} < 0$, determines the hardness, or in other words, the resolving power of the interaction. 
The negative four-momentum transfer squared is defined as:
\begin{equation}
  Q^{2}  = -q^{2} = -(k -k')^{2}
\label{eq:q2}
\end{equation}
and the invariant mass $W$ of the hadronic final state $X$ is given by:
\begin{equation}
  W^{2}  =  (q + P)^{2} .
\label{eq:w2}
\end{equation}
Interactions at HERA are denoted {\it elastic} if the proton remains
intact, {\it quasi-elastic} if the proton dissociates into a low-mass
hadronic system,  or {\it inelastic} if the proton breaks up.
Deep inelastic scattering (DIS) events are characterised as having
large momentum transfer $Q^{2} \gg m_{p}^{2}$ and are highly
inelastic, $W \gg m_p$, where $m_p$ is the mass of the proton.

%%%

The fraction of the proton momentum carried by the struck parton 
is given by the quantity {\it Bjorken} $x$, where:
\begin{equation}
  x = \frac{Q^2}{2 P \cdot q}
  \label{eq:x}
\end{equation}
under the assumption of massless quarks.
The {\it inelasticity} of the interaction, $y$, is given by:
\begin{equation}
  y  = \frac{q \cdot P}{k \cdot P}
  \label{eq:y}
\end{equation}
and is equal to the fraction of the incident electron momenta
carried by the exchange boson in the rest frame of the proton.
Both $x$ and $y$ take values between $0$ and $1$ since they describe
fractions of momenta, as defined above.
The above quantities are related by:
\begin{equation}
  Q^2 \approx sxy ,
  \label{eq:qsxy}
\end{equation}
resulting in a maximum squared four-momentum exchange
equal to the centre of mass energy squared $s$.  
%%%

The value of $Q^{2}$ is a measure of the {\it virtuality} of the exchange boson: for $Q^{2} \approx 0$,
the photon is considered to be almost real, and in such interactions the initial state electron
scatters under very small angles.
An interaction at higher $Q^{2}$ represents a more energetic exchange, resulting in a higher resolution
of the parton participating in the interaction: one of the three valence quarks,
a sea quark or a gluon.
Hence the exchanged boson acts as a probe for the determination of the structure of the
proton and depending on the scattering angle and energy of the outgoing lepton, different
$Q^2$ and $x$ ranges can be investigated, thus examining the electromagnetic and weak charge
distribution inside the proton.
\begin{figure}[!h]
\centerline{\includegraphics[width=1.0\columnwidth]{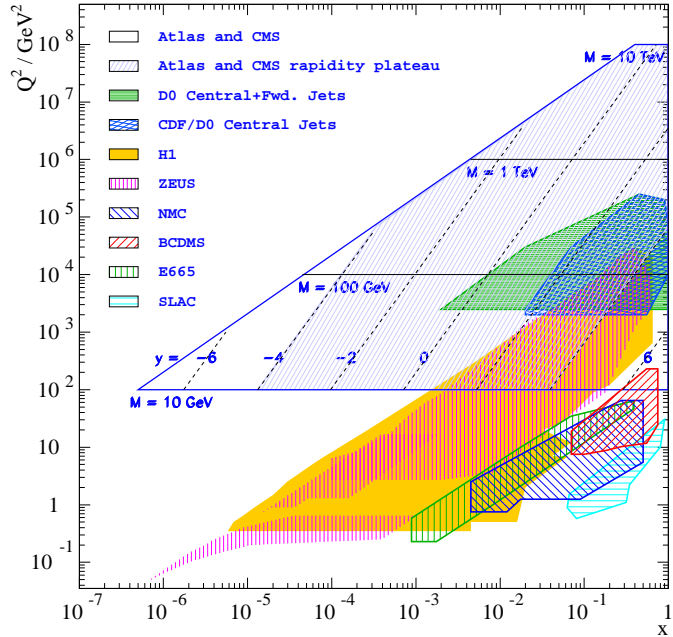}}
\caption{Regions of phase space in the $x$-$Q^2$ kinematic plane covered by several
  collider and fixed target experiments.}
\label{fig:kinematicplane}
\end{figure}

%%%

The advent of the HERA collider in 1992 made it possible to explore a much wider region in $x$ and $Q^{2}$
than that previously accessible at fixed target experiments, with measurements possible down to
$10^{-5}$ in $x$ and low $Q^2$, in particular in the non-perturbative
region and up to a $Q^{2}$ of $10^{4}$~GeV$^{2}$ in the valence (high-$x$) region.
The kinematic coverage of the HERA experiments in the $x$-$Q^{2}$ plane is shown in
figure~\ref{fig:kinematicplane}, compared to that of several fixed target DIS experiments as well as
the phase space covered by the hadron-hadron collision experiments at both the Tevatron ($p\bar{p}$)
and the LHC ($pp$). 
By exploiting QCD factorisation~\cite{Collins:1989gx} and utilising
the DGLAP~\cite{Gribov:1972ri,Altarelli:1977zs,Dokshitzer:1977sg} parton evolution scheme, the
HERA parton distribution functions (PDFs) derived from H1 and ZEUS measurements across a
large range in $x$ can be used as input to calculate predictions for the LHC at much higher
values of $Q^{2}$.
We will return to these measurements and the calculation and impact of the HERA PDFs
in section~\ref{sec:sm}.

\section{The HERA collider at DESY}
\label{sec:hera}

\begin{figure}[t]
  \centerline{\includegraphics[width=1.0\columnwidth]{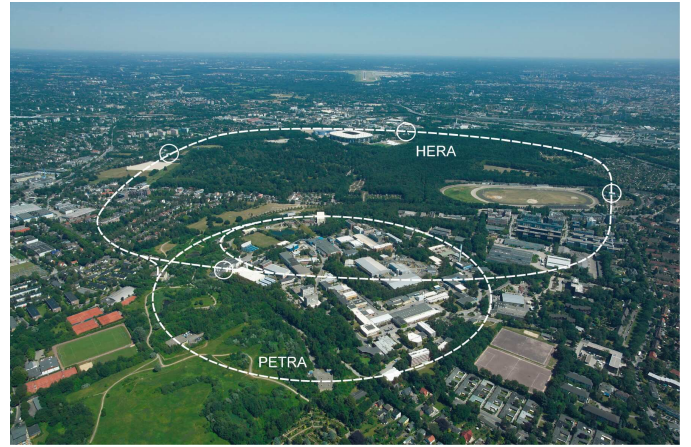}}
  \vspace{0.5cm}
  \centerline{\includegraphics[width=1.0\columnwidth]{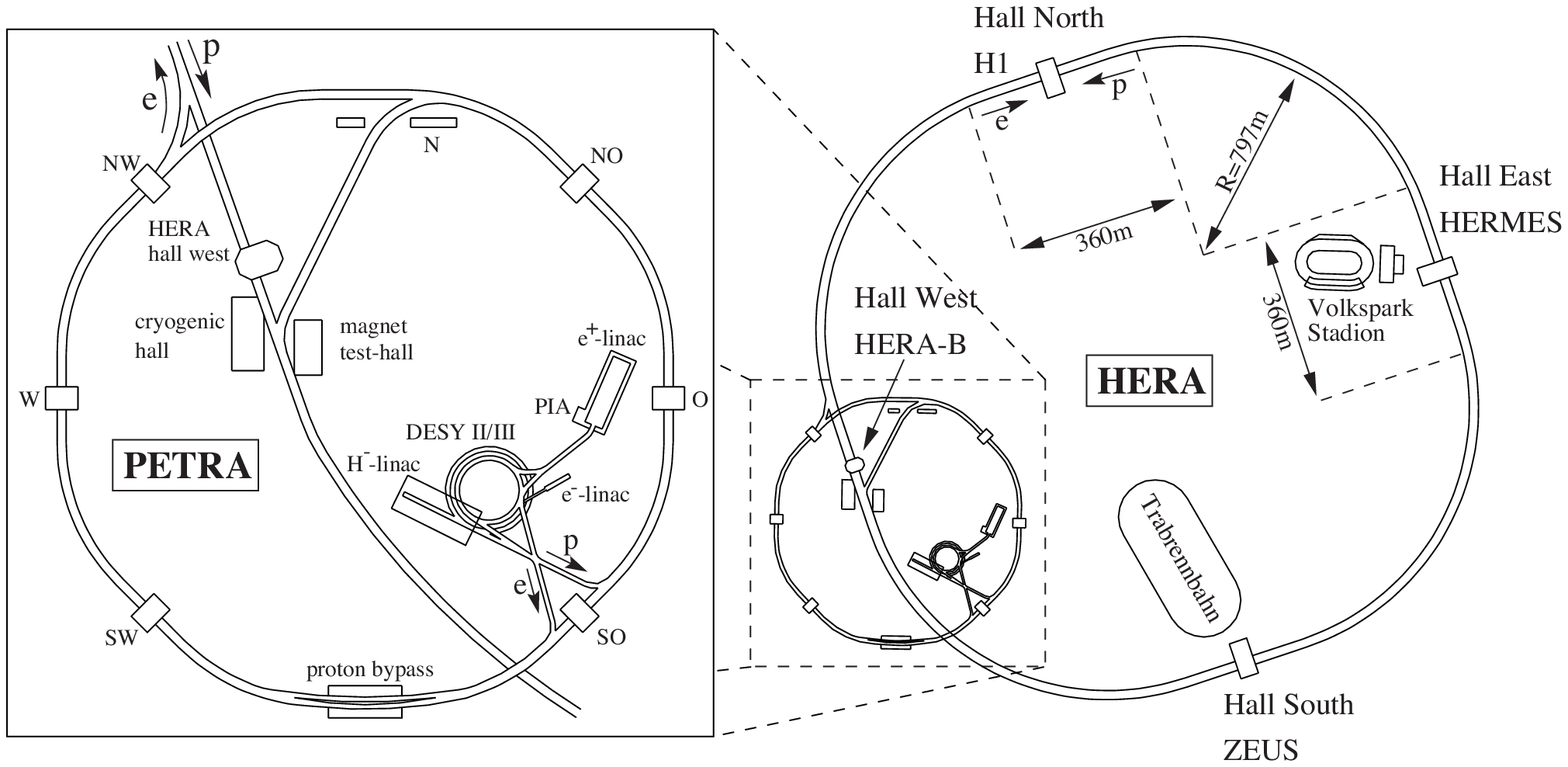}}
  \caption{The DESY research centre in Hamburg, Germany. In the upper photograph, the dashed circles
    show the path of the underground ring accelerators PETRA and HERA; the four large halls
    containing the HERA detectors are indicated by the smaller, solid circles. The lower diagram details the
    system of pre-accelerators employed in order to produce the proton and electron beams that were
    brought into collision at the H1 and ZEUS experiments.}
  \label{fig:heralayout}
\end{figure}

HERA (Hadron Electron Ring Anlage)~\cite{heraacc} is so far the only lepton-proton collider
in the world to have been constructed\footnote{The conceptual design report of the proposed
LHeC project, a machine to collide a high energy electron beam with the hadron beams of the
LHC, is now available~\cite{AbelleiraFernandez:2012cc}.}. 
It was located at the DESY (Deutsches Electronen Synchrotron) laboratory, as pictured in
the upper half of figure~\ref{fig:heralayout}, and was in operation during the years 1992 to 2007.
The HERA machine accelerated and brought into collision $27.6$~GeV electrons or positrons with
$920$~GeV protons\footnote{The proton beam energy was $820$~GeV from 1992-1997,
resulting in a centre of mass energy of $300$~GeV. The data recorded during this period amounts
to less than $10$\% of the total integrated luminosity yield.}, resulting in a centre of mass energy
of $319$~GeV.
At the time, this energy represented more than an order of magnitude increase with respect to
the previous fixed-target experiments and consequently a new and wider kinematic region
was accessible for the first time at HERA.
In the final data taking period the proton beam was accelerated to lower energies, first
$460$~GeV and then $575$~GeV, in order to provide data used for a direct measurement
of the longitudinal structure function $F_{L}$ (see section \ref{sec:sm}).

The $6.3$~km long HERA tunnel is located between $15$ and $30$ metres below ground, in which
electrons and protons were accelerated in two counter-rotating rings.
Four experiments were located in experimental halls around the HERA ring: 
Electron-proton collisions occurred at two interaction points, one in the North Hall where the H1
experiment~\cite{Abt:1996hi,Abt:1996xv} was located, the other in the South Hall where the ZEUS
experiment~\cite{ZEUSdetector} could be found.
The HERMES experiment~\cite{Ackerstaff:1998av} in the East Hall studied the spin structure
of the nucleon using collisions of the lepton beam on an internal polarised gas target.
The  HERA-B experiment~\cite{Hartouni:1995cf,HERAB:2000aa} in the West Hall was built to use
collisions of the proton beam halo with a wire target in order to produce $B$-mesons for the study
of CP violation in the $B-\bar{B}$ system.
The layout of the HERA ring and the system of pre-accelerators at DESY is illustrated in the
lower half of figure~\ref{fig:heralayout}.

%%%

The proton beam began as negative hydrogen ions (H$^{-}$) accelerated in a linear accelerator
to $50$~MeV.
The electrons were then stripped off the H$^{-}$ ions to obtain protons, which were injected
into the proton synchrotron DESY III, accelerated up to $7.5$~GeV, and transferred to the
PETRA ring, where they were accelerated to $40$~GeV.
The protons were then finally injected in three shots into the HERA proton storage ring, which is
made up of superconducting magnets with a maximum field of $4.65$~T, where they were then
accelerated to the nominal beam energy of $920$~GeV.  

%%%

The electron (positron) pre-acceleration chain began in a linear accelerator, LINAC I (LINAC II),
where the leptons were accelerated up to $450$~MeV.
The leptons were then injected into the electron synchrotron DESY II, accelerated to $7$~GeV
and, similarly to the protons, transferred to the PETRA ring, where they reached an energy
of $14$~GeV.
Injection transfer into the HERA ring followed, where they were accelerated to the nominal
lepton-beam energy of $27.6$~GeV using conventional magnets with a maximum field
of $0.165~$T.

%%%

Up to $210$ bunches of leptons and protons were accelerated in the HERA ring, spaced
at $96$~ns intervals.
Only $175$ bunches were typically used for collisions, where the remainder were used as {\it pilot}
bunches to study background rates arising from interactions of the beams with residual
gas in the beam-pipe.
When the proton bunches were compressed by HERA during acceleration, small secondary
or {\it satellite} bunches were formed, separated from the main bunch by up to $8$~ns.

%%%

\begin{figure}[t]
\centerline{\includegraphics[width=1.0\columnwidth]{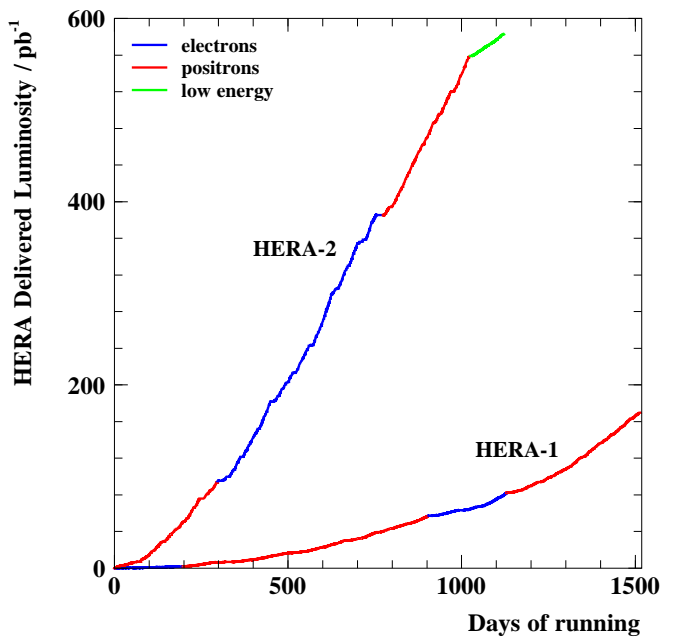}}
\caption{A summary of the integrated luminosity delivered by the HERA collider during the HERA~I
  (1992-2000) and HERA~II (2003-2007) phases. The different electron and positron running periods
  are indicated, as well as the data taken at lower proton beam energies in 2007.}
\label{fig:lumi}
\end{figure}

The data taking at HERA may be divided into two distinct periods: HERA~I, which was from
1994 until 2000, and HERA~II, from 2003 until 2007.
A luminosity upgrade \cite{heraupgrade} of the machine took place between the two
data taking periods and brought an observed increase in the luminosity delivered to the
experiments from $1.5~\times~10^{31}$ cm$^{-2}$~s$^{-1}$ in the HERA~I phase up to a
peak value of $5.0~\times~10^{31}$ cm$^{-2}$~s$^{-1}$, achieved during HERA~II $e^{-}p$
running.
The integrated luminosity delivered by the HERA accelerator is shown in
figure~\ref{fig:lumi}.

%%%

The integrated luminosity collected for analysis by H1 and ZEUS amounts to
about $0.5$~fb$^{-1}$ per experiment.
This is less than the delivered integrated luminosity, as quality conditions are applied
to the data used for analysis, such as requirements on the high voltage status of the
various detector subsystems (see section~\ref{sec:det}).
The luminosity is measured by both experiments from the rate of the  well understood QED
Bethe-Heitler process $ep \rightarrow ep \gamma$.
As the photon is emitted almost collinear to the incoming electron, it is detected using
devices located close to the beam line beyond the main detectors.
A photon detector~\cite{suszyski,levonian,Ahmed:1995cf} is employed by H1 and the ZEUS
experiment uses two independent systems, a photon
calorimeter~\cite{lumi1,zfp:c63:391,app:b32:2025} and a magnetic
spectrometer~\cite{Helbich:2005qf}.
A recent analysis~\cite{Aaron:2012kn} of Compton scattering events provided an
alternative and improved measurement of the luminosity recorded by the H1 experiment.
The integrated luminosities of the data sets\footnote{Note that for
  some analyses presented in this review the integrated luminosity
  may vary from this table. For example, some searches do not require
  a good polarisation measurement and this results in a higher
  luminosity yield from HERA~II. In such cases, the integrated
  luminosity of the analysed datasets is given in the text.} are detailed in table~\ref{tab:datasets}.

Another feature of the HERA~II upgrade was the use of a longitudinally polarised lepton beam.
As the lepton beam circulated in HERA it naturally became transversely polarised via the
Sokolov-Ternov effect~\cite{sokolov-ternov1,sokolov-ternov2}.
The typical polarisation build-up time for the HERA accelerator was approximately 40 minutes.
At HERA~II, spin rotators installed on either side of the H1 and ZEUS detectors changed the
transverse polarisation of the beam into longitudinal polarisation and back again.
The lepton beam polarisation was measured using two independent polarimeters, the transverse
polarimeter (TPOL)~\cite{tpol} and the longitudinal polarimeter (LPOL)~\cite{lpol}.
Both devices exploited the spin-dependent cross section for Compton scattering of circularly
polarised photons off positrons to measure the lepton beam polarisation~\cite{Sobloher:2012rc}.

%%%

During the HERA~II period, the machine was run in both left handed and right handed polarisation
modes, where the average beam polarisation is given by:

\begin{equation}
  \mathcal{P}_{e} = (N_{R} - N_{L})/(N_{R} + N_{L}) 
  \label{eq:pe}
\end{equation}
and $N_{R}$ ($N_{L}$) is the number of right (left) handed leptons in the beam.
Accordingly, four distinct data sets were recorded by the experiments at HERA~II, by running
with either left or right handed polarised electrons or positrons.
The luminosity weighted average polarisations of the HERA~II data sets are also given in
table~\ref{tab:datasets}.

\begin{table}[h]
  \renewcommand{\arraystretch}{1.3}
\caption{Integrated luminosity $\mathcal{L}$ and luminosty-weighted average
  lepton beam polarisation $\mathcal{P}_{e}$ of the H1 and ZEUS
  data. The centre of mass energy $\sqrt{s}$ of the data is $319$~GeV,
  except for the 1994-1997 data, which is $301$~GeV.}
\label{tab:datasets}
\begin{tabular*}{1.0\columnwidth}{cccccc}
\hline
\multicolumn{2}{l}{Data set} & \multicolumn{2}{c}{H1} & \multicolumn{2}{c}{ZEUS} \\
& & $\mathcal{L}$ [pb$^{-1}$] & $\mathcal{P}_{e}$ [\%] & $\mathcal{L}$ [pb$^{-1}$] & $\mathcal{P}_{e}$ [\%] \\
\hline
1994-1997 & $e^{+}p$ & $36$ & ~~~~$0$ & $48$ & ~~~~$0$\\ 
1998-1999 & $e^{-}p$ & $16$ & ~~~~$0$ & $16$ & ~~~~$0$\\ 
1999-2000 & $e^{+}p$ & $65$ & ~~~~$0$ & $63$ & ~~~~$0$\\
\hline
HERA II & $e^{+}p$ & $98$ & $+32$ & $91$ & $+32$\\
            &                & $82$ & $-38$ & $68$ & $-37$\\
\hline
HERA II &$e^{-}p$ & $46$ & $+37$ & $71$ & $+29$\\
             &             & $103$ & $-26$ & $99$ & $-27$\\
\hline
\end{tabular*}
\end{table}

\section{Particle physics detectors}
\label{sec:det}

\begin{figure}[t]
\centerline{\includegraphics[width=0.85\columnwidth]{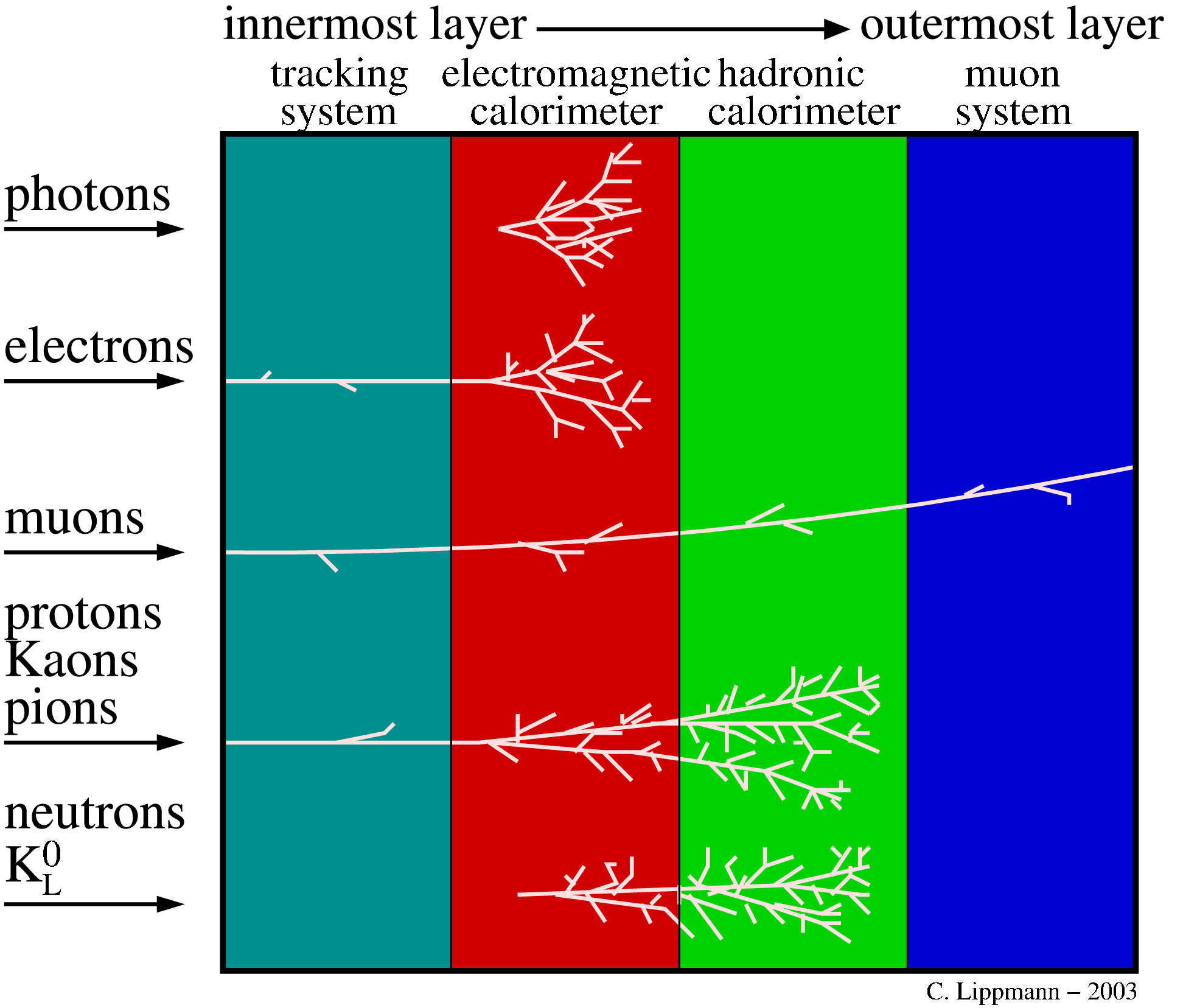}}
\caption{Typical components of a particle physics detector. Different
types of particle leave different signatures in the detector, allowing particle
identification to be performed~\cite{Lippmann:2011bb}.}
\label{fig:pidparticles}
\end{figure}

A general purpose particle physics experiment is normally composed of
a series of different detectors surrounding the interaction region,
which is the nominal point where the two counter-rotating beams of
particles are brought into collision.
Each detector identifies and measures particles produced in the interaction by taking
advantage of their different properties.
Figure~\ref{fig:pidparticles} shows a general layout of a particle physics experiment:
a tracking device in a magnetic field is surrounded by an electromagnetic and an hadronic
calorimeter, and finally by muon detectors.
Particles produced in the interaction region traverse these detectors sequentially from
the collision point outwards, leaving different signatures depending on their physics
properties.

%%%

The first detector surrounding the interaction region is a tracking system, which records
the passage of charged particles.
This system may be comprised of multiple detectors, such as drift chambers and silicon trackers.
Charged particles leave their energy in these detectors via ionisation, and due to the low material
budget only a small fraction of the particle energy is lost in such devices.
In general neutral particles are not seen as tracks, but photons may convert into an electron-positron pair
which can then be detected in the tracker.
Tracks are reconstructed by detecting the particle energy lost along their path inside the detector.
A magnetic field, which is parallel to the beam direction, allows the particle momentum and charge to be
determined, based on the curvature of the tracks.
The characteristic pattern of energy loss in the tracker may also be used to perform particle identification,
allowing the separation of electrons, pions, kaons and protons at low momenta.

%%%

Moving outwards from the interaction region, the tracking system is enclosed inside a calorimeter.
Calorimeters measure both neutral and charged particles by completely degrading them and absorbing
all of the energy they deposit.
A typical design is a \emph{sampling calorimeter}, in which layers of active material such as scintillator
or liquified noble gas are interleaved by layers of absorber such lead or depleted uranium.
A calorimeter is also able to determine if a particle has electromagnetic or hadronic interaction by examining
the pattern of the energy loss.
This allows the separation of electron and photons from other particles which undergo hadronic interactions.
A calorimeter is designed such that electrons and photons leave all of their energy in the
electromagnetic section, which is closer to the interaction region and typically many
radiation lengths deep.
The electromagnetic interactions occur rapidly with the nuclei in the absorbing layers via the
bremsstrahlung and pair-production processes.
As electrons and photons cannot be distinguished using the calorimeter alone, this is done by spatially
matching the electromagnetic energy deposit to a track in the tracking system.
Hadrons penetrate more deeply in the detector, leaving energy deposits not only in the electromagnetic
calorimeter but also in the hadronic section, where they interact strongly with the nuclei of the absorbing
layers, elastically and inelastically, resulting in a shower composed of secondary hadrons.
The characteristic length of the hadronic shower, the {\it interaction length}, is much longer than
the radiation length for the same material and the shower, which also contains an electromagnetic
component, is typically much broader than a purely electromagnetic interaction, allowing the two
shower types to be distinguished.
A full discussion on the separation of electromagnetic and hadronic showers
can be found elsewhere~\cite{Kogler:2011zz}.

%%%

Muons do not interact like hadrons via the strong force and do not radiate via bremsstrahlung as much as
electrons due to their heavier mass, losing their energy only via ionisation and behaving like
a minimum ionising particle.
They therefore penetrate beyond the main calorimeters and additional, dedicated muon detectors are
installed as the outermost layer of the detector.
In order to make a muon momentum measurement, information from these outermost detectors is typically
matched to a track measured in the tracking system.

%%%

A detailed description of the H1 and ZEUS detectors can be found
elsewhere~\cite{Abt:1996hi,Abt:1996xv,ZEUSdetector}.
Both detectors are described by a right handed  cartesian coordinate system $(x,y,z)$
with the nominal interaction point defined at the origin, $+x$ pointing towards the
centre of the ring, $+y$ pointing vertically upwards and $+z$ in the direction of the
incoming proton beam (also referred to as the {\it forward} direction).
The corresponding spherical coordinate system $(r,\theta,\phi)$ is defined so that
$\theta = 0^{\circ}$ is in the proton direction and consequently $\theta = 180^{\circ}$
is in the electron ({\it backward}) direction.

%%%

\begin{figure}[h]
  \centerline{\includegraphics[width=1.0\columnwidth]{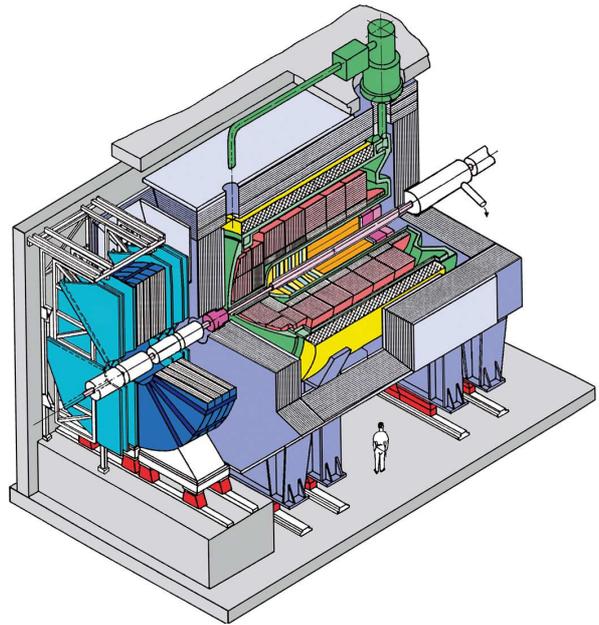}}
  \vspace{1cm}
  \centerline{\includegraphics[width=1.0\columnwidth]{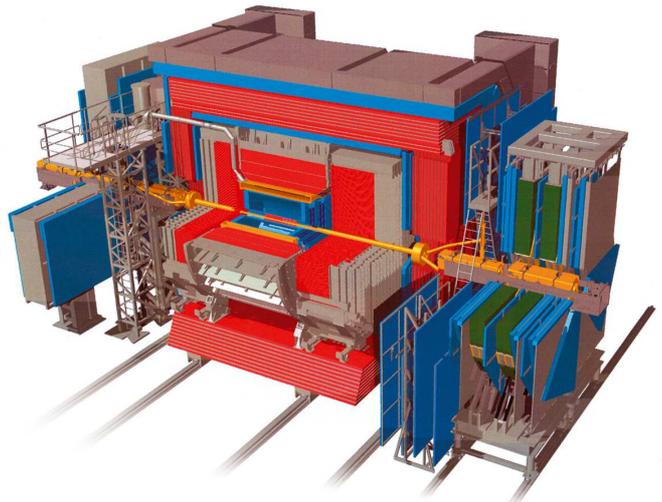}}
  \caption{The H1 (top) and ZEUS (bottom) multi-purpose detectors employed at HERA.
    The proton beam enters the H1 (ZEUS) detector from the right (left).}
  \label{fig:detectors}
\end{figure}

The H1 and ZEUS detectors are illustrated in figure~\ref{fig:detectors}.
Both designs feature tracking detectors closest to the beam pipe, which runs
through the centre, surrounded by electromagnetic and hadronic calorimetry, which
is enclosed in a muon system.
Two striking features can be seen in the design of both the H1 and ZEUS detectors.
Firstly, almost complete coverage is achieved around the interaction point.
This not only allows particles produced in the $ep$ interaction to be almost
completely contained, but also results in a reliable calculation of any
net transverse momentum imbalance in an event, which is important if the resulting
final state includes neutrinos.
Secondly, both designs are very asymmetric,  corresponding to the large
asymmetry in the beam energies.
Thus the backward region of each detector is mainly dedicated to the detection
of the scattered electron, whereas the forward region contains more
instrumentation and a deeper coverage.
More details on the main H1 and ZEUS detector components are given in the
following sections.

%%%

\subsection{The H1 detector}
\label{sec:det-h1}

The central and forward tracking detectors, which cover the regions
\mbox{$20^\circ < \theta < 160^\circ$} and \mbox{$7^\circ < \theta < 25^\circ$}
respectively, are used to measure charged particle trajectories and to reconstruct
the interaction vertex.
These detectors are enclosed together with the Liquid Argon (LAr)
calorimeter inside a superconducting
magnetic coil with a field strength of $1.16$~T parallel to the $z$ axis.
From the curvature of charged particle trajectories in the magnetic field, the central
tracking system provides transverse momentum measurements with a
resolution of
$\sigma_{P_T}/P_T = 0.002 P_T /~{\rm GeV} \oplus~0.015$.
The charge of the particle may also be ascertained from the direction of the curvature,
and an accurate measurement of the tracks in an event provides spatial information on
the interaction vertex.
The central and forward tracking detectors are complimented at the very centre of the H1 detector
by the central~\cite{Pitzl:2000wz}, backward~\cite{Eick:1996gv} and (during HERA~II only)
forward silicon trackers, providing improved $z$-resolution and polar angle measurements.
Additional track reconstruction is provided in the backward region by the backward drift
chamber (backward proportional chamber) during the HERA~I (HERA~II) running period.

%%%

The LAr calorimeter~\cite{Andrieu:1993kh} covers the polar angle range
\mbox{$4^\circ < \theta < 154^\circ$} with full azimuthal ($\phi$) acceptance and is composed of
two sections: an electromagnetic calorimeter (EMC) and a hadronic calorimeter (HAC).
The passive layers of the EMC are formed from 2.4mm thick lead plates, whereas the
HAC uses 16mm thick plates of stainless steel; liquid argon forms the common sampling
layer for both the EMC and the HAC.
Charged particles produced in the shower ionise the argon atoms and the resulting
electrons are converted to a signal and read out.
The LAr is a {\it non-compensating} calorimeter, resulting in a $30$\% loss of the
initial hadronic energy to the showering process. To account for this, an offline software
technique is employed \cite{compensate}.
The energies of electromagnetic showers are measured in the LAr calorimeter with a precision of
$\sigma (E)/E \simeq 11\%/ \sqrt{E/\mbox{GeV}} \oplus 1\%$ and hadronic
energy deposits with $\sigma (E)/E \simeq 50\%/\sqrt{E/\mbox{GeV}} \oplus 2\%$,
as determined in test beam measurements~\cite{Andrieu:1993tz,Andrieu:1994yn}.
A lead-scintillating~fibre calorimeter\footnote{This device was installed in 1995, replacing a
lead-scintillator sandwich calorimeter~\cite{Abt:1996xv}.}~(SpaCal)~\cite{Appuhn:1996na} covering the
backward region $153^\circ < \theta < 178^\circ$ completes the measurement of
charged and neutral particles.
Like the LAr calorimeter, it is divided into electromagnetic and hadronic sections, although its primary
function is the detection of electrons scattered through low angles.
Both sections consist of scintillating fibres, which form the sampling material, embedded in
a lead matrix absorber.
Charged particles produced by showering in the lead cause the fibres to scintillate, and
the resulting light is recorded using photomultipliers.
The response to electrons is given by
$\sigma (E)/E = 7\%/ \sqrt{E/\mbox{GeV}} \oplus 1\%$ and
$\sigma (E)/E = 13\%/ \sqrt{E/\mbox{GeV}} \oplus 4\%$ for the
electromagnetic~\cite{Nicholls:1995di} and
hadronic~\cite{Appuhn:1996xh} sections, respectively.

%%%

Two muon detection systems are used by H1.
The return yoke of the magnetic coil is the outermost part of the detector and is equipped with
limited streamer tubes (LSTs), forming the central muon detector,
covering the range $4^\circ < \theta < 171^\circ$. 
When the streamer occurs within the LST, pulses generated on the strips are combined by
the central muon trigger to reconstruct the muon track.
In the central (barrel) region, at least $2$ hits in the inner $4$ layers are required.
In the endcaps, which complete the iron shell around the detector, at least $3$/$5$ hits are required.
The signals from the central muon system are combined with tracking information from the central
tracking detector to form the muon momentum measurement.
In the very forward region \mbox{($3^\circ < \theta < 17^\circ$)} a set of six double layers of drift
chambers, three either side of a central $1.5$~T toroidal magnet, is used to detect muons and measure
their momenta.
Each layer is divided into octants and a total of $1520$ drift chambers are used in the detector,
varying in length from $0.40$m to $2.40$m.
The field of the main H1 solenoid has little effect at such low $\theta$ angles and so the
toroidal magnet is used to bend the path of traversing muons, enabling an independent
momentum measurement.

%%%

The H1 detector contains approximately $270,000$ readout channels, which
combined with the HERA bunch crossing frequency of $96$~ns (equivalent
to an event rate of $\approx 10$ MHz) provides a potential rate of data flow that is
too high for the detector components and electronics employed to process.
To cope with this, a multi-level trigger system is employed. The first level
comprises triggers built using about $200$ different {\it subtriggers}, each using
basic information from different parts of the detector: calorimeter energies,
tracks multiplicities and so on.
The level one subtriggers fired by more common physics processes are only accepted by the
central trigger a fraction of the time using a technique called {\it prescaling}: for
a trigger with a prescale of $p$, only 1 in $p$ events are kept.
Level two triggers add additional information and combine existing triggers, often utilising
different subdetectors. There are two components: a topological trigger, which employs
pattern recognition using a 2D projection (or {\it topology}) of the event in $\theta$ and
$\phi$ and a neural network based system.
During HERA~II particle identification was employed at level three, using the
Fast Track Trigger (FTT).
Finally, the level four trigger runs on a filter farm and provides more detailed
reconstruction and selection of tracks and clusters.
The level 4 farm runs at up to $50$~Hz, at which rate the H1 data events are written out
for analysis.

\subsection{The ZEUS detector}
\label{sec:det-zeus}

Charged particles are detected by ZEUS using the Central Tracking Detector
(CTD) \cite{nim:a279:290,npps:b32:181,nim:a338:254}, the Microvertex Detector
(MVD) \cite{nim:a581:656} and the straw-tube tracker (STT)~\cite{nim:a535:191}.
The CTD and the MVD are located in a magnetic field of $1.43$~T, provided by a thin
superconducting solenoid.
The CTD consists of $72$ cylindrical drift chamber layers, organised in nine
superlayers covering the polar-angle region $15^\circ < \theta  < 164^\circ$.
The MVD silicon tracker consists of a barrel (BMVD) and a forward (FMVD) section.
The BMVD provides polar angle coverage for tracks with three measurements in the
range $30^\circ < \theta < 150^\circ$.
The FMVD extends the polar-angle coverage in the forward region down to $7^\circ$.
The STT covers the polar-angle region $5^\circ < \theta < 23^\circ$ and consists of $48$ sectors
of two different sizes.
Each sector, which is trapezoidal in shape and subtends an azimuthal angle of $60^\circ$, contains
$192$ (small sector) or $264$ (large sector) straws of diameter $7.5$~mm, arranged into three layers.
Six sectors form a superlayer, and a particle passing through the complete STT traverses eight superlayers,
which are rotated around the beam direction at angles of $30^\circ$ or $15^\circ$ to each other. 

%%%

A high-resolution, uranium-scintillator calorimeter\\
(CAL) \cite{nim:a309:77,nim:a309:101,nim:a321:356,nim:a336:23} is employed by ZEUS, consisting
of three parts: the forward (FCAL), the barrel (BCAL) and the rear (RCAL) calorimeter, covering $99.7$\%
of the solid angle around the nominal interaction point.
Each part is subdivided transversely into towers and longitudinally into one electromagnetic section
(EMC) and either one (RCAL) or two (BCAL and FCAL) hadronic sections (HAC).
The relative energy resolutions of the CAL, as measured under test-beam conditions, are
$\sigma(E)/E = 18\%/ \sqrt{E/\mbox{GeV}}$ for electrons and
$\sigma(E)/E = 35\%/ \sqrt{E/\mbox{GeV}}$ for hadrons.
The timing resolution of the CAL is better than $1$~ns for energy deposits exceeding $4.5$~GeV.
Pre-sampler detectors~\cite{Bamberger:1996hi,zeus-pres2} are mounted in front of the CAL, consisting
of scintillator tiles matching the calorimeter towers to measured signals from particle showers created
by interactions in the material located between the interaction point and the calorimeter.
The RCAL is instrumented with a layer of $3\times 3~{\rm cm^2}$ silicon-pad detectors at a depth of
$3.3$ radiation lengths forming the hadron-electron separator~\cite{Dwurazny:1988fj}, which is used
to improve the electron angle measurement.

%%%

The ZEUS muon system consists of a central barrel (BMUON) and rear (RMUON) tracking
detectors~\cite{nim:a333:342}, in addition to the forward (FMUON) tracking detectors~\cite{ZEUSdetector}.
The BMUON ($34^\circ < \theta < 135^\circ$) and RMUON ($135^\circ < \theta < 171^\circ$) are comprised of
LSTs, located behind the CAL, and inside and outside the magnetised iron yoke surrounding the CAL.
The FMUON is made up of six planes of LSTs and four planes of drift chambers covering the angular region
$5^\circ < \theta < 32^\circ$.
Whereas the central and rear muon systems use the magnetic field of the iron yoke, two iron toroids
with a field strength of $1.6$~T provide an independent muon momentum measurement in the
forward direction.
Muons are also detected by ZEUS in the sampling Backing Calorimeter (BAC)~\cite{nim:a313:126}, which consists
of $5200$ proportional drift chambers, typically $5~{\rm m}$ long with a wire spacing of $1~{\rm cm}$.
The BAC is located within the magnetised iron yoke covering the CAL
and is equipped with analogue and digital readouts for energy and
tracking measurements, respectively.
The digital information from the hit wires allows the reconstruction of muon trajectories in two dimensions
($x-y$ in the barrel, $y-z$ in endcaps) with an accuracy of a few mm.

%%%

Similarly to H1, ZEUS also employs a multi-level trigger system.
The first level, which deals with the very high $10$~MHz rate, handles simple event level information,
such as calorimeter energy deposits or the number of tracks in the event.
At the second level more refined information is available, such as the arrival time of particles in the calorimeter,
which is used to achieve an efficient separation between $ep$ collision events and background.
At the third level, part of the offline reconstruction software is run
in order to select signal events.

\section{Particle identification and event reconstruction}
\label{sec:pid}

The design and concept of the HERA detectors are driven by the physics they are required to measure
and the resulting particle identification and event reconstruction methods employed by the H1 and ZEUS
experiments are similarly defined.
This section describes the identification of the particles and event level quantities used in the analyses
presented in this review.
A description of the various kinematic reconstruction methods, which require at least one of the
components detailed below, is given in the following section.

%%%

{\bf Electrons} are identified as compact, isolated energy clusters in the electromagnetic
part of the calorimeters.
Electron candidates are also required to have an associated
track, with a distance of closest approach (DCA) to the calorimetic
cluster typically less than $12$cm.
Further requirements on the electron track are often applied in the central region of the detector
where coverage is more complete, such as a minimum measured transverse momentum, $P_{T}$, and
a minimum radial starting position measured with respect to the nominal interaction point.
The electron cluster is required to be clean, such that the energy in a cone of radius~$1$
in pseudorapidity\footnote{The pseudorapidity is defined as $\eta= -\log{\tan (\theta/2)}$.}-azimuth
($\eta-\phi$) space around the electron cluster is limited to a small fraction of the electron energy.
Inefficient regions between calorimeter modules, where an electron may pass through the electromagnetic
section and into the hadronic section without interaction, are excluded using fiducial volume cuts.
The electron energy $E_{e}$ and polar angle $\theta_{e}$ are
determined from the calorimeter cluster; the azimuthal
angle $\phi_{e}$ is determined from the track.
A sample of NC events with a well contained hadronic final state is
used to perform a calibration of the electron energy, where the
measured electron energy in the calorimeter, $E_{e}$, is compared to
that determined via the ``double angle method'', $E_{DA}$, as described
in Section~\ref{sec:kinemethods}.
Further details on the electron calibration performed by H1 can be
found in \cite{Aaron:2012qi} and references therein.

%%%

{\bf Photons} are identified using the same criteria as electrons concerning the isolation of the 
electromagnetic cluster.
Conversely to electrons however, a track veto is applied, so that any photon candidates
with an associated track are rejected.
For example, in the case of H1, a minimum track-cluster DCA of 12 cm
and no tracks within a cone of radius $0.5$ in $\eta-\phi$ space
around the cluster are required.

%%%

{\bf Muons} are identified using a wide range of detector components, where the muon reconstruction
algorithms require a track in the tracking system, a minimum ionising particle (m.i.p.) energy deposit
in the electromagnetic and hadronic calorimeter, which is of the order of $1-2$~GeV, and a signal in the
outer muon detectors.
Muons used in the analyses presented in this review typically have a high enough transverse
momentum to traverse the tracking detectors and calorimeters, and reach the muon detectors, either central or forward,
which are furthest from the interaction point. 
The muon momentum $P_{T}^{\mu}$ and its angular variables $\theta_{\mu}$ and $\phi_{\mu}$ are given by the
associated track.
Muons with a transverse momenta lower than a few GeV do not reach the outer muon detectors and are usually 
stopped in the calorimeter, but still may be identified as muons by the characteristic calorimeter pattern.
Given the multi-detector nature of muon identification, a series of muon classes or grades are usually employed,
depending on which information is available.
These grades typically use track quality and DCA arguments, as well as requirements on the number of hit
in the muon system as well as matching between the measurements available.

%%%

{\bf Jets} are narrow cones of hadrons or other particles produced from the hadronisation of a quark or a gluon.
In the detector, they are reconstructed as clusters of energy in the electromagnetic and hadronic calorimeter, which
are recognised as coming from a collimated particle flow.
Tracking information can also be used at particle momenta for which the resolution of the tracking detector is
better that that of the calorimeter.

%%%

The features of jet in a hadronic final state are closely related to those of the partons originating them.
However, jets are complex objects which are not uniquely defined in QCD and whose definition relies on the
algorithms used to reconstruct them~\cite{Salam:2009jx}.
These come essentially in two types: \emph{cone algorithms}, in which a jet is defined as a cone of radius $R$ in the
$\eta-\phi$ plane, and \emph{clustering algorithms}, in which particles (or energy deposits) are assigned to jets
iteratively according to whether a given energy-angle resolution variable $y_{ij}$ exceeds a fixed resolution
parameter $y_{\rm cut}$.
Clustering algorithms are more reliable in hadron-hadron and lepton-hadron collisions, as they are not affected from
ambiguities related to the presence of overlapping jets in multi-jet events.

%%%

At H1 and ZEUS, jets are reconstructed using the $k_{T}$ clustering
algorithm~\cite{Catani:1993hr}, which uses the
relative transverse momentum $k_{T}$ between the particles as
a resolution variable to identify jets.
This algorithm is infrared and collinear safe at any order in QCD, can
be used with the same procedure both on theoretical calculations and
on experimental data and treats multi-jet events without ambiguities.
A jet is usually kinematically identified by its transverse energy $E_{T}^{\rm jet}$ and its angular variables
$\eta_{\rm jet}$ and $\phi_{\rm jet}$.
Typical values used in the $k_{T}$ clustering algorithm are $R < 1$ and a minimum $E_{T}^{\rm jet} = 5$~GeV.
Later publications~\cite{Andreev:2014wwa} have also employed the {\it anti}$-k_{T}$
algorithm; discussions on the merits of various jet clustering
algorithms are available elsewhere~\cite{Cacciari:2008gp}.

%%%

All identified leptons are excluded from the {\bf inclusive hadronic
  final state} and any energy around the lepton in a cone of radius of $0.5$
in $\eta-\phi$ space for electrons and radius $1.0$ for muons is also
typically excluded.
The cone is larger for muons as they tend to deposit energy in both parts of the calorimeter, whereas electrons are
generally stopped by the electromagnetic section.
Calibration of the hadronic final state is performed using a large NC sample by comparing the transverse momentum
of the calibrated electron, $P_{T}^{e}$ to that of the inclusive
hadronic final state, $P_{T}^{h}$.
The $P_{T}$ balance,  $P_{T}^{h} / P_{T}^{e}$ should be equal to $1$ in intrinsically balanced NC events and the data
are adjusted iteratively until they are in agreement with the MC
simulation.
The calibration procedure may involve additional steps, in particular
in the treatment of jets, see for example~\cite{Kogler:2011zz}.

%%%

The {\bf missing transverse momentum}, $P_{T}^{\rm miss}$, is
calculated using the vector sum of all identified particles and a
significant value of this quantity may indicate the presence of a
neutrino, or neutrinos, in the event.
The vector $\vec{p}_T^{\rm miss}$ is derived from the total visible
hadronic momentum vector, $\vec{p}_T$, by
$\vec{p}_T^{\rm miss}  = -\vec{p}_T$, where:
\begin{eqnarray}
\vec{p}_T & = & \left(P_x, P_y\right) \nonumber \\
& = & \left(\sum\limits_{i} E_i \sin \theta_i \cos \phi_i\;,\; \sum\limits_{i} E_i \sin \theta_i \sin \phi_i \right).
 \label{eq:ptmiss}
\end{eqnarray}
For example, in CC events a significant imbalance of transverse
momenta of measured final state particles is observed. As a consequence, the $P_x$ and $P_y$ components 
of $\vec{p}_T^{\rm miss}$ are non zero, and they can be attributed to the outgoing neutrino.
Many of the final states examined in this review feature neutrinos,
and as such $P_{T}^{\rm miss}$ is employed in such analyses.
A reliable measurement of this quantity is made possible via the
near hermetic coverage of the H1 and ZEUS detectors around
their respective interactions points.
A related quantity used in several H1 analyses is $P^{\rm calo}_{T}$,
the net transverse momentum calculated from all reconstructed
particles measured in the calorimeter.
This quantity reflects the missing transverse momentum as seen
by the trigger.
For events containing high energy muons, where only the (relatively
small) muon energy deposited in the calorimeter is included,
$P_T^{\rm calo} \simeq P_T^{h}$,
otherwise $P_{T}^{\rm calo}~=~P_{T}^{\rm miss}$.

\section{Kinematic reconstruction methods}
\label{sec:kinemethods}

The reconstruction of the scattered electron together with that
of the hadronic final state is of particular importance at HERA as
it allows the determination of the kinematic variables introduced in
Section~\ref{sec:intro}.
Indeed, one of the salient features of the H1 and ZEUS experiments
is the possibility to determine NC event kinematics from the
scattered electron or from the hadronic final state, or using a
combination of the two.
In the case of NC DIS, exclusively using the scattered
electron to determine the event kinematics results in a less model
dependent analysis, which is easier to interpret theoretically.
Resolution, measurement accuracy and effects due to the radiation of
photons by the incoming electron influence the choice of the
reconstruction method in a given kinematic region.
For a broader discussion of the reconstruction methods presented
in this section, see for example~\cite{Aaron:2012qi,Derrick:1996hn}.

%%%

For NC scattering, the ``electron method'' is favoured, where the
inelasticity and the negative four-momentum transfer squared are
calculated from the scattered electron energy $E_{e}$ and polar
angle $\theta_{e}$ as:
\begin{subequations}
\begin{equation}
  Q^{2}_{e} = \frac{{P_{T}^{e}}^{2}} {1 - y_{e}}\,,~~~y_{e} = 1-\frac{\Sigma_{e}}{2 E_{e}^{0}}\,,~~~x_{e} = \frac{Q^{2}_{e}}{s y_e}\,, 
  \label{eq:emeth}
\end{equation}
where $P_{T}^{e} = E_{e}\sin\theta_{e}$ is the electron transverse
momentum, $\Sigma_e = E_{e}(1-\cos\theta_{e}) = E_{e} - P_{z}^{e}$,
and $E_{e}^{0}$ is the energy of the initial state electron.
Using this method, the negative four-momentum transfer
squared may also be expressed as:
\begin{equation}
  Q^{2}_{e} = 4E_{e}^{0}E_{e}\cos^{2} \frac{\theta_{e}}{2}\,.
  \label{eq:emeth2}
\end{equation}
\end{subequations}

In the case of CC events, as the scattered neutrino is not
detected only the information on the hadronic final state can be
used for the reconstruction of the event kinematics.
The ``hadron method'' or ``Jacquet-Blondel method''~\cite{yjb} uses
similar relations to those defined in eqation~\ref{eq:emeth}, obtained
exclusively from the reconstructed hadronic final state:
\begin{equation}
  Q^2_{h} = \frac{{P_{T}^{h}}^2} {1 - y_{h}}\,,~~~y_{h} = \frac{\Sigma_{h}}{2 E_{e}^{0}}\,,~~~x_{h} = \frac{Q^{2}_{h}}{s y_{h}}\,,
  \label{eq:jbmeth}
\end{equation}
where $\Sigma_h = \sum_i{(E^{h}_i-p^{h}_{z,i})}$ is the total hadronic
$E_{h}-P_{z}^{h}$, summed over all reconstructed hadronic final state
particles $i$, and $P_{T}^{h}$ is transverse momentum of the inclusive
final state.
A combination of $P_{T}^{h}$ and $\Sigma_{h}$ defines the hadronic
scattering angle, $\gamma_{h}$, where:
\begin{equation}
  \tan \frac{\gamma_h}{2} = \frac{\Sigma_h}{P_{T}^{h}}
  \label{eq:thh}
\end{equation}
which, within the Quark Parton Model (QPM)~\cite{Callan:1969uq}
corresponds to the direction of the struck quark.

%%%

The ``sigma method''~\cite{Bassler:1994uq} makes use of both electron
and hadronic final state variables and equations~\ref{eq:emeth} and
\ref{eq:jbmeth} are modified as:
\begin{equation}
  Q^2_{\Sigma}=\frac{P^2_{T,e}}{1-y_{\Sigma}}\,,~~~y_{\Sigma}  = \frac{\Sigma_h} {E-P_z}\,, ~~~x_{\Sigma} = \frac{Q^2_{\Sigma}}{s y_{\Sigma}}\,.
\label{eq:sigmameth}
\end{equation} 
The total $E-P_{z}$ of the event, $\delta$, is defined as:
\begin{eqnarray}
\delta \equiv E-P_{z} & = & E_{e} (1-\cos{\theta_{e}}) + \sum\limits_{i}{(E^{h}_i - p^{h}_{z,i})} \nonumber \\
& = & \Sigma_{e} + \Sigma_{h}
\label{eq:epz}
\end{eqnarray}
and for events with no photon radiation from the incoming electron
$\delta = 2E_{e}^{0} = 55$~GeV.

%%%

The ``e-sigma method''~\cite{Bassler:1997tv} is most often used for
the reconstruction of NC kinematics and is an optimum combination
of the two methods, providing good resolution whilst minimising
radiative effects~\cite{Heinemann:1999ry}:
\begin{equation}
Q^2_{e\Sigma} = Q^2_e\,,~~~y_{e\Sigma} =  \frac{Q^2_e}{s x_{\Sigma}}\,,~~~x_{e\Sigma} = x_{\Sigma}\,.
\label{eq:esigmameth}
\end{equation}

%%%

Finally, the  ``double angle method'' \cite{standa,hoegerda} is used to
reconstruct the event kinematics from the electron and hadronic scattering angles:
\begin{subequations}
\label{eq:dameth}
\begin{eqnarray}
y_{DA}& =& \frac{\tan{(\gamma_h/2)}}{\tan{(\theta_e/2)} + \tan{(\gamma_h/2)}}\,,  \nonumber \\
Q^2_{DA} & = & 4 {E_{e}^{0}}^{2} \cdot  
\frac{\cot{(\theta_e/2)}}{\tan{(\theta_e/2)} + \tan{(\gamma_h/2)}}\,,  \\
x_{DA} & = & \frac{Q^2_{DA}}{s y_{DA}}\,. \nonumber
\end{eqnarray}
In this method, the energy of the scattered electron may be
calculated via:
\begin{equation}
  E_{DA} = \frac{2E_{e}^{0} \sin{\gamma_{h}}}{\sin{\gamma_{h}} +
    \sin{\theta_{e}} - \sin{(\gamma_{h} + \theta_{e})}}\,,
\end{equation}
\end{subequations}
which is used in the calibration of the electron energy as described
in section \ref{sec:pid}.
This method is largely insensitive to hadronisation and is, to first
order, independent of the detector energy scales.
However, the hadronic angle is less well-determined than the electron
angle due to particle loss in the beampipe, and an additional
correction may be applied~\cite{Derrick:1996hn}.

\section{Physics simulation}
\label{sec:mc}

To study different physics models and in particular compare these models with experimental data, stochastic techniques
are employed.
These techniques, which use random numbers and probability distributions, are termed {\it Monte Carlo} (MC) methods.
Simulation is an essential tool for physics analysis, contributing to a better understanding of the data and of the
detector response to physics events.
Moreover, the theoretical models implemented in MC programs may be tested by comparing the prediction
from the simulation to what is observed in the real data.
The simulation of physics events at HERA, much like at other particle physics colliders, can be broken down into
three discreet steps: event generation, detector simulation, and finally reconstruction of the simulated events.
These steps are briefly described in the following.

\subsection{Event generation}
\label{sec:mcgen}

Firstly, QCD MC event event generators are used, which employ the factorisation
theorem~\cite{Collins:1988ig} to describe the $ep$ hard scatter, characterised by an
associated scale allowing the collision to be factorised into separate stages.
Event generators produce all final state partons for a given interaction,
using all relevant diagrams and parton density functions.
The {\it hard sub-process} is the main feature of the event, and is the interaction of a parton
extracted from the proton and the photon (or a photon constituent in resolved photon events).
This process can be calculated in a fixed order perturbative expansion since it involves
a hard scale $\mu$.
{\it Hadronisation} is the process in which colourless hadrons are formed starting from
coloured partons produced in the hard scatter.
It is a non-perturbative phenomenon which is modelled by the simulation programs
using phenomenological inputs.
The main hadronisation models available include the cluster model (for
example, as done by HERWIG~\cite{Corcella:2000bw}) and the Lund string
model~\cite{Andersson:1983ia} (for example, as done by PYTHIA~\cite{Sjostrand:2006za} 
and JETSET~\cite{jetset}).
In processes involving charged and coloured objects the topology of an event can be strongly
influenced by the emission of gluons and photons by the initial or final state.
These perturbative corrections are usually modelled by the so called {\it parton shower method}, where
the radiation is simulated by an arbitrary number of branchings of one parton into two, such as
$e \rightarrow e\gamma$, $q \rightarrow qg$, $q \rightarrow q\gamma$ or $g \rightarrow q \bar q$.
A final consideration is the beam remnant, which comprises the remainder of the initial state particles, following the hard scatter and any initial and/or final state radiation.
If this remnant is coloured, it will be necessarily connected to the rest of the event and needs to be fragmented
and reconstructed coherently.
The interactions of any unstable partons produced (mainly quarks and gluons) are further simulated until
only long lived stable particles exist.
The simulated event then consists of a list of 4-vectors, describing
the final state particles.

\subsection{Detector simulation}
\label{sec:mcsim}

The output of the event generator comprises a list of particles produced in the hard scattering, as well as the 
particles produced from the parton shower, as explained in the previous section.
In real data the only available information from an electron-proton scattering is the signal the particles
produced in the collision leave in the various detector sub-components as they pass through the detector.
A full detector simulation is therefore performed in order to also describe this at the MC level.
The passage of particles through the detector is simulated with the GEANT3~\cite{geant} package.
GEANT provides a description of all detector components, including the composite material,
as well as the shapes and relative positions.
The program traces the passage of a particle through the whole detector, simulating its response whilst taking
into account the relevant physics processes such as energy loss, multiple scattering and particle
decays in flight.
After the detector response has been simulated, the trigger logic as implemented in the
data taking is added to the simulation.
The simulated event now resembles a set of hits on wires, energy deposits in the calorimeters,
signals in the muons chambers and so on, mimicking the traces left in the detector by a real
$ep$ collision event.

\subsection{Reconstruction of simulated events}
\label{sec:mcrec}

As a final step, the same reconstruction program used for the data is applied to simulated events.
This program reconstructs the event variables, like particle momenta and energies, treating
the data and the Monte Carlo in the same way.
All the information coming from the different detector sub-components are taken as input by
the reconstruction program.
MC simulated data are thus identical to real data, with the addition of the generator level information.
This information and the  difference between the two levels of simulation also provide a method of correcting detector acceptances and resolution effects in the data.

\section{Standard Model physics at HERA}
\label{sec:sm}

The cross sections of inclusive NC and CC interactions, which are measured at HERA with
high precision, are one of the most important ingredients for the determination of the
proton parton distribution functions.
The kinematic region in $x$ and $Q^2$ covered by the HERA experiments is shown
in figure~\ref{fig:kinematicplane} and comprises a significant part of the $x$ region of
interest for the LHC experiments.
As a consequence, taking advantage of QCD factorisation and the use of DGLAP
equations~\cite{Gribov:1972ri,Altarelli:1977zs,Dokshitzer:1977sg,Gribov:1972rt,Lipatov:1974qm},
the PDFs extracted from the HERA data can be used as input for
cross section determination at the LHC.

%%%

The H1 and ZEUS collaborations published~\cite{Abramowicz:2015mha} a combination of their
NC and CC cross section measurements extracted from the full data sample collected at HERA.
The details of how the combination was performed, in particular regarding the treatment of the
systematic uncertainties, are described in detail elsewhere~\cite{Abramowicz:2015mha}.
For the aims of this paper it is sufficient to say that the cross sections
measured by the two collaborations are combined using a $\chi^2$
minimisation method, which takes into account both the statistical and
systematic uncertainties of the data.
In particular, a distinction is made between the {\it correlated} and
{\it uncorrelated} uncertainties among the different points (bins) of the analysis.
In addition to the clear improvement on the statistical uncertainty
obtained by combining the data, an improvement on the systematics
uncertainties is also obtained.
Intuitively, correlated systematic uncertainties that affect the
measurement in one direction for H1, and in the other for ZEUS, can be
significantly reduced in the data combination.
The fact that the detectors and the analysis techniques are different
is fully exploited, where these differences translate into systematic
uncertainties affecting the data in a different way.
In this sense, the detectors are used to ``cross calibrate'' each other.
These arguments are explained in full mathematical
detail in the papers illustrating the method
employed~\cite{Aaron:2009bp,Glazov:2005rn}.
It is also worthwhile pointing out also that this method allows the
consistency of the data of the two experiments to be checked in a
model-independent way, as the main assumption done in the data
combination is  that there is a single true value of the cross section
corresponding to each data point and each
process, NC or CC $e^+p$ or $e^-p$ scattering.

%%%

\subsection{Neutral Current measurements}
\label{sec:ncmeas}
Cross sections for NC DIS interactions have been published~\cite{Abramowicz:2015mha} for
$0.045 \le Q^2 \le 50000~{\rm GeV^2}$, in a large phase space region $6\cdot 10^{-7} \le x \le 0.65$
for values of inelasticity $0.005 \le y \le 0.95$ (see figure~\ref{fig:kinematicplane}).
Covering the very low $Q^2$ regions required special experimental techniques.
The lowest-$Q^2$ data, $Q^2 > 0.045~{\rm GeV}^2$, were collected during the HERA~I data
taking period with the ZEUS detector using special tagging devices~\cite{Breitweg:2000yn}.
The $Q^2$ range between $0.2~{\rm GeV}^2$ and $1.5~{\rm GeV}^2$ was covered using
special HERA~I runs, in which the interaction vertex position was shifted forward, bringing
backward-scattered electrons with small scattering angles into the
acceptance of the detectors~\cite{Aaron:2009bp,Breitweg:1998dz,Adloff:1997mf}.
The $Q^2 > 1.5~{\rm GeV^2}$ region was covered with HERA~I and HERA~II
data in different configurations.

In performing the analyses, three main different $Q^2$ regions were
considered, as the analysis methods differ substantially among them.
At $\sqrt{s} = 318~{\rm GeV}$, the high $Q^2$ region is defined for
$Q^2$ between $150$ and $30000~{\rm GeV}^2$, whereas the low $Q^2$ region
comprises the range $2 < Q^2 < 120~{\rm GeV}^2$.
The very-low $Q^2$ region, populated only with data collected at $\sqrt{s} \le 300~{\rm GeV}$,
is defined for $Q^2 < 2~{\rm GeV}^2$.
As a breakdown of perturbative QCD (pqCD) is expected for $Q^2$
approaching $1~\rm{GeV^2}$, the data of this last region cannot be
compared to predictions from pQCD.

%%%

The most interesting region for BSM searches is the high $Q^2$ region,
in particular at very low $x$, where the cross section is lower and
the precision of the measurements is worse so that new phenomena
can hide in the large SM background.
In general, searches for deviations from the SM are performed at the limits of
the accessible kinematic regions.
Therefore, the analysis for the high $Q^2$ analysis is reported here as
representative. The selection techniques of the many other analyses
included in the combined data are described in detail
elsewhere (\cite{Abramowicz:2015mha} and references therein).

%%%

NC DIS events are generated with the DJANGOH~\cite{django} MC simulation program,
which is based on HERACLES~\cite{heracles} for the electroweak calculation and
LEPTO~\cite{lepto} for the hard matrix element calculation.
The colour dipole model as implemented in ARIADNE~\cite{ariadne} is used to
generate higher order QCD dynamics.
The JETSET program is used to simulate the hadronisation process
in the 'string-fragmentation' model.

%%%

Events are selected by requiring the DIS electron to be reconstructed in the final state.
The electron reconstruction is based on an algorithm combining information from the
calorimeter energy deposits and tracks measured in the central tracking detectors.
The electron is required to have an energy $E^\prime_e > 10~{\rm GeV}$ and to be
isolated from other energy deposits in the calorimeter.
If the electron is found in the acceptance region of the tracking detectors, a track matched
to the energy deposit in the calorimeter is required.
The matching is performed considering the distance of closest approach between the track
extrapolated to the calorimeter surface and the energy cluster position.
A matched track is not required if the electron emerged outside the acceptance of the
tracking detectors.

%%%

The most important background in the NC DIS analysis comes from photoproduction
events, when an energy deposit in the calorimeter associated to a charged track is wrongly
identified as the scattered DIS electron.
In order to suppress this background, a cut on the total $E - P_{Z}$
as defined in equation~\ref{eq:epz} is used in the event selection, by
requiring events to have $38 < \delta < 65~{\rm GeV}$.
As already mentioned in section~\ref{sec:kinemethods}, conservation of energy and
longitudinal momentum implies that $\delta = 2E_e = 55~{\rm GeV}$, if all final-state
particles are detected and perfectly measured.
Undetected particles that escape down the forward beampipe have a negligible effect on $\delta$.
However, particles lost down the backward beampipe could lead to a substantial reduction in $\delta$.
This is the case for photoproduction events, where the electron emerges at a very small scattering angle,
or for events in which an initial-state bremsstrahlung photon is emitted.
In order to further reduce the photoproduction background, the selected NC events are required
to have $y < 0.9$.
For the puposes of this review, it is worth noting that this
requirement cuts out part of the kinematic region interesting for BSM searches. 

%%%

Further cuts are applied in order to suppress the backgrounds from beam-gas and cosmic interactions.
After all cuts, the remaining background contamination estimated using photoproduction MC is about $0.2\%$.

%%%

\begin{figure}[!h]
\centerline{\includegraphics[width=1.0\columnwidth]{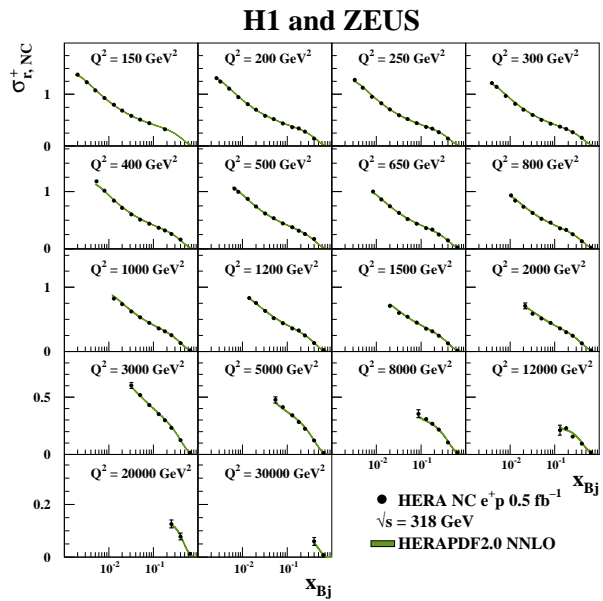}}
\caption{The combined high $Q^2$ HERA inclusive NC $e^+ p$ reduced cross sections at $\sqrt{s} = 318~{\rm GeV}$, plotted as a function of $x$ at fixed $Q^2$, with overlaid predictions of the HERAPDF2.0 NNLO. The bands represent the total uncertainties on the predictions.}
\label{fig:h1zeus_eplusp_doublediff_highQ2}
\end{figure}

\begin{figure}[!h]
\centerline{\includegraphics[width=1.0\columnwidth]{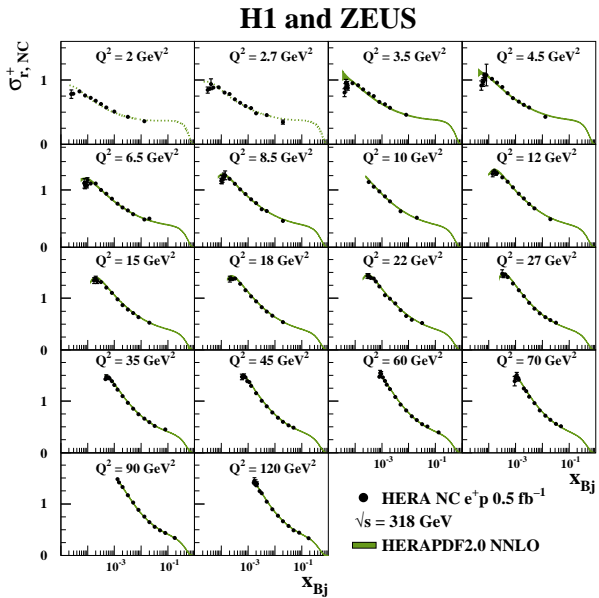}}
\caption{The combined low $Q^2$ HERA inclusive NC $e^+ p$ reduced cross sections at $\sqrt{s} = 318~{\rm GeV}$, plotted as a function of $x$ at fixed $Q^2$, with overlaid predictions of the HERAPDF2.0 at NNLO order in pQCD. The bands represent the total uncertainties on the predictions. Dotted lines indicate extrapolation into kinematic regions not included in the fit.}
\label{fig:h1zeus_eplusp_doublediff_lowQ2}
\end{figure}

Cross sections are extracted and compared to theoretical predictions from pQCD, based on the PDF set HERAPDF2.0,
which will be introduced in section~\ref{sec:herapdf2.0}. 
The reduced cross section $\tilde\sigma^{\pm}_{NC}$ is defined in terms of the inclusive NC cross section as:
\begin{eqnarray}
\sigma^{\pm}_{r,NC}&  =&   \frac{d^2\sigma(e^\pm p)}{dx dQ^2} \cdot \frac{Q^4 x}{2\pi \alpha^2 Y_+}
\end{eqnarray}
where the fine-structure constant, $\alpha$, is defined at scale zero
and $Y_\pm = 1 \pm (1-y)^2$.
The combined high $Q^2$ inclusive NC $e^{+}p$ reduced cross sections at
$\sqrt{s} = 318~{\rm GeV}$, as extracted from the combined HERA data, are shown in 
figure~\ref{fig:h1zeus_eplusp_doublediff_highQ2}.
The cross sections are shown as a function of $x$ in different $Q^2$ bins and are compared to the HERAPDF2.0
predictions~\cite{Abramowicz:2015mha} at next-to-next-to-leading (NNLO).
The combined low $Q^2$ HERA inclusive NC $e^{+}p$ reduced cross sections
at $\sqrt{s} = 318~{\rm GeV}$ are shown in figure~\ref{fig:h1zeus_eplusp_doublediff_lowQ2}.
The data description by the SM prediction is generally good, except for the turnover of the cross
section at low $x$ and low $Q^2$.

\begin{figure}[!h]
\centerline{\includegraphics[width=1.0\columnwidth]{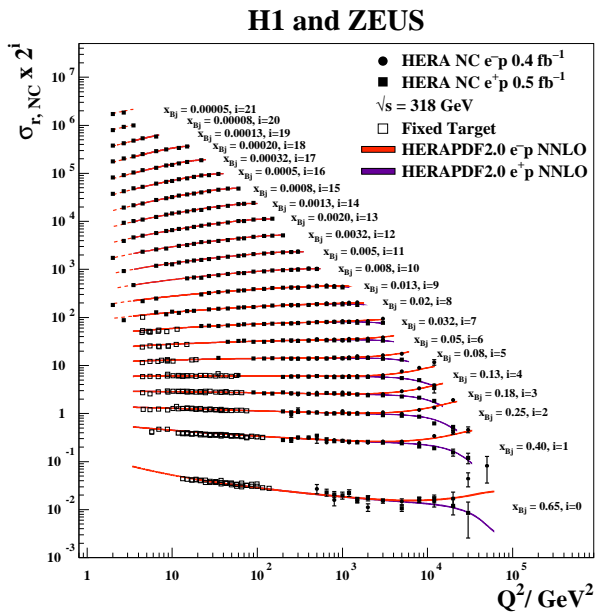}}
\caption{The combined HERA data for the inclusive NC $e^+ p$ and $e^- p$ reduced cross 
sections together with fixed target data~\cite{Benvenuti:1989rh,Arneodo:1996qe} and the predictions of HERAPDF2.0 NNLO~\cite{Abramowicz:2015mha}. The bands represent the total uncertainty on the predictions.}
\label{fig:h1zeus_combnchera}
\end{figure}

The combined reduced cross section as a function of $Q^2$ at different $x$ values is shown in figure~\ref{fig:h1zeus_combnchera}.
In the high $Q^2$ region, the  $e^-p$ cross section is significantly larger than the  $e^+p$.
This is due to the parity-violating component of the NC interactions, namely to the exchange
of a $Z^{0}$ boson between the electron and the proton.
This component is suppressed at low $Q^2$ due to the large mass of the $Z^{0}$ boson and
becomes relevant only for $Q^2 \sim M_{Z^{0}}^2$. 

In figure~\ref{fig:h1zeus_combnchera} the HERA data are compared to the results of
fixed-taget experiments, which populate the region of lower $Q^2$ and higher $x$ values.
The HERA and the fixed-target results are in good agreement, as shown by the comparison
with theoretical predictions based on the HERA data alone~\cite{Abramowicz:2015mha}.
The level of agreement between the two fixed-target experiment results shown in the 
figure~\cite{Benvenuti:1989rh,Arneodo:1996qe} was investigated in~\cite{Arneodo:1996qe}
and found to be good, taking into account the normalisation uncertainties of the data 
and the systematic uncertainties quoted in the analyses.
The HERA data significantly extend the region where the reduced cross section can be measured
with very good precision. 

%%%

Scaling violations, as predicted by the DGLAP equations,
are also clearly visible in figure~\ref{fig:h1zeus_combnchera},
where in the low $x$ region the reduced cross section is not independent of $Q^2$,
but shows a slope that becomes steeper with decreasing $x$. 
These violations are due to QCD effects not present in the na\"ive QPM model.
The large kinematic range covered by HERA clearly demonstrates the scaling
violations and allows them to be used for the extraction of the gluon density in the proton. 
On the other hand, it has to be noted that the gluon content of 
the proton cannot be measured directly at HERA. 
This means that phenomenological assumptions are needed in the extraction of the gluon 
density from the data, introducing an important source of uncertainty in the extraction
of the proton PDFs.

%%%

The HERA inclusive NC DIS cross sections are an important input to the determination of the
proton structure functions.
The double-differential cross section of the electron-proton NC DIS as
a fucntion of $x$ and $Q^2$ can be expressed in terms of the proton
structure functions $F_2$, $F_3$ and $F_L$, as~\cite{Abramowicz:2015mha}:
\begin{eqnarray}
\frac{d^2\sigma(e^\pm p)}{dx dQ^2} & = & \frac{2\pi\alpha^2}{xQ^4} 
\left\{ Y_+ F_2 \mp  Y_-xF_3 - y^2 F_L\right\}.
\label{eq:disxsection}
\end{eqnarray}
It follows that from the reduced cross section, $\tilde\sigma^{\pm}_{NC}$,
the proton structure functions can be constrained:
\begin{eqnarray}
\sigma^{\pm}_{r,NC}
%&  =&   \frac{d^2\sigma(e^\pm p)}{dx dQ^2} \cdot \frac{Q^4 x}{2\pi \alpha^2 Y_+} =  \nonumber \\ 
%& = & F_2(x,Q^2) \mp \frac{Y_-}{Y_+}xF_3(x,Q^2) - \frac{y^2}{Y_+}F_L(x,Q^2) 
& = & F_2 \mp \frac{Y_-}{Y_+}xF_3 - \frac{y^2}{Y_+}F_L.
\label{eq:nc_reduced}
\end{eqnarray}
The structure functions  $F_2$, $F_3$ and $F_L$ are process dependent.
$F_3$ is non-zero only for weak interactions and is generated by parity-violating
interactions, i.e. by the exchange of a $W$ boson in CC DIS
or a $Z^{0}$ boson in NC DIS between the electron and the proton.
Therefore, for $Q^2 \ll M_{Z^{0}}^2$:
\begin{eqnarray}
\sigma^{\pm}_{r,NC}
& = & F_2 \mp - \frac{y^2}{Y_+}F_L
\label{eq:f2_fl}
\end{eqnarray}

The $F_L$ term is related to the longitudinally polarised virtual boson exchange process. This
term vanishes at lowest order QCD but has been predicted~\cite{Altarelli:1978tq} to
be non-zero when including higher order QCD terms. The contribution of $F_L$ to the reduced
NC cross section is relevant only for values of $y$ larger than approximately $0.5$.

%%%

A direct measurement of $F_L$ has been performed at HERA by determining the
reduced cross section, $\sigma^{\pm}_{r,NC}$, at different values of $\sqrt{s}$
by reducing the proton beam energy from $E_p = 920~{\rm GeV}$ to $E_p = 460~{\rm GeV}$ and
$E_p = 575~{\rm GeV}$, as described in section~\ref{sec:hera}. 
This method had been previously used to extract $F_L$ in fixed-target 
experiments~\cite{Benvenuti:1989rh,Arneodo:1996qe,Aubert:1982ts,Whitlow:1990gk}.
From equation~\ref{eq:qsxy}, $y= Q^2/sx$ and therefore the cross
sections can be measured at the same values of $x$ and $Q^2$ but at
different values of $y$, allowing an experimental separation between $F_2$ and $F_L$
in equation~\ref{eq:f2_fl}.
The sensitivity to $F_L$ is increased by measuring the cross sections in the high $y$ region, but in
this region the electron energy is low and the background from photoproduction is large.
The separation between NC DIS and the photoproduction background is therefore one of the
main challenges of the high $y$ analysis. 

%%%

\begin{figure}[!h]
  \centerline{\includegraphics[width=1.0\columnwidth]{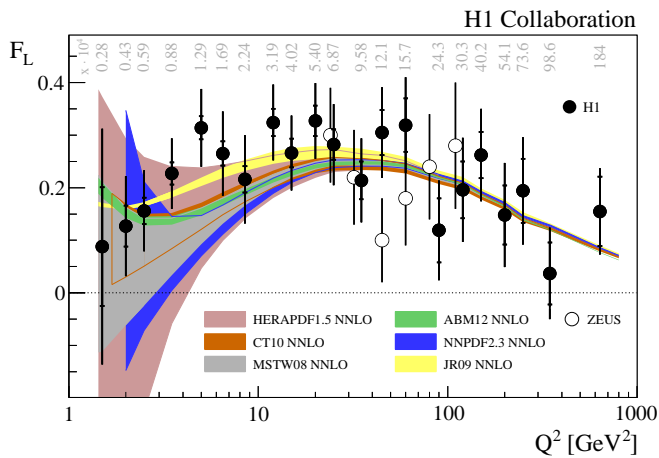}}
  \caption{The proton structure function $F_L$ averaged over $x$ at different $Q^2$ (solid points).
    The average value of $x$ for each $Q^2$ is given above each data point. The inner error bars
    represent the statistical uncertainties, the full error bars include the statistical and systematic
    uncertainties added in quadrature, including all correlated and uncorrelated uncertainties.
    The data are compared to NNLO predictions from a selection of PDF sets as indicated
    (\cite{Andreev:2013vha} and references therein).}
  \label{fig:h1_fl}
\end{figure}

The structure function $F_L$ as measured by the H1~\cite{Andreev:2013vha} and
ZEUS~\cite{Chekanov:2009na} collaborations is shown in figure~\ref{fig:h1_fl}.
The data are compared with several QCD predictions at NNLO, which describe the
measurements reasonably well. 

A precise direct measurement of $F_L$ would allow, according to the
Altarelli-Marinelli relation~\cite{Altarelli:1978tq}, a direct extraction of the
gluon density in the proton, which is constrained indirectly via scaling violations.
The $F_L$ measurement was used~\cite{Andreev:2013vha} to perform a gluon density
extraction based on a NLO approximation, and the agreement with the gluon as determined
from scaling violations was found to be reasonably good.
Although the scaling violations allow a much better constraint of the gluon in the proton,
the comparison with the direct extraction from the $F_L$ measurement provides an 
independent check of its validity. Although more at a qualitative than at 
a quantitative level, this measurement represents an 
improvement on the knowledge of the gluon density in the proton.

%%%

From equation~\ref{eq:nc_reduced} it follows that the structure function $xF_3$ can be extracted
from the difference between the $e^+p$ and $e^-p$ reduced cross sections: 
\begin{eqnarray}
xF_3 = \frac{Y_+}{2Y_-}(\sigma^-_{\rm r,NC} - \sigma^+_{\rm r,NC}). 
\label{eq:xF3}
\end{eqnarray}
It is useful to consider the simplified picture as expressed in the QPM, where $F_L=0$.
In the QPM, $xF_3$ is directly related to the valence quark
distributions in the proton, which, assuming symmetry between the
quarks and anti-quarks in the sea, can be expressed as:
\begin{eqnarray}
xu_v & = & xU - x\bar{U}, \nonumber \\
xd_v & = & xD -x \bar{D};
\end{eqnarray}
where $xU$, $x\bar{U}$, $xD$ and $x\bar{D}$ represent the sums of parton distributions for up-type and down-type quarks and anti-quarks, respectively. Below the $b$-quark mass threshold, they are related to the quark distributions as follows:
\begin{eqnarray}
xU = xu+xc, & & x\bar{U} = x\bar{u}+x\bar{c}, \nonumber \\
xD = xd+xs, & & x\bar{D} = x\bar{d}+x\bar{s},
\end{eqnarray} 
where $xs$ and $xc$ are the strange- and charm-quark distributions. 

\begin{figure}[!h]
\centerline{\includegraphics[width=1.0\columnwidth]{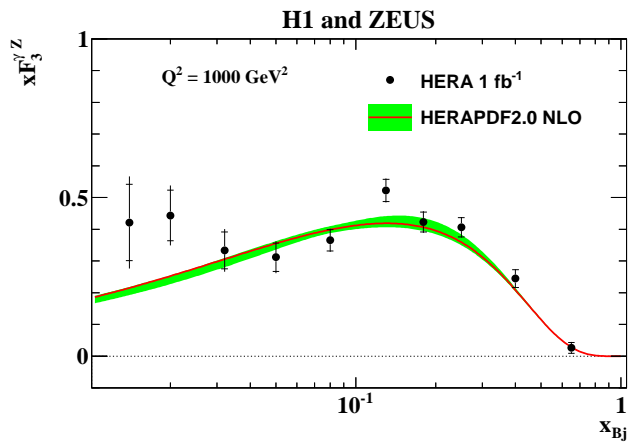}}
\caption{The structure function $xF_3^{\gamma Z}$ averaged over $Q^2 \ge  1000~{\rm GeV^2}$ at the scale $Q^2 = 1000~{\rm GeV^2}$ together with the prediction from HERAPDF2.0 NLO. The band represents the total uncertainty on the prediction.}
\label{fig:h1zeus_xF3}
\end{figure}

In the HERA kinematic regime the dominant contribution to $xF_3$ comes
from its photon-$Z^{0}$ exchange interference term, $xF_3^{\gamma Z}$.
In the QPM, this is expressed by the relation:
\begin{eqnarray}
xF_3^{\gamma Z} & \approx & \frac{x}{3}(2u_v + d_v),
\end{eqnarray}
which shows that $xF_3$ is directly related to the valence quark distributions in the proton.
The measurements of $xF_3^{\gamma Z}$ therefore make it possible to determine the behaviour
of the valence quark distributions at low $x$.
The structure function $xF_3^{\gamma Z}$ as extracted from the HERA combined NC DIS cross
sections is shown in figure~\ref{fig:h1zeus_xF3} and compared to pQCD predictions from HERAPDF2.0.

%%% 

\subsection{Charged Current measurements}
\label{sec:ccmeas}

The CC DIS process $ep \rightarrow \nu X$  was measured at HERA for the first time in 1993. 
The $Q^2$ dependence of the CC DIS cross section is sensitive to the
$W$ boson mass, which enters
the CC propagator. One of the aims of the first measurements of CC DIS was the determination 
of this mass, which was extracted from a fit to the differential CC DIS cross section as a function of 
$Q^2$, leaving it as a free parameter of the fit.
The fact that the obtained value was consistent to the mass of the $W$ boson as measured
by the hadron colliders at that time demonstrated the presence of the
$W$ propagator~\cite{Ahmed:1994fa,Derrick:1995us}.
The measurement of CC DIS interactions in $ep$ collisions,
$e^+ p \rightarrow \bar{\nu}X~ (e^-p \rightarrow \nu X)$, provides a complementary
view with respect to NC DIS for the understanding of the proton structure and the SM.
While NC DIS is mediated by the exchange of photons and $Z^{0}$ bosons and is sensitive
to all quark flavours, only down-type quarks and up-type antiquarks
(down-type antiquarks and up-type quarks) contribute at leading order to
$e^+p$ ($e^-p$) CC DIS.
Thus this process is a powerful probe of flavour-specific PDFs, as
described in more detail in the following. 
Although part of this information could already be obtained from the previous
fixed-target neutrino experiments, NC and CC DIS became accessible at
the same machine for the first time at HERA, and in an extended kinematic region.

%%%

The precision of the measurement of the CC cross sections at HERA has been significantly
improved by combining the data of the two experiments H1 and ZEUS.
Cross sections for CC interactions have been published~\cite{Abramowicz:2015mha}
for $200 \le Q^2 \le 50000~{\rm GeV^2}$ and $1.3 \cdot 10^{-2} \le x \le 0.40$ at values of
$y$ between $0.037$ and $0.76$.

As described for NC in the previous section, CC DIS events are generated using the
DJANGOH MC simulation program, interfaced with LEPTO for the hard matrix calculation
and HERACLES for electroweak effects.
A set of NC DIS events generated with DJANGOH is used to estimate the NC contamination in the CC sample
while the photoproduction background was simulated using HERWIG or
PYTHIA.
The GRAPE~\cite{Abe:2000cv} and EPVEC~\cite{Baur:1991pp} MC programs
are used to simulate the background from lepton pair and single $W$ boson production, respectively.

%%%

In CC DIS,  the struck quark gives rise to one or more jets of hadrons
and the energetic final state neutrino escapes detection, leaving a
large imbalance in the transverse momentum observed in the
detector. Therefore, CC DIS events are selected by requiring a large
missing transverse momentum in the event, $P_{T}^{\rm miss}$, as
defined in equation~\ref{eq:ptmiss}.

%%%

Backgrounds to CC DIS arise from high $E_{T}$ events in which the finite energy resolution
of the calorimeter or energy that escapes detection leads to significant missing transverse momentum.
Non-$ep$ events such as beam-gas interactions, beam-halo muons or cosmic rays can also
cause substantial imbalance in the measured transverse momentum and constitute additional
sources of background.
Moreover, single $W$ boson production events and di-lepton events could also constitute a
potential background in case the lepton(s) are poorly reconstructed.

%%%

CC events are selected in the kinematic region $Q^2 > 200~{\rm GeV^2}$ and $y < 0.9$, with
large missing transverse momentum, $P_{T}^{\rm miss}>~12~{\rm GeV}$, and a
primary vertex reconstructed in the nominal interaction region.
The vertex requirement significantly reduces the non-$ep$ background.
Background from NC events with a poorly measured scattered electron or hadronic jets,
leading to significant missing transverse momentum, are removed by rejecting events
with an isolated electromagnetic cluster in the calorimeter and a
longitudinal balance $\delta > 30~{\rm GeV}$.
Other cleaning cuts are used to reduce the remaining backgrounds, based on for example
the number of tracks fitted to the vertex compared to the total, or on the azimuthal collimation
of the energy flow in the event, using the ratio $V_{\rm ap}/V_{\rm p}$~\cite{Adloff:1999ah}, where 
$V_{\rm p}$ and $V_{\rm ap}$ are the transverse energy flow parallel and antiparallel
to the hadronic final state $\vec{p}_{T}^{h}$, respectively, and are determined from the
transverse momentum vectors $\vec{p}_{T}^{i}$ of all the particles $i$ which belong
to the hadronic final state according to:
\begin{eqnarray}
  V_{\rm p} = \sum\limits_{i} \frac{\vec{p}_{T}^{h} \cdot \vec{p}_{T}^{i}}{P_{T}^{h}}~~~~{\rm for}~~~~\vec{p}_T^{h} \cdot \vec{p}_{T}^{i} >0\\
  V_{\rm ap} = \sum\limits_{i} \frac{\vec{p}_{T}^{h} \cdot \vec{p}_{T}^{i}}{P_{T}^{h}}~~~~{\rm for}~~~~\vec{p}_T^{h} \cdot \vec{p}_{T}^{i} <0
\label{eq:vpvap}
\end{eqnarray}

\begin{figure}
  \centerline{\includegraphics[width=1.0\columnwidth]{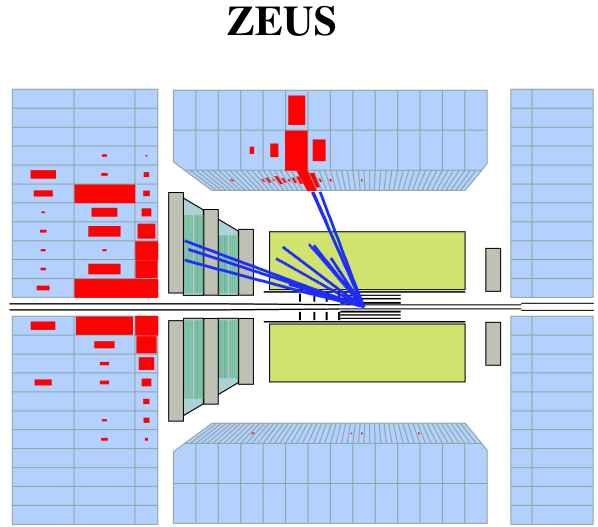}}
  \caption{A charged current event with $Q^2 = 53 060~{\rm GeV^2}$ and $x = 0.59$
    observed in the ZEUS data. The transverse momentum imbalance in the event
    is visible in the figure, as most of the activity appears in the
    upper part of the detector.}
  \label{fig:cc_event}
\end{figure}

After all cuts, the contamination from NC DIS and non-$ep$ events is found to be
negligible, while other backgrounds from photoproduction, single $W$ boson
production and lepton pairs contribute up to $20\%$ in the lowest-$Q^2$ and lowest $x$ bins. 
One of the selected CC events, as seen by the ZEUS detector, is shown in figure~\ref{fig:cc_event}.

\begin{figure}[!h]
\centerline{\includegraphics[width=1.0\columnwidth]{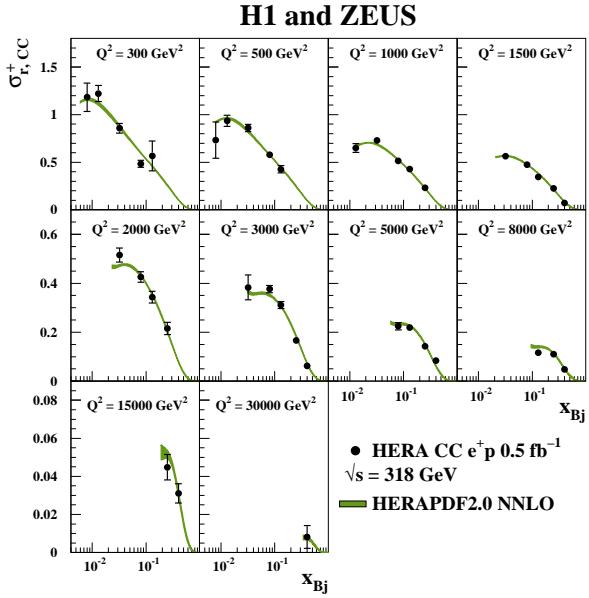}}
\caption{The combined HERA inclusive CC $e^+p$ reduced cross sections
  at $\sqrt{s} = 318~{\rm GeV}$ with overlaid predictions from
  HERAPDF2.0 NNLO. The bands represent the total uncertainties
  on the predictions.}
\label{fig:h1zeus_cceplus}
\end{figure}

Similarly to NC, cross sections are extracted and compared to the theoretical predictions
from pQCD, based on HERAPDF2.0.
The reduced cross sections for unpolarised CC $e^\pm p$ scattering are defined as:
\begin{eqnarray}
\sigma^\pm_{r,{\rm CC}} & = & \frac{2\pi x}{G_F^2}\bigg[ \frac{M_W^2+Q^2}{M_W^2} \bigg]^2 
\frac{d^2\sigma^{e^\pm p}_{\rm CC}}{dxdQ^2}.
\end{eqnarray}
The combined inclusive CC reduced cross sections at $\sqrt{s} = 318~{\rm GeV}$, as
extracted from the combined HERA data, are shown in figure~\ref{fig:h1zeus_cceplus}.
The precise predictions describe the CC cross sections well.
The CC data are in general less precise than the NC data.

\begin{figure}[!h]
\centerline{\includegraphics[width=1.0\columnwidth]{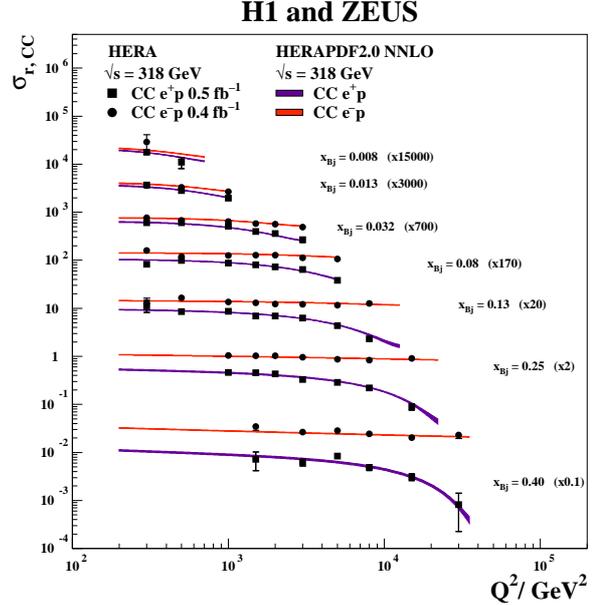}}
\caption{The combined HERA data for inclusive CC $e^+p$ and $e^-p$ reduced cross sections at $\sqrt{s} = 318~{\rm GeV}$ with overlaid predictions of HERAPDF2.0 NNLO. The bands represent the total uncertainty on the predictions.}
\label{fig:h1zeus_ccvsq2}
\end{figure}

The combined inclusive CC reduced cross sections as a function of $x$
in different $Q^2$ bins are shown in figure~\ref{fig:h1zeus_ccvsq2}.
The difference between the $e^- p$ and the $e^+ p$ cross sections
can be intuitively understood at leading order by considering the
valence quark composition of the proton and the charge of the
exchanged vector boson in $e^- p$ and $e^+ p$ interactions and is
explained more formally below.
The data are well described by pQCD predictions. 
%

%%%

The cross sections of CC DIS also provide an important input for the determination of the proton structure. 
In analogy to equation~\ref{eq:nc_reduced}, CC structure functions are defined such that:
\begin{eqnarray}
\sigma^\pm_{r,{\rm CC}} & = & \frac{Y_+}{2}W^{\pm}_2 \mp \frac{Y_-}{2}xW^{\pm}_3 -\frac{y^2}{2}W^{\pm}_L
\end{eqnarray}
In the QPM~\cite{Callan:1969uq}, $W_L^{\pm}=0$, while $W_2^{\pm}$ and
$xW_3^{\pm}$ are expressed as sum and differences of quark and
anti-quark distributions, depending on the charge of the lepton beam:
\begin{eqnarray}
W_2^+ \approx x\bar{U} +xD, & &  xW_3^+ \approx xD - x \bar{U}, \nonumber \\ 
W_2^- \approx xU +x\bar{D}, & &  xW_3^- \approx xU - x \bar{D},
\end{eqnarray}
and consequently:
\begin{eqnarray}
\sigma^+_{r,{\rm CC}} & \approx & (x\bar{U}+(1-y)^2xD), \nonumber \\
\sigma^-_{r,{\rm CC}} & \approx & (xU+(1-y)^2x\bar{D}). 
\label{eq:sigmacc_epluseminus}
\end{eqnarray}
The combination of NC and CC DIS measurements therefore makes it possible to determine both the
combined sea-quark distributions, $x\bar{U}$ and $x\bar{D}$, and the
valence quark distributions, $xu_v$ and $xd_v$.

Equation~\ref{eq:sigmacc_epluseminus} also provides the formal
explanation for the numerical difference
between the cross section for $e^- p$ and $e^+ p$ collisions (see figure~\ref{fig:h1zeus_ccvsq2}),
which is due to the presence of the helicity factor $(1-y)^2$.
As this factor multiplies the valence quark distribution in the $e^+ p$ cross section
formula, it results in a suppression of the cross section at high $y$.
For the $e^- p$ cross section, the $(1-y)^2$ helicity factor multiplies the anti-quark distribution,
which is part of the sea and is already suppressed at high $Q^2$ and
high $x$, and contributes little compared to the quark term.

%%% 

\subsection{Electroweak unification of NC and CC DIS}

\begin{figure}[!h]
\centerline{\includegraphics[width=1.0\columnwidth]{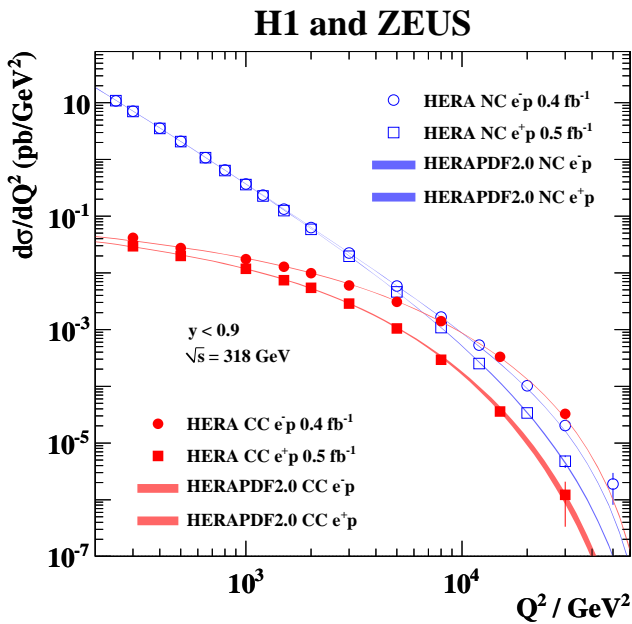}}
\caption{The combined HERA NC and CC $e^- p$ and $e^+ p$ cross sections
 together with predictions from HERAPDF2.0 NLO. The bands represent the total uncertainty on the predictions.}
\label{fig:h1zeus_ncccvsq2}
\end{figure}

The H1 and ZEUS combined NC and CC cross sections as a function of $Q^2$ are shown
in figure~\ref{fig:h1zeus_ncccvsq2}. 
The NC processes dominate the cross section at low $Q^2$.
This is due to the fact that in the low $Q^2$ region the major contribution to the
cross section is given by the exchange of a photon between the electron and the proton;
$W$ or $Z^{0}$ exchange is suppressed due to the high mass of the
vector boson, which enters the propagator term, $Q^2/(Q^2+M_{W/Z}^2)$.
At higher $Q^2$, of the order of the mass squared of the $W$ or $Z$ boson,
$Q^2 \simeq 10000~{\rm GeV}^2$, CC and NC processes are equally important, 
providing experimental evidence of electroweak unification.
It can also be seen that the $e^{-}p$ cross section is higher than the $e^{+}p$ data
for both NC and CC at higher values of $Q^2$ due to the couplings of the
quarks in the proton to the exchanged boson.
The data are compared with the predictions of HERAPDF2.0 at next-to-leading order
(NLO), which provide a good description of the measurements.

%%%

\subsection{HERAPDF2.0}
\label{sec:herapdf2.0}

The combined HERA data are used as sole input for the extraction
of the PDF set HERAPDF2.0~\cite{Abramowicz:2015mha}.
The framework already established for the extraction of
the HERAPDF1.0~\cite{Aaron:2009aa} is used.
The $x$ and $Q^2$ dependences of the NC and CC DIS cross sections
were used to determine the free parameters of the assumed shape of the 
parton distribution functions at a given value of $Q^2$.

%%%

As pQCD is not applicable below $Q^2$ of the order of the inverse of the proton radius squared, an
intrinsic uncertainty is present in performing the QCD fits, due to the choice of
the scale at which to start the DGLAP evolution.
In the analysis performed to extract the HERAPDF2.0 PDF set, to safely
remain in the kinematic region where pQCD is expected to be applicable, only cross sections
for $Q^2$ starting from $Q^2_{\rm min} = 3.5~{\rm GeV^2}$ are used in the analysis.
The $Q^2$ range of the cross sections entering the fit is therefore $3.5 \le Q^2 \le 50000~{\rm GeV^2}$.
The corresponding $x$ range is $0.651 \cdot 10^{-4} \le x \le 0.65$. In addition to experimental
uncertainties, model and parameterisation uncertainties are also considered. 

%%%

To extract the proton PDFs, predictions from pQCD are fitted to the data.
These predictions are obtained by solving the DGLAP evolution
equations~\cite{Gribov:1972ri,Altarelli:1977zs,Dokshitzer:1977sg,Gribov:1972rt,Lipatov:1974qm}
at LO, NLO and NNLO in the $\bar{MS}$ scheme~\cite{Fanchiotti:1992tu,Giele:1998gw}.
The DGLAP equations yield the PDFs at all scales $\mu^2_{\rm f}$ and $x$, if they are provided as functions of
$x$ at some starting scale, $\mu^2_{\rm f_0}$.
This is chosen to be $\mu^2_{\rm f_0} = 1.9~{\rm GeV^2}$ as for HERAPDF1.0~\cite{Aaron:2009aa}, since in
the used formalism $\mu^2_{\rm f_0}$ has to be lower than the charm-quark mass parameter squared.
The renormalisation and factorisation scales were chosen to be $\mu^2_{\rm r} = \mu^2_{\rm f} = Q^2$.  

The values of the charm and beauty mass parameters are chosen after performing
$\chi^2$ scans of NLO and NNLO pQCD fits to the HERA inclusive data and the H1 and ZEUS charm
and beauty data.
The procedure is described in detail elsewhere~\cite{Abramowicz:1900rp}. 

%%%

\begin{figure}[!h]
\centerline{\includegraphics[width=1.0\columnwidth]{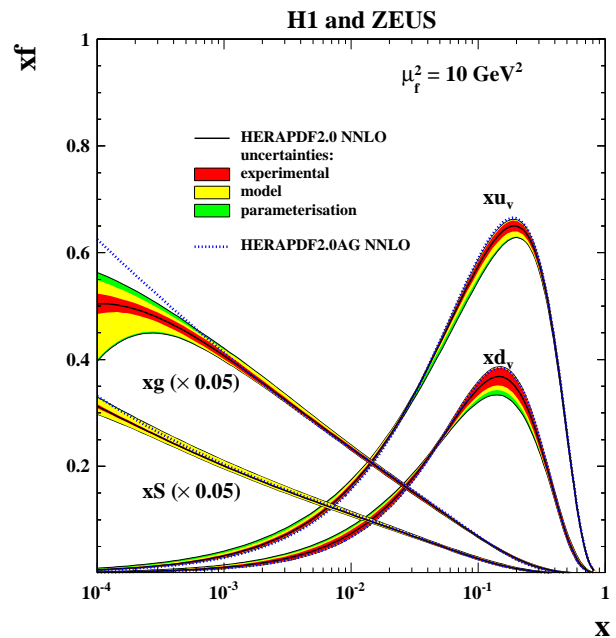}}
\caption{The parton distribution functions $xu_v$, $xd_v$, $xS = 2x( {\bar U} + {\bar D})$ and $xg$ of HERAPDF2.0 NNLO at $\mu^2_{\rm f} = 10~{\rm GeV^2}$. The gluon and sea distributions are scaled down by a factor 20. The experimental, model and parameterisation uncertainties are shown. The dotted lines represent HERAPDF2.0AG NNLO with the alternative gluon parameterisation, see~\cite{Abramowicz:2015mha}.}
\label{fig:h1zeus_herapdf2.0}
\end{figure}

A detailed description of the HERAPDF2.0 PDFs is beyond the scope of this paper and
further information is given in the publication~\cite{Abramowicz:2015mha}.
Here as an example the quark and gluon distributions are shown in figure~\ref{fig:h1zeus_herapdf2.0}
at a scale $Q^2=10~{\rm GeV}^2$.
The gluon distribution domainates the low $x$ region, while at high $x$, as expected,
the valence quark distribution are prevalent, and the $u$-type quarks are twice as much as the $d$-type,
due to the QED couplings, which are proportional to the quark electrical charge.

HERAPDF2.0 has small experimental uncertainties due to the high precision and
coherence of the input data, and makes precise predictions which describe the
input data well, as can be seen also in the examples reported above.
The precision data on the inclusive $ep$ scattering from the H1 and ZEUS experiments
are one of the main legacies of HERA.

%%% 

\subsection{Helicity structure of the SM and limits on the right-handed {\boldmath $W$} boson}

The presence of a polarised electron beam during the second phase of the data taking period at
HERA also provides the opportunity to investigate the helicity structure of the Standard Model.
The most striking evidence can be seen in CC DIS processes.
The electroweak Born-level cross section for the CC reaction,
$e^+ p \rightarrow \bar \nu X$ ($e^-p \rightarrow \nu X$), with a longitudinally
polarised positron beam can be expressed as a function of the charged current
structure functions $W_2^{\pm}$, $W_3^{\pm}$ and $W_L^{\pm}$~\cite{dis_book}:
\begin{eqnarray}
\frac{d^2\sigma^{CC}(e^\pm p)}{dxdQ^2} =  & (1\pm \mathcal{P}_e)&  \frac{G_F^2}{2\pi x} \bigg( \frac{M_W^2} {M_W^2+Q^2}\bigg)^2  \nonumber \\
\bigg[ \frac{Y_+}{2}W_2^{\pm}-& \frac{Y_{-}}{2}xW_3^{\pm}-&\frac{y^2}{2}W_L^{\pm}\bigg],
\label{eq:ccsigma_pol}
\end{eqnarray}
where, $\mathcal{P}_e$ is the polarisation of the lepton beam, as given in equation~\ref{eq:pe}.

\begin{figure}[!h]
\centerline{\includegraphics[width=1.0\columnwidth]{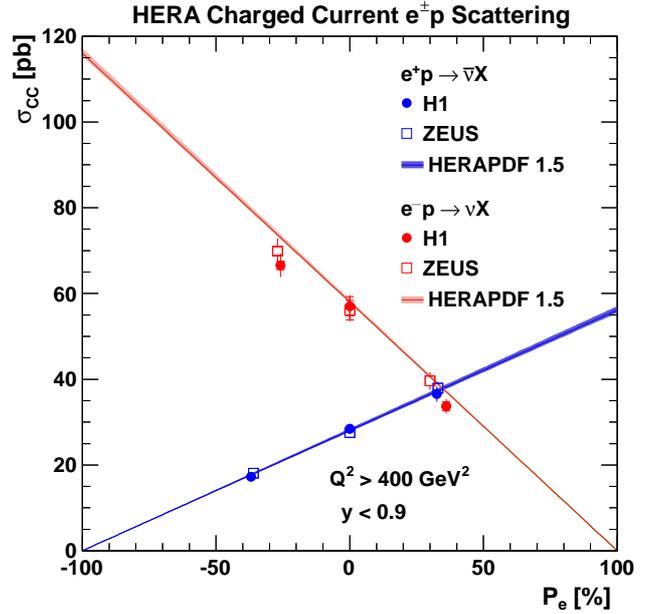}}
\caption{The measured CC DIS cross sections versus the lepton beam polarisation as
  measured by the H1 and ZEUS collaborations. The prediction from
  HERAPDF1.5 is also shown.}
\label{fig:h1zeus_ccvspol}
\end{figure}

It can be seen from equation~\ref{eq:ccsigma_pol} that the cross section for the process
$e^-p \rightarrow \nu X$ ($e^+p \rightarrow \bar \nu X$) has a linear dependence on the beam
polarisation, and vanishes in the SM for $P_{e^-} = 1$ ($P_{e^+} = -1$).
The dependence of the CC cross section on the polarisation as measured by the H1 and
ZEUS collaborations is shown in figure~\ref{fig:h1zeus_ccvspol}, compared to an earlier iteration
of HERAPDF, namely HERAPDF1.5~\cite{Radescu:2013mka,CooperSarkar:2011aa}. 
The data are well described by the SM predictions.

According to the SM, the CC $e^+p$ DIS cross section becomes zero for a fully negatively polarised
positron beam, a non-zero cross section at $P_e=-1$ might point to the existence of a right-handed
$W$ boson, $W_R$, and right-handed neutrinos, $\nu_R$.
The H1 and ZEUS data have been used to constrain the mass of such
boson~\cite{Aaron:2012qi,Collaboration:2010xc}, assuming the coupling strength and propagator
dependence on the mass of the boson to be the same as in SM CC interactions,
and the outgoing right-handed neutrinos to be light.
The ZEUS collaboration obtains $M_{W_R} > 198~{\rm GeV}$ at $95\%$~confidence level
(CL)~\cite{Collaboration:2010xc}, while the H1 collaboration gives a $95\%$~CL limit
of $M_{W_R} > 214~{\rm GeV}$ and $M_{W_R} > 194~{\rm GeV}$ for $e^-p$ and $e^+p$
collisions, respectively~\cite{Aaron:2012qi}.

%%%

The HERA results shown in this section confirm the excellent agreement between the
data and the predictions of the SM.
However, there are parts of the phase space which are not measured with very high precision,
or types of processes for which the cross sections as predicted by the SM are very low and
would not manifest themselves in significant changes to the NC and CC inclusive cross sections.

%%%

Searches for rare processes and BSM physics at HERA focus therefore on processes with
striking features but low SM cross sections, like particle production at high transverse momenta,
or specific model-predicted signatures like SUSY, or heavy resonances like leptoquarks,
and in general on deviations from SM cross sections in the regions which are less well measured,
such as high $Q^2$ and high $x$.

\section{Contact interactions}
\label{sec:ci}

The high $Q^{2}$ NC DIS interactions $e^{\pm}p \rightarrow e^{\pm}X$
at HERA described in the previous section provide the means to search
for new physics beyond the SM at short distances, using the concept of
four-fermion contact interactions (CI).
As opposed to $s$-channel {\it direct} searches, where the mass of any
resonant particle is limited to the available centre of mass energy
via $M_{X} =\sqrt{x s}$, the search for CI relies upon the {\it
  indirect} effects of the interference of the SM photon and $Z$ boson
field with the field of any new particle produced.
Therefore, CI may manifest as deviations from the SM expectation in
the measured differential cross section $d\sigma/dQ^{2}$ and any
observed deviation may be related to new heavy particles with masses
$M_{X}$ much larger than the electroweak scale.
In the low energy limit $\sqrt{s} \ll M_{X}$ such phenomena can be
described by an effective four-fermion CI model, and different
implementations of such a model are described in the following.

%%%

The most general chiral invariant Lagrangian for NC four-fermion CI in
$ep$ scattering may be written
as~\cite{Eichten:1983hw,Ruckl:1983hz,Ruckl:1983ag,haberl}:
\begin{equation}
\mathcal{L} = \sum\limits_{q}\sum\limits_{a,b = L,R} =
\eta^{q}_{ab}(\bar{e}_{a}\gamma_{\mu}e_{a}) (\bar{q}_{b}\gamma^{\mu}q_{b}),
\label{eq:ci-lagrangian}  
\end{equation}
where $\eta^{q}_{ab}$ are the CI coupling coefficients, $a$ and $b$
indicate the left-handed and right-handed fermion helicities and the
first sum is over all quark flavours.
In the kinematic region of interest at HERA the valence ($u$ and $d$)
quarks dominate.

%%%

For general models of fermion compositeness or substructure the CI
coupling coefficients are defined as:
\begin{equation}
\eta^{q}_{ab} = \epsilon^{q}_{ab}\frac{4\pi}{\Lambda^{2}}
\label{eq:ci-general}
\end{equation}
where $\Lambda$ is the compositeness scale and the coefficients
$\epsilon^{q}_{ab}$ describe the chiral structure of the coupling and
may take the values $\pm 1$ or $0$, depending on whether the coupling
represents pure left-handed (L), right-handed (R), or vector (V) and
axial-vector (A) scenarios.
Hence, depending on the model and the sign of the coefficients, the
new physics process interferes either constructively or destructively
with the SM.

%%%

Leptoquarks are hypothetical colour triplet scalar or vector bosons
carrying both lepton and baryon number, and appear naturally in
extensions of the SM which aim to unify the lepton and quark sectors.
For leptoquark masses much larger than the probing scale at HERA,
$M_{LQ} \gg \sqrt{s}$, the coupling $\lambda$ is related to the CI
coupling coefficients via:
\begin{equation}
\eta^{q}_{ab} = \epsilon^{q}_{ab}\frac{\lambda^{2}}{M^{2}_{\rm LQ}}
\label{eq:ci-lqs}
\end{equation}

The classification of leptoquarks is discussed in
section~\ref{sec:lq}, and follows the Buchm\"{u}ller-R\"{u}ckl-Wyler
(BRW) model \cite{Buchmuller:1986zs} where the coefficients
$\epsilon^{q}_{ab}$ depend on the leptoquark type \cite{schrempp}
and take values $0$, $\pm 0.5$ , $\pm 1$ or $\pm 2$.

%%%

\begin{figure*}
\centerline{\includegraphics[width=1.40\columnwidth]{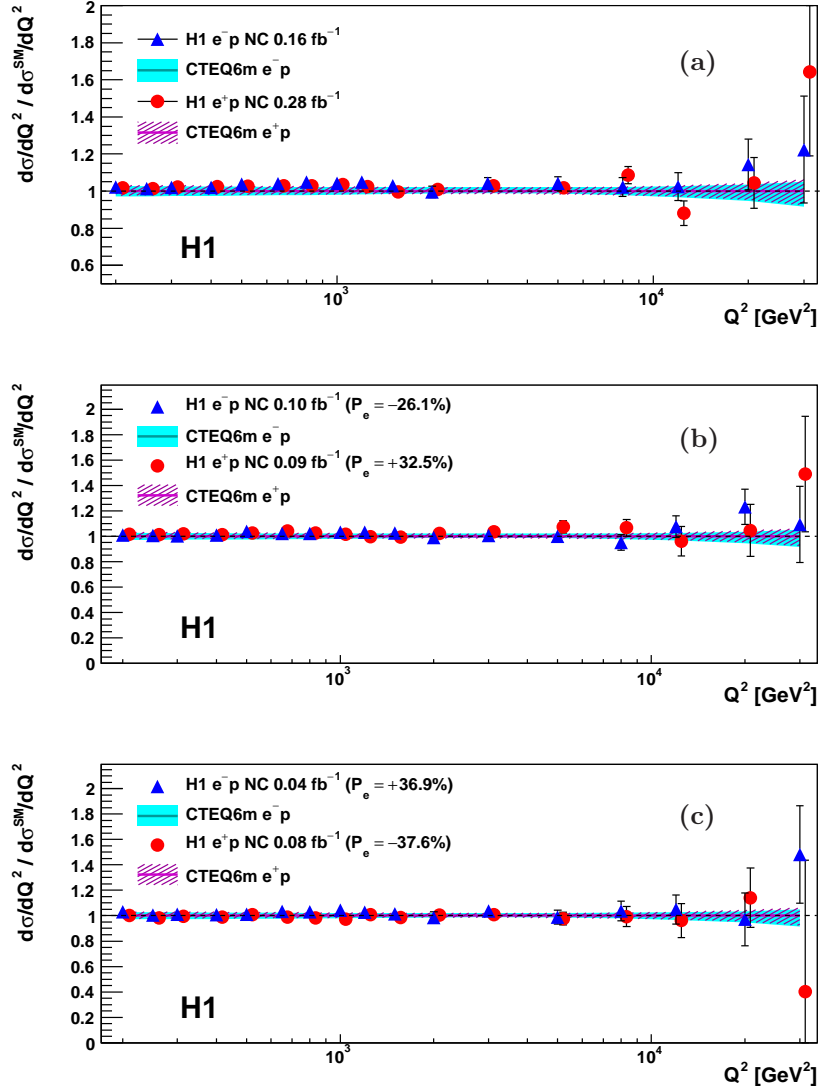}}
 \begin{picture} (0.,0.)
  \setlength{\unitlength}{1.0cm}
    \put (12.0,14.0){\bf\normalsize (a)}
    \put (12.0,9.0){\bf\normalsize (b)}
    \put (12.0,4.0){\bf\normalsize (c)}
 \end{picture}
\caption{The ratio of the measured cross section to the SM prediction
  determined using the CTEQ6m PDF set for $e^{+}p \rightarrow
  eX$ and $e^{-}p \rightarrow eX$ scattering. Figure (a) shows the
  full H1 data with an average longitudinal polarisation of
  $\mathcal{P}_{e} \sim 0$. Figures (b) and (c) show the polarised H1
  data divided into the different lepton charge and polarisation data
  sets. The error bars represent the statistical and uncorrelated
  systematic errors added in quadrature. The bands indicate the PDF
  uncertainties of the SM cross section predictions.}
\label{fig:ci_analysis}
\end{figure*}

A proposed explanation as to why the gravitational force is so much
weaker than the other fundamental forces, the so called
{\it hierarchy problem}, is the existence of large extra
dimensions~\cite{ArkaniHamed:1998rs,Antoniadis:1998ig,ArkaniHamed:1998nn}.
In such a model it is proposed that whilst particles, including strong
and electroweak bosons, are confined to four dimensions, gravity can
propagate into additional $n$ spatial dimensions, which are compactified
with a radius $R$.
The theory predicts a new gravitational scale $M_{S}$, related to the
Planck scale by $M^{2}_{P} \sim R^{n}M^{2+n}_{S}$, which for $R \sim
1$~mm and $n = 2$ can be of order TeV.
Consequently, at high energies the strengths of the gravitational
and electroweak interactions may be comparable~\cite{Chekanov:2003pw}.
The addition of the graviton-photon interference and graviton-$Z$
interference to the SM cross section is described in~\cite{Adloff:2000dp}.
The graviton-exchange contribution to the total $eq \rightarrow eq$
scattering cross section can thus be
described~\cite{Giudice:1998ck,Cheung:1999qh} as a contact interaction
with an effective coupling strength:
\begin{equation}
\eta_{G} = \frac{\lambda}{M^{4}_{S}}
\label{eq:ci-extradim}
\end{equation}
where the coupling $\lambda$ is conventionally set to $\pm 1$.

%%%

Finally, the observation of finite size effects such as the measurement
of the electroweak charge distribution of fermions would provide a
clear manifestation of quark substructure.
Such effects are typically described by a standard form factor in the
$eq \rightarrow eq$ scattering cross section:
\begin{equation}
f(Q^{2}) = 1 - \frac{\langle R^{2} \rangle}{6} Q^{2}
\label{eq:ci-quarkradius}
\end{equation}
where $\langle R^{2} \rangle$ is the mean squared radius of the
electroweak charge distribution of the quark, or {\it quark radius}.
This form factor modifies the $Q^{2}$ dependence of the SM cross
section similarly to the above CI models.

%%%

Several results on searches for contact interactions have been
published by both H1 and ZEUS using partial HERA
data sets~\cite{Chekanov:2003pw,Adloff:2000dp,Breitweg:1999ssa,Adloff:2003jm}.
%i
In this review we focus on the H1 result~\cite{Aaron:2011mv} that uses
their full data sample, corresponding to an integrated
luminosity of $446$~pb$^{-1}$ as described in table~\ref{tab:datasets}.
As some of the models described above exhibit chiral sensitivity, the
longitudinally polarised data sets are analysed separately.

%%%

To investigate contact interactions the measured H1 NC differential
cross sections $d\sigma/dQ^{2}$ at high four-momentum
transfer squared $Q^{2} > 200$~GeV$^{2}$ are compared to the SM
prediction.
The standard H1 NC event selection is employed~\cite{Aaron:2012qi}.
For the analysis of contact interactions, the SM prediction at
high $Q^2$ and high $x$ are of special relevance. The use of a PDF set
to determine the SM prediction in this kinematic region based mostly
on the HERA data is not appropriate, as the SM predictions would
also include the effects from any potential contact interactions
present in the data.
Therefore, the PDF set CTEQ6m PDF~\cite{Pumplin:2002vw} is used. 
The CTEQ6m set was obtained by fitting several experimental data sets.
At high $x$ this PDF is mostly constrained by fixed target experiments
and also by $W$-boson production and jet data from the Tevatron
experiments, which are not sensitive to possible $eq$ contact
interaction processes.
CTEQ6m does include early $e^\pm p$ scattering data at high $Q^2$ from
the H1 and ZEUS experiments, but as the $e^+p$ ($e^-p$) data sets
analysed here are 6 (10) times larger the residual correlations
between the HERA data and the CTEQ6m PDF are small and are
neglected in the following.
The CTEQ6m parton densities can be therefore be regarded as unbiased
with respect to possible contact interaction effects.
CTEQ6m is chosen as it describes many experimental data and in
particular, the HERA data in the region $Q^2 < 200~{\rm GeV}^2$, which
are not used in this analysis. The analysis was also verified using an
alternative PDF not based on HERA high $Q^2$ data, a dedicated H1 PDF
set, obtained from a next-to-leading order QCD fit to the H1
data~\cite{Aaron:2009bp} with $Q^2 < 200~{\rm GeV}^2$, excluding the
high $Q^2$ data used in this analysis.
Both the SM expectation and limits derived using the dedicated H1 PDF
agree well with those obtained using the CTEQ6m PDF within the uncertainties.
The ratio of the measured cross sections to those from the SM
prediction is shown in figure~\ref{fig:ci_analysis}, where a good
agreement is observed both in the full data set and each of the
polarised data sets from the HERA~II.

%%%

A quantitive test is performed to evaluate the impact of each of the
different CI models on the SM prediction and the level of agreement to
the data using a $\chi^{2}$ minimisation fitting function~\cite{Adloff:2003jm}.
The fit to the experimental and theoretical cross sections
fully takes into account all statistical and systematic uncertainties.

The H1 data are found to be consistent with the SM expectation based
on the CTEQ6m PDF, yielding a $\chi^{2}/dof = 16.4/17 (7.0/17)$ for
the $e^{+}p$ ($e^{-}p$) data, where $17$ is the number of $Q^2$ bins.
For each of the CI models described above, the effective scale
parameters and couplings associated to the new physics scale are
determined by a fit to the differential NC cross section.
All scale parameters are found to be consistent with the SM and limits are
calculated at the $95\%$~CL using the frequentist
method as described in the H1 HERA~I publication~\cite{Adloff:2003jm}.

\begin{table}
  \renewcommand{\arraystretch}{1.3}
\caption{Lower limits from H1 at the $95\%$~CL on the compositeness scale
  $\Lambda$. The $\Lambda^{+}$ limits correspond to the upper signs and
  the $\Lambda^{-}$ limits correspond to the lower signs of the chiral
  coefficients $\epsilon^{q}_{LL}$,$\epsilon^{q}_{LR}$, $\epsilon^{q}_{RL}$ and
$\epsilon^{q}_{RR}$, where the CI coupling coefficient is given by $\eta^{q}_{ab}
= \epsilon^{q}_{ab} 4\pi/\Lambda^{2}$.}
\label{tab:h1generalcompositeness}
\begin{tabular*}{1.0\columnwidth}{@{\extracolsep{\fill}} l r r r r c c}
\hline
\multicolumn{7}{@{\extracolsep{\fill}} l}{\bf H1 Search for General
  Compositeness ({\boldmath ${\mathcal L} = 446~{\rm pb}^{-1}$})}\\
\hline
Model & $\epsilon^{q}_{LL}$ & $\epsilon^{q}_{LR}$ & $\epsilon^{q}_{RL}$&$\epsilon^{q}_{RR}$& $\Lambda^{+}$ [TeV] & $\Lambda^{-}$ [TeV]\\
\hline
LL & $\pm 1$ & 0 & 0 & 0 & 4.2 & 4.0\\
LR & 0 & $\pm 1$ & 0 & 0 & 4.8 & 3.7\\
RL & 0 & 0 & $\pm 1$ & 0 & 4.8 & 3.8\\
RR & 0 & 0 & 0 & $\pm 1$ & 4.4 & 3.9\\
VV & $\pm 1$& $\pm 1$ & $\pm 1$ & $\pm 1$ & 5.6 & 7.2\\
AA & $\pm 1$& $\mp 1$ & $\mp 1$ & $\pm 1$ & 4.4 & 5.1\\
VA & $\pm 1$& $\mp 1$ & $\pm 1$ & $\mp 1$ & 3.8 & 3.6\\
LL+RR & $\pm 1$& 0 & 0 & $\pm 1$ & 5.3 & 5.1\\
LR+RL & 0 & $\pm 1$ & $\pm 1$ & 0 & 5.4 & 4.8\\
\hline
\end{tabular*}
%\end{tabular}}
\end{table}

Lower limits on the compositeness scale $\Lambda$ in the context of
the general CI model are presented in
table~\ref{tab:h1generalcompositeness}.
The results are presented for nine scenarios, which differ in their
chiral structure as determined by the CI coupling coefficients $\eta^{q}_{ab}$.
Depending on the model and the sign of the coefficients, limits on
$\Lambda$ are obtained in the range $3.6$~TeV ($\Lambda^{-}_{VA}$) to
$7.2$~TeV ($\Lambda^{-}_{VV}$).
In a similar analysis by ZEUS~\cite{Chekanov:2003pw} using their HERA
I data alone, limits of $3.3$~TeV and $6.2$~TeV are found for
$\Lambda^{-}_{VA}$ and $\Lambda^{-}_{VV}$, respectively.
For comparison, in a recent ATLAS analysis where the background process
to contact interactions arises from Drell-Yan production $qq \rightarrow
\ell\ell$, limits are derived at the $95\%$~CL on the LL, LR and RR
scenarios in the range $15.4$~TeV and $26.3$~TeV~\cite{Aad:2014wca}.

%%%

Within the model of leptoquark-type contact interactions lower limits on
$M_{LQ}/\lambda$ at the $95\%$~CL are presented in
table~\ref{tab:h1heavyleptoquarks}, for each of the $14$ leptoquarks
in the BRW model.
The observed limits in this analysis are in the range $M_{LQ}/\lambda > 0.41$ -
$1.86$~TeV.
Leptoquarks coupling to $u$ quarks are probed with higher sensitivity,
corresponding to more stringent limits than those coupling to $d$
quarks due to the valence quark content of the proton
and the QED couplings of the electron to the quarks, which reflect the
electric charge of the different quarks.
For a Yukawa coupling of electromagnetic strength, $\lambda = 0.3$,
scalar and vector leptoquark masses up to $0.33$~TeV and $0.56$~TeV
are excluded, respectively.

%%%

\begin{figure}[t]
\centerline{\includegraphics[width=1.0\columnwidth]{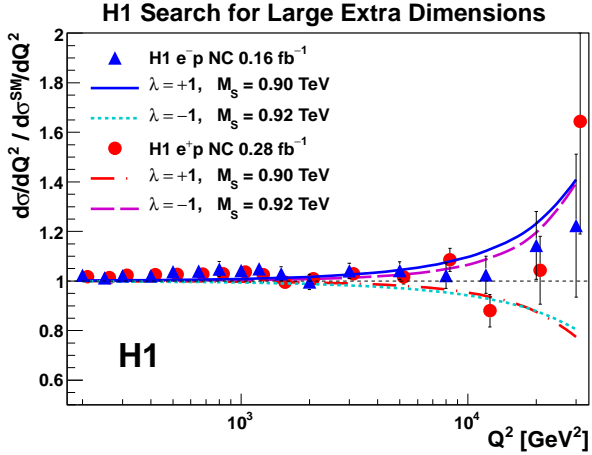}}
\caption{The measured NC cross section ${\rm d}\sigma/{\rm d}Q^{2}$ normalised to
  the SM expectation. H1 $e^{\pm}p$ scattering data are compared with
  curves corresponding to $95\%$~CL exclusion limits obtained from the
  full H1 data on the gravitational scale, $M_{S}$, for both positive
  ($\lambda = +1$) and negative ($\lambda = -1$) couplings. The error
  bars represent the statistical and uncorrelated systematic errors
  added in quadrature.}
\label{fig:h1-led}
\end{figure}

The $S^{L}_{0}$ and $\tilde{S}^{L}_{1/2}$ leptoquarks have identical
quantum numbers to down and up squarks, respectively.
The couplings associated to these leptoquarks therefore correspond to the
Yukawa couplings $\lambda'_{11k}$ and $\lambda'_{1j1}$ in the
framework of ${R}_{p}$ violating supersymmetry and the limits on
these leptoquarks presented in table~\ref{tab:h1heavyleptoquarks}
are also applicable to the the ratio
 $M_{\tilde{q}}/\lambda'$~\cite{Aaron:2011mv}.
Dedicated leptoquark searches performed at HERA are discussed in the
following sections.

Lower limits in a model with large extra dimensions on the
gravitational scale $M_{S}$ in $4+n$ dimensions are derived assuming a
positive or negative coupling.
For a $\lambda = +1$ ($\lambda = -1$) coupling mass scales $M_{S} <
0.90$~TeV ($M_{S} < 0.92$~TeV) are excluded at the $95\%$~CL.
The corresponding cross section predictions normalised to the SM
expectation are compared to the $e^{\pm}p$ data in figure~\ref{fig:h1-led}.
Since the advent of the LHC and the higher energy centre of mass data
that it provides these limits have been surpassed, for example by
ATLAS, where limits are set on $M_{S} > 3.2$~TeV~\cite{Aad:2014wca}.

\begin{table}
  \renewcommand{\arraystretch}{1.3}
  \caption{Lower limits from H1 at the $95\%$~CL on $M_{LQ}/\lambda$ for $14$ scalar (S)
    and vector (V) leptoquarks, where L and R denote the lepton
    chirality and the subscript ($0$, $1/2$, $1$) is the weak
    isospin. In this case the CI coupling coefficient is $\eta^{q}_{ab} =
    \epsilon^{q}_{ab} \lambda^{2}/M^{2}_{LQ}$. For each leptoquark type,
    the relevant coefficients $\epsilon^{q}_{ab}$ and fermion number $F = L + 3B$ are
    indicated. Leptoquarks with identical quantum numbers except for
    weak hypercharge are distinguished using a tilde, for example
    $V^{R}_{0}$ and $\tilde{V}^{R}_{0}$.}
  % Quantum numbers and helicities refer to $e^{-}q$ and $e^{-}\bar{q}$ states.}
  \label{tab:h1heavyleptoquarks}
  \begin{tabular*}{1.0\columnwidth}{@{\extracolsep{\fill}} l c c c c}
\hline
\multicolumn{5}{@{\extracolsep{\fill}} l}{\bf H1 Search for Heavy
    Leptoquarks ({\boldmath ${\mathcal L} = 446~{\rm pb}^{-1}$})}\\
\hline
LQ & $\epsilon^{u}_{ab}$ & $\epsilon^{d}_{ab}$ & $F$ & $M_{LQ}/\lambda$ [TeV]\\
\hline
$S^{L}_{0}$ & $\epsilon^{u}_{LL} = + \frac{1}{2}$ & & 2 & 1.10\\
$S^{R}_{0}$ & $\epsilon^{u}_{RR} = + \frac{1}{2}$ & & 2 & 1.10\\
$\tilde{S}^{R}_{0}$ & & $\epsilon^{d}_{RR} = + \frac{1}{2}$ & 2 & 0.41\\
$S^{L}_{1/2}$ & $\epsilon^{u}_{LR} = - \frac{1}{2}$ & & 0 & 0.87\\
$S^{R}_{1/2}$ & $\epsilon^{u}_{RL} = - \frac{1}{2}$ & $\epsilon^{d}_{RL} = - \frac{1}{2}$ & 0 & 0.59\\
$\tilde{S}^{L}_{1/2}$ & & $\epsilon^{d}_{LR} = - \frac{1}{2}$ & 0 & 0.66\\
$S^{L}_{1}$ & $\epsilon^{u}_{LL} = + \frac{1}{2}$ & $\epsilon^{d}_{LL} = + 1$ & 2 & 0.71\\
\hline
$V^{L}_{0}$ & & $\epsilon^{d}_{LL} = -1 $ & 0 & 1.06\\
$V^{R}_{0}$ & & $\epsilon^{d}_{RR} = -1$ & 0 & 0.91\\
$\tilde{V}^{R}_{0}$ & $\epsilon^{u}_{RR} = -1$ & & 0 & 1.35\\
$V^{L}_{1/2}$ & & $\epsilon^{d}_{LR} = + 1$& 2 & 0.51\\
$V^{R}_{1/2}$ & $\epsilon^{u}_{RL} = +1$ & $\epsilon^{d}_{RL} = +1$ & 2 & 1.44\\
$\tilde{V}^{L}_{1/2}$ & $\epsilon^{u}_{LR} = + 1$ & & 2 & 1.58\\
$V^{L}_{1}$ & $\epsilon^{u}_{LL} = - 2$ & $\epsilon^{d}_{LL} = - 1$ & 0 & 1.86\\
\hline
\end{tabular*}
\end{table}

Finally, an upper limit at $95\%$~CL on the quark radius $R_{q} < 0.65
\cdot 10^{-18}$~m is found,  assuming point-like leptons. The same
limit from ZEUS is $R_{q} < 0.81 \cdot 10^{-18}$~m, which is derived
in a similar analysis using their HERA~I data set~\cite{Chekanov:2003pw}.

\section{Leptoquarks}
\label{sec:lq}

\begin{table*}
\caption{The $14$ LQ types of the Buchm\"uller-R\"uckl-Wyler
    classification in the commonly used notation.
    The LQ subscripts refer to the weak isospin and the superscripts refer to the lepton
    chirality. The spin $J$, fermion number $F$ and charge $Q$ of each leptoquark is
    indicated, as well as the dominant resonant production process in $ep$ scattering
    and possible decay modes, the corresponding coupling and the branching ratio
    to charged leptons $\beta_{\ell}$.}
\label{tab:lqs}
\begin{small}
\setlength{\extrarowheight}{3pt}
\centerline{\begin{tabular*}{0.66\textwidth}{@{\extracolsep{\fill}} c c c c r@{\hskip 0.2cm} l@{\hskip 0.2cm} l c c}
\hline
LQ type & $J$ & $F$ & $Q$ & \multicolumn{3}{c}{Production and} & Coupling & $\beta_{\ell}$ \\
&  &  & & \multicolumn{3}{c}{decay modes} & & \\
       \hline
       %scalar F=2, e-
       & & & & & & $\ell^{-}u$ & $\lambda_L$ & $1/2$  \\
       \up{$S^L_0$} & \up{$0$} & \up{$2$} & \up{$-1/3$} & \up{$e^{-}_Lu_L$} &
       \up{$\rightarrow \bigg\{ $} & $\nu_\ell d$ & $-\lambda_L$ & $1/2$  \\
       \hline
       $S^R_0$ & $0$ & $2$ & $-1/3$ & $e^{-}_Ru_R$ & $\rightarrow$ & $\ell^{-}u$
       & $\lambda_R$ & $1$ \\     
       \hline
       $\tilde{S}^R_0$ & $0$ & $2$ & $-4/3$ & $e^{-}_Rd_R$ & $\rightarrow$ & $\ell^{-}d$
       & $\lambda_R$ & $1$ \\
       \hline
       & & & & & & $\ell^{-}u$ & $-\lambda_L$ & $1/2$  \\
       $S^L_1$ & $0$ & $2$ & \up{$-1/3$} & \up{$e^{-}_Lu_L$} &
       \up{$\rightarrow \bigg\{ $} & $\nu_\ell d$
       & $-\lambda_L$ & $1/2$ \\ 
       & & & $-4/3$ & $e^{-}_Ld_L$ & $\rightarrow$ & $\ell^{-}d$ &
       $-\sqrt{2}\lambda_L$ & $1$  \\
       \hline
       %vector F=2, e-
       $V^L_{1/2}$ & $1$ & $2$ & $-4/3$ & $e^{-}_Ld_R$ & $\rightarrow$ & $\ell^{-}d$
       & $\lambda_L$ & $1$ \\
       \hline
       & & & $-1/3$ & $e^{-}_Ru_L$ & $\rightarrow$ & $\ell^{-}u$ & $\lambda_R$ & $1$  \\
       \up{$V^R_{1/2}$} & \up{$1$} & \up{$2$} & $-4/3$ & $e^{-}_Rd_L$ &
       $\rightarrow$ & $\ell^{-}d$ & $\lambda_R$ & $1$  \\
       \hline
       $\tilde{V}^L_{1/2}$ & $1$ & $2$ & $-1/3$ & $e^{-}_Lu_R$ & $\rightarrow$ &
       $\ell^{-}u$ & $\lambda_L$ & $1$ \\
       \hline
       \hline
       %vector F=0, e+
       & & & & & & $\ell^{+}d$ & $\lambda_L$ & $1/2$  \\
       \up{$V^L_0$} & \up{$1$} & \up{$0$} & \up{$+2/3$} & \up{$e^{+}_Rd_L$} &
       \up{$\rightarrow \bigg\{ $} & $\bar{\nu}_\ell u$ & $\lambda_L$ & $1/2$  \\
       \hline
       $V^R_0$ & $1$ & $0$ & $+2/3$ & $e^{+}_Ld_R$ & $\rightarrow$ & $\ell^{+}d$
       & $\lambda_R$ & $1$ \\     
       \hline
       $\tilde{V}^R_0$ & $1$ & $0$ & $+5/3$ & $e^{+}_Lu_R$ & $\rightarrow$ & $\ell^{+}u$
       & $\lambda_R$ & $1$ \\
       \hline
       & & & & & & $\ell^{+}d$ & $-\lambda_L$ & $1/2$  \\
       $V^L_1$ & $1$ & $0$ & \up{$+2/3$} & \up{$e^{+}_Rd_L$} &
       \up{$\rightarrow \bigg\{ $} & $\bar{\nu}_\ell u$
       & $\lambda_L$ & $1/2$ \\ 
       & & & $+5/3$ & $e^{+}_Ru_L$ & $\rightarrow$ & $\ell^{+}u$ &
       $\sqrt{2}\lambda_L$ & $1$  \\
       \hline
       %scalar F=0, e+        
       $S^L_{1/2}$ & $0$ & $0$ & $+5/3$ & $e^{+}_Ru_R$ & $\rightarrow$ & $\ell^{+}u$
       & $\lambda_L$ & $1$ \\
       \hline
       & & & $+2/3$ & $e^{+}_Ld_L$ & $\rightarrow$ & $\ell^{+}d$ & $-\lambda_R$ & $1$  \\
       \up{$S^R_{1/2}$} & \up{$0$} & \up{$0$} & $+5/3$ & $e^{+}_Lu_L$ &
       $\rightarrow$ & $\ell^{+}u$ & $\lambda_R$ & $1$  \\
       \hline
       $\tilde{S}^L_{1/2}$ & $0$ & $0$ & $+2/3$ & $e^{+}_Rd_R$ & $\rightarrow$ &
       $\ell^{+}d$ & $\lambda_L$ & $1$ \\
       \hline
     \end{tabular*}}
  \end{small}
\end{table*}

\begin{figure}
\begin{center}
\includegraphics[width=0.98\columnwidth]{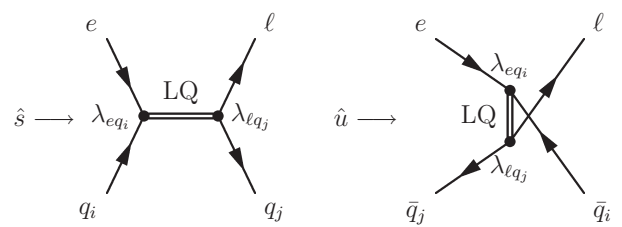}
\caption{Leptoquark production in $ep$ collisions: $s$-channel resonant production (left) and $u$-channel virtual exchange (right) with subsequent decay to a lepton-quark pair.}
\label{fig:lq_diag}
\end{center}
\end{figure}

The electron-proton collisions at HERA provide a unique opportunity
to search for new particles coupling directly to a lepton and a quark.
Leptoquarks (LQs) are an example of such particles, colour-triplet bosons
which appear in many extensions of the SM~\cite{Pati:1974yy,Georgi:1974sy,Langacker:1980js,Schrempp:1984nj,Wudka:1985ef,Dimopoulos:1979es,Dimopoulos:1979sp,Farhi:1979zx,Farhi:1980xs,Nilles:1983ge,Haber:1984rc}.
They can be produced at HERA directly via $s$-channel resonant production,
or indirectly via $u$-channel virtual LQ exchange, as shown in figure \ref{fig:lq_diag}.
The coupling at the electron-quark-LQ vertex is described by a dimensionless
parameter $\lambda$.
LQs carry both lepton ($L$) and baryon ($B$) quantum numbers
and the fermion number $F = L+3B$ is conserved.  
The $s$-channel production dominates for LQ masses lower than the centre of
mass energy, while for masses larger than $\sqrt{s}$ both the $s$- and
the $u$-channels, as well as the interference with SM processes, are important. 
In the first case, LQs appear as peaks in the spectra of the final state lepton-jet
invariant mass. In the second case, heavy LQ exchange could lead to measurable
low-energy exchange and so to deviations of the measured cross sections with
respect to SM predictions.
For LQ masses lower than $\sqrt{s}$, $F=0$ ($F=2$) LQs dominate
in $e^+p$ ($e^-p$) collisions, while for higher masses both $e^+p$ and $e^-p$
have similar sensitivity to all LQ types.

\begin{figure*}
\centerline{\includegraphics[width=1.60\columnwidth]{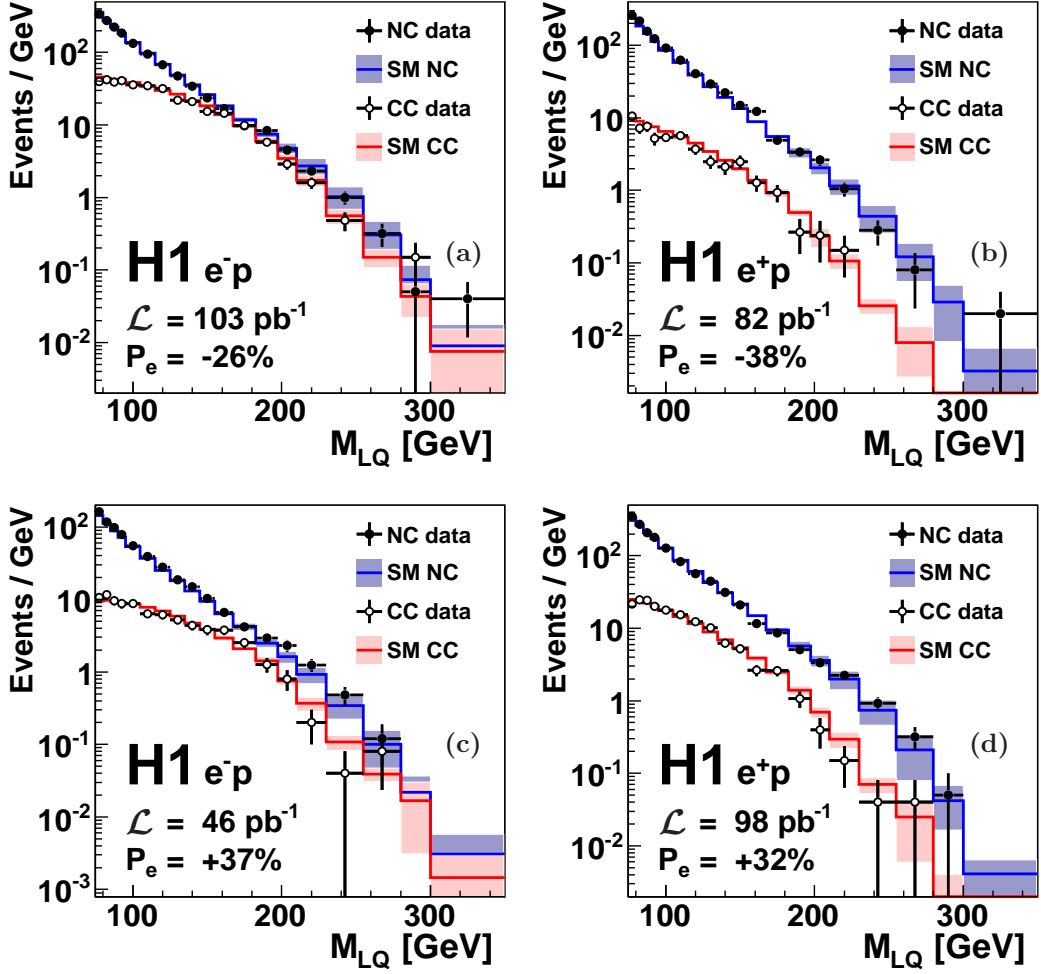}}
  \begin{picture} (0.,0.)
  \setlength{\unitlength}{1.0cm}
    \put (8.0,10.0){\bf\normalsize (a)}
    \put (14.9,10.0){\bf\normalsize (b)}
    \put (8.0, 3.5){\bf\normalsize (c)}
    \put (14.9,3.5){\bf\normalsize (d)}
 \end{picture}
\caption{The reconstructed leptoquark mass in the search for first generation leptoquarks in the 2003-2007 H1 data, which was taken with a polarised lepton beam. The left-handed electron data (a) and left-handed positron data (b) are shown in the top row; the right-handed electron data (c) and right-handed positron data (d) are shown in the bottom row. The luminosity $\mathcal{L}$ and average longitudinal lepton polarisation $\mathcal{P}_{e}$ of each data set is indicated. The NC (solid points) and CC (open points) data are compared to the SM predictions (histograms), where the shaded bands indicate the total SM uncertainties.}
\label{fig:lq-massdistns}
\end{figure*}

%%%

In the framework of the phenomenological Buchm\"uller-R\"uckl-Wyler (BRW)
model~\cite{Buchmuller:1986zs}, LQs are classified into $14$ types~\cite{schrempp}
with respect to the quantum numbers spin $J$, weak isospin $I$ and chirality
$C$ (left-handed $L$, right-handed, $R$).
The $14$ different LQs are detailed in table~\ref{tab:lqs}.
Scalar ($J=0$) LQs are denoted as $S_{I}^{C}$ and vector ($J=1$) LQs are denoted
$V_{I}^{C}$ in the following.
LQs with identical quantum numbers except for weak hypercharge are distinguished
using a tilde, for example $V_0^R$ and $\tilde{V}_0^R$.

%%%

Whereas all $14$ LQs couple to electron-quark pairs, four of the left-handed LQs,
namely $S_0^L$, $S_{1}^L$, $V_0^L$ and $V_{1}^L$, may also decay to a
neutrino-quark pair.
The branching ratio to charged leptons is given by:
\begin{equation}
\beta_\ell\!=\!\Gamma_{\ell q}/(\Gamma_{\ell q}+\Gamma_{\nu_\ell q}) 
\label{eq:lq-beta}  
\end{equation}
where \mbox{$\Gamma_{\ell q}$} (\mbox{$\Gamma_{\nu_\ell q}$}) denotes the partial
width for the LQ decay to the charged lepton $\ell$ (neutrino $\nu_\ell$) and a quark $q$.
The branching fraction of decays into a neutrino-quark pair is then given by
$\beta_{\nu_\ell} =1-\beta_{\ell}$.
In particular, for $S_0^L$ and $V_0^L$ the branching fraction of decays into an
electron-quark pair is predicted\footnote{In the case of $S_1^L$ and $V_1^L$ LQs,
which are superpositions of two states, this branching ratio is less trivial.}
by the model to be $\beta_\ell = 0.5$.

%%%

If the lepton number is conserved, a LQ can decay only into a quark and a
first generation lepton: an electron, positron or electron-neutrino, and is
termed a {\it first generation leptoquark}.
If the lepton number is not conserved, the production of {\it second} and {\it third generation
leptoquarks} is possible, leading to a final state containing a second
or third generation lepton together with the quark.

%%%

The BRW model assumes lepton number conservation, although a general
extension of this model allows for the decay of LQs to final states containing a
quark and a lepton of a different flavour, that is a muon or tau lepton, introducing
lepton flavour violation (LFV).
Non-zero couplings $\lambda_{eq_i}$ to an electron-quark pair and $\lambda_{\mu q_j}$
($\lambda_{\tau q_j}$) to a muon(tau)-quark pair are assumed.
The indices $i$ and $j$ represent quark generation indices, such
that $\lambda_{eq_i}$ denotes the coupling of an electron to a quark of
generation $i$, and $\lambda_{\ell q_j}$ is the coupling of the outgoing lepton
(where $\ell = \mu$ or $\tau$) to a quark of generation $j$.
When second and third generation LQs are considered, the branching ratio $\beta$
to a given charged lepton flavour $LQ \rightarrow \mu (\tau) q$ is calculated as:
\begin{equation}
  \beta = \beta_{\ell} \times \beta_{LFV}
\label{eq:lfv1}
\end{equation}
where
\begin{equation}
  \beta_{LFV}=\frac{\Gamma_{\mu(\tau)q}}{\Gamma_{\mu(\tau)q}+\Gamma_{eq}}\\
\label{eq:lfv2}
\end{equation}
and
\begin{equation}
  \Gamma_{\ell q}=m_{{\rm LQ}}\lambda^2_{\ell q}\times 
  \begin{cases}
    \frac{1}{16\pi} & \text{scalar LQ}\\
    \frac{1}{24\pi} & \text{vector LQ}
  \end{cases}
\label{eq:lfv3}
\end{equation}
where \mbox{$\Gamma_{\ell q}$} denotes the partial LQ decay width for the decay
to a lepton \mbox{$\ell=e,\mu,\tau$} and a quark $q$.
Assuming lepton universality, and that only one LFV transition is possible, $\beta_{LFV} = 0.5$.
An overview of this extended model for the LQ coupling to $u$ and $d$ quarks is provided
elsewhere~\cite{Aktas:2007ji}.

\subsection{Searches for first generation leptoquarks}
\label{sec:lq1}

Searches for first generation leptoquarks have been regularly
performed by both H1 and ZEUS during the HERA data
taking~\cite{Adloff:1999tp,Aktas:2005pr,Breitweg:2000sa,Breitweg:2000ks,Chekanov:2003af},
and two papers published in 1997 using up to $20$~pb$^{-1}$ of the
first data taken revealed a handful of outstanding events observed by both
collaborations at masses above and around
$200$~GeV~\cite{Adloff:1997fg,Breitweg:1997ff}.

%%%

The final H1 and ZEUS first generation leptoquark searches are
summarised in this section, which between them include the
full $1$~fb$^{-1}$ of HERA
data~\cite{Aaron:2011qaa,Abramowicz:2012tg}.
The polarisation $\mathcal{P}_{e}$ of the lepton beam in the HERA~II
data taking period is exploited to enhance the chiral sensitivity of
leptoquarks by analysing the positively and negatively polarised data
samples separately. 

%%%

First generation leptoquarks lead to final states which are similar to those of
NC and CC DIS, in that the LQ decays into either an electron or neutrino
and a quark.
As the experimental signature of LQ production is similar to that of
NC and CC events, the data selection closely follows that used in
the standard inclusive DIS analyses performed by H1 and ZEUS
as described in section~\ref{sec:sm}.
The NC and CC DIS processes are modelled using the DJANGOH
generator and the smaller photoproduction contribution is estimated
using PYTHIA.

%%%

In the NC-like event topology $ep \rightarrow eX$, the H1
analysis~\cite{Aaron:2011qaa} is performed in the region $Q^{2} > 133$~GeV$^2$,
$E_{e'} > 11$~ GeV and $0.1 < y < 0.9$, where the electron reconstruction
method is used (see equation~\ref{eq:emeth}).
The ZEUS analysis~\cite{Abramowicz:2012tg} employs both the
electron and double angle methods (see equation~\ref{eq:dameth}) and the
phase space is defined as $Q^{2} > 2500$~GeV$^2$, $E_{e'} > 10$~GeV,
$x > 0.1$ and $y_{e} < 0.95$.
An identified hadronic jet with $P_{T} > 15$~GeV in the region $|\eta| <3$ is
also required by the ZEUS analysis.
Background from neutral hadrons or photons misidentified as leptons is suppressed
by requiring a charged track to be associated to the lepton candidate and both analyses
use additional cuts on the total $E - P_{z}$ in the event to reduce the residual
photoproduction background.

%%%

In the CC like event topology $ep \rightarrow \nu X$, events are selected in the H1
analysis by requiring significant missing transverse momentum $P_{T}^{\rm miss}>12$~GeV,
which is due to the undetected neutrino, in the inelasticity region $0.1 <  y < 0.85$.
The ZEUS analysis phase space is $P_{T}^{\rm miss}>22$~GeV, $Q^{2} > 700$~GeV$^{2}$,
$y < 0.9$.
An identified hadronic jet with $P_{T} > 10$~GeV in the region $|\eta| <3$ is also required.
Photoproduction background is suppressed in both the H1 and ZEUS analyses by
exploiting the correlation between $P_{T}^{\rm miss}$ and the ratio
$V_{\rm ap}/V_{\rm p}$~\cite{Adloff:1999ah}.
Both H1 and ZEUS use the hadronic reconstruction method described in
equation~\ref{eq:jbmeth} for the CC like event topology.

%%%

The lepton scattering angle in the lepton-jets centre of mass frame, $\theta^{*}$,
is used in the ZEUS analysis to isolate a potential leptoquark signal
further.
The decay of a scalar resonance results in a flat distribution in $\cos\theta^{*}$,
while SM events show an approximately $1/(1 - \cos \theta^{*})^{2}$
dependence~\cite{Breitweg:2000sa}.
In a final step a cut of $\cos\theta^{*} < 0.4$ is applied by ZEUS in their analysis of
both the NC and CC topologies.

%%%

\begin{figure}[h]
  \centerline{\includegraphics[width=0.98\columnwidth]{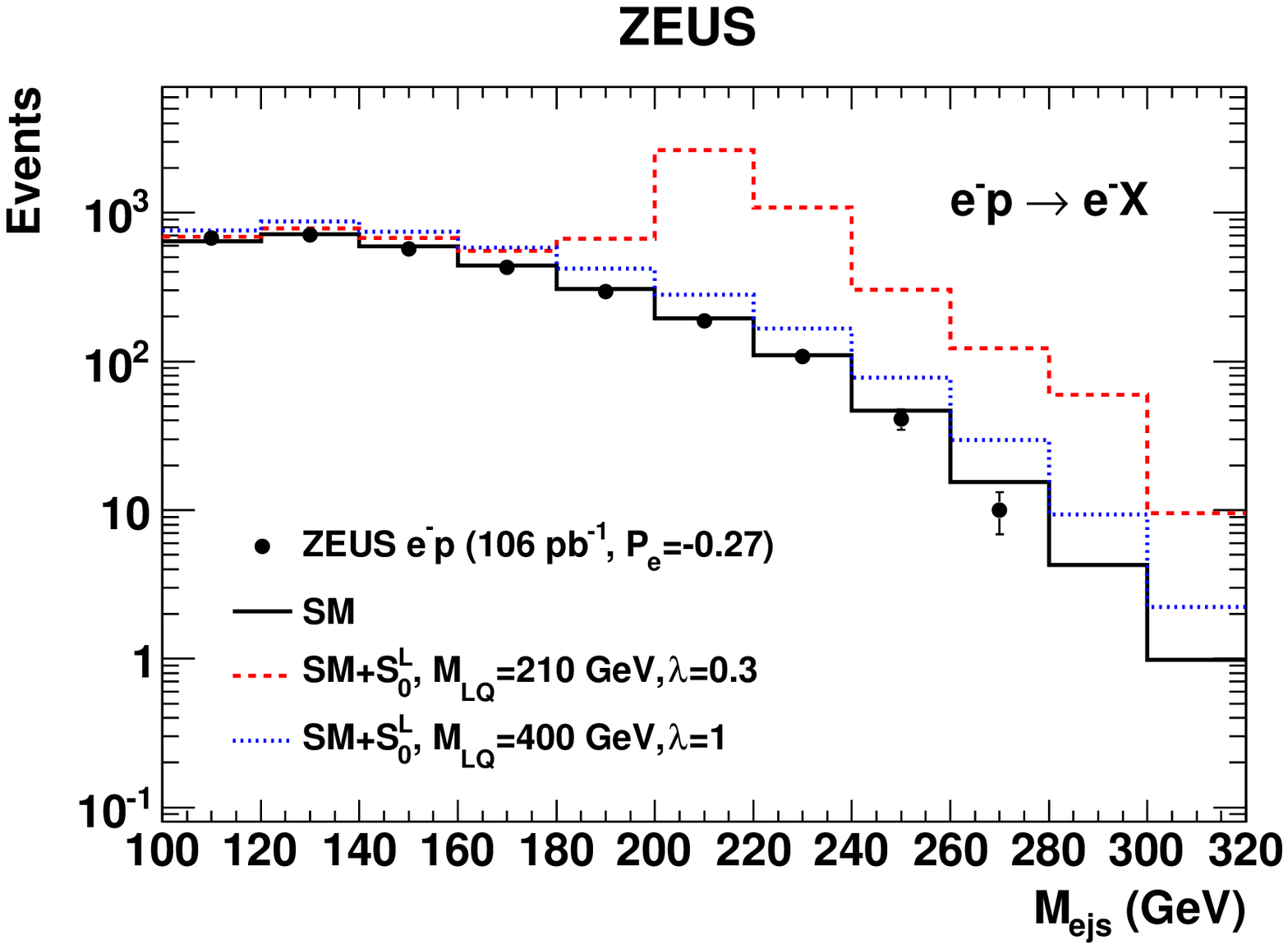}}
  \centerline{\includegraphics[width=0.98\columnwidth]{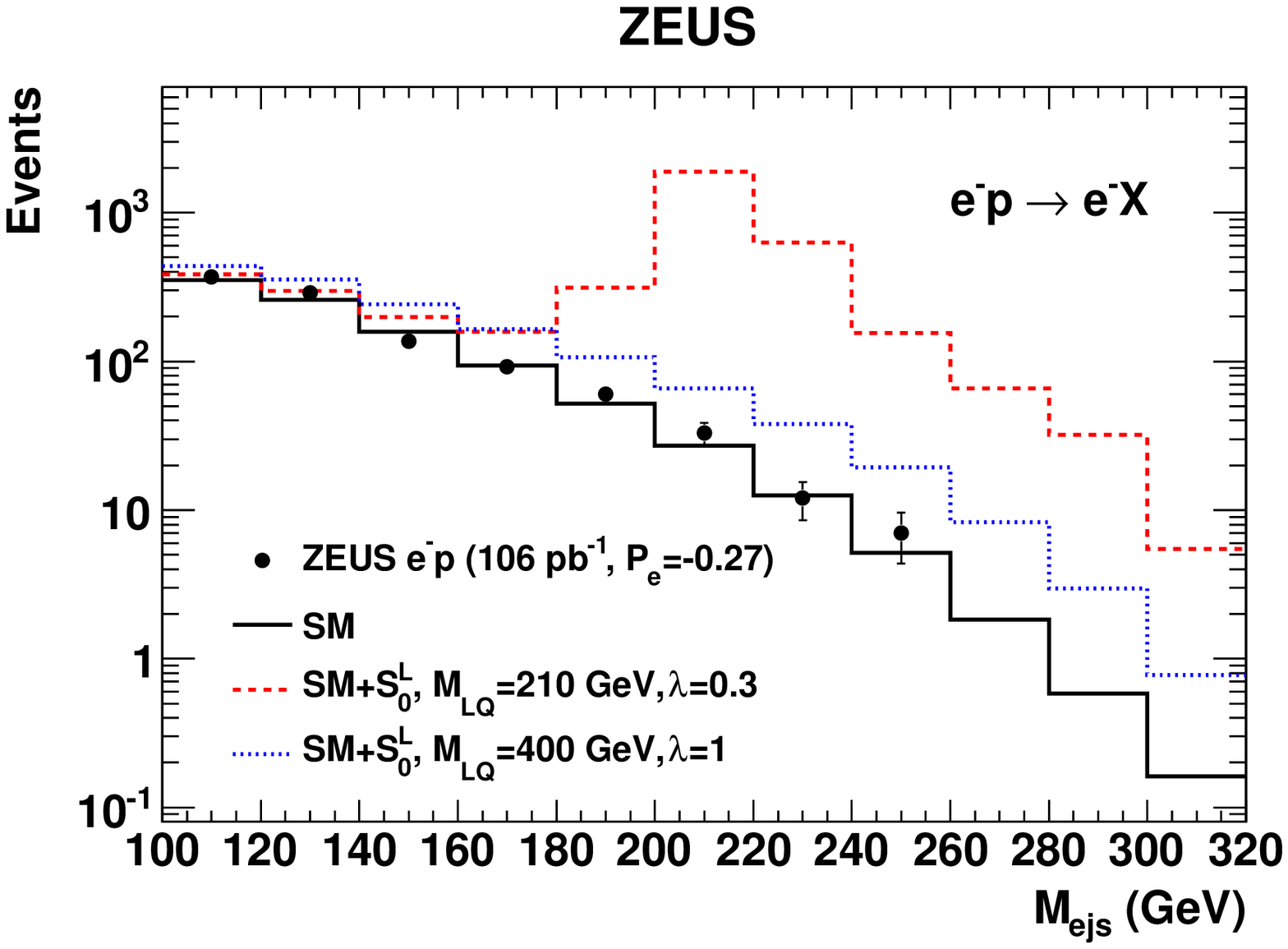}}
  \caption{Top: The reconstructed invariant mass, $M_{ejs}$, distribution in the
    $e^{-}p \rightarrow e^{-}X$ topology for the left-handed $e^{-}p$ ZEUS sample (dots),
    compared to the SM prediction (solid histogram) and to the predictions of the BRW-based
    LQ model including a $S^{L}_{0}$ LQ state with a mass of $210$~GeV with a coupling strength
    $\lambda=0.3$ (dashed histogram) and a mass of $400$~GeV and a coupling $\lambda=1.0$
    (dotted histogram). Bottom: the same distribution after the additional analysis cut
    on $\cos\theta^{*} < 0.4$ is applied.}
\label{fig:lq-zeuscostheatcomp}
\end{figure}

The leptoquark mass is reconstructed by H1 as:
\begin{equation}
  M_{\rm LQ} =\sqrt{Q^2/y}
\label{eq:mlq-h1}
\end{equation}
and uses the measured kinematics of the scattered electron
(hadronic final state) in the analysis of NC (CC) topologies.
In the ZEUS analysis, the LQ mass is reconstructed from the momentum and
energy of the jet and the lepton:
\begin{equation}
M_{ljs} = \sqrt{E^2_{ljs}- \vec{p}_{ljs}^2}, 
\label{eq:mlq-zeus}
\end{equation}
where $E^2_{ljs}$ is the sum of the energies of the jet and the lepton, and
$\vec{p}_{ljs}^2$ is the vector sum of their momenta.

%%%

In the analysis of their complete data sets a good agreement between the
data and the SM is observed by both H1 and ZEUS and no evidence for
LQs at HERA is observed.
The SM expectation is dominated by DIS processes in all event samples,
with small additional contributions from photoproduction.
%

%%%

%%%

Mass spectra of the four H1 HERA~II data sets taken with a longitudinally polarised
lepton beam are shown in figure~\ref{fig:lq-massdistns}, where both the NC-like
and CC-like event samples are presented~\cite{Aaron:2011qaa}.
The shape and normalisation of all samples are well described.
Similarly good agreement is observed in the equivalent mass distributions
in the ZEUS analysis~\cite{Abramowicz:2012tg}.

%%%

Figure~\ref {fig:lq-zeuscostheatcomp} shows as an example the left-handed
$e^{-}p$ ZEUS data without (top) and with (bottom) the $\cos\theta^{*} < 0.4$
analysis cut applied.
In both cases good agreement is seen between the data and the SM NC prediction.
Also shown are two predictions for a $S^{L}_{0}$ LQ state with a mass of $210$~GeV
and coupling strength $\lambda=0.3$ and a mass of $400$~GeV and a coupling
strength $\lambda=1.0$. 

%%%

\begin{figure*}
\centerline{\includegraphics[width=0.98\columnwidth]{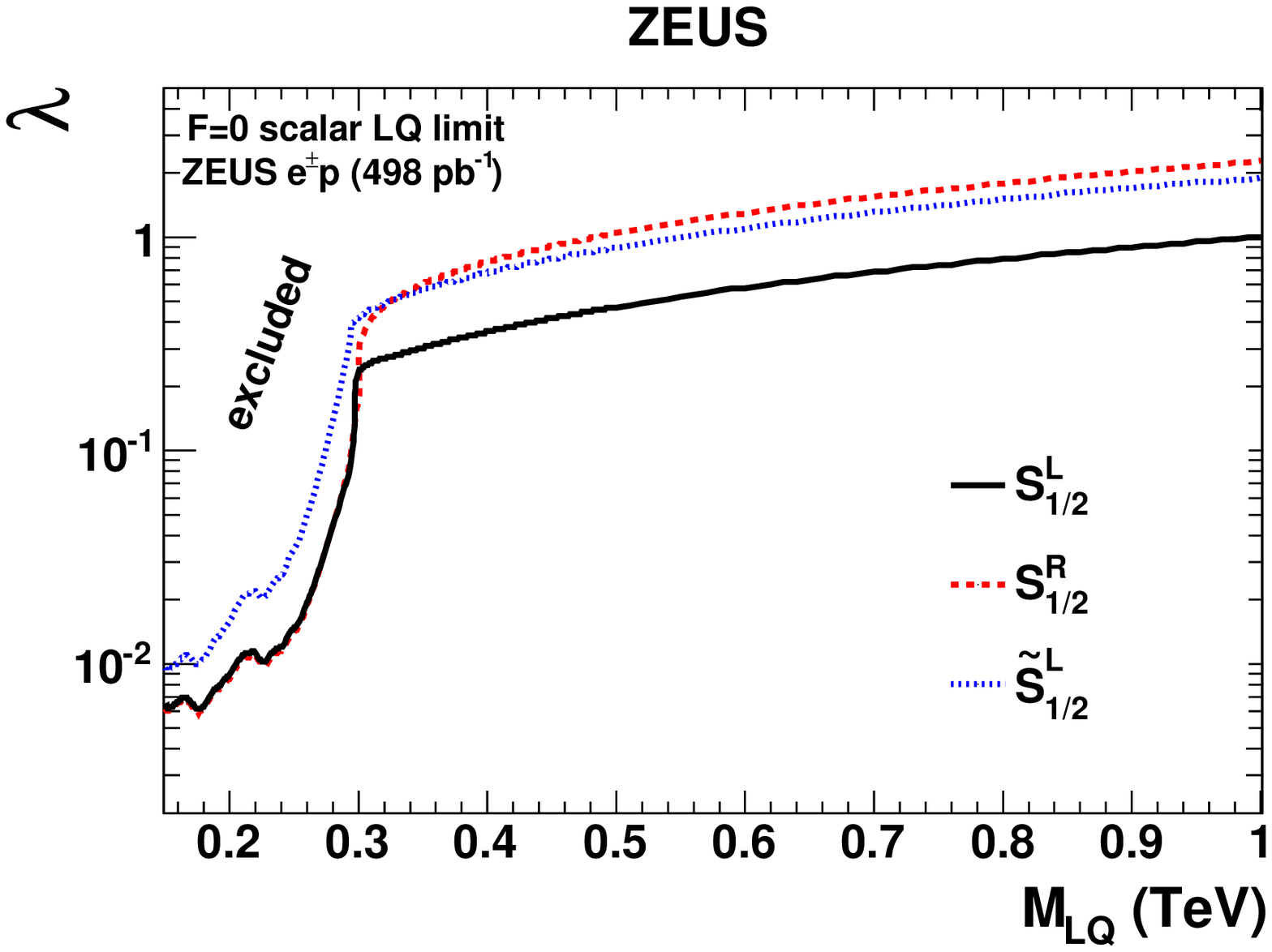}\includegraphics[width=0.98\columnwidth]{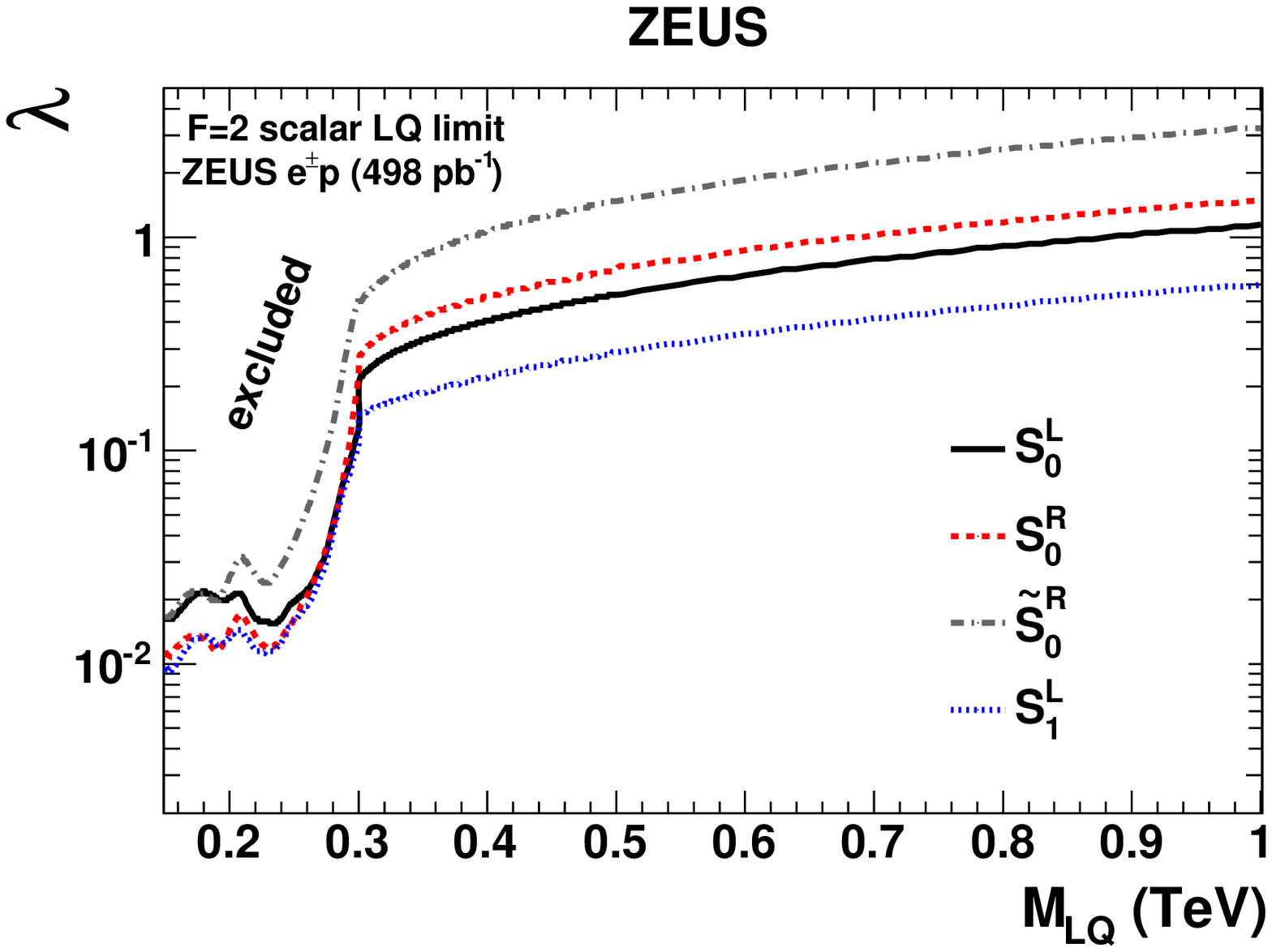}}
\centerline{\includegraphics[width=0.98\columnwidth]{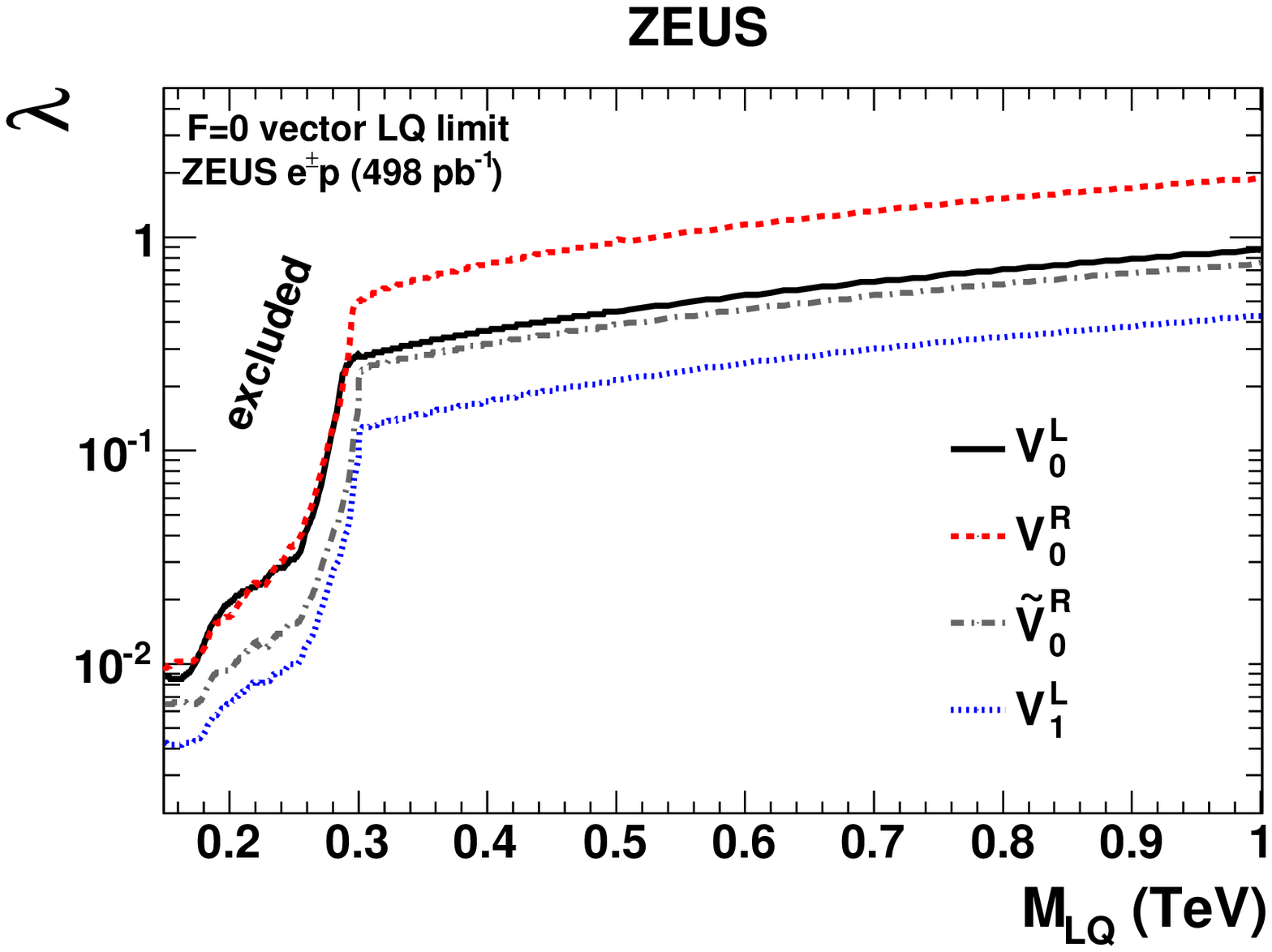}\includegraphics[width=0.98\columnwidth]{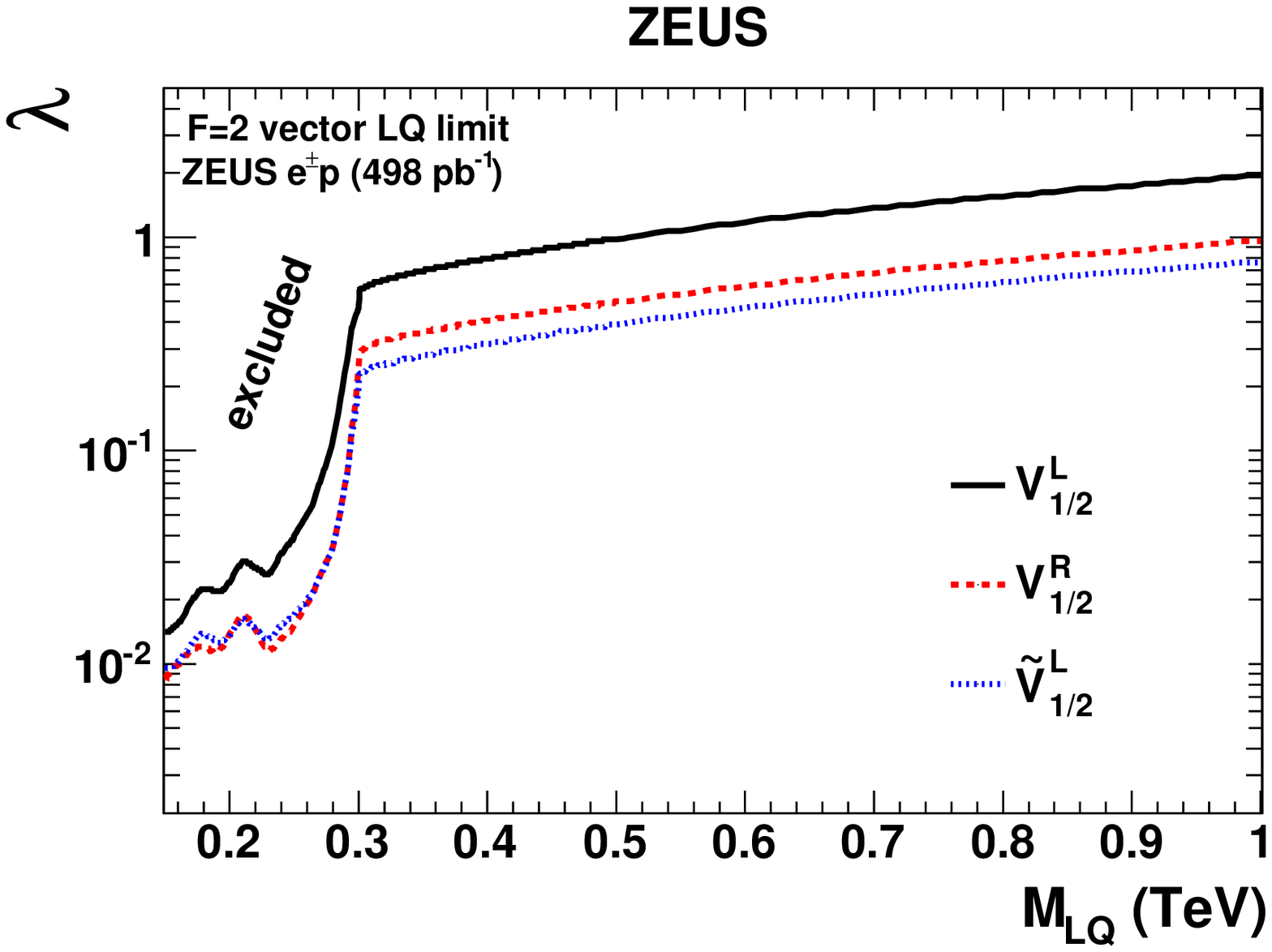}}
  \begin{picture} (0.,0.)
  \setlength{\unitlength}{1.0cm}
    \put (3,8.2){\bf\normalsize (a)}
    \put (11.8,8.2){\bf\normalsize (b)}
    \put (3,2.3){\bf\normalsize (c)}
    \put (11.8,2.3){\bf\normalsize (d)}
 \end{picture}
\caption{Exclusion limits from ZEUS for the 14 leptoquarks in the framework of the Buchm\"uller, R\"uckl and
    Wyler model on the coupling $\lambda$ as a function of leptoquark mass for the scalar leptoquarks
    with $F = 0$ (a) and $F = 2$ (b) and the vector leptoquarks with $F = 0$ (c) and $F = 2$ (d).
    Domains above the curves are excluded at $95\%$~CL.}
\label{fig:lq-zeuslimits}
\end{figure*}

\begin{figure*}
\centerline{\includegraphics[width=0.98\columnwidth]{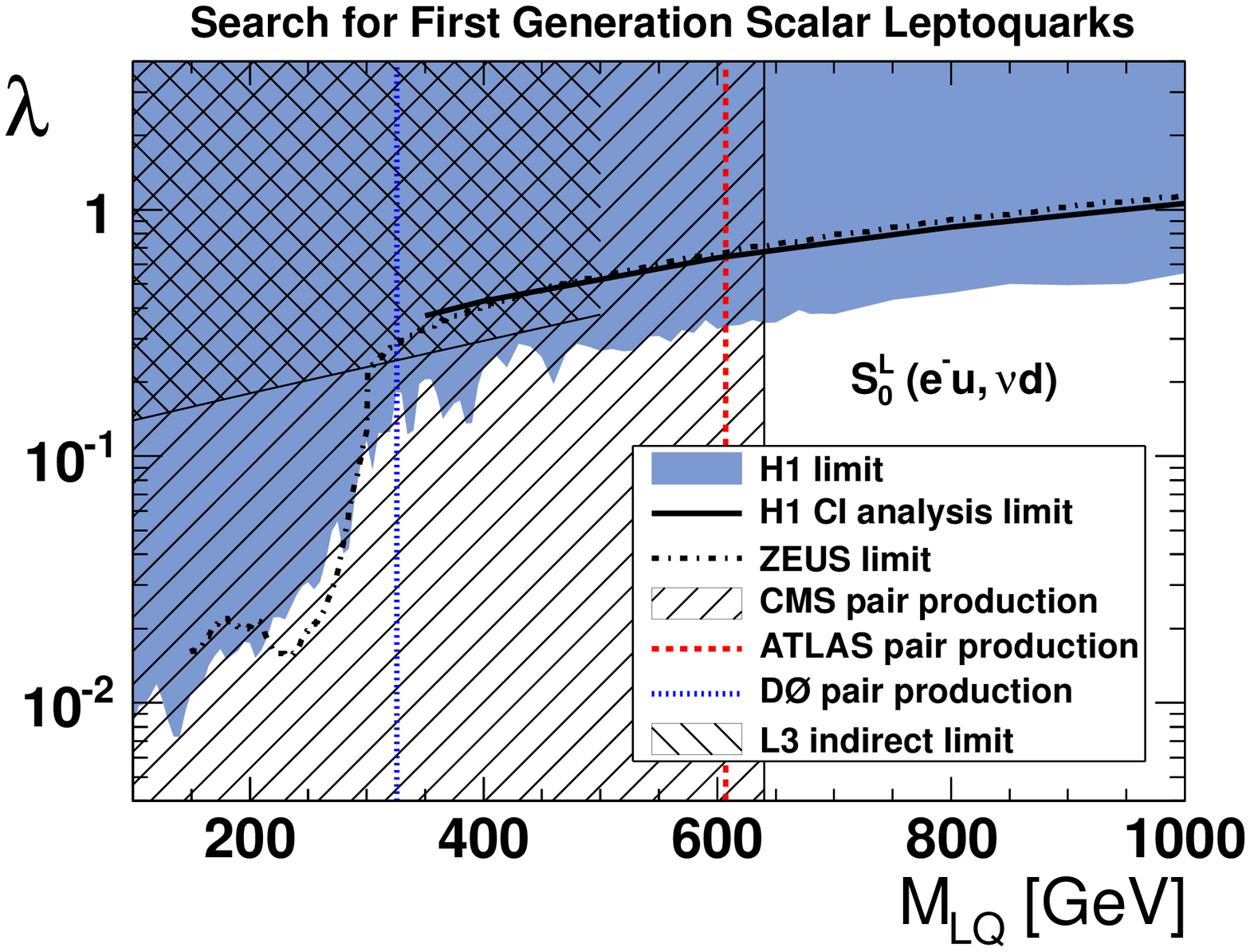}\includegraphics[width=0.98\columnwidth]{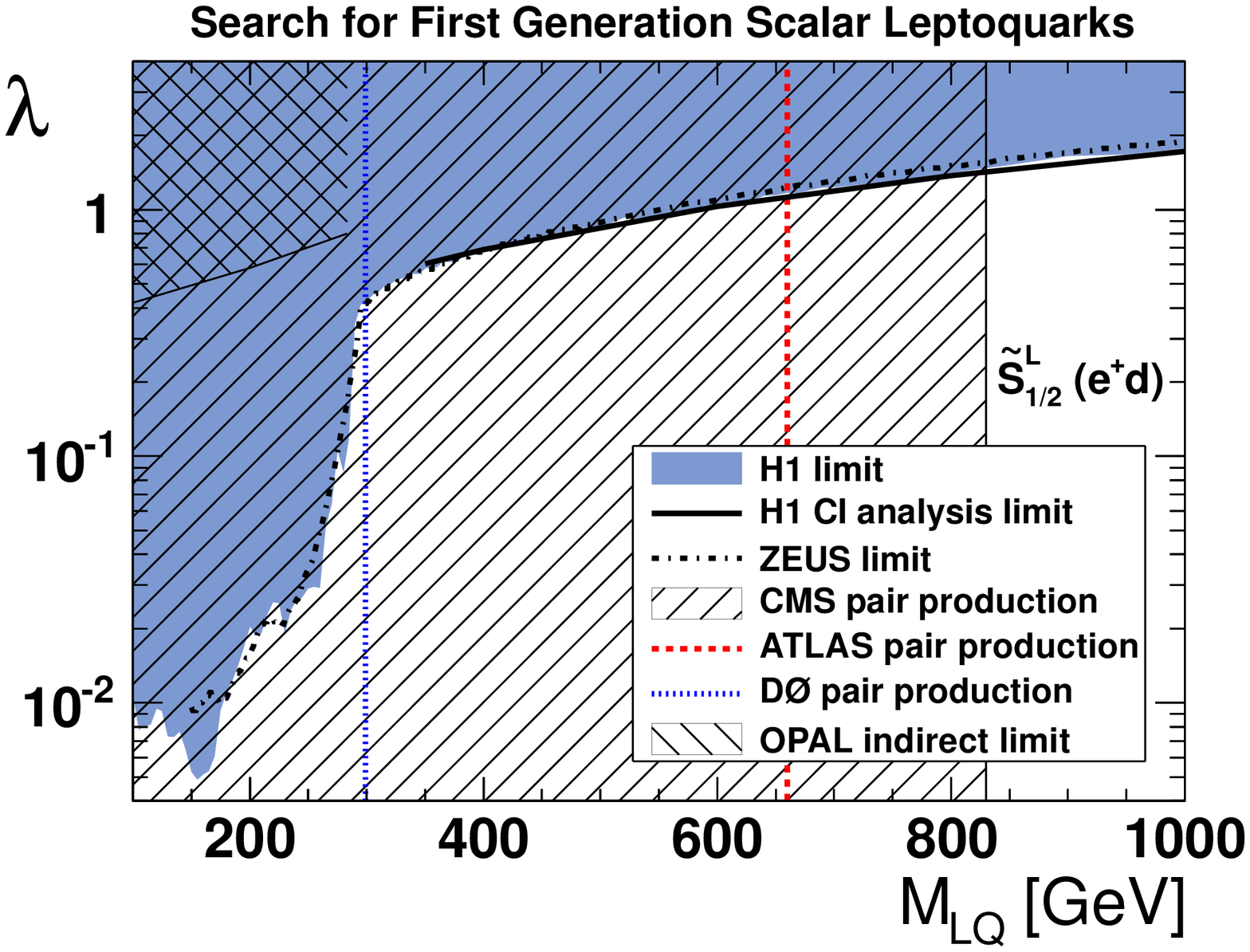}}
\caption{Exclusion limits from H1 and ZEUS in the framework of the Buchm\"uller, R\"uckl and Wyler model
  on the coupling as a function of leptoquark mass for the $S^{L}_{0}$ (left) and  $\tilde{S}^{L}_{1/2}$ (right)
  leptoquarks, which have branching fraction to charged leptons of $\beta_{\ell} = 0.5$
  and $\beta_{\ell} = 1.0$, respectively. Domains above the curves and to the left of the vertical lines
  are excluded at $95\%$~CL. Published limits from the Tevatron (D{\O}), LEP (L3 and OPAL) and the LHC experiments
  (CMS and ATLAS, $\sqrt{s} = 7$~TeV data) are also shown for comparison, as well as constraints on LQs with masses
  above 350 GeV from the H1 contact interaction (CI) analysis.}
\label{fig:lq-limitcomparison}
\end{figure*}

%%%

Since no evidence for LQ production is observed in any of the NC or CC data samples,
the data are used to set constraints on leptoquarks coupling to first generation fermions.
To estimate the LQ signal, an event reweighting technique is applied to the SM MC events.
The reweighting procedure uses the CTEQ5~\cite{Lai:1999wy} PDF for the LQ signal. 
For the limit analysis, the data are studied in bins by H1 (ZEUS) in the $M_{\rm LQ}-y$ ($M_{ljs}-\cos\theta^{*}$)
plane, where the NC and CC data samples with different lepton beam charge and polarisation
are kept as distinct data sets.
Limits are determined from a statistical analysis which uses the method of fractional event
counting~\cite{Bock:2004xz}, taking into account the polarisation and systematic uncertainties,
the most relevant of which are the electromagnetic and hadronic energy scales and the
PDF uncertainty.
Full details on the limit calculation employed can be found in the individual
publications~\cite{Aaron:2011qaa,Abramowicz:2012tg}.

%%%

Upper limits from ZEUS on the coupling $\lambda$ obtained at $95$\%~CL are
shown as a function of leptoquark mass in figure~\ref{fig:lq-zeuslimits},
displayed as groups of scalar and vector LQs for both $F = 2$ and $F = 0$.

Limits corresponding to LQs coupling to a $u$ quark are better compared to those for
LQs coupling to the $d$ quark only, as expected from the larger $u$ quark density
in the proton.
Corresponding to the steeply falling parton density function for high values of $x$,
the LQ production cross section decreases rapidly and exclusion limits from HERA
are less stringent towards higher LQ masses.
For LQ masses near the kinematic limit of $319$~GeV, the limit corresponding
to a resonantly  produced LQ turns smoothly into a limit on the virtual effects of
both an off-shell $s$-channel LQ process and a $u$-channel LQ exchange.

%%%

The presented limits extend beyond those from previous leptoquark and
contact interaction analyses by H1~\cite{Adloff:2003jm,Aktas:2005pr} and
ZEUS~\cite{Chekanov:2003pw,Chekanov:2003af} based on smaller HERA data sets.
For a coupling of electromagnetic strength $\lambda =\sqrt{4\pi\alpha_{\rm em}} = 0.3$,
LQs produced in $ep$ collisions decaying to an electron-quark or a neutrino-quark pair are
excluded at $95\%$~CL up to leptoquark masses between $290$~GeV ($\tilde{S}_{0}^{R}$)
and $699$~GeV ($V_{1}^{L}$), depending on the leptoquark type.
Similar limits are derived by H1, where LQs decaying to an electron-quark
or a neutrino-quark pair are excluded at $95\%$~CL up to leptoquark masses
between $277$~GeV ($V_{0}^{R}$) and $800$~GeV ($V_{0}^{L}$),
depending on the leptoquark type.

%%%

Within the framework of the BRW model, the $\tilde{S}_{1/2}^{L}$ LQ
decays exclusively to an electron-quark pair, resulting in a branching fraction
for decays into charged leptons of $\beta_{\ell} = 1.0$, whereas the $S_{0}^{L}$
LQ also decays to neutrino-quark, resulting in $\beta_{\ell} = 0.5$.
The H1 and ZEUS limits on $\tilde{S}_{1/2}^{L}$ and $S_{0}^{L}$ from the analysis
of the full HERA data are compared to those from other experiments in
figure~\ref{fig:lq-limitcomparison}
The H1 limits at high leptoquark mass values are also compared with those
obtained in the contact interaction analysis described in section~\ref{sec:ci},
which is based on single differential NC cross sections
${\rm d}\sigma/{\rm d}Q^2$ measured using the same data.
The additional impact of the CC data can be seen in the case of the $S_{0}^{L}$ LQ,
where a stronger limit is achieved in this analysis, whereas for the $\tilde{S}_{1/2}^{L}$
LQ the two analyses result in a similar limit.

%%%

Indirect limits from searches for new physics in $e^+e^-$ collisions at LEP by the
OPAL~\cite{Abbiendi:1998ea} and L3~\cite{Acciarri:2000uh} experiments
are indicated, as well as the limits from the D{\O}~\cite{Abazov:2009ab,Abazov:2011qj}
experiment at the Tevatron and from searches by the ATLAS~\cite{Aad:2011ch} and
CMS~\cite{Chatrchyan:2012vza} experiments based on their $\sqrt{s} = 7$~TeV data.
The limits from hadron colliders are primarily based on searches for LQ pair-production
and are independent of the coupling $\lambda$.
For example, for a leptoquark mass of $640$~GeV, the H1 LQ analysis rules out the
$S_{0}^{L}$ LQ for coupling strengths larger than about $0.35$.

%%%

At the time of writing in winter 2015-16, results
from the $\sqrt{s} = 8$~TeV data LHC are appearing, extending these
limits into the kinematic regime beyond
$1$~TeV~\cite{Aad:2015caa} for $\beta = 1.0$ leptoquarks.
Limits from the LHC experiments so far refer only to scalar leptoquarks.

\subsection{Search for second and third generation leptoquarks}
\label{sec:lfv}

The introduction of lepton flavour violation~\cite{Barbieri:2011ci} to leptoquark models would mean
that the processes $ep \rightarrow \mu X$ or $ep \rightarrow \tau X$, mediated
by the exchange of a second or third generation leptoquark, would be observable at
HERA with final states containing a muon or the decay products of a tau lepton
back-to-back in the transverse plane with a hadronic system $X$.
Searches for such signatures have been previously performed at HERA by both
H1 and ZEUS using HERA~I data~\cite{Aktas:2007ji,Adloff:1999tp,Chekanov:2005au}.
The analysis performed by H1 using their complete $\sqrt{s} = 319$~GeV data set
(see table~\ref{tab:datasets}) is described in the following.
The event selections for both the searches are described below; full details can be
found in the publication~\cite{Aaron:2011zz}.

%%%

Leptoquarks with couplings to first and second generation leptons may
decay to a muon and a quark, therefore event topologies with an isolated, high transverse
momentum muon back-to-back to a hadronic system in the transverse plane are selected.
An initial sample of events with muons and jets is selected by requiring at least one
$P_{T}^{\mu} > 8$~GeV muon in the polar angular range $10^{\circ} < \theta_{\mu} < 120^{\circ}$
and at least one jet.
Events with isolated muons are then selected; the angular distance, 
$D=\sqrt{(\Delta\eta)^2+(\Delta\phi)^2}$, of the muon to the nearest track and
to the nearest jet are required to be greater than $0.5$ and $1.0$, respectively.
Limitations on the calorimetric energy in a cone around the muon are also
introduced~\cite{Aaron:2011zz}.
To reduce the muon-pair SM background exactly one isolated muon is required,
as expected in LFV LQ signal events.

%%%

NC DIS background is suppressed by applying a cut on the calorimetric
momentum imbalance, $P_{T}^{\rm calo}>25$~GeV,
and by rejecting events with identified isolated electrons.
The back-to-back event topology in the azimuthal plane is also exploited to
remove the SM background: the difference between the azimuthal angle of the
hadronic system and the muon, the acoplanarity $\Delta\phi_{\mu-X}$,
is required to be greater than $170^{\circ}$.
Further SM background reduction is achieved by exploiting the calorimetric
energy  imbalance $V_{\rm ap}/V_{\rm p}<0.3$ and the overall longitudinal
balance of the event $\delta > 40$~GeV.

\begin{figure*}
\centerline{\includegraphics[width=0.495\textwidth]{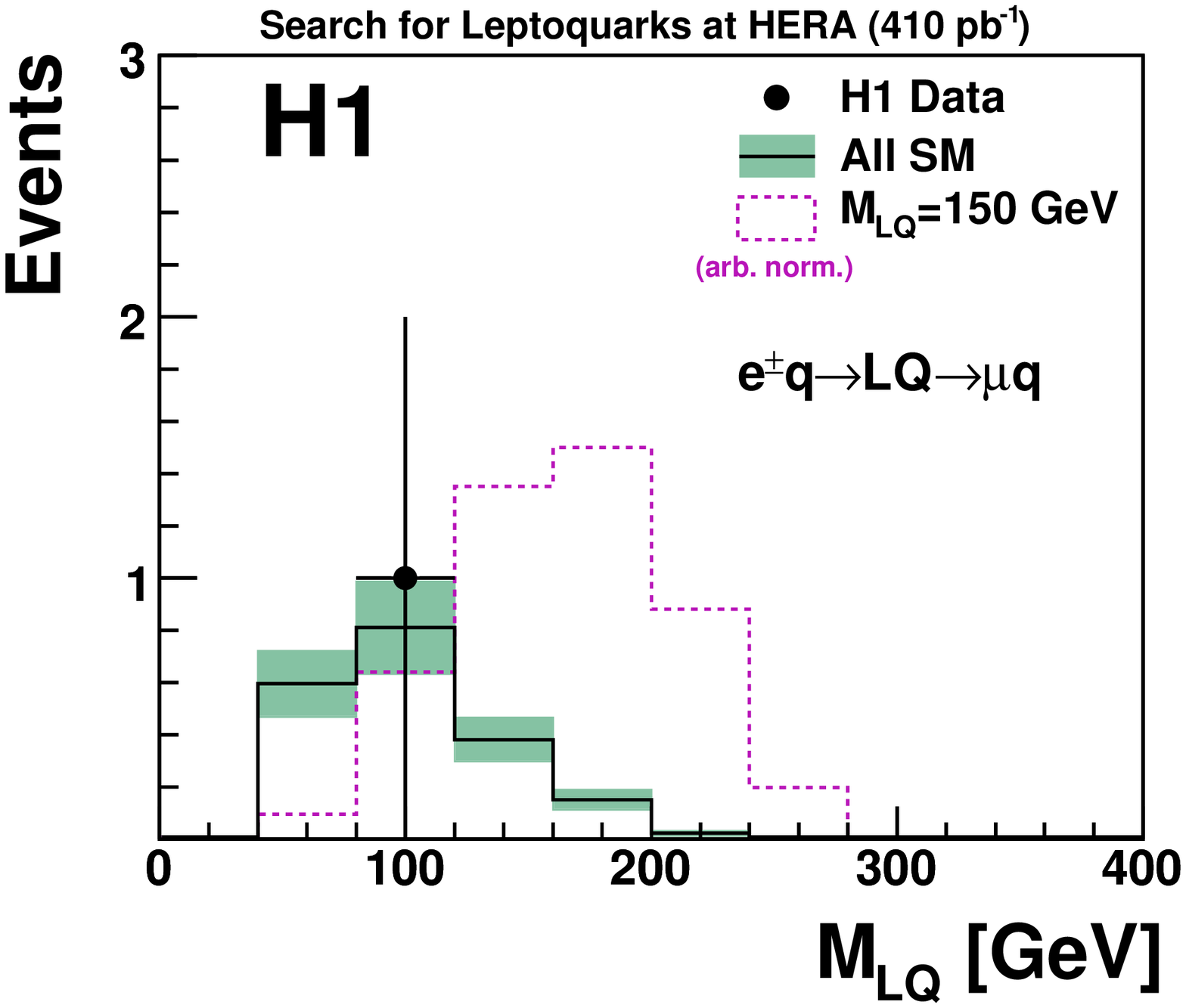}\includegraphics[width=0.495\textwidth]{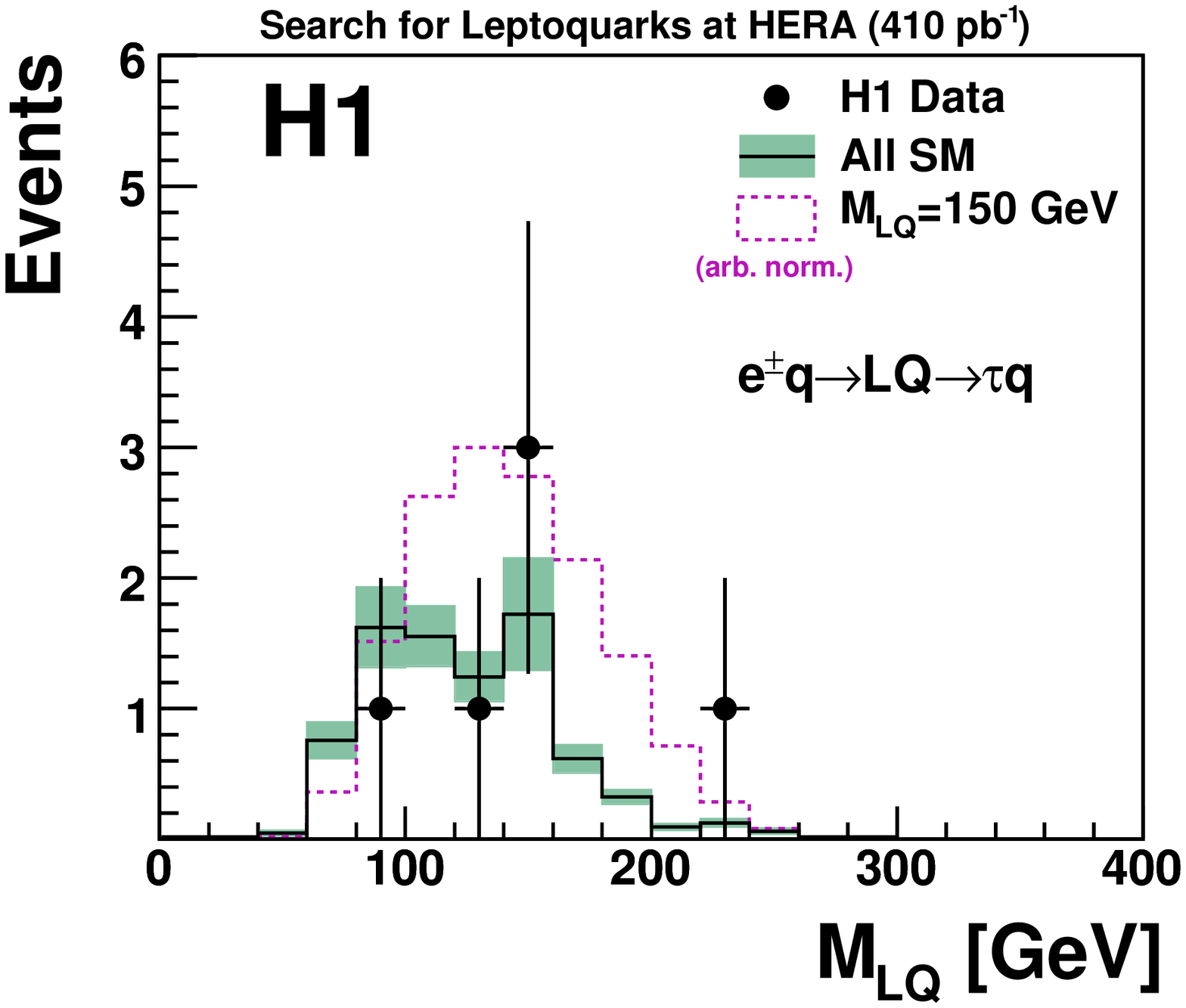}}   
  \caption{The reconstructed leptoquark mass in the H1 search for
    $ep \rightarrow \mu X$~(left) and $ep \rightarrow \tau X$~(right)
    events. The data are the points and the total uncertainty on the
    SM expectation (open histogram) is given by the shaded band. The
    dashed histogram indicates the LQ signal with arbitrary
    normalisation for a leptoquark mass of $150$~GeV.}
  \label{fig:lfv-massplots}
\end{figure*} 
\begin{figure*}
\centerline{\includegraphics[width=0.495\textwidth]{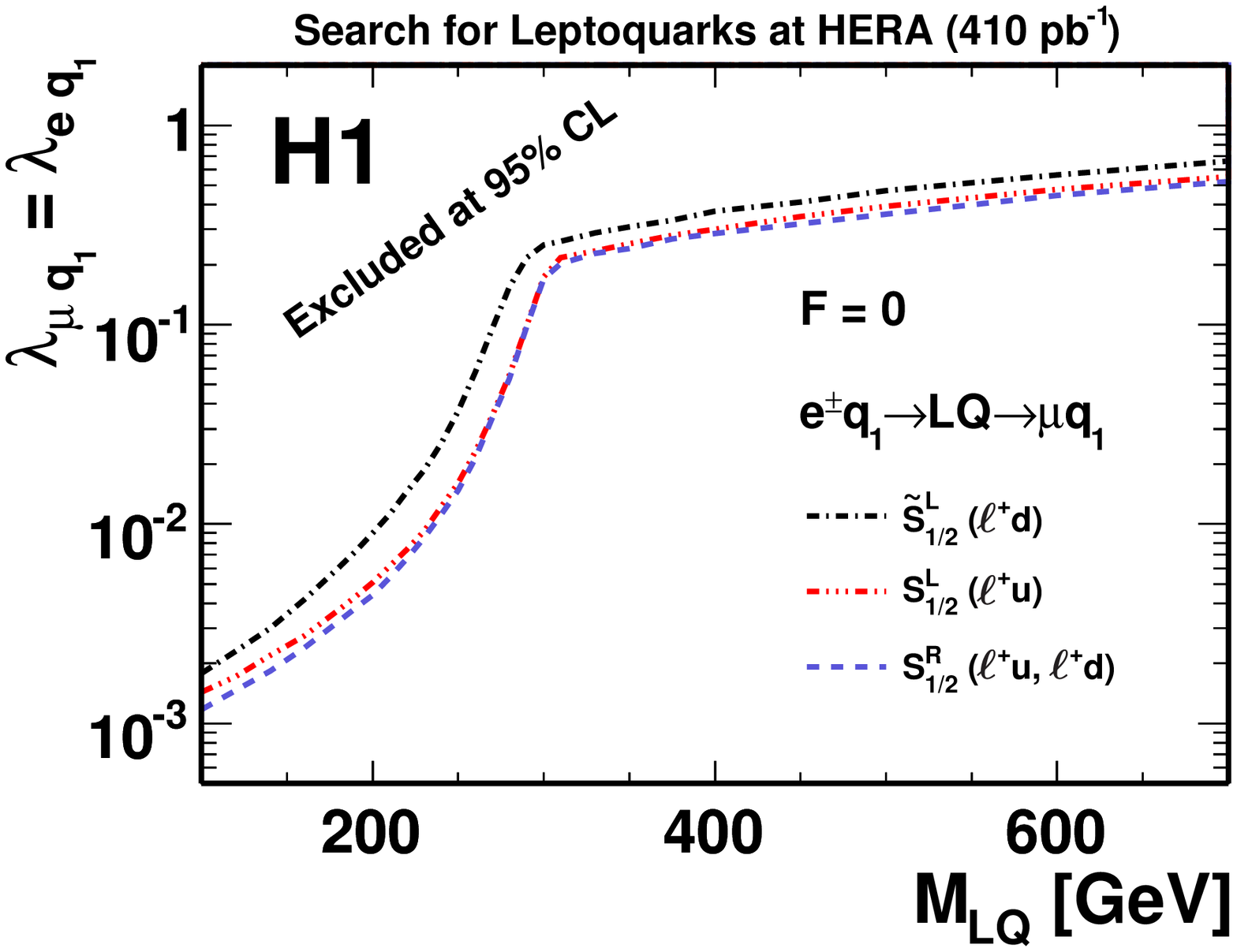}\includegraphics[width=0.495\textwidth]{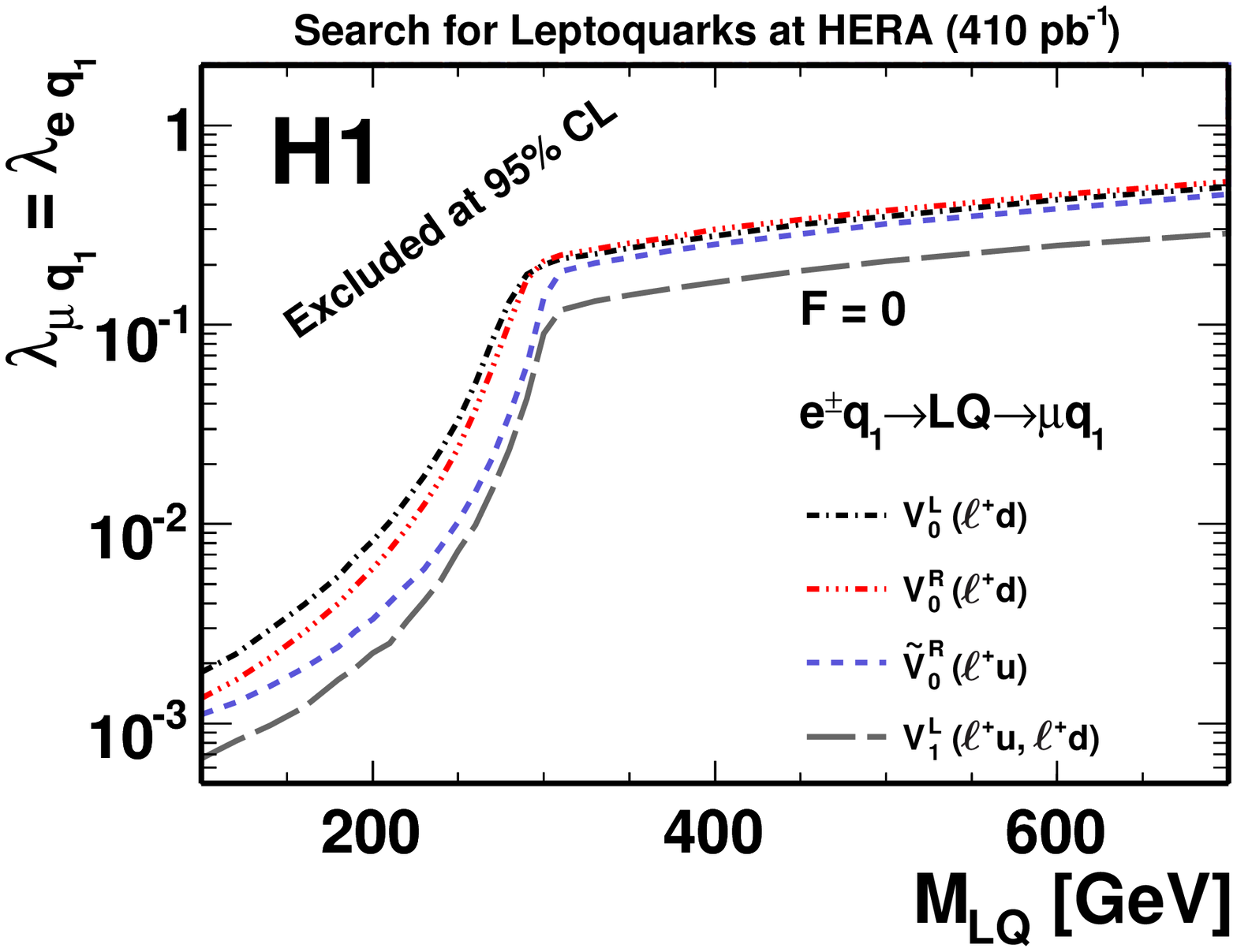}}
\centerline{\includegraphics[width=0.495\textwidth]{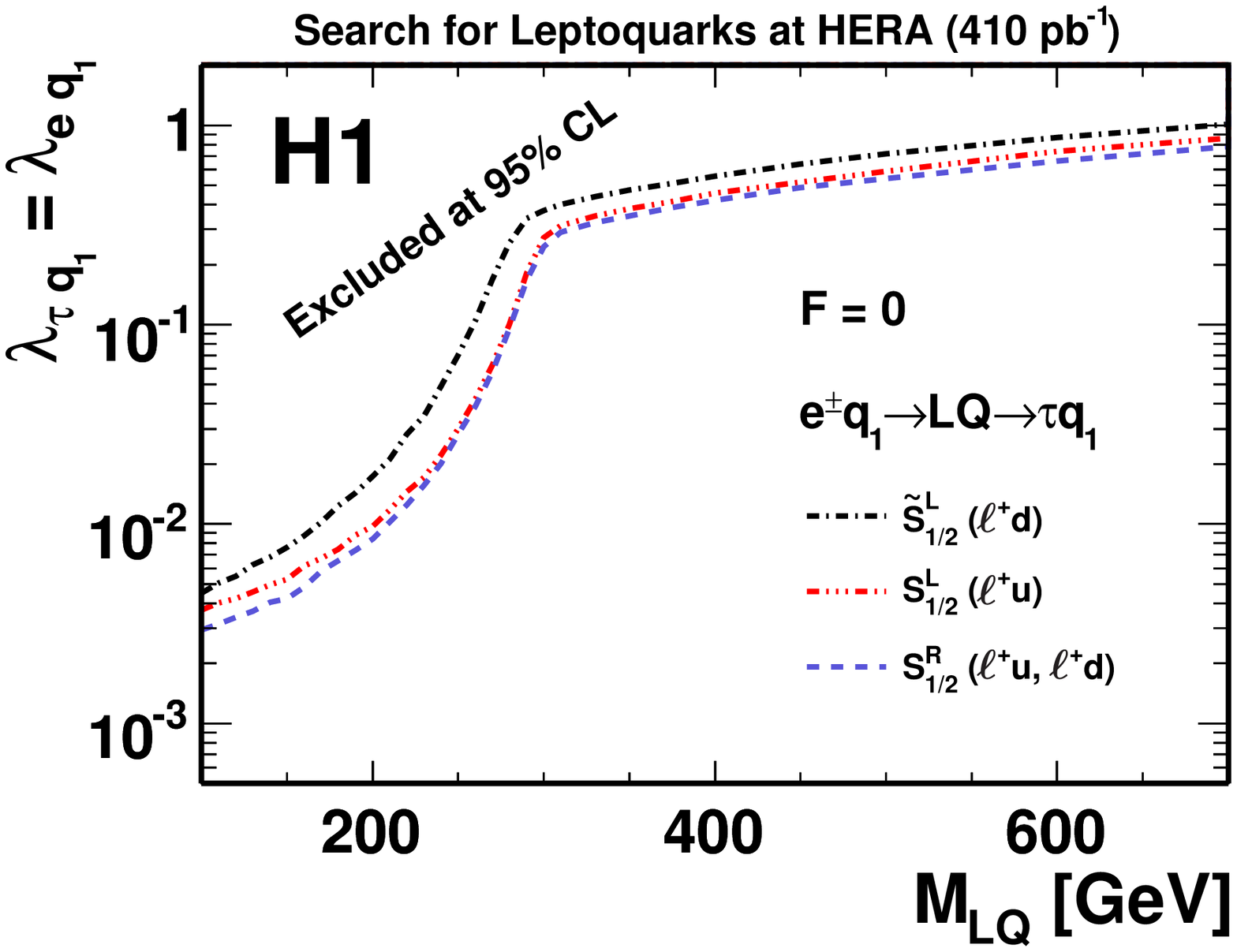}\includegraphics[width=0.495\textwidth]{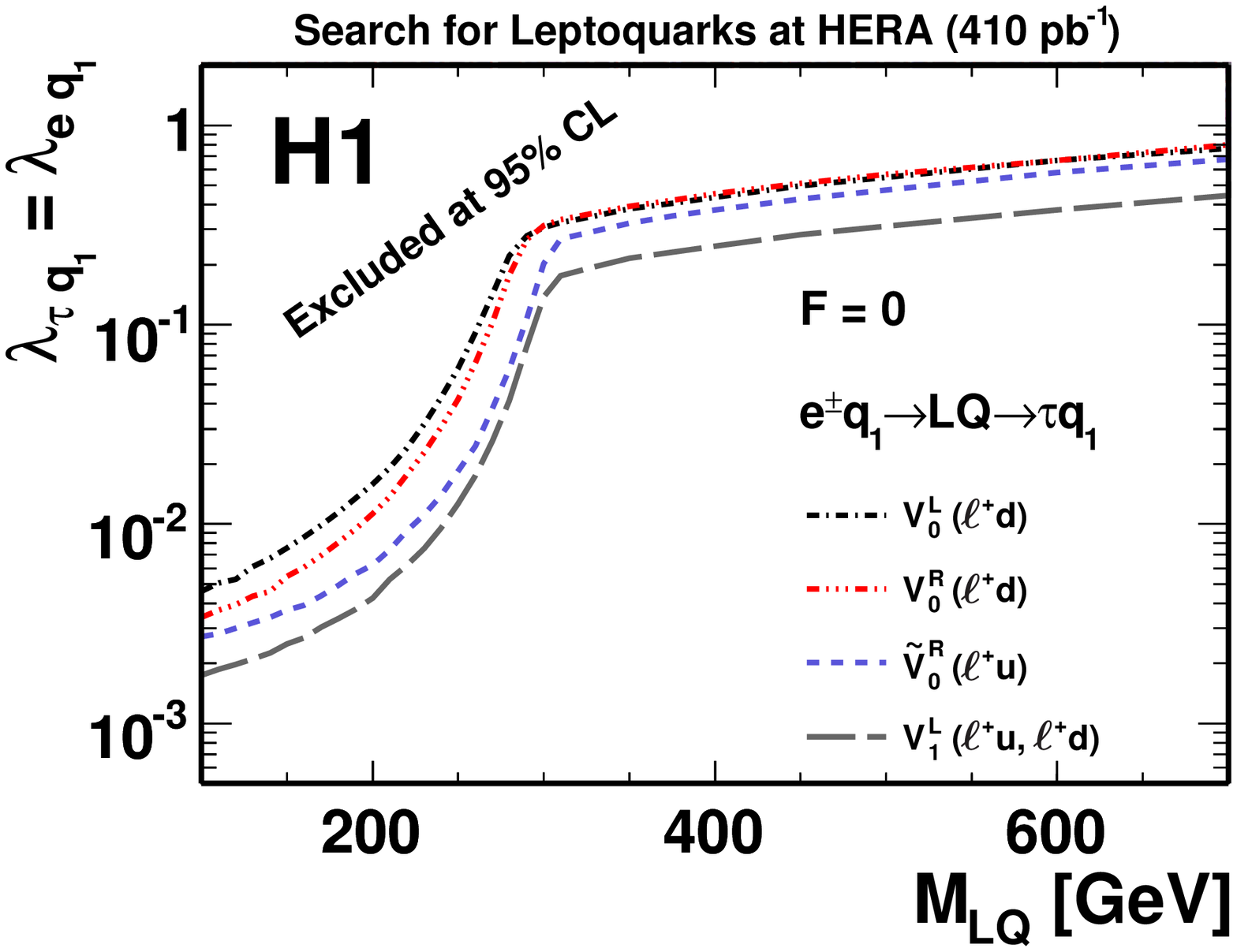}}
  \begin{picture} (0.,0.)
    \setlength{\unitlength}{1.0cm}
    \put (3.4,8.8){\bf\normalsize (a)}
    \put (13,8.8){\bf\normalsize (b)}
    \put (3.4,2.4){\bf\normalsize (c)}
    \put (13,2.4){\bf\normalsize (d)}
  \end{picture}
  \caption{Exclusion limits on the coupling constants $\lambda_{\ell q_{1}} = \lambda_{eq_{1}}$
    as a function of leptoquark mass $M_{\rm LQ}$ for $F = 0$ leptoquarks:
    limits on second generation ($\ell = \mu$) scalar (a) vector (b) LQs;
    limits on third generation ($\ell = \tau$) scalar (c) vector (d) LQs.
    Regions above the lines are excluded at $95\%$~CL. The notation $q_1$ indicates that only
    processes involving first generation quarks are considered. The parentheses after the LQ name
    indicate the fermion pairs coupling to the LQ, where pairs involving anti-quarks are not shown.}
  \label{fig:lfv-limits}
\end{figure*}

%%%

After all selection cuts, in the analysis of $\mu X$ final states one event is observed in the
data which compares well to the SM prediction of $2.0 \pm 0.4$.
The largest contribution comes from muon-pair events, which are modelled by
GRAPE.

%%%

Leptoquarks with couplings to first and third generation leptons may decay to a tau and a quark
and in the search for third generation leptoquarks, tau leptons are identified using the muonic
and one-prong hadronic decays of the tau\footnote{The electronic decays of the tau
$\tau\rightarrow e \nu_{e} \nu_{\tau}$ result in final states very difficult to distinguish
from SM NC DIS, where the missing transverse momentum is aligned with
the electron. As such, they are not considered in the analysis.}.
Hadronic tau decays, both one-prong and three-prong, are described in
more detail in section~\ref{sec:mlep-ditau}.
Muonic tau decays, $\tau\rightarrow \mu\nu_\mu\nu_\tau$, result in final states similar to the
high $P_{T}$ muon signatures described above and the same selection is therefore applied in that channel.

%%%

The one-prong hadronic decay of the tau leads to a high $P_T$, narrow ``pencil-like''
jet, so that the typical LFV signal event topology is a di-jet event.
An initial di-jet event sample for the analysis of this decay channel is formed by selecting
events with at least two jets in the polar angle range
$5^{\circ} < \theta_{\rm jet} < 175^{\circ}$ and with
$P_{T}^{\rm jet 1} > 20$~GeV and $P_{T}^{\rm jet 2} > 15$~GeV.
The undetected neutrinos from tau lepton decays result in an overall $P_{T}$ imbalance
and therefore a minimum missing transverse momentum $P_{T}^{\rm miss} > 12$~GeV is
required.

%%%
%
A tau jet is characterised by a narrow energy deposit in the calorimeter and
a low track multiplicity within the identification cone of the jet.
Tau jets are identified in the di-jet sample, where the candidates are
required to be in the polar angle range $20^{\circ} < \theta_{\rm jet} < 120^{\circ}$
and to have a maximum jet radius $R_{\rm jet}$ of $0.12$~\cite{Aktas:2006fc}, where
\begin{equation}
  R_{\rm jet}=\frac{1}{E_{\rm jet}}\sum_{h}E_{h}\sqrt{\Delta\eta({\rm jet},h)^{2}+\Delta\phi({\rm jet},h)^{2}},
\label{eq:taujet}
\end{equation}
and $E_{\rm jet}$ is the total jet energy.
The sum runs over all hadronic final state particles in jets with energy $E_{h}$.
At least one track with $P_{T}$ larger than $2$~GeV not associated with an identified
electron or muon is required within the jet radius of the tau jet.

A single tau jet is required in the final sample, with selection requirements including
isolation from from tracks and other jets by an angular distance $D > 1.0$, a track multiplicity
of one in a cone of radius $R = 1.0$ around the jet axis and a maximum of $90\%$ of the
jet energy in the EMC to reject purely electromagnetic jets.

%%%

Further cuts are then applied to reduce the remaining SM background: the hadronic transverse
momentum $P_{T}^{h}$ is required to be larger than $30$~GeV and the acoplanarity between the
tau jet and hadronic system in the transverse plane $\Delta\phi_{\tau - X}$ is required to be
greater than $160^{\circ}$.
Analogous to the muon channel, a cut of $\delta > 40$~GeV is also applied
to exploit the longitudinal balance of the event.

%%%

After all selection cuts, in the analysis of $\tau X$ final states where the tau lepton decays
hadronically, $6$ events are observed in the data, in good agreement with the SM prediction of 
$8.2 \pm 1.1$.
The main SM contribution is from remaining NC DIS events, which
are modelled by the RAPGAP~\cite{Jung:1993gf} event generator.

%%%

The reconstructed leptoquark-candidate mass in the search for $ep \rightarrow \mu X$ and
$ep \rightarrow \tau X$ events is shown in figure~\ref{fig:lfv-massplots}, compared to
the SM prediction and an example LQ signal with arbitrary normalisation. 
The LQ kinematics are reconstructed using the double angle method~(equation~\ref{eq:dameth}).
The direction of the detected lepton and the hadronic final state are used to reconstruct
the Bjorken scaling variable $x$ and subsequently the LQ mass following equation~\ref{eq:mlq-h1}.

%%%

As the observed number of events is in agreement with the SM prediction and
therefore no evidence for LFV is found, the results of the search are
interpreted in terms of exclusion limits on the mass and the coupling of LQs
mediating LFV.
The LQ production mechanism at HERA involves a non-zero coupling to the
first generation fermions \mbox{$\lambda_{eq_{i}} > 0$}.
For the LFV leptoquark decay, it is assumed that only one of the couplings
$\lambda_{\mu q_{j}}$ and $\lambda_{\tau q_{j}}$ is non-zero and that
$\lambda_{eq_{i}} = \lambda_{\mu q_{j}}(\lambda_{\tau q_{j}})$.
A modified frequentist method with a likelihood ratio as the test statistic
is used to combine the individual data sets and the $ep \rightarrow \tau X$
search channels~\cite{Barate:2003sz}.
In order to avoid the need to generate many signal MC samples at each leptoquark
mass, coupling and branching ratio, a weighting technique is used to provide
predictions across the full range of LQ production parameters~\cite{Aktas:2007ji}.
As in the search for first generation leptoquarks, the lepton beam polarisation
enters the limit calculation for the HERA~II data.

Figure~\ref{fig:lfv-limits} shows the $95\%$~CL upper limits on the
couplings to the first generation quarks, $\lambda_{\mu q_{1}}$ and
$\lambda_{\tau q_{1}}$, for $F = 0$ LQs as a function of the mass of the
LQ leading to LFV in $ep$ collisions.
These limits extend beyond those from H1 and ZEUS based on
the HERA~I data alone~\cite{Aktas:2007ji,Adloff:1999tp,Chekanov:2005au}.
Similar limits are found for $F = 2$ LQs and these can be found in the
H1 publication along with limits involving heavier quark
flavours~\cite{Aaron:2011zz}.

For $\lambda = \sqrt{4 \pi \alpha_{\rm em}} = 0.3$, LFV leptoquarks
produced in $ep$ collisions decaying to a muon-quark or a tau-quark
pair are excluded at $95\%$~CL up to leptoquark masses of
$712$~GeV and $479$~GeV, respectively.
In both cases, the LQ with the highest sensitivity is $V_{1}^{L}$.
As described above, the most appropriate value for the branching ratio 
when comparing to results from hadron colliders is $\beta=0.5$ or less
(see equations~\ref{eq:lfv1}-\ref{eq:lfv3}).
In a search for pair production of second generation scalar
leptoquarks, the ATLAS collaboration rules out leptoquark masses at
$95\%$~CL less than about $850$~GeV for
$\beta=0.5$~\cite{Aad:2015caa}.
The CMS collaboration has recently performed searches for third
generation scalar LQ pairs decaying to
$t+\tau$~\cite{Khachatryan:2015bsa} and 
$b+\tau$~\cite{Khachatryan:2014ura} final states, where the best limit 
for $\beta=0.5$ is $M_{\rm LQ} > 560$~GeV at $95\%$~CL.

\section{Multi-lepton production at high transverse momentum}
\label{sec:mlep}

\begin{figure*}
\centerline{\includegraphics[clip,width=0.80\textwidth]{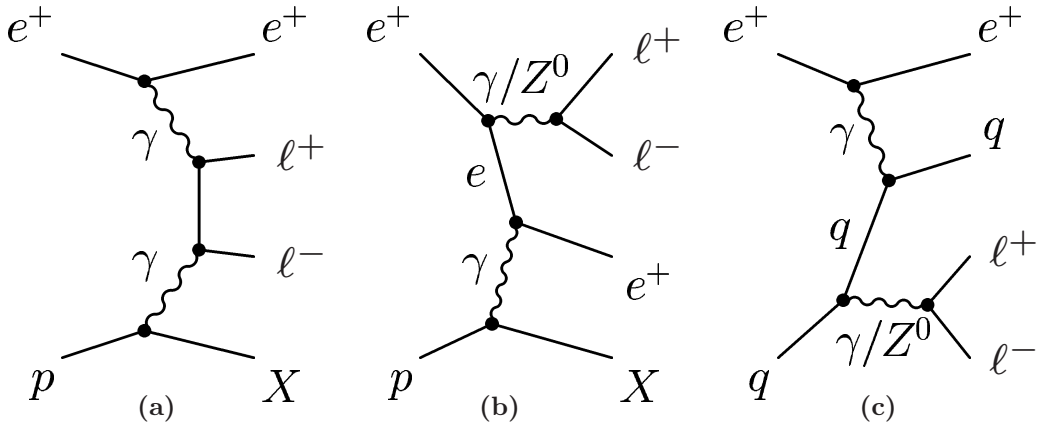}}
\begin{picture} (0.,0.)
  \setlength{\unitlength}{1.0cm}
  \put(4,0.7){\bf\normalsize (a)}
  \put(8.5,0.7){\bf\normalsize (b)}
  \put(13.5,0.7){\bf\normalsize (c)}
\end{picture}
\caption{Lepton pair production in $e^{+}p$ collisions at HERA. Examples of Feynman diagrams
   are shown for the photon-photon interaction (a) and $\gamma$/$Z^{0}$
   conversion (b) and (c). The hadronic final state can
   be a proton (elastic production) or a higher mass system
   (quasi-elastic and inelastic production).}
\label{fig:multilepfeyn}
\end{figure*}

Multi-lepton final states may be produced in electron-proton
collisions at HERA, proceeding mainly via photon-photon 
$\gamma \gamma \rightarrow \ell^{+} \ell^{-}$
interactions~\cite{Vermaseren:1982cz}, as shown in
figure~\ref{fig:multilepfeyn}.
The dominant contribution, shown in figure~\ref{fig:multilepfeyn}(a),
is from lepton pair production via the interaction of two
photons radiated from the incident electron and proton.
Lepton pairs may also originate from internal conversion of a
photon or a $Z^{0}$ boson, radiated either from the incident
electron line (figure~\ref{fig:multilepfeyn}(b)) or from the quark line
(figure~\ref{fig:multilepfeyn}(c)).

%%%

As this is a purely QED (Quantum Electrodynamic) process,
the cross section is precisely calculable in the SM.
Multi-lepton events are simulated using the GRAPE event generator,
which includes all leading order electroweak matrix elements.
The final state leptons provide a clean event signature, which is well
described by the SM and the investigation of the high-mass,
high-$P_{T}$ regions, where the SM expectation is low, may reveal
some signal of new physics.

%%%

The following section concerns final states with electrons and muons.
Studies of tau-pair production are described in
section~\ref{sec:mlep-ditau}.

\subsection{Events with electron and muon final states}
\label{sec:multi-elmu}

Measurements of both multi-electron~\cite{Aktas:2003jg} and muon
pair \cite{Aktas:2003sz} production at high transverse momentum have
previously been performed by H1 using their HERA~I data set.
Since then, both H1 and ZEUS have published~\cite{Aaron:2008jh,Chekanov:2009cv}
their final results on the search for multi-lepton events,
using the full available statistics of the complete HERA data set.
The integrated luminosity in the H1 analysis corresponds to
$463$~pb$^{-1}$, of which $285$~pb$^{-1}$ are from $e^{+}p$ collisions
and $178$~pb$^{-1}$ from $e^{-}p$ collisions.
In the ZEUS analysis, the integrated luminosity corresponds to
$480$~pb$^{-1}$, of which $278$~pb$^{-1}$ are from $e^{+}p$ collisions
and  $202$~pb$^{-1}$ from $e^{-}p$ collisions.

%%%

The analysis strategy and event selection used in the H1 and ZEUS
analyses are similar.
First, electron or muon candidates are identified in a wide angular
region using a loose selection criteria.
Then, at least two central ($20^\circ < \theta_{\ell} < 150^\circ$)
lepton candidates are required in the event.
Electrons are identified in the polar angle range  $5^\circ < \theta_{e} <
175^\circ$ with an energy greater $E_{e} > 5$~GeV.
This energy threshold is raised to $10$~GeV in the region
$5^\circ < \theta_{e} < 20^\circ$ in the H1 analysis and in the region
$5^\circ < \theta_{e} < 150^\circ$ in the ZEUS analysis.
Muons are identified in the polar angle range $20^\circ <
\theta_{\mu} < 160^\circ$ and are required to have transverse 
momentum $P_{T}^{\mu}>2$~GeV. 

%%%

The final event selection requires at least two central lepton
candidates, one with $P_{T} > 10~{\rm GeV}$ and the other with $P_{T}>
5~{\rm GeV}$.
Additional leptons identified according to the above criteria may also
be present in the event.
The leptons are required to be isolated from each other by an angular distance
$D >0.5$ in the $\eta-\phi$ plane.
According to the number and flavours of the identified leptons, the
events are classified into mutually exclusive topologies containing up
to $4$ leptons.
%%%

The main source of SM background in each analysis depends on the
number and flavour of the identified leptons in the event sample.
NC DIS and QED Compton scattering (QEDC, $ep \to e\gamma X$)
constitute a significant background only for event topologies in which
two leptons, one of which is an electron, are found in the final
state.
The NC DIS background is modelled in the H1 (ZEUS) analysis using the
RAPGAP (DJANGOH) event generator.
The smaller QEDC background contribution is modelled using
WABGEN~\cite{Berger:1998kp} (GRAPE) in the H1 (ZEUS) analysis. 
Background contributions to the SM expectation are negligible in
events in which two muons or more than two leptons are observed.

%%%

\begin{figure*}
  \centerline{
    \includegraphics[clip,width=0.90\columnwidth]{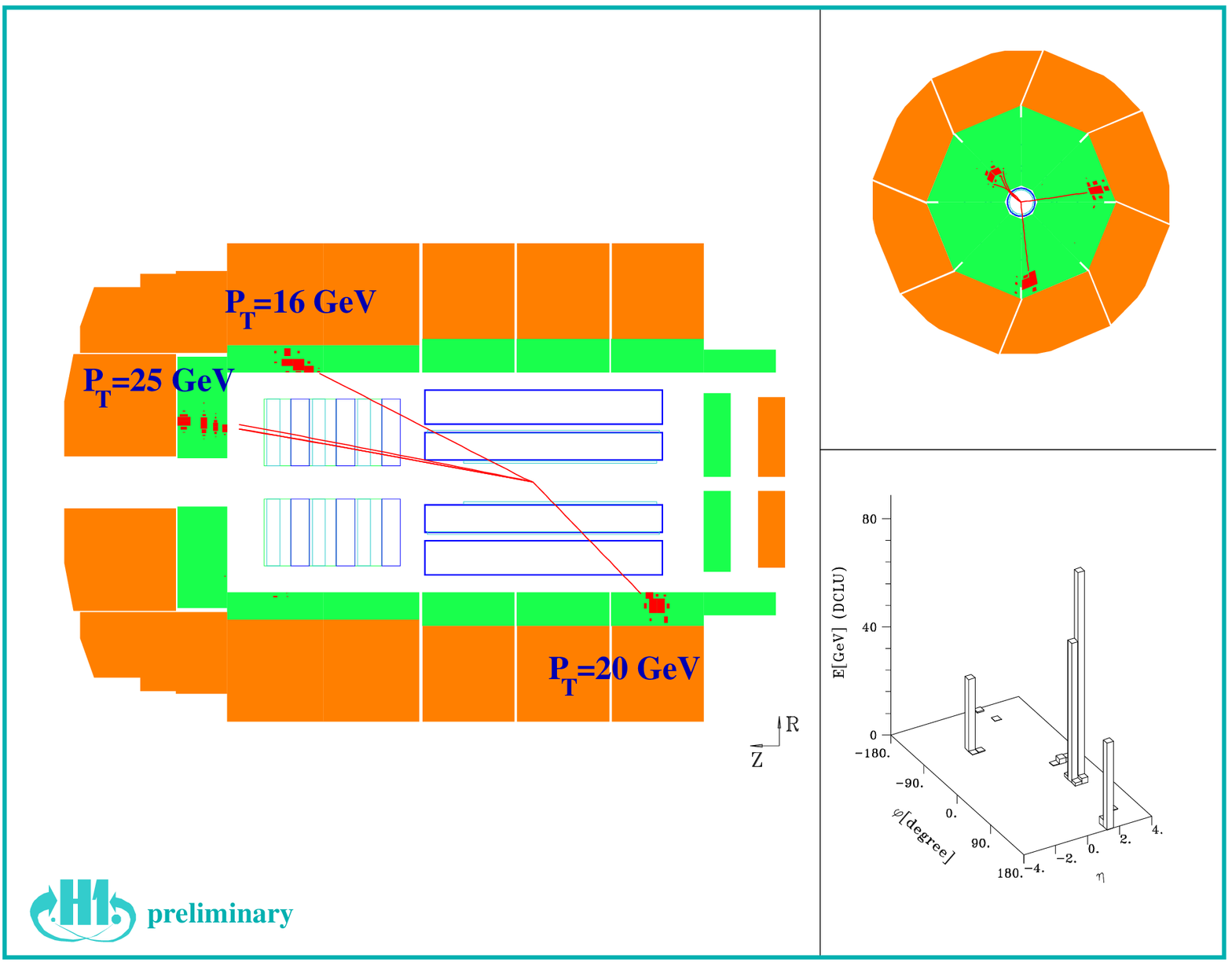}
    \hspace{1cm}
    \includegraphics[width=0.90\columnwidth]{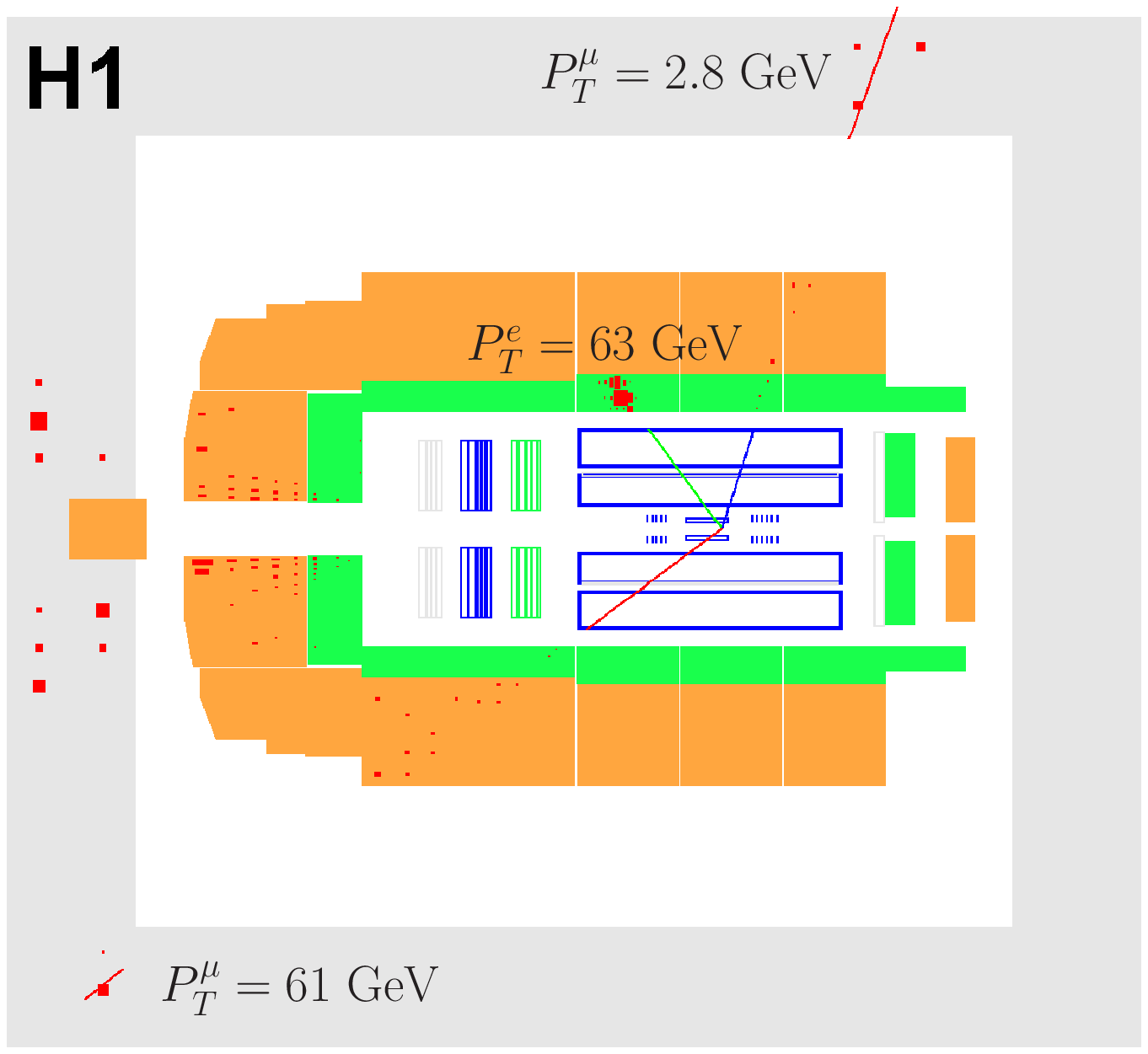}
  }
  \vspace{1cm}
  \centerline{
    \includegraphics[width=0.90\columnwidth]{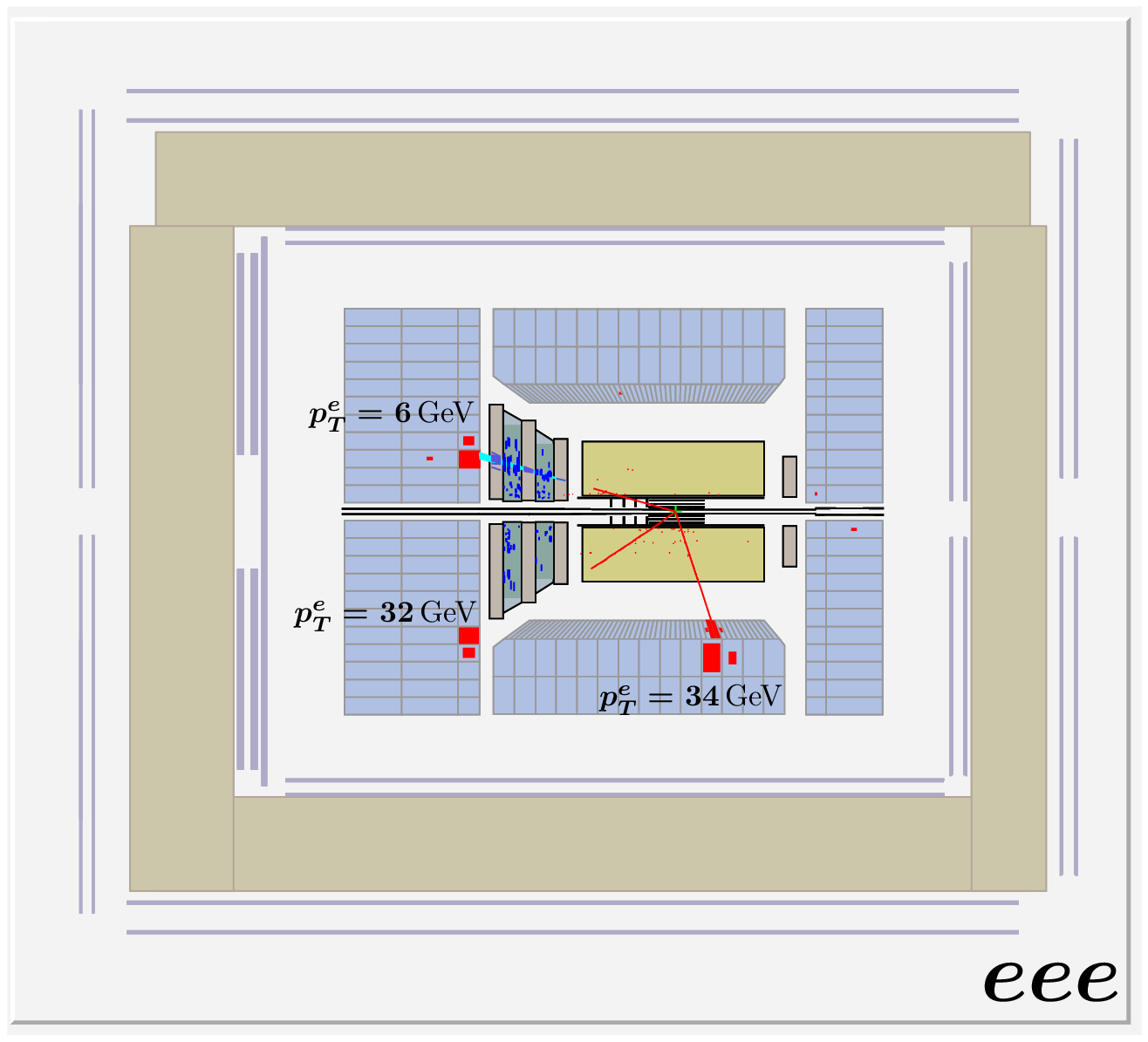}
    \hspace{1cm}
    \includegraphics[width=0.90\columnwidth]{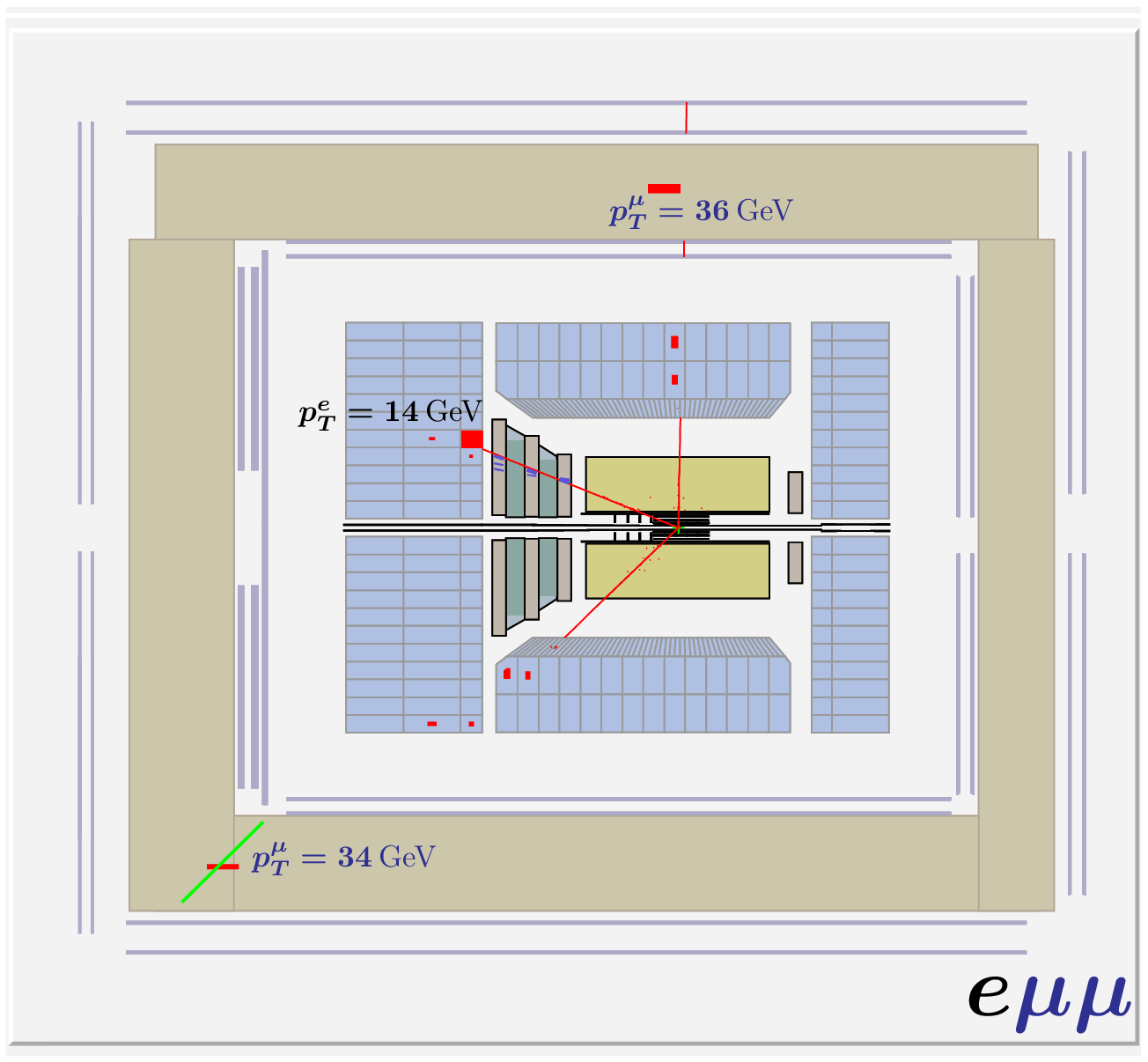}
  }
  \caption{Events observed in the H1 (top) and ZEUS (bottom)
    multi-lepton analyses. The measured transverse momentum of the
    leptons is indicated. Top left: an event observed by H1 in the
    HERA~I data containing three electrons. The invariant mass of the
    two highest $P_{T}$ electrons is measured as $M_{12}
    =118$~GeV. Top right: an $e\mu\mu$ event observed in the H1 HERA~II
    data where $M_{12} = 127$~GeV, formed by the electron and one of
    the muons. Bottom left: a three electron event observed by ZEUS
    with $M_{12} =113$~GeV. Bottom right: an  $e\mu\mu$ event observed
    by ZEUS where the muon pair form an invariant mass $M_{12} = 77$~GeV.}
  \label{fig:mlep-events}
\end{figure*}

\begin{figure*}
  \centerline{
    \includegraphics[width=0.7\columnwidth]{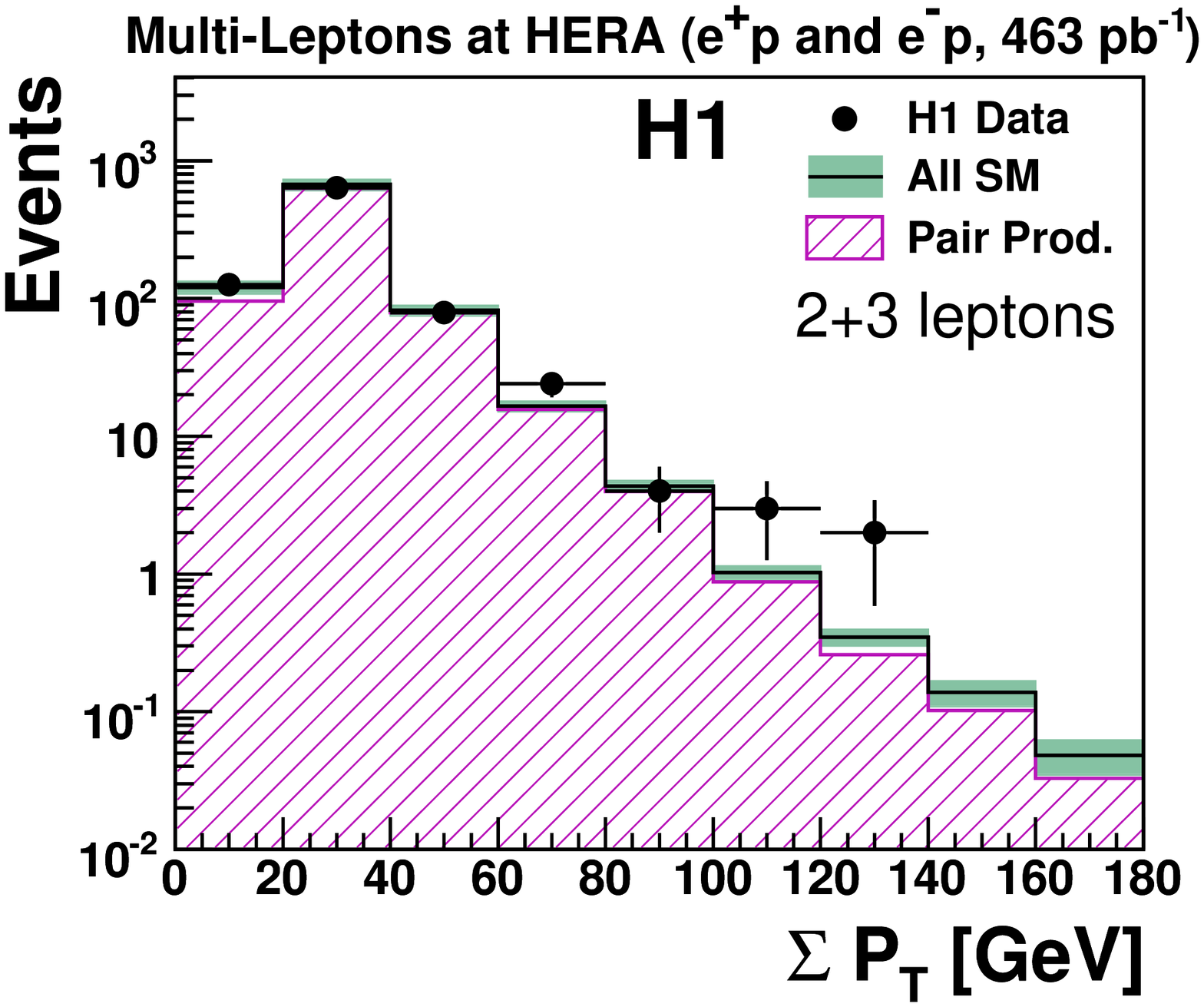}
    \includegraphics[clip,width=1.30\columnwidth]{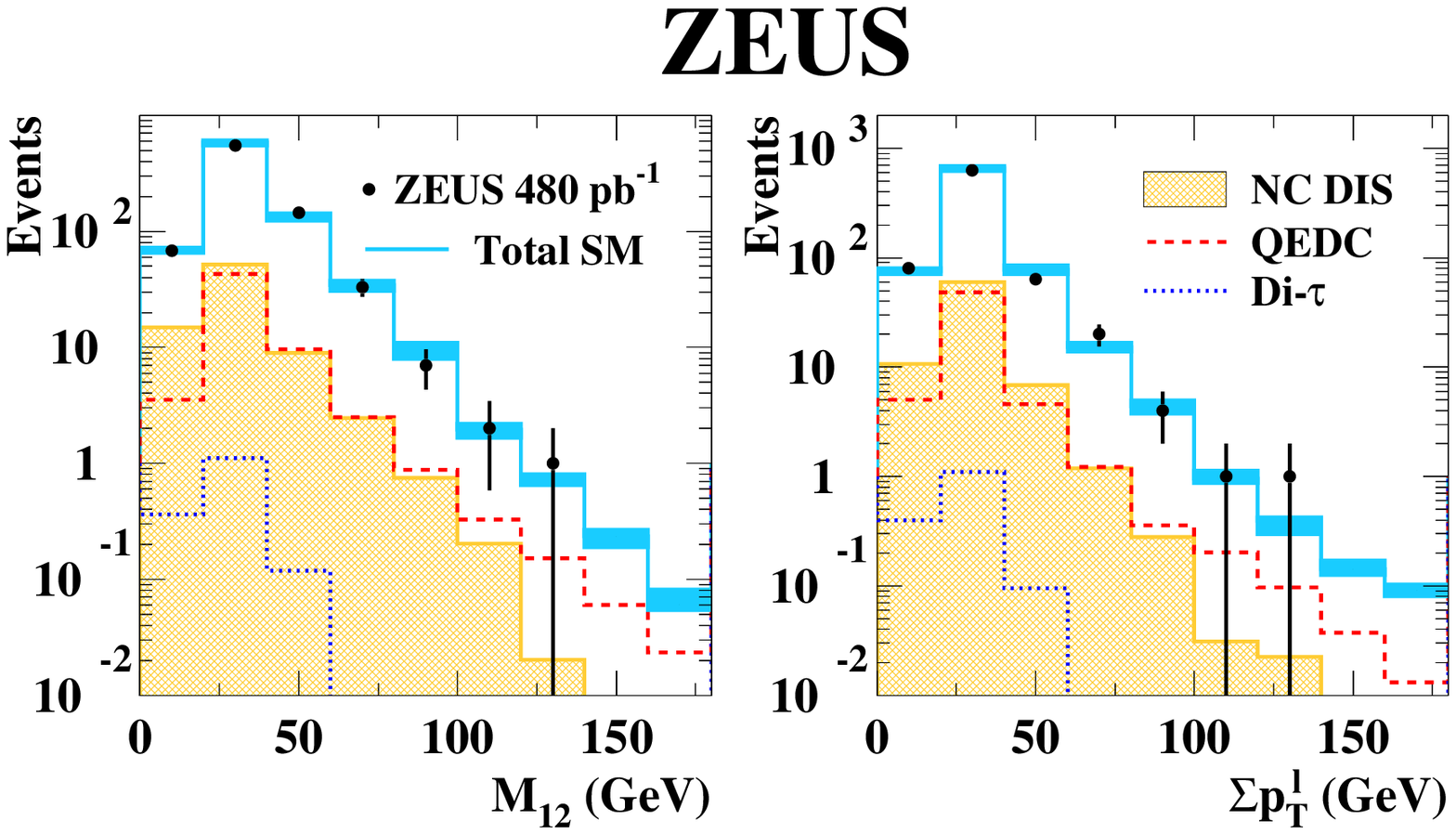}
  }
  \caption{The distribution of the scalar sum of the transverse
    momenta of all the leptons in the multi-lepton final states from
    the H1 (left) and ZEUS (right) analyses, as well as the invariant mass of the two
    highest-$P_{T}$ leptons in the ZEUS analysis (centre), for the
    complete data sets and for all lepton topologies combined.
    The points correspond to the observed data events and the histogram to
    the SM expectation. The total uncertainty on the SM expectation is
    given by the shaded band. The individual components of the SM are as
    indicated within the figures.}
  \label{fig:mlep-individualsplots}
\end{figure*}

The number of selected events in the H1 analysis~\cite{Aaron:2008jh} in
the various topologies are compared to SM predictions in
table~\ref{tab:mlepyields}, where a good agreement is observed for
all event samples.
In the table, the number of events from genuine pair production is shown,
together with the number of events from NC DIS and QED Compton, which
constitute the most significant source of background.
Events with high invariant mass of the two highest $P_T$ leptons
$M_{12}$ are observed in the data, where the SM expectation is low.
All events with $M_{12} > 100$~GeV are seen in the $e^{+}p$ data only.
Two of the high mass events observed by H1 are shown in
figure~\ref{fig:mlep-events}.
A high scalar sum of the lepton transverse momenta
$\sum P_{T}$~GeV, which is summed over all identified leptons in the
event, may be an indication of new physics and five events are seen in
the H1 data, compared to a SM expectation of $1.60\pm 0.20$.
All events are seen in the $e^+p$ data, where $0.96\pm 0.12$
are expected from the SM.
The $\sum P_{T}$~GeV distribution from the H1 analysis is shown in
figure~\ref{fig:mlep-individualsplots}.

\begin{table*}
  \renewcommand{\arraystretch}{1.3}
  \caption{Observed and predicted multi-lepton event yields for the different event
    topologies in the H1, ZEUS and combined analysis. The total SM
    event yield is given by the sum of the signal, lepton
    pair production in $\gamma \gamma$ interactions, and the
    background, mainly from NC DIS and QEDC events.
    For the combined analysis, performed in a common phase space,
    the event yields shown for the $\gamma\gamma$ subsamples are
    those used in the cross section measurement.
    The uncertainties on the predictions include model
    uncertainties and experimental systematic uncertainties added in
    quadrature. The limits on the background estimations are quoted at
    $95\%$~CL.}
  \label{tab:mlepyields}
  \centerline{
    \begin{tabular*}{1.0\textwidth}{@{\extracolsep{\fill}} c c c c c} 
      \hline
      \multicolumn{5}{@{\extracolsep{\fill}} l}{\bf Searches for Multi-lepton Events at HERA}\\
      \hline
      \multicolumn{5}{@{\extracolsep{\fill}} l}{\bf H1 Analysis ({\boldmath ${\mathcal L} = 463~{\rm pb}^{-1}$})}\\
      \hline                                        
      Event sample & Data & Total SM & Pair production & NC DIS + QEDC\\
      \hline                      
      $ee$ & $368$ & $390 \pm 46$ & $332 \pm 26$  & $58 \pm 30$ \\ 
      $\mu\mu$ & $201$   & $211 \pm 32$ & $211 \pm 32$ &  $< 0.005$ \\
      $e\mu$ & $132$   & $128 \pm 9$~~ & $118 \pm 8$~~ & $\!\!\!10.0 \pm 2.5$ \\
      $eee$ & $73$ & $70 \pm 7$ & $69.8 \pm 7.0$ & $0.2 \pm 0.1$ \\    
      $e\mu\mu$ & $97$ & $102 \pm 14$  & $102 \pm 14$  &  $< 0.005$    \\   
      $ee\mu$   & $4$  & ~~$1.43 \pm 0.26$ & ~~$1.18 \pm 0.20$ & $0.25 \pm 0.14$ \\ 
      $eeee$   &  $1$  & ~~$0.33 \pm 0.07$ & ~~$0.33 \pm 0.07$ & $< 0.005$  \\ 
      % \hline
      % $(\gamma\gamma)_{e}$ & $146$  & $138 \pm 12$ & $135 \pm 11$ & $3.0 \pm 1.0$\\ 
      %    % 
      % $(\gamma\gamma)_{\mu}$ & $163$ & $162 \pm 24$ & $162 \pm 24$ & $< 0.005$  \\  
      %    % 
      \hline     
      % \multicolumn{5}{c}{ZEUS analysis}\\
      \multicolumn{5}{@{\extracolsep{\fill}} l}{\bf ZEUS Analysis ({\boldmath ${\mathcal L} = 480~{\rm pb}^{-1}$})}\\
      \hline                                        
      Event sample & Data & Total SM & Pair production & NC DIS + QEDC\\
      \hline                      
      $ee$ & $545$ & $563^{+29}_{-37}$ & $429^{+21}_{-29}$  & $134 \pm 11$  \\
      $\mu\mu$ & $93$   & $106 \pm 12$ & $106 \pm 12$ &  $< 0.5$\\
      $e\mu$ & $46$   & $42 \pm 4$ & $37^{+3}_{-4}$ & $4.5 \pm 2$ \\
      $eee$ & $73$ & $75^{+5}_{-4}$ & $73^{+4}_{-5}$ & $<4$ \\    
      $e\mu\mu$ & $47$ & $48 \pm 5$  & $48 \pm 5$  &  $< 0.5$ \\   
      $eeee$   &  $1$  & $0.9^{+0.5}_{-0.1}$ & $0.6 \pm 0.1$ & $< 1.4$ \\ 
      $ee\mu\mu$   & $2$  & $0.5^{+0.3}_{-0.1}$ & $0.4 \pm 0.1$ & $<0.5$ \\ 
      % \hline
      % $(\gamma\gamma)_{e}$ & $166$  & $185^{+8}_{-14}$ &  $183^{+8}_{-14}$ & $2.8 \pm 1.2$ \\
      %    % 
      % $(\gamma\gamma)_{\mu}$ & $72$ & $85^{+9}_{-10}$ & $85^{+9}_{-10}$ & $< 0.5$ \\  
      %    % 
      \hline                      
      \multicolumn{5}{@{\extracolsep{\fill}} l}{\bf Combined H1 and ZEUS Analysis ({\boldmath ${\mathcal L} = 0.94~{\rm fb}^{-1}$})}\\
      \hline                                        
      Event sample & Data & Total SM & Pair production & NC DIS + QEDC\\
      \hline                      
      $ee$ & $873$ & $895 \pm 57$ & $724 \pm 41$  & $171 \pm 28$ \\ 
      $\mu\mu$ & $298$   & $320 \pm 36$ & $320 \pm 36$ &  $< 0.5$ \\
      $e\mu$ & $173$   & ~~$167 \pm 10$~~ & $152 \pm 9$~~ & ~$15 \pm 3$~ \\
      $eee$ & $116$ & $119 \pm 7$~~ & ~$117 \pm 6$~~~ & $< 4$ \\    
      $e\mu\mu$ & $140$ & $147 \pm 15$  & $147 \pm 15$  &  $< 0.5$    \\   
      \hline
      $(\gamma\gamma)_{e}$ & $284$  & $293 \pm 18$ & $289 \pm 18$ & ~$4 \pm 1$\\ 
      $(\gamma\gamma)_{\mu}$ & $235$ & $247 \pm 26$ & $247 \pm 26$ & $< 0.5$  \\  
      \hline
    \end{tabular*}
  }
\end{table*}

Similarly for the ZEUS analysis~\cite{Chekanov:2009cv}, the number of
selected events in the various topologies studied are compared to the
SM prediction in table~\ref{tab:mlepyields}, where a good agreement
is observed for all event samples.
The $\sum P_{T}$  and $M_{12}$ distributions of all topologies
combined are shown in figure~\ref{fig:mlep-individualsplots}, where
the shape and normalisation are well described by the SM predictions.
Some interesting events are also observed in the ZEUS data at
high-$M_{12}$ and high-$\sum{P_T}$: in particular, two events are
present in the data with $\sum{P_T} >100$~GeV compared to a SM prediction
of $1.56 \pm 0.15$.
Two of the highest mass events observed by ZEUS are shown in
figure~\ref{fig:mlep-events}.

%%%

A combination of the H1 and ZEUS multi-lepton analyses has also been
performed, using a data sample with a total integrated luminosity
of $0.94$~fb$^{-1}$~\cite{Aaron:2009ad}.
Five of the final state topologies are combined,
namely $ee$, $\mu\mu$, $e\mu$, $eee$ and $e\mu\mu$.
The combination is performed in a common phase space, corresponding
to a tightening of the selection cuts of the two experiments. 
For example, in the H1 analysis the electron energy threshold in the
central region was raised from $5$~GeV to $10$~GeV.
Both the number of the observed events and the cross sections for multi-lepton
production measured by the two experiments were combined.
This allows a better sensitivity to rare processes in the high
$M_{12}$ and high $\sum P_{T}$ regions to be achieved and an improved
precision of the measured cross sections.

%%%

\begin{figure}
\begin{center}
  \includegraphics[width=0.90\columnwidth]{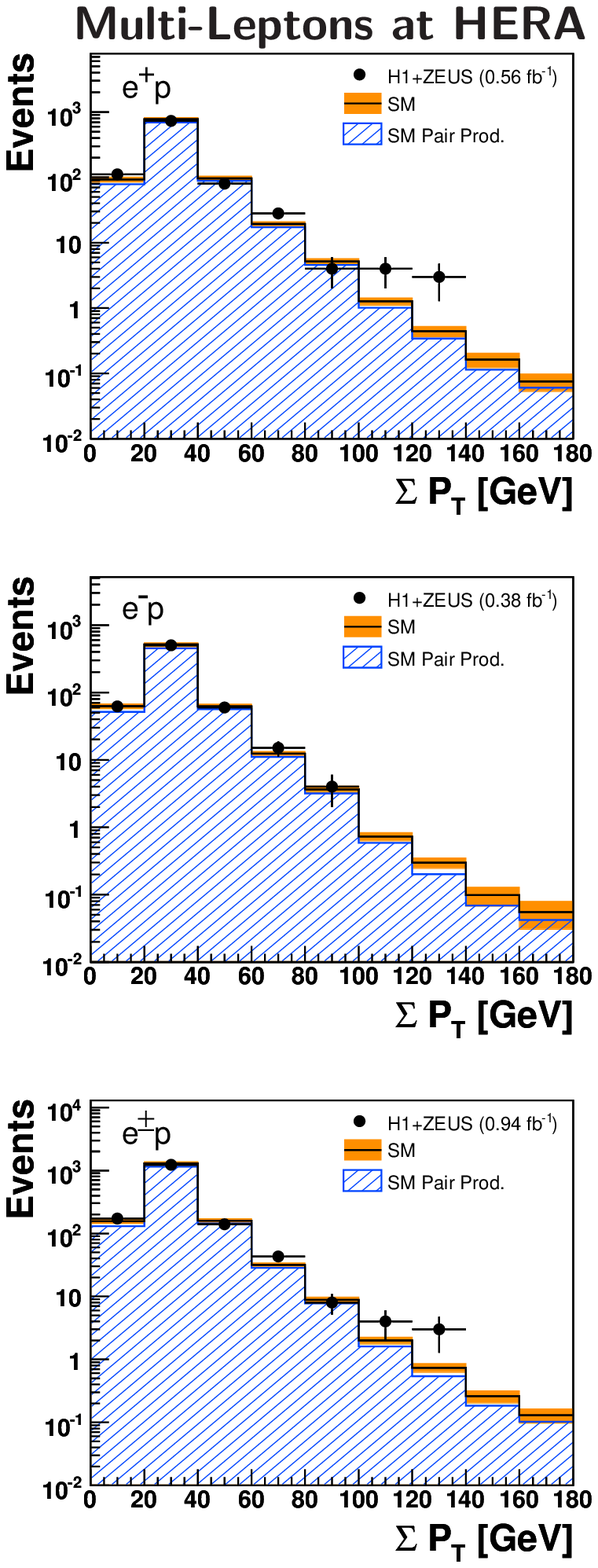}
  \caption{The scalar sum of the transverse momentum $\sum{P_T}$ for combined
    di-lepton and tri-lepton event topologies for $e^{+}p$ (top),
    $e^{-}p$ (middle) data and for all data (bottom) in the combined
    H1 and ZEUS analysis.
    The points correspond to the observed data events and the histogram to
    the SM expectation. The total uncertainty on the SM expectation is
    given by the shaded band. The component of the SM expectation arising
    from lepton pair production is given by the striped histogram.}
  \label{fig:h1zeus-mlepsumpt}
\end{center}
\end{figure}

The event yields of the combined analysis are shown in
table~\ref{tab:mlepyields}, where a good agreement is observed with
the SM.
The $\sum{P_T}$ distributions for the full combined $e^{\pm}$ data, as
well as separately for the $e^+p$ and $e^-p$ data, are shown in
figure~\ref{fig:h1zeus-mlepsumpt}.
In general, a good agreement is found between the data and the SM
predictions.
For $\sum P_{T}>100$~GeV, seven data events are observed in total,
compared to $3.13\pm0.26$ expected from the SM.
These seven events were all recorded in $e^+p$ data, for which the 
SM expectation is $1.94\pm 0.17$.
Events are observed in all topologies with $M_{12}>100~{\rm
  GeV}$, as detailed in table~\ref{tab:h1zeusmlepyieldsM100}.
Both experiments recorded these events in $e^+p$ collisions
only~\cite{Aaron:2009ad}.

\begin{table*}
  \renewcommand{\arraystretch}{1.3}
  \caption{Observed and predicted multi-lepton event yields for
    masses $M_{12} > 100$~GeV for the different event topologies in
    the H1 and ZEUS combined analysis. Event yields are
    shown for all data and divided into $e^+p$ and $e^-p$ collisions. The
    uncertainties on the predictions include model uncertainties and
    experimental systematic uncertainties added in quadrature. The
    limits on the background estimations correspond to the selection
    of no event in the simulated topology and are quoted at $95\%$~CL.}
  \label{tab:h1zeusmlepyieldsM100}
  \centerline{
    \begin{tabular*}{1.0\textwidth}{@{\extracolsep{\fill}} c c c c c}
      \hline
      \multicolumn{5}{@{\extracolsep{\fill}} l}{\bf Multi-lepton Events at HERA with \boldmath{$M_{12}>100$}~GeV}\\
      \hline
      \multicolumn{5}{@{\extracolsep{\fill}} l}{\bf Combined H1 and ZEUS Analysis} \\
      \hline                                        
      {\bf \boldmath $e^{+}p$ collisions (${\mathcal L} = 0.56$ fb$^{-1}$)} & & & & \\
      %\hline                                        
      Event sample & Data & Total SM & Pair production & NC DIS + QEDC \\
      \hline                                        
      $ee$  & $4$ & $1.68 \pm 0.18$ & $0.94 \pm 0.11$  & $0.74 \pm 0.12$ \\ 
      $\mu\mu$  & $1$  & $0.32 \pm 0.08$ & $0.32 \pm 0.08$ &  $< 0.01$  \\ 
      $e\mu$  & $1$ & $0.40 \pm 0.05$ & $0.39 \pm 0.05$ & $< 0.02$  \\ 
      $eee$ & $4$ & $0.79 \pm 0.09$ & $0.79 \pm 0.09$ &  $< 0.03$    \\	
      $e\mu\mu$ & $2$ & $0.16 \pm 0.04$ & $0.16 \pm 0.04$ &  $< 0.01$    \\
      \hline                                        
      {\bf \boldmath $e^{-}p$ collisions (${\mathcal L} = 0.38$ fb$^{-1}$)} & & & & \\
      %\hline                                        
      Event sample & Data & Total SM & Pair production & NC DIS + QEDC \\
      \hline                                        
      $ee$  & $0$ & $1.25 \pm 0.13$ & $0.71 \pm 0.11$ & $0.54 \pm 0.08$ \\ 
      $\mu\mu$  & $0$  & $0.23 \pm 0.10$ & $0.23 \pm 0.10$ &  $< 0.01$  \\ 
      $e\mu$  & $0$ & $0.26 \pm 0.03$ & $0.25 \pm 0.03$  & $< 0.02$  \\ 
      $eee$ & $0$ & $0.49 \pm 0.07$ & $0.49 \pm 0.07$ &  $< 0.03$    \\	
      $e\mu\mu$  & $0$ & $0.14 \pm 0.05$ & $0.14 \pm 0.05$ &  $< 0.01$    \\
      \hline  				      
      {\bf \boldmath $e^{\pm}p$ collisions (${\mathcal L} = 0.94$ fb$^{-1}$)} & & & & \\
      %\hline                                        
      Event sample & Data & Total SM & Pair production & NC DIS + QEDC \\
      \hline
      $ee$  & $4$ & $2.93 \pm 0.28$ & $1.65 \pm 0.16$  & $1.28 \pm 0.18$ \\ 
      $\mu\mu$  & $1$  & $0.55 \pm 0.12$ & $0.55 \pm 0.12$ &  $< 0.01$  \\ 
      $e\mu$  & $1$   & $0.65 \pm 0.07$ & $0.64 \pm 0.06$ & $< 0.02$  \\ 
      $eee$   & $4$ & $1.27 \pm 0.12$ & $1.27 \pm 0.12$ &  $< 0.03$    \\    
      $e\mu\mu$   & $2$ & $0.31 \pm 0.06$ & $0.31 \pm 0.06$ &  $< 0.01$    \\
      \hline
    \end{tabular*}
  }
\end{table*}

The cross sections for lepton pair production were also measured by
both collaborations~\cite{Aaron:2008jh,Chekanov:2009cv}  in the
photoproduction regime, in which the virtuality $Q^2$ of the photon
emitted by the beam lepton is low.
Photoproduction events were selected by requiring a total
$E-P_{z}$ of $\delta < 45$~GeV, which singles out events in which
the scattered lepton is lost in the beam pipe and corresponds to phase
space cuts of $Q^2 < 1$~GeV$^2$ and $y< 0.82$. 
The event yield in this cross section phase space is also presented
for the combined analysis in table~\ref{tab:mlepyields} in the
rows marked $(\gamma\gamma)_{e}$ and $(\gamma\gamma)_{\mu}$.
The cross section is evaluated in each bin $i$ using the formula
\begin{equation}
  \sigma_i = \frac{N_i^{\rm{data}}-N_i^{\rm{bgr}}}{ {\cal L} \cdot A_i},
  \label{eq:xsection}
\end{equation}
where $N_i^{\rm{data}}$ is the number of observed events in bin $i$,
$N_i^{\rm{bgr}}$ the expected contribution from background processes
in bin $i$, ${\cal L}$ the integrated luminosity of the data and $A_{i}$
is the signal acceptance in bin $i$ and is calculated using GRAPE.

\begin{figure*}
  \centerline{\includegraphics[width=1.8\columnwidth]{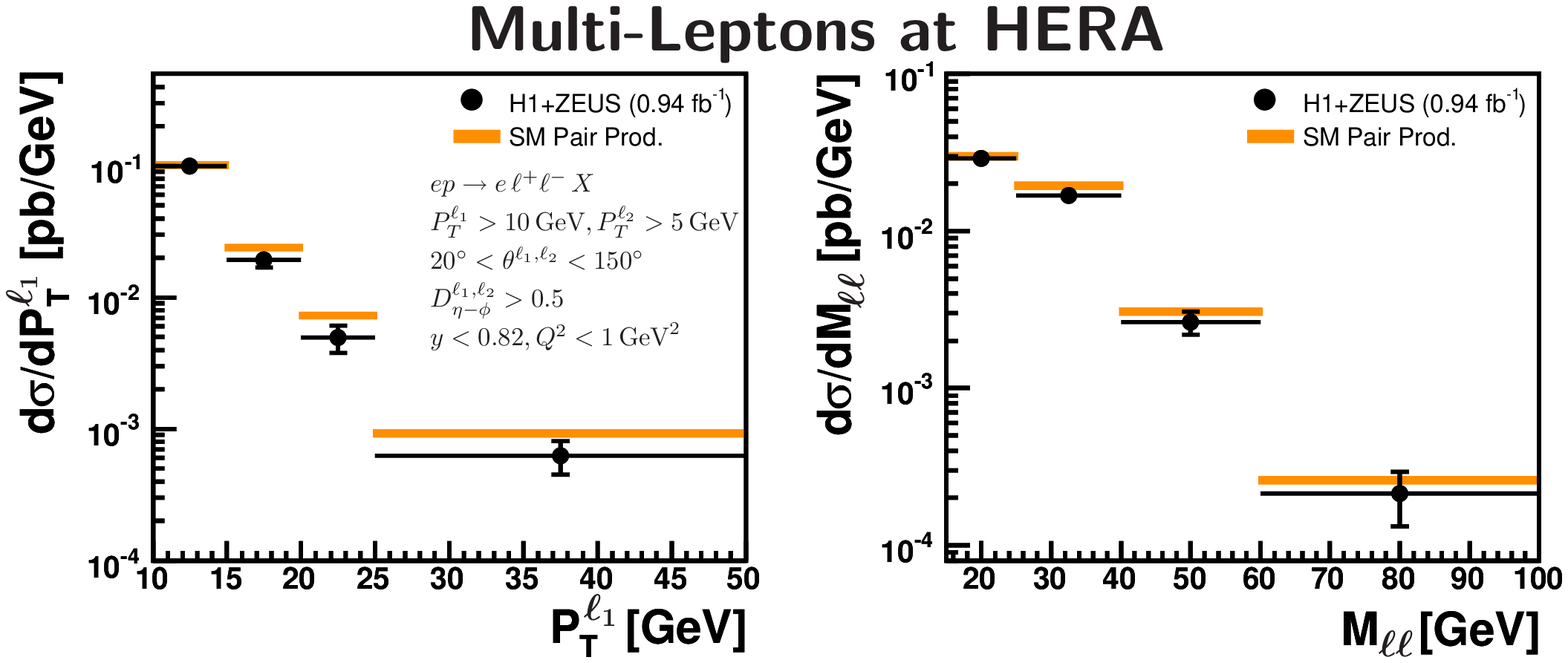}}
  \caption{The cross section for lepton pair photoproduction in a
    restricted phase space as a function of the leading lepton
    transverse momentum $P_T^{\ell_1}$ (left) and the invariant mass of
    the lepton pair $M_{\ell\ell}$ (right). The total error bar is shown,
    representing the statistical and systematic uncertainties added in
    quadrature, which is dominated by the statistical. The bands
    represent the one standard deviation uncertainty in the SM
    prediction, dominated by the photon-photon process.}
  \label{fig:h1zeus-mlepxsec}
\end{figure*}

%%%
%
The combined H1-ZEUS cross section measurement is evaluated
using a weighted mean of the values measured by the two
collaborations~\cite{Aaron:2009ad}.
%
%For $ep \rightarrow ee^{+}e^{-}X$ events, the mean signal
%acceptances in the H1 and %ZEUS experiments are 45% and 60%,
% respectively. In case of  $ep \rightarrow e\mu^{+}\mu^{-}X$ events,
% it is 60% for H1 and 30% for ZEUS.
%
The total visible electron pair production cross section for the
process $ep \rightarrow ee^{+}e^{-}X$ is measured
in the restricted phase space as:
\[
 \sigma_{e^{+}e^{-}} = 0.68~\pm~0.04~({\rm stat.})\pm~0.03~({\rm sys.})~{\rm pb,}
\]
where the first uncertainty is statistical and the second systematic.
The total visible muon pair production cross section for the process
$ep \rightarrow e\mu^{+}\mu^{-}X$ is measured as:
\[
 \sigma_{\mu^{+}\mu^{-}} = 0.63~\pm~0.05~({\rm stat.})\pm~0.06~({\rm sys.})~{\rm pb.}
\]
As the SM cross sections for $e^+e^-$ and $\mu^+\mu^-$ production in
$\gamma\gamma$ interactions are expected to be the same, the electron
and muon pair production cross sections given above are combined into
a single visible lepton pair production cross section of:
\[
 \sigma_{\ell^{+}\ell^{-}} = 0.66~\pm~0.03~({\rm stat.})\pm~0.03~({\rm sys.})~{\rm pb,}
\]
which is in good agreement both with the SM prediction
from GRAPE of $0.69 \pm 0.02$~pb, as well as the individual H1 and ZEUS
measurements~\cite{Aaron:2009ad}.
The differential cross sections for lepton pair photoproduction as a
function of the transverse momentum of the leading lepton,
$P_T^{\ell_1}$, and of the invariant mass of the lepton pair, $M_{\ell\ell}$,
are also measured and found to be in good agreement with SM
predictions, as shown in figure~\ref{fig:h1zeus-mlepxsec}.

\subsection{Tau-pair production}
\label{sec:mlep-ditau}

The dominant tau-pair production mechanism is the same as for
multi-electron and multi-muon production, 
$\gamma \gamma \rightarrow \tau^+ \tau^-$, as illustrated
in figure~\ref{fig:multilepfeyn}.
Tau particles can decay into leptons, $\tau \rightarrow \mu \nu_\mu \nu_\tau$,
or $\tau \rightarrow e \nu_e \nu_\tau$, or into hadrons,
$\tau \rightarrow h \nu_\tau$, which happens about $2/3$ of the time.
Therefore, whilst the search of tau-pair production has the same
physical motivation of that of multi-leptons, it is performed with
different experimental techniques due to the different experimental
signature of the tau lepton.

%%%

The tau-pair decay is classified as {\it leptonic} if both
tau leptons decay leptonically, {\it semi-leptonic} if one tau lepton decays
hadronically and one leptonically, and {\it hadronic} if both tau leptons
decay into hadrons.
From the experimental point of view, leptons from a tau decay
cannot be distinguished from prompt electron or muon production,
as the accompanying neutrino is not detected and hence events in which
the two taus decay to leptons of the same flavour are rejected when
examining tau-pair production.

In the hadronic tau decay, due to the mass and charge of the tau lepton, 
most likely only one or three charged hadrons are
produced\footnote{The branching ratio for tau decays into more than three
charged hadrons is small (about $2\%$).}.
These decays modes are  referred to as \emph{1-prong}
(tau decay branching ratio $49\%$) and \emph{3-prong}
(tau decay branching ratio $14\%$), respectively.
The hadronic decay products of the tau lepton look like a collimated jet,
featuring only one or three tracks, which are very close to each other.
A $\tau$ jet can therefore be distinguished from QCD jets based on its
shape and on the charge and multiplicity of its tracks.
Neural networks and multivariate analysis techniques are typically employed
to separate identified tau jet candidates from the large QCD background.
One of the main differences between the analysis of tau-pair production
and other di-lepton analyses is that a significant part of the tau-lepton
momentum is carried away by the tau neutrino.
This result in the tau decay products having a significantly lower $P_{T}$
compared to the original tau lepton.

%%%

The H1~\cite{Aktas:2006fc} and ZEUS~\cite{Levy:2010aa} collaborations have
both studied tau-pair production looking at the leptonic ($e\mu$),
semi-leptonic ($e~{\rm jet}$ or $\mu~{\rm jet}$) and hadronic (${\rm jet}~{\rm jet}$)
decay modes.
The tau-pair signal is modelled using the GRAPE event generator,
which is also used to model the background expectation from other di-leptons.
In order to reduce the significant SM background contribution
from NC DIS and photoproduction events, the analyses are restricted to
elastic or quasi-elastic tau-pair production, $ep \rightarrow eX \tau^+ \tau^-$,
where $X$ is the proton or a resonant state.
This is effectively done by vetoing any additional objects in the final state beyond
the tau decay products and the scattered electron, which may also be present.

%%%

The main background contribution after applying the above elasticity requirements
is due to exclusive diffractive events, both in DIS and
photoproduction, which is modelled using the RAPGAP event generator. 
Non-diffractive DIS background is modelled by H1 (ZEUS) using RAPGAP
(DJANGOH) and photoproduction is modelled by both experiments using
PYTHIA.

\begin{table}
  \renewcommand{\arraystretch}{1.3}
  \caption{Observed and predicted event yields for the different event
    topologies in the H1 and ZEUS di-tau analyses. The total MC expectation
    includes the sum of tau-pair production, NC DIS and photoproduction,
    as well as electron and muon pair production. The experimental
    systematic uncertainties are quoted on the total MC expectations.}
  \label{tab:ditauyields}
    \begin{tabular*}{1.0\columnwidth}{@{\extracolsep{\fill}} c c c c} 
      \hline
      \multicolumn{4}{@{\extracolsep{\fill}} l}{\bf Tau-pair Production at HERA}\\
      \hline
      \multicolumn{4}{@{\extracolsep{\fill}} l}{\bf H1 Analysis ({\boldmath ${\mathcal L} = 106~{\rm pb}^{-1}$})}\\
      \hline                                        
      Tau decay & Data & Total SM & Tau-pair\\
      &  & & production\\
      \hline                      
      $e\mu$ & $7$ & $2.9 \pm 0.4$ & $56\%$\\ 
      $e~{\rm jet}$ & $2$   & $6.3 \pm 0.9$ & $47\%$\\
      $\mu~{\rm jet}$ & $10$   & $7.0 \pm 1.3$ & $85\%$\\
      ${\rm jet}~{\rm jet}$ & $11$ & $11.0 \pm 2.0$ & $50\%$\\    
      \hline                                        
      Total & $30$ & $27.1 \pm 4.1$  & $59\%$\\   
      \hline     
      \multicolumn{4}{@{\extracolsep{\fill}} l}{\bf ZEUS Analysis ({\boldmath ${\mathcal L} = 334~{\rm pb}^{-1}$})}\\
      \hline                                        
      Tau decay & Data & Total SM & Tau-pair\\
      &  & & production\\
      \hline                                        
      $e\mu$ & $4$ & $3.6^{+1.3}_{-0.3}$ & $3.0^{+0.3}_{-0.2}$\\ 
      $e~{\rm jet}$ & $7$ & $8.8^{+1.8}_{-0.8}$ & $5.3^{+0.3}_{-0.2}$\\ 
      $\mu~{\rm jet}$ & $4$ & $8.0^{+2.2}_{-1.2}$ & $5.9^{+0.5}_{-0.5}$\\ 
      ${\rm jet}~{\rm jet}$ & $10$ & $14.4^{+2.2}_{-3.5}$ & $9.0^{+0.4}_{-0.3}$\\ 
      \hline                                        
      Total & $25$ & $34.8^{+3.9}_{-3.8}$ & $23.2^{+0.7}_{-0.7}$\\ 
      \hline                      
    \end{tabular*}
\end{table}

\begin{figure*}
\centerline{\includegraphics[clip,width=0.9\columnwidth,angle=270]{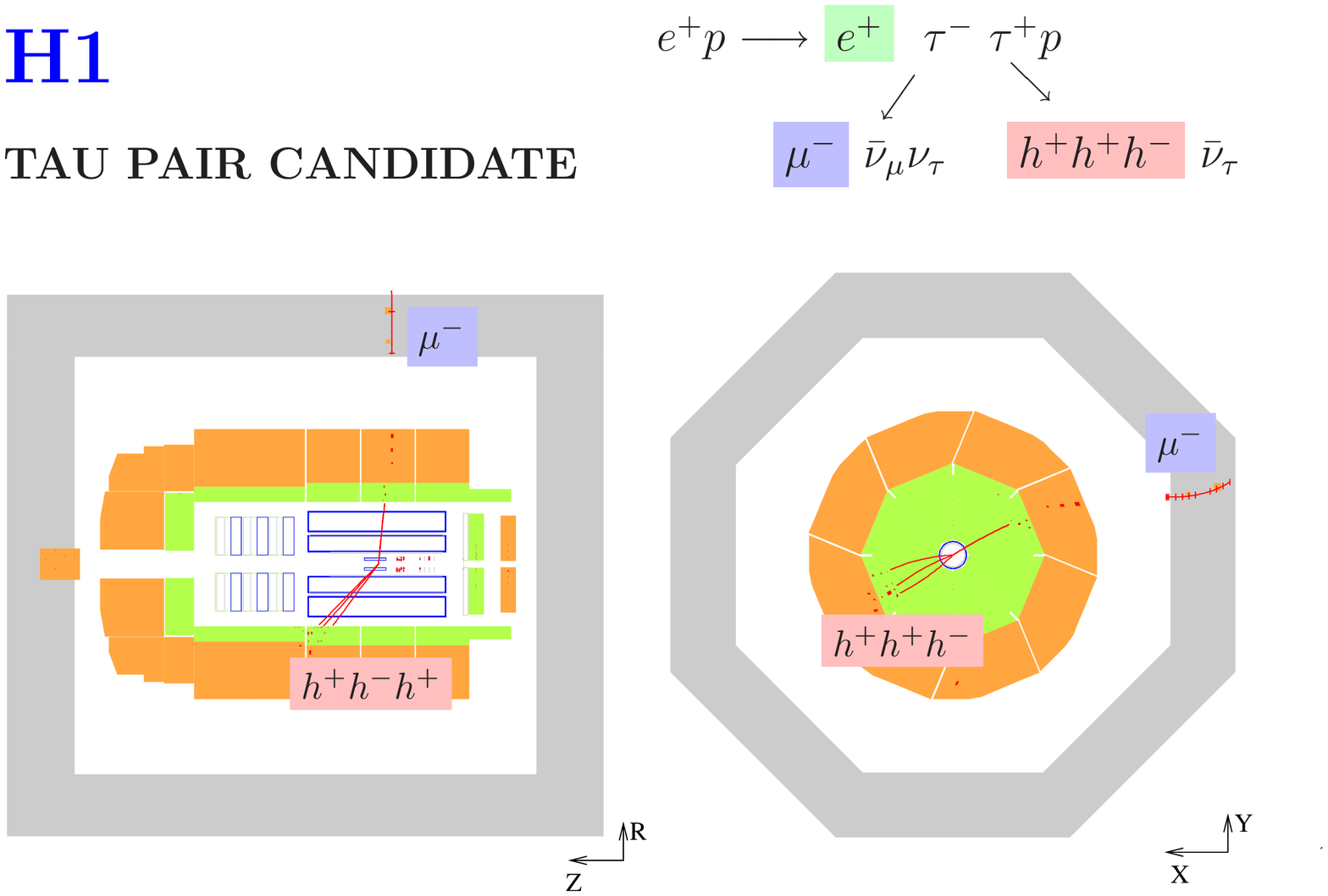}}
\caption{A tau-pair event observed in the H1 detector, where one tau lepton decays leptonically
  to a muon and the other tau lepton decays to three charged hadrons, a so called 3-prong decay.}
\label{fig:ditau_H1_event}
\end{figure*}

The H1 analysis of tau-pair production is based on their HERA~I
data sample~\cite{Aktas:2006fc}, corresponding to an
integrated luminosity of $106$~pb$^{-1}$.
Electrons are selected with $E_{e} > 5$~GeV and  $P_{T}^{e} > 3$~GeV,
and muons are required to have  $P_{T}^\mu > 2$~GeV.
Both electrons and muons are reconstructed in the central region of the detector,
$20^\circ < \theta_{\ell} < 140^\circ$, and are required to be isolated from jets of at
least one unit in angular distance $D=\sqrt{(\Delta\eta)^2+(\Delta\phi)^2}$.
Jets composed of one or three tracks with  $E_T^{\rm jet} > 2$~GeV and
in the region $20^\circ < \theta_{\rm jet} < 120^\circ$ are reconstructed by a
dedicated algorithm, which uses multiple neural networks to discriminate
between tau jets and the QCD background~\cite{Aktas:2006fc}.

%%%

In total, $30$ tau-pair events are selected in the H1 analysis, compared to a
SM prediction of $27.1 \pm 4.1$, of which $16.0\pm3.4$ are expected from
$\gamma \gamma \rightarrow \tau^+ \tau^-$ events.
The number of events collected in the different tau-pair topologies are
summarised in table~\ref{tab:ditauyields}.
The distributions of the visible invariant mass, as well as the polar angle
and transverse momentum of the identified tau candidates were also examined
and show no deviation from the SM predictions. 
A tau-pair event in the final H1 sample is shown
in figure~\ref{fig:ditau_H1_event}, where one of the tau leptons decays into a muon and
the other undergoes a 3-prong hadronic decay.
In this case, the scattered electron is not detected in the event.

%%%

The ZEUS analysis of tau-pair production is based on their HERA~II
data sample~\cite{Levy:2010aa}, corresponding to an integrated
luminosity of $334$~pb$^{-1}$.
Electrons are selected with $P_{T}^{e} > 2$~GeV, in the polar angle region
$17^\circ < \theta_{e} < 160^\circ$,  isolated from the rest other calorimetric energy
deposits by a distance $D> 0.8$.
Muons are selected as tracks in the central detectors matched to segments in the muon 
chambers, with $P_T^{\mu} > 2$~GeV and in the polar angle region
$34^\circ < \theta_{\mu} < 157^\circ$.
The muon track is required to be separated by a distance $D>1.0$ from any other
track in the event. 
Tau jets with at least one track and $E_T^{\rm jet} > 5~{\rm GeV}$ are reconstructed in
the region $|\eta^{\rm jet}| < 2$.
A multivariate discrimination technique~\cite{Carli:2002jp} is used to discriminate tau jets
from QCD jets. 

%%%

\begin{figure*}
  \centerline{\includegraphics[width=1.8\columnwidth]{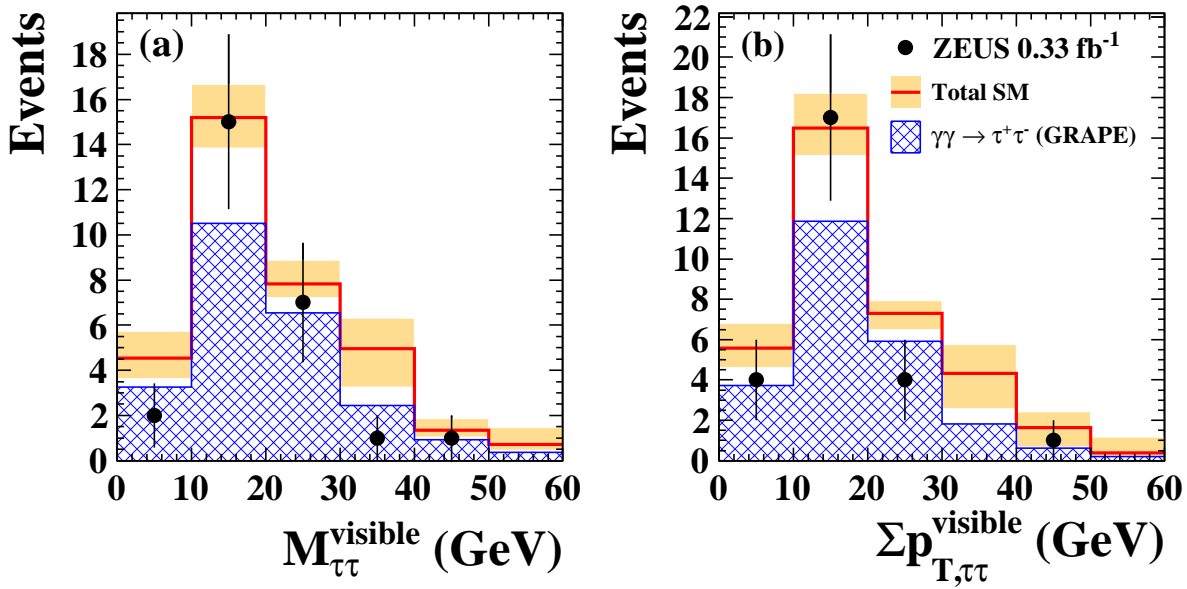}}
  \caption{Distribution of ZEUS tau-pair candidates as a function of (a) the visible invariant mass
    of the tau pair, $M^{\rm visible}_{\tau \tau}$, and (b) the scalar sum of transverse momenta of the
    two tau candidates, $\sum p_{T,\tau \tau}^{\rm visible}$. The data (points) are compared with the
    predictions of the sum of the MC expectations. The tau-pair expectation is given by the
    hatched histogram. The shaded band represents the systematic uncertainty on the SM expectation.}
  \label{fig:ditau_zeus_distr}
\end{figure*}

A total of 25 events are observed in the data, compared to a SM expectation of
$34.8^{+3.9}_{-3.8}$, of which $23.2^{+0.7}_{-0.7}$ are from tau-pair production.
The events are classified into the different tau decays topologies in
table~\ref{tab:ditauyields}.
Figure~\ref{fig:ditau_zeus_distr} shows the visible invariant mass,
$M^{\rm visible}_{\tau \tau}$, calculated from the two tau candidates,
and the scalar sum of their visible transverse momenta, $\sum P_{T,\tau \tau}^{\rm visible}$.
No event with a visible mass $M^{\rm visible}_{\tau \tau}$ greater
than $50$~GeV is observed in the data.
The highest visible-mass candidate observed in the ZEUS data, found in the
$e-\mu$ topology, has $M^{\rm visible}_{\tau \tau} = 49$~GeV.
The SM prediction describes the data well and no excess is observed in either
the high mass or high $\sum P_{T,\tau \tau}^{\rm visible}$ regions.

%%%

H1 performs a measurement of the cross section for the elastic production of $\tau^{+}\tau^{-}$
pairs in the kinematic region $P_{T}^{\tau} > 2$~GeV and $20^{\circ} < \theta_{\tau} < 140^{\circ}$.
The cross section is calculated according to equation~\ref{eq:xsection}, where the acceptance is
calculated using GRAPE.
The H1 measured visible cross section for elastic tau pair production $ep \rightarrow ep\tau^{+}\tau^{-}$
integrated over the kinematic phase space defined above is:
\[
 \sigma_{\tau^{+}\tau^{-}} = 13.6~\pm~4.4~({\rm stat.})\pm~3.7~({\rm sys.})~{\rm pb,}
\]
where the first uncertainty is statistical and the second systematic.
The result is in good agreement with the SM expectation from GRAPE of $11.2 \pm 0.3$~pb.
A similar cross section measurement is performed by ZEUS in the kinematic region
$P_{T}^{\tau} > 5$~GeV and $17^{\circ} < \theta_{\tau} < 160^{\circ}$.
The visible measured cross section in the ZEUS phase space is:
\[
\sigma_{\tau^{+}\tau^{-}} = 3.3~\pm~1.3~({\rm stat.})\pm~2.0~({\rm sys.})~{\rm pb,}
\]
which is in reasonable agreement with the SM expectation $5.7 \pm
0.2$~pb, as evaluated using GRAPE.

\subsection{Search for doubly-charged Higgs production}
\label{sec:higgs}

\begin{figure}
  \begin{center}
  \includegraphics[width=0.45\columnwidth]{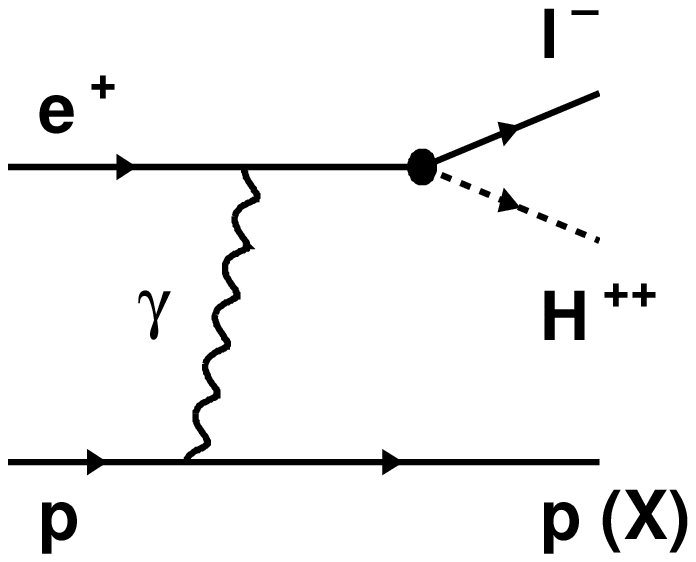}\\
  \includegraphics[width=0.45\columnwidth]{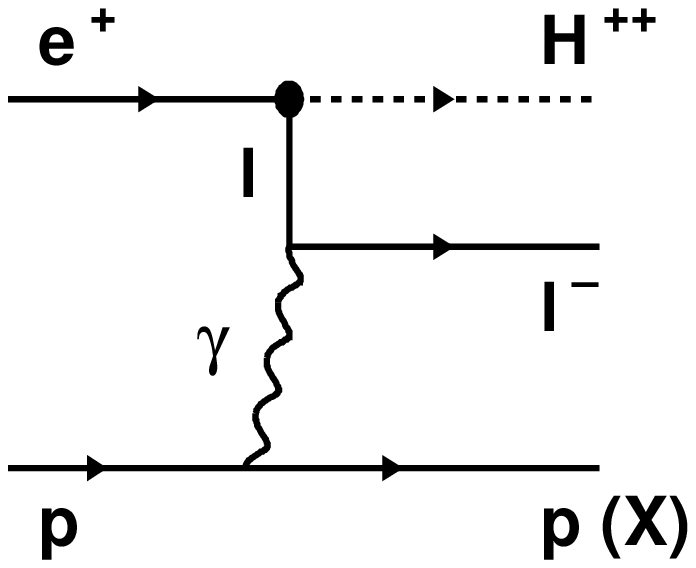}
  \includegraphics[width=0.45\columnwidth]{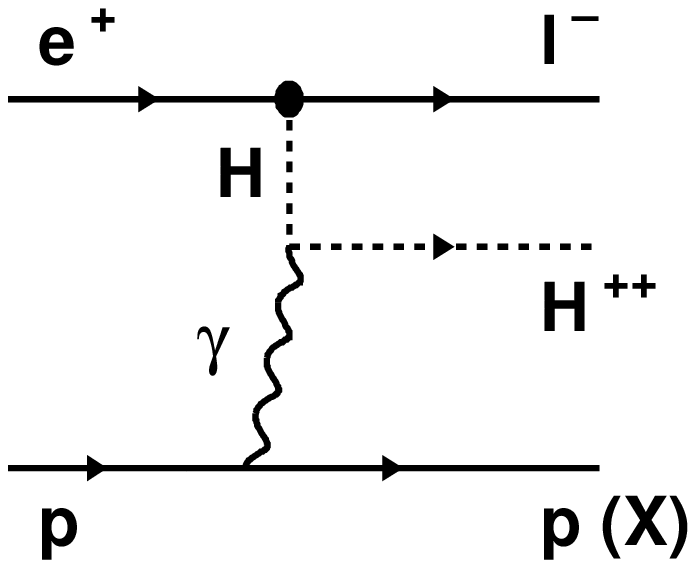}
  \caption{Diagrams for the single production of a doubly-charged
    Higgs boson in $e^{+}p$ collisions at HERA via a coupling
    $h_{e\ell}$.} 
  \label{fig:doublehiggs-feyn}
\end{center}
\end{figure}

The production in $ep$ collisions of a single doubly-charged Higgs boson
$H^{\pm\pm}$ could be a source of events containing multiple high
$P_{T}$ leptons at HERA~\cite{Accomando:1993ar}.
Figure~\ref{fig:doublehiggs-feyn} shows the possible diagrams for
single $H^{++}$ production in $e^{+}p$ collisions at HERA.
Following the observed excess of high mass events observed in
multi-electron events in the H1 HERA data~\cite{Aktas:2003jg}, these
events were investigated in this context~\cite{Aktas:2006nu}.
The compatibility of these events with a hypothetical doubly-charged
Higgs coupling to $ee$ is addressed and further searches for a
$H^{\pm\pm}$ boson coupling to $e\mu$ and $e\tau$ are performed.

%%%

The $ee$ analysis is based on the multi-electron event selection
described in section~\ref{sec:mlep} and~\cite{Aktas:2003jg},
where a like-charge requirement~\cite{Aktas:2006nu} is added to the
two highest electrons (positrons), which are assigned to the hypothetical
$H^{--}$ ($H^{++}$).
After the addition of the charge requirement, $3$ events are observed
in the data with $M_{ee} > 65$~GeV, in agreement with the SM
expectation of $2.45 \pm 0.11$.
Only one of the high mass events seen in the original multi-electron
analysis~\cite{Aktas:2003jg} survives the additional selection
requirements.

%%%

Events with one electron and one muon with minimal transverse
momenta of $P_{T}^{e} > 10$~GeV and $P_{T}^{\mu} > 5$~GeV are selected
to investigate the $H^{\pm\pm} \rightarrow e\mu$ decay. 
Electrons are selected in the polar angle range $20^\circ < \theta_{e}
< 140^\circ$ to reduce the NC DIS background, and muons in the range
$10^\circ < \theta_{\mu} < 140^\circ$, which is more extended into the
forward region to increase the efficiency for high $H^{\pm\pm}$
masses.
The same charge criteria is applied to the $e\mu$ final states as used
in the $ee$ search. 
For $M_{e\mu} > 65$~GeV one event is observed, compared to a SM
expectation of $4.17 \pm 0.44$.

%%%

The search for a $H^{++}$ boson decaying into $e\tau$ is
performed in three final states, depending on whether the tau-lepton
decays into an electron, a muon or hadronically\footnote{The search
  for the $e\tau$ decay is only performed for using the larger
  $e^{+}p$ data set, so here only a $H^{++}$ is considered.}.
Events are selected which contain either two electrons ($ee$), or
an electron and a muon ($e\mu$), or an electron and a hadronic
tau-jet ($eh$).
The two leptons, or the electron and the hadronic tau-jet (see 
section~\ref{sec:mlep-ditau}) are required to be in the angular range
$20^{\circ} < \theta < 140^{\circ}$, and have an angular separation
of $D > 2.5$ in pseudorapidity-azimuth. A minimum $P_{T}$ of
$10$~($5$)~GeV is required for the leading (second) lepton or hadronic tau-jet.
A significant amount of missing transverse and longitudinal momentum is
expected due to the neutrinos produced in the tau lepton decay.
Events in the $ee$ class are required to have a missing transverse
momentum $P_{T}^{\rm miss} > 8$~GeV.
This is increased to $P_{T}^{\rm miss} > 11$~GeV for the $eh$ class and additional
cuts are applied on the longitudinal balance of the event to reduce
large NC DIS background~\cite{Aktas:2006nu}.
Finally, events are rejected if the track associated with one of the
Higgs decay product candidates has a negative charge,
which is opposite to that of the incoming lepton beam.
Following all $e\tau$ selection cuts, only one event is observed
in the data (in the $eh$ class), compared to a SM expectation of
$2.1 \pm 0.5$.

%%%

Overall a good agreement is observed with the SM predictions and no
evidence for a doubly charged Higgs boson $H^{\pm\pm}$ is found in the
data.
Upper limits on the coupling $h_{e\ell}$ are derived at $95\%$~CL
using a modified frequentist approach~\cite{Junk:1999kv} taking into
account systematic uncertainties and assuming that the $H^{\pm\pm}$
couples with $100\%$ branching ratio to only $ee$, $e\mu$ or $e\tau$. 

The H1 limits on $h_{ee}$ are not competitive to those set by the OPAL
experiment~\cite{Abbiendi:2003pr}.
The $e\mu$ ($e\tau$) analysis allows masses below $141$~GeV
($112$~GeV) to be ruled out at $95\%$~CL for a coupling
$h_{e\mu}$ ($h_{e\tau}$) of electromagnetic strength $h_{e\ell} =
0.3$, which extends beyond the limits from
LEP~\cite{Abbiendi:2001cr,Abdallah:2002qj,Achard:2003mv}.
More recent analyses from hadron colliders, which
investigate $H^{\pm\pm}$ pair production and are independent of the
coupling strength, have pushed these limits further, first beyond the
$200$~GeV regime at the Tevatron (for example, an analysis of
$e\mu$ final states from CDF, $M_{H^{\pm\pm}}
>210$~GeV~\cite{Aaltonen:2011rta})
and later at the LHC where the most stringent limits are from
ATLAS~\cite{Aad:2014hja,ATLAS:2014kca} and currently exclude masses
up to $468$~GeV ($400$) in an analysis of $e\mu$ ($e\tau$) final
states.
The limits from the H1 remain unique in that they
derived from a search for single $H^{\pm\pm}$ production.

\section{Events with isolated leptons and missing transverse momentum}
\label{sec:isolep}

Events containing high energy charged leptons (electron, muon or tau-leptons)
together with missing transverse momentum produced in high energy particle
collisions are interesting as they may be a signature of BSM physics.
When such an event containing an isolated muon was observed by H1
in the first $4$~pb$^{-1}$ of $e^{+}p$ data, a detailed investigation
into the potential physics source was performed~\cite{firstH1event}.
After this initial H1 event was discovered during the visual
scanning of high $Q^{2}$ events with large $P_{T}^{\rm miss}$, which was
routinely done during data taking for CC DIS analysis, such events would
regularly appear throughout HERA~I and II data taking.

%%%

\begin{figure*}
  \centerline{
    \includegraphics[width=0.80\columnwidth]{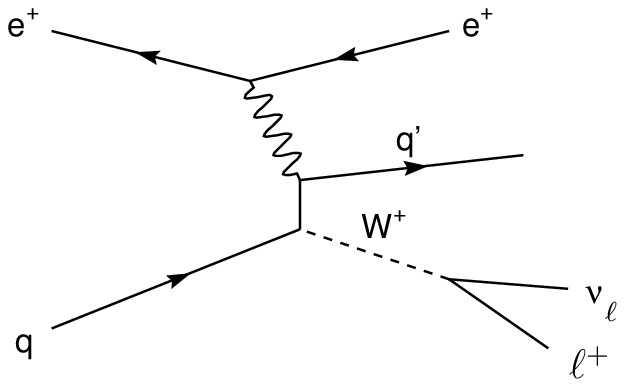}
    \hspace{0.4cm}
    \includegraphics[width=0.85\columnwidth]{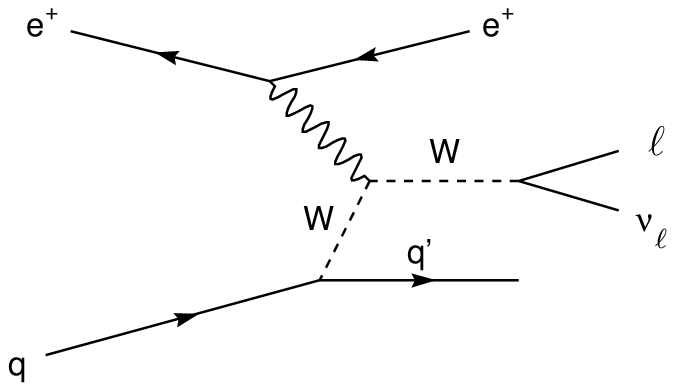}
  }
  \vspace{1.5cm}
  \centerline{
    \includegraphics[width=0.80\columnwidth]{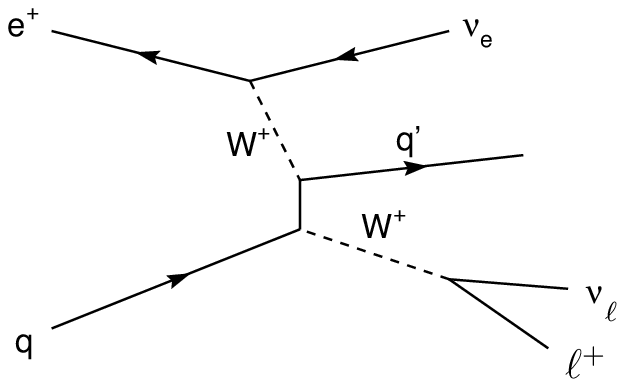}
    \hspace{0.4cm}
    \includegraphics[width=0.85\columnwidth]{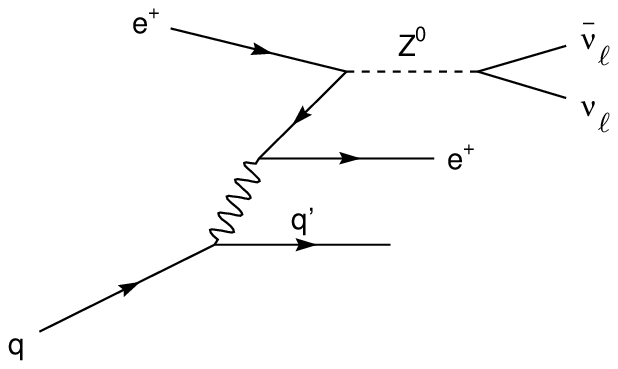}
  }
  \begin{picture} (0.,0.) 
    \setlength{\unitlength}{1.0cm}
    \put (4.8,6){\bf\normalsize  (a)} 
    \put (13,6){\bf\normalsize  (b)} 
    \put (4.8,0.5){\bf\normalsize  (c)} 
    \put (13,0.5){\bf\normalsize  (d)} 
  \end{picture} 
  \caption{Diagrams of processes at HERA which produce an isolated lepton together
    with missing transverse momentum in the final state:
    Figure (a): $ep~\rightarrow~eW(\rightarrow\ell\nu)X$; figure (b): $W$
    production via the $WW\gamma$ triple gauge boson coupling;
    figure (c): $ep~\rightarrow~\nu~W(\rightarrow\ell\nu)X$;
    figure (d): $ep~\rightarrow~eZ(\rightarrow\nu\bar{\nu})X$. The diagrams are
    shown for $e^{+}p$ collisions.}
  \label{fig:isolep-feynman}
\end{figure*}

In this analysis, processes are considered {\it signal} if they produce events
containing a genuine isolated lepton and genuine missing transverse
momentum in the final state.
Within the SM, single $W$ boson production with subsequent
leptonic decay $W\rightarrow \ell\nu$, as illustrated in
figures~\ref{fig:isolep-feynman}(a)-(c), is the main SM process at HERA that
produces events with high energy isolated leptons and missing
transverse momentum.
The inclusive hadronic state, which results
primarily\footnote{In principle, the beam remnant may also contribute to the 
hadronic final state, although in the case of $W$ production this is typically
low $P_{T}$ and continues down the beampipe following the $e^{\pm}p$ collision.}
from the hadronisation of the struck quark $q'$, is denoted by $X$.
The SM prediction for $W$ production via $ep\rightarrow eWX$ 
is modelled by both H1 and ZEUS using the EPVEC event generator, which
employs the full set of LO diagrams~\cite{Baur:1991pp}, including $W$
production via the $WW\gamma$ triple gauge boson coupling
as illustrated in figure \ref{fig:isolep-feynman}(b).
This prediction is corrected to NLO by applying a reweighting to
the LO cross section dependent on the transverse momentum and rapidity
of the $W$ boson~\cite{Diener:2003df,Diener:2002if,Nason:1999xs,Spira:1999ja}.
The NLO corrections range from $30\%$ at low $W$ transverse
momentum (resolved photon interactions) to around $10\%$ at high 
$W$ transverse momentum (direct photon interactions).

%%%

Two further processes contribute to signal events.
Firstly, the equivalent charged current $W$ production process $ep\rightarrow \nu
WX$, as illustrated in figure~\ref{fig:isolep-feynman}(c), which is calculated
at LO with EPVEC and contributes an additional $7\%$ to the predicted 
signal cross section.
Secondly, $Z^{0}$ production with subsequent decay to neutrinos, as illustrated
in figure~\ref{fig:isolep-feynman}(d), where the outgoing electron is
the isolated lepton in the event and genuine missing transverse momentum
is produced by the $Z^{0}$ decay neutrinos.
This process, which is only relevant when the identified isolated lepton
in the final state is an electron, is also calculated using EPVEC and found
to contribute less than $3\%$ to the predicted signal cross section.

%%%

Final states with isolated electrons or muons and missing transverse
momentum, referred to as the electron channel and muon channel
respectively, are discussed in the following.
Analyses of events with isolated tau leptons and missing
transverse momentum are covered in section~\ref{sec:isolep:singletau}.

\subsection{Events with isolated electrons or muons}
\label{sec:isoelmu}

Events with isolated electrons or muons and missing transverse
momentum have been observed at HERA, and results from the analyses
have been regularly published by both
collaborations~\cite{Adloff:1998aw,Andreev:2003pm,Breitweg:1999ie,Chekanov:2003yt},
culminating in the final results using the complete HERA data
set~\cite{Aaron:2009wp,Chekanov:2008gn}.
The integrated luminosity in the H1 analysis corresponds to
$474$~pb$^{-1}$, of which $291$~pb$^{-1}$ are from $e^{+}p$ collisions
and $183$~pb$^{-1}$ from $e^{-}p$ collisions.
In the ZEUS analysis, the integrated luminosity corresponds to
$504$~pb$^{-1}$, of which $296$~pb$^{-1}$ are from $e^{+}p$ collisions
and  $208$~pb$^{-1}$ from $e^{-}p$ collisions.

%%%

The H1 and ZEUS event selections are very similar.
Lepton candidates identified by H1 (ZEUS) are required to lie
within the polar angle range $5^{\circ} < \theta_{\ell} < 140^{\circ}$
($15^\circ<\theta_{\ell}<120^\circ$) and to have transverse
momentum, $P_{T}^{\ell} > 10$~GeV.
The lepton is also required to be isolated with respect to jets
and other tracks in the event using the distances in
$\eta$--$\phi$ space to the nearest jet $D_{\rm jet}>1.0$ and
nearest track $D_{\rm track}>0.5$.
A large transverse momentum imbalance $P_{T}^{\rm miss} >12$~GeV
is required and a cut on $P_{T}^{\rm calo}>12$~GeV is also imposed to
ensure a high trigger efficiency.
As muons deposit little energy in the calorimeter 
$P_{T}^{\rm calo} \simeq P_{T}^{X}$ in events with isolated muons, and
therefore the $P_{T}^{\rm calo}$ requirement effectively acts as a cut on
the hadronic transverse momentum $P_{T}^{X} > 12$~GeV in the muon channel.
To ensure that the two final states are exclusive, events in the
electron channel must contain no isolated muons.

%%%

SM background processes contribute to the analysis mainly through
misidentification or mismeasurement.
NC DIS events ($ep \rightarrow eX$), in which genuine isolated
high $P_{T}$ electrons are produced, form a significant background
in the electron channel search when fake $P_{T}^{\rm{miss}}$
arises from mismeasurement.
The NC DIS background is modelled in the H1 (ZEUS) analysis using the
RAPGAP (DJANGOH) event generator.
Charged current (CC) DIS events ($ep \rightarrow \nu_{e}X$), in which
there is real $P_{T}^{\rm{miss}}$ due to the escaping neutrino,
contribute to the background when fake isolated electrons or muons are
observed and is modelled by both H1 and ZEUS using DJANGOH.
Lepton pair production ($ep \rightarrow e \ell^{+}\ell^{-}X$)
contributes to the background via events where one lepton escapes
detection and/or measurement errors cause apparent missing transverse
momentum and is modelled by both H1 and ZEUS using the GRAPE event
generator.

%%%

In order to reject the NC background contribution in the electron
channel, further cuts are applied on the calorimetric energy
imbalance, $V_{\rm ap}/V_{\rm p}$~\cite{Adloff:1999ah}, and the
longitudinal momentum imbalance, $\delta_{\rm miss} =2E^{0}_{e}-
\delta$, where $\delta$ is the total $E-P_{z}$ in the event as defined
in equation~\ref{eq:epz} and $E^{0}_{e}$ is the electron beam energy.
In the case of H1, the cut on $\delta_{\rm miss}$ is only performed if the event contains
exactly one electron, which has the same charge as the beam lepton.
A cut on the difference in azimuthal angle between the lepton and the direction
of the hadronic system, $\Delta \phi_{\ell-X} <160^{\circ}$, is used
to reject NC DIS background, which has a back--to--back topology.
Further background rejection in the electron channel is achieved using
${\zeta}^{2}_{e}=4 E_{e}E^{0}_{e} \cos^2 \theta_e/2$, where $E_{e}$ is
the energy of the final state electron\footnote{For NC events,
where the scattered electron is identified as the isolated high
transverse momentum electron, ${\zeta}^{2}_{e}$ is equal
to the four momentum transfer squared $Q^{2}_{e}$, as calculated via
electron method (see section~\ref{sec:kinemethods}).}.
Lepton pair background is removed from the muon channel by rejecting
azimuthally balanced events using $V_{\rm ap}/V_{\rm p}$ and by requiring
$\Delta\phi_{\mu-X} < 170^{\circ}$, as well as by rejecting events
with two or more isolated muons.
Finally, the lepton--neutrino transverse mass:
\begin{equation}
  M_{T}^{\ell\nu} = \sqrt{(P_T^{\rm miss} + P_T^{\ell})^2 - (\vec{p}_T^{\rm miss} + \vec{p}_T^{\ell})^2}
  \label{eq:mt}
\end{equation}
is required to be larger than $10$~GeV in in the H1 analysis to further reject NC
(lepton pair) background in the electron (muon) channel.
Full details of the event selections can be found in the individual
publications~\cite{Aaron:2009wp,Chekanov:2008gn}.

%%%

A total of $53$ events are observed in the H1 analysis, in good
agreement with the SM prediction of $54.1 \pm 7.4$, which is
dominated by the expectation from signal processes of
$40.4 \pm 6.3$~\cite{Aaron:2009wp}.
The data event yield is made up of $39$ events in the electron channel and
$14$ events in the muon channel, compared to a SM prediction of
$43.1 \pm 6.0$ and $11.0 \pm 1.8$ respectively.
Fewer events are observed in the muon channel due to the $P_{T}^{X}$
cut, which is applied due to the $P_{T}^{\rm calo}$ requirement of the
trigger (see section \ref{sec:pid}).

%%%

\begin{figure}
    \includegraphics[width=0.98\columnwidth]{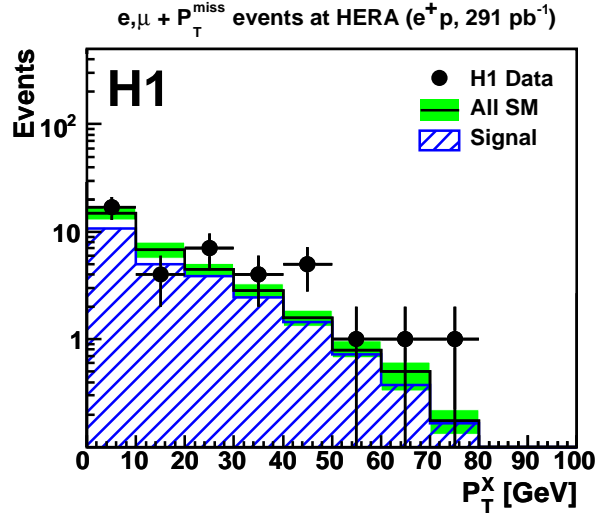}
  \caption{The hadronic transverse momentum $P_{T}^{X}$ distribution
    for the electron and muon channels combined from the H1 analysis 
    of the $e^{+}p$ data sample. The data (points) are compared to the SM
    expectation (open histogram). The signal component of the SM
    expectation, dominated by real $W$ production, is shown as the
    striped histogram. The total uncertainty on the SM expectation is
    shown as the shaded band.}
  \label{fig:isolep-h1ptx}
\end{figure}

For large transverse momentum, $P_{T}^{X}> 25$~ GeV, which is atypical
of SM $W$ production, a total of 18 events are found in the data,
compared to a SM prediction of $13.6 \pm 2.2$, of which $17$ are
found in $e^{+}p$ collisions compared to a SM prediction of $8.0 \pm 1.3$.
The $P_{T}^{X}$ distribution for the combined H1 electron and muon
channels is shown for the $e^{+}p$ data sample in figure~\ref{fig:isolep-h1ptx}. 
The observation of an excess of data events over the SM prediction,
but only in the $e^{+}p$ data, is also a feature of earlier H1
publications~\cite{Adloff:1998aw,Andreev:2003pm}.

%%%

\begin{figure*}
  \centerline{\includegraphics[width=0.80\textwidth]{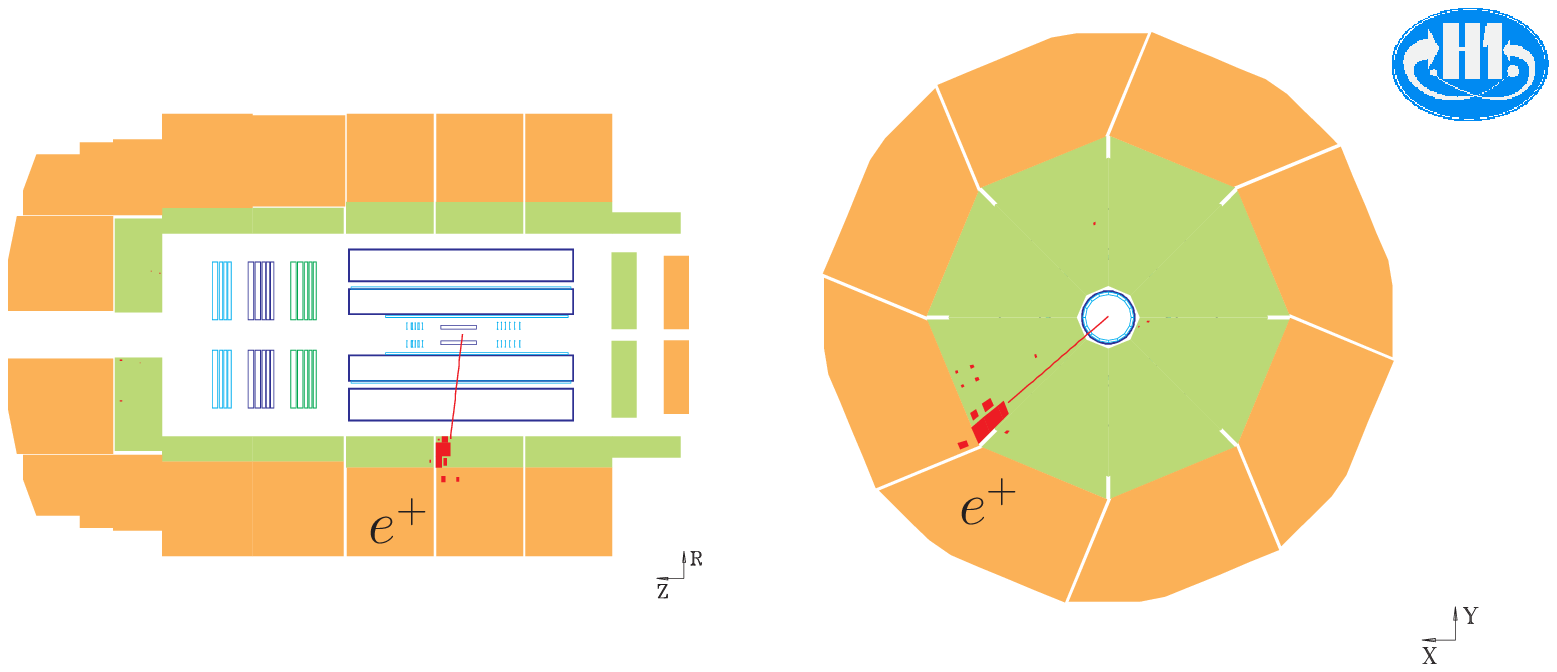}}
  \vspace{0.5cm}
  \centerline{\includegraphics[width=0.80\textwidth]{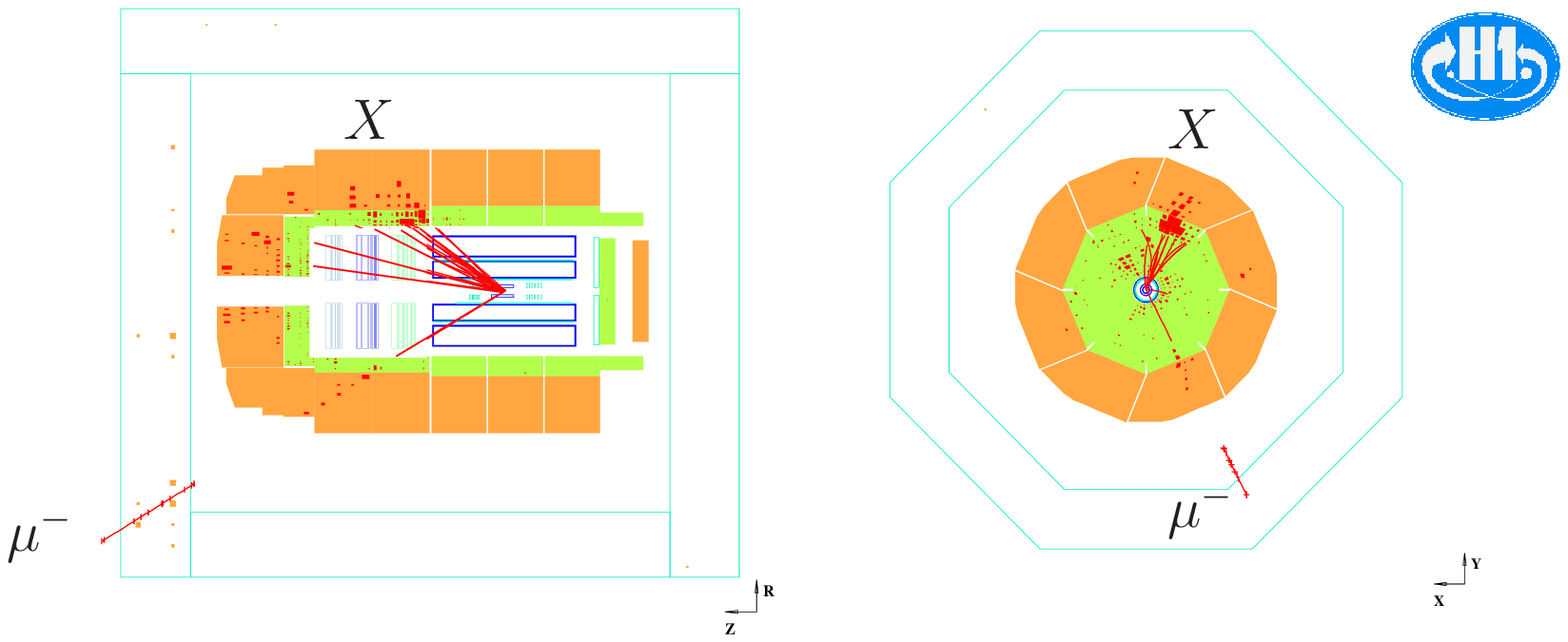}}
  \vspace{0.5cm}
  \centerline{\includegraphics[width=0.80\textwidth]{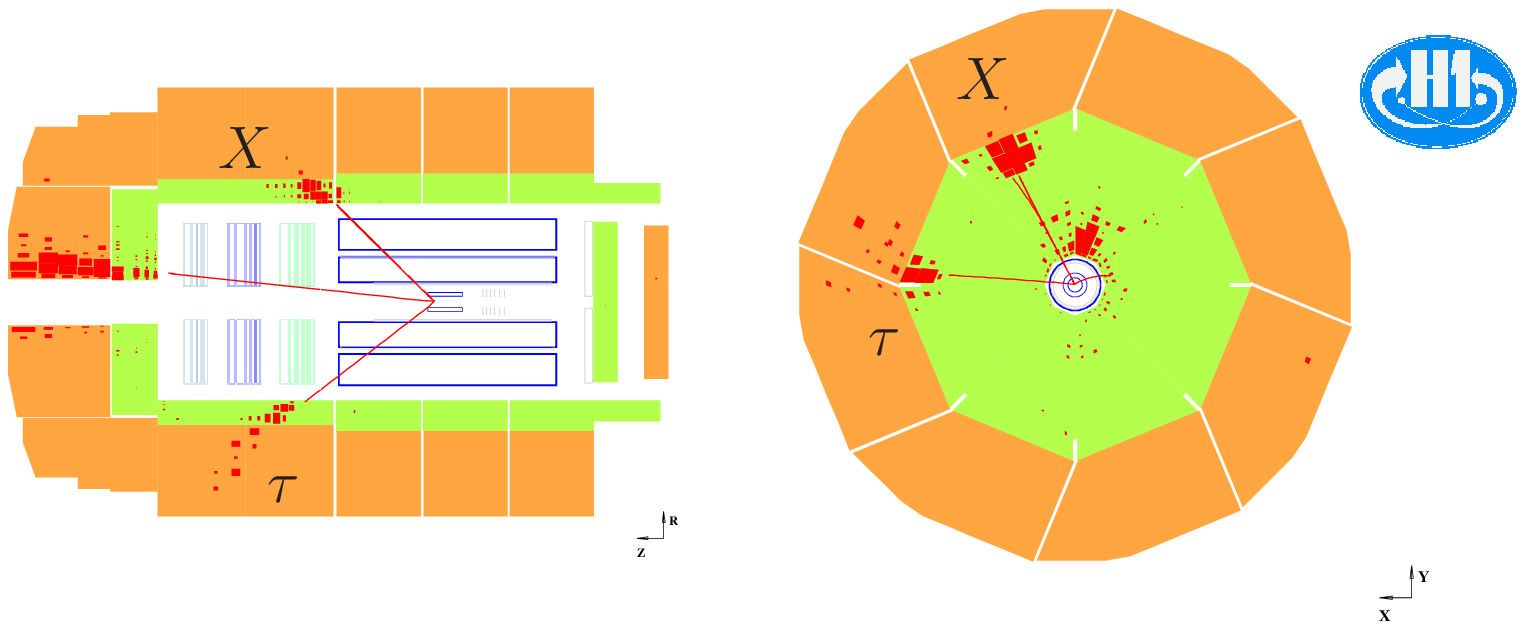}}
  \caption{Event displays in the H1 search for events with and
    isolated lepton and missing transverse momentum. Top: An electron event with missing
    transverse momentum and no visible hadronic final state, typical of
    low $P_{T}$ single $W$ production. Middle: An isolated muon event, with missing transverse
    momentum and a prominent hadronic final state. Bottom: An event with
    an isolated tau lepton candidate, missing transverse momentum and a
    prominent hadronic final state.}
  \label{fig:isolep-eventdisplays}
\end{figure*}

A typical $W$ production event observed in the H1 analysis is
displayed in figure~\ref{fig:isolep-eventdisplays} (top), featuring a
single, isolated electron and large missing transverse momentum, which
is is clearly visible in the azimuthal plane.
A further event from the H1 analysis is displayed in
figure~\ref{fig:isolep-eventdisplays} (middle), this time featuring
an isolated muon, large missing transverse momentum and a prominent
hadronic final state with large $P_{T}^{X}$.

%%%

\begin{table}
  \renewcommand{\arraystretch}{1.3}
  \caption{Results of the ZEUS search for events with isolated electrons
    and missing transverse momentum. The number of observed events is
    compared to the SM prediction. The fraction of the SM expectation
    arising from $W$ production is also given.
    The quoted errors contain statistical and systematic
    uncertainties added in quadrature.}
  \label{tab:zeusiolepelmu}
    \begin{tabular*}{1.0\columnwidth}{@{\extracolsep{\fill}} c c c c} 
      \hline
      \multicolumn{4}{@{\extracolsep{\fill}} l}{\bf Search for Events with an Isolated Electron or }\\[-4pt]
      \multicolumn{4}{@{\extracolsep{\fill}} l}{\bf Muon and Missing Transverse Momentum at HERA}\\
      \hline
      \multicolumn{4}{@{\extracolsep{\fill}} l}{\bf ZEUS Analysis}\\
      \hline
      \multicolumn{4}{@{\extracolsep{\fill}}l}{\bf \boldmath $e^{+}p$ collisions (${\mathcal L} = 296$ pb$^{-1}$)}\\
      & Data & Total SM & $W$ production\\
      \hline
      ~~~~~~~$P_T^X < 12$~GeV &  $7$ & $12.6 \pm 1.7$ & $68\%$\\
      $12<P_T^X<25$~GeV & $7$ & ~$6.2 \pm 0.9$ & $75\%$\\
      ~~~~~~~$P_T^X > 25$~GeV & $6$ & ~$7.4 \pm 1.0$ & $79\%$\\
      \hline
      \multicolumn{4}{@{\extracolsep{\fill}}l}{\bf \boldmath $e^{-}p$ collisions (${\mathcal L} = 208$ pb$^{-1}$)}\\
      & Data & Total SM & $W$ production \\
      \hline
      ~~~~~~~$P_T^X < 12$~GeV &  $9$ & $11.3 \pm 1.5$ & $54\%$\\
      $12<P_T^X<25$~GeV & $6$ & ~$5.1 \pm 0.7$ & $67\%$\\
      ~~~~~~~$P_T^X > 25$~GeV & $5$ & ~$5.5 \pm 0.8$ & $75\%$\\
      \hline
      \multicolumn{4}{@{\extracolsep{\fill}}l}{\bf \boldmath $e^{\pm}p$ collisions (${\mathcal L} = 506$ pb$^{-1}$)}\\
      & Data & Total SM & $W$ production\\
      \hline
      ~~~~~~~$P_T^X < 12$~GeV &  $16$ & $23.9 \pm 3.1$ & $61\%$\\
      $12<P_T^X<25$~GeV & $13$ & $11.2 \pm 1.5$ & $71\%$\\
      ~~~~~~~$P_T^X > 25$~GeV & $11$ & $12.9 \pm 1.7$ & $77\%$\\
      \hline
    \end{tabular*}
\end{table}

The results of the ZEUS analysis are summarised in
table~\ref{tab:zeusiolepelmu}, for the electron and muon
channels combined.
In the full phase space a total of 40 events are observed in the data
broadly in agreement with the SM prediction~\cite{Chekanov:2008gn}.
For large hadronic transverse momentum, $P_{T}^{X}> 25$~ GeV, 
$11$ events are found in the data, compared to a SM prediction of
$12.9 \pm 1.7$.
Unlike the H1 analysis, no excess is observed in this region, where
for the ZEUS $e^{+}p$ data $6$ events are seen compared to a
SM prediction of $7.4 \pm 1.0$.

%%%

A combination of the H1 and ZEUS isolated lepton and missing
transverse momentum analyses has also been performed, using the full
HERA data sample corresponding to an integrated luminosity
of $0.98$~fb$^{-1}$ comprising $0.39$~fb$^{-1}$ of $e^{-}p$ collisions and
$0.59$~fb$^{-1}$ of $e^{+}p$ collisions.
A study of the selection efficiency for signal processes found the
H1 and ZEUS analyses to be compatible in the kinematic region where they
are directly comparable~\cite{Diaconu:2006qs}.
Nevertheless, in order to coherently combine the results from the two
experiments, a common phase space is established~\cite{Aaron:2009ab}.
The lepton polar angle acceptance is the main difference in the
common phase space with respect to the H1 analysis, where
the more restricted range of the ZEUS analysis is used,
$15^{\circ} < \theta_{\ell} < 120^{\circ}$.
Additionally, the more restrictive cuts on
$\delta_{\rm miss}$ and $V_{\rm ap}/V_{\rm p}$ are taken from the ZEUS
analysis~\cite{Chekanov:2008gn}.
The minimum lepton--neutrino transverse mass and electron multiplicity
requirements are taken from the H1 analysis~\cite{Aaron:2009wp}.

%%%

The results of the combined H1 and ZEUS analysis are summarised in
table~\ref{tab:isolep-h1zrates}.
The signal contribution, mainly from real $W$ production, is seen to
dominate the total SM expectation in all data samples.
A total of 81 events are observed in the data, compared to a SM
expectation of $87.8 \pm 11.0$.
At large hadronic transverse momentum $P_{T}^{X} > 25$~GeV a total of
$29$ events are observed in the $e^{\pm}p$ data compared to a
SM prediction of $24.0 \pm 3.2$.
In the $e^{+}p$ data alone, $23$ events are observed with $P_{T}^{X} >$ 25~GeV
compared to a SM prediction of $14.0 \pm 1.9$.
Seventeen of these data events are the same events observed in the
standard H1 analysis~\cite{Aaron:2009wp}, but now compared to a
lower SM expectation of $6.7 \pm 0.9$ in the more restricted
common phase space.
Whilst this intriguing excess in the $e^{+}p$ analysis remains in the
common phase space, it also remains a feature only seen in the
H1 data.

%%%

\begin{figure*}
  \centerline{
    \includegraphics[width=0.475\textwidth]{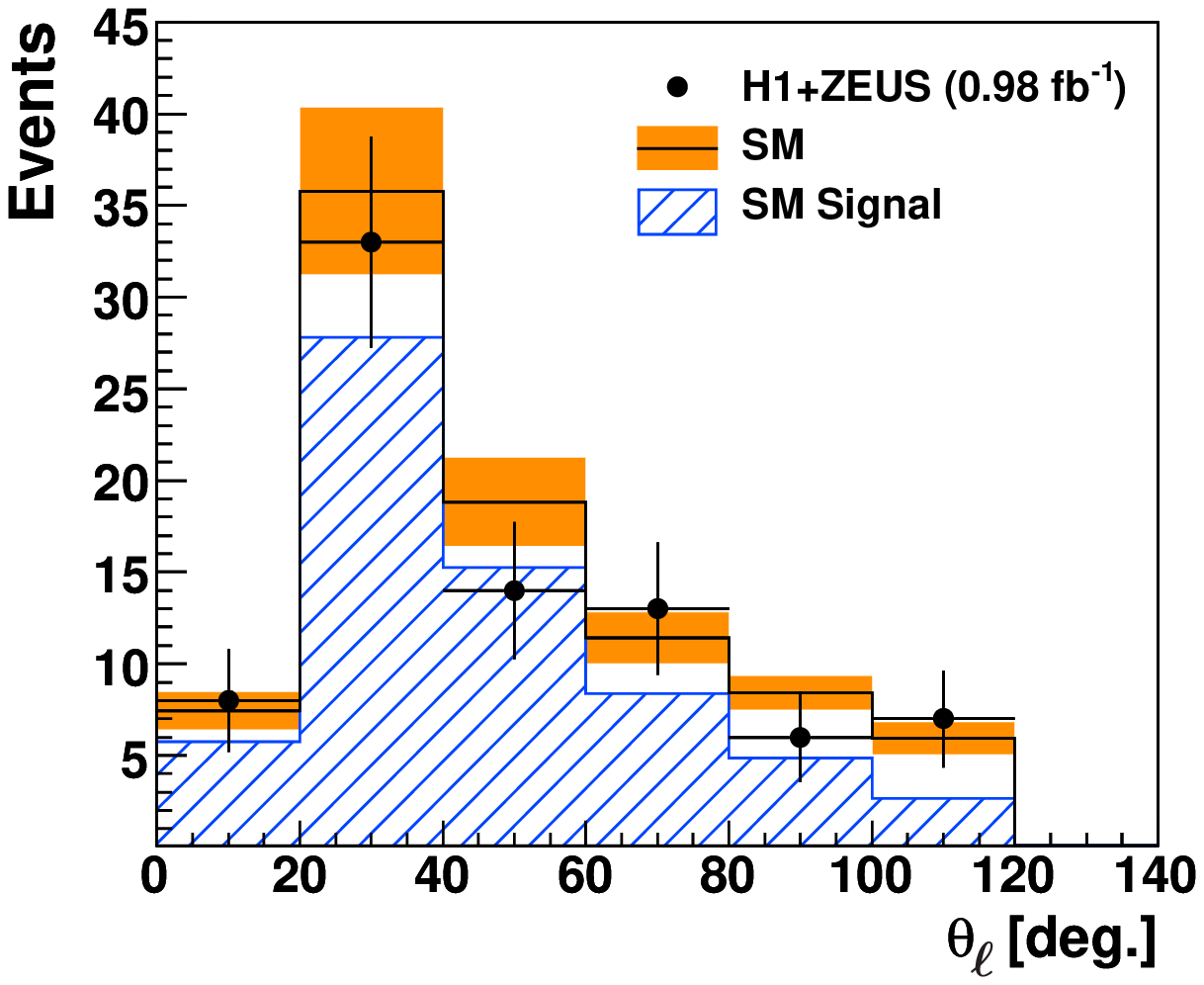}
    \includegraphics[width=0.475\textwidth]{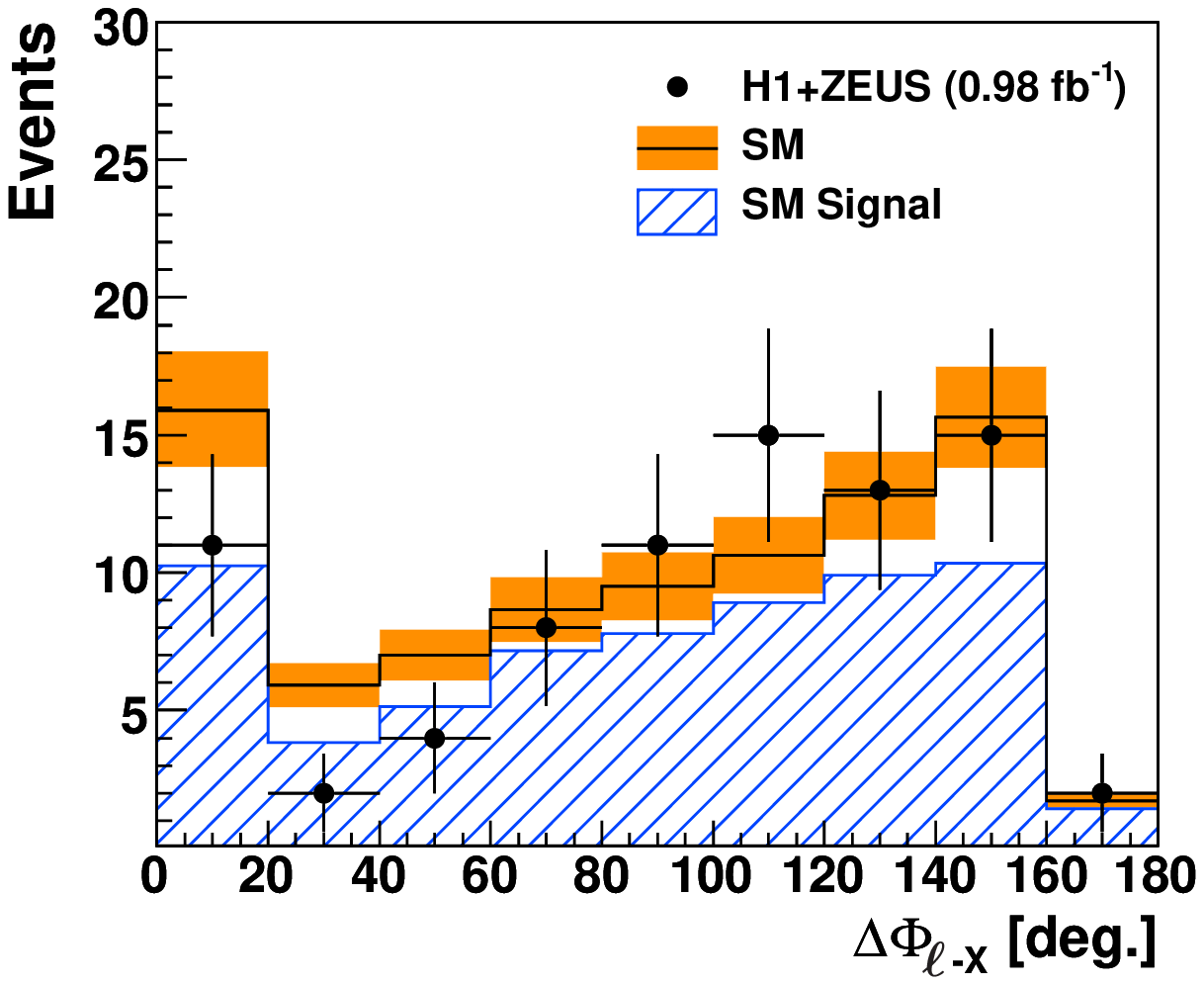}
  }
  \vspace{-0.2cm}
  \centerline{
    \includegraphics[width=0.475\textwidth]{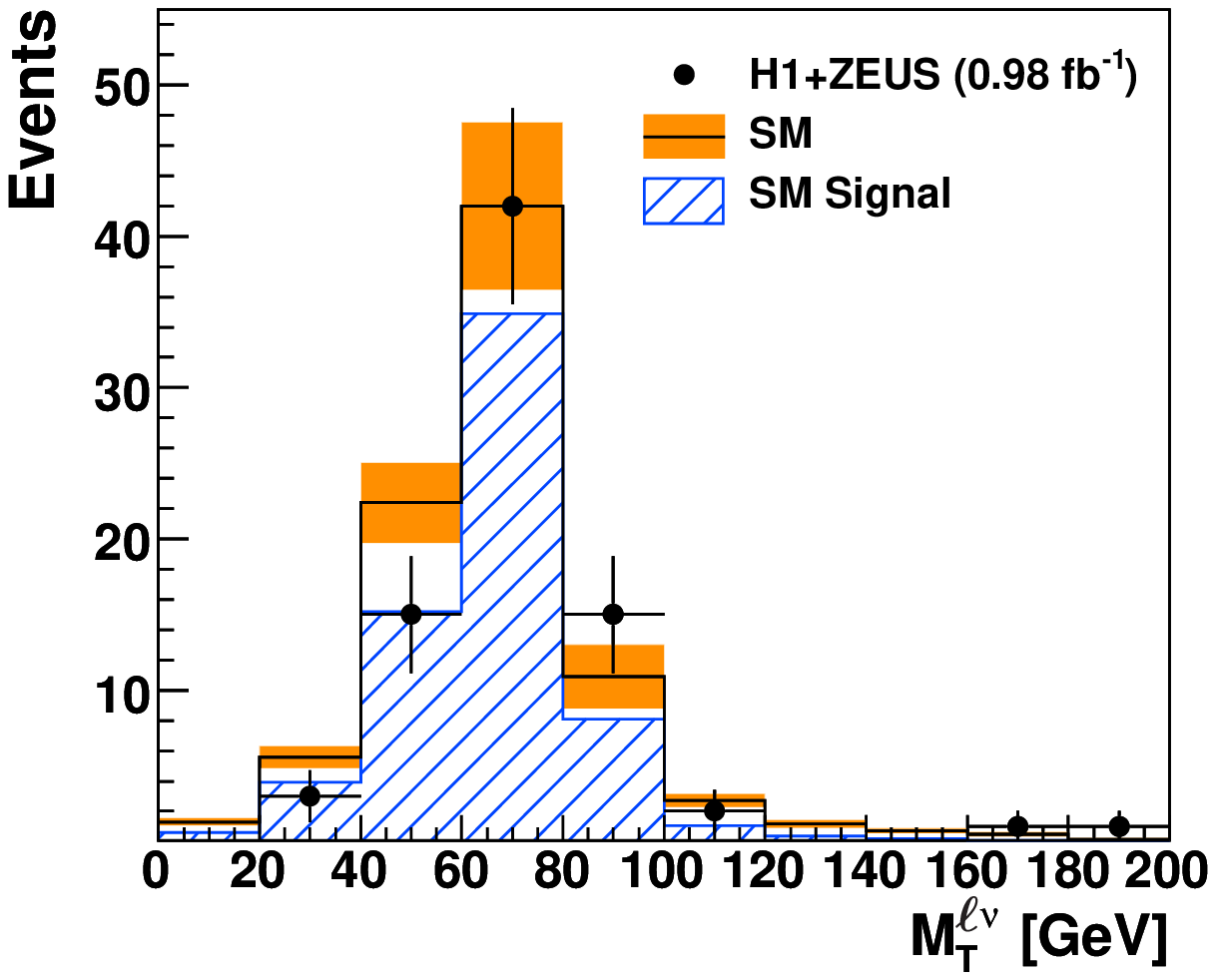}
    \includegraphics[width=0.475\textwidth]{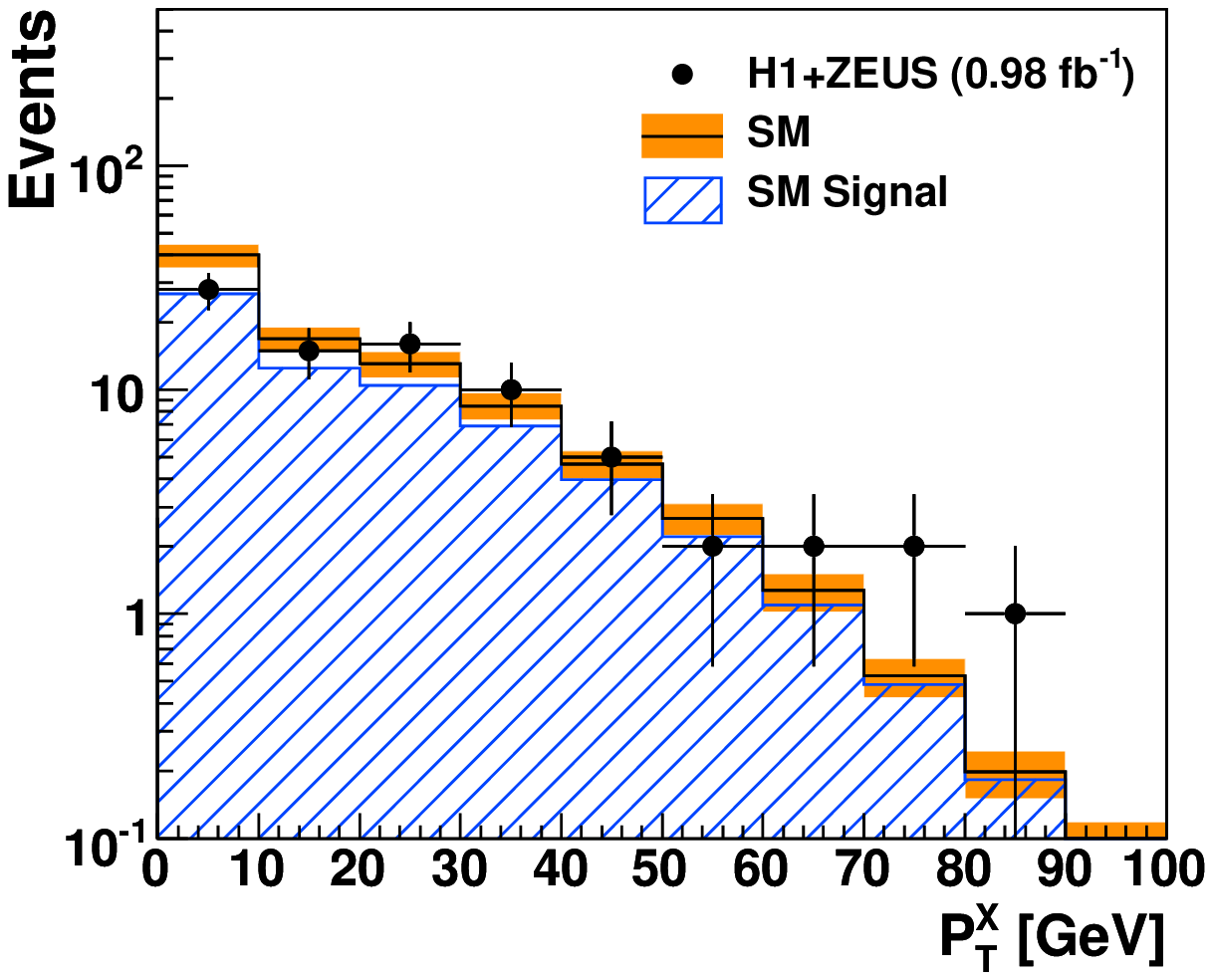}
  }
  \vspace{-0.2cm}
  \centerline{
    \includegraphics[width=0.475\textwidth]{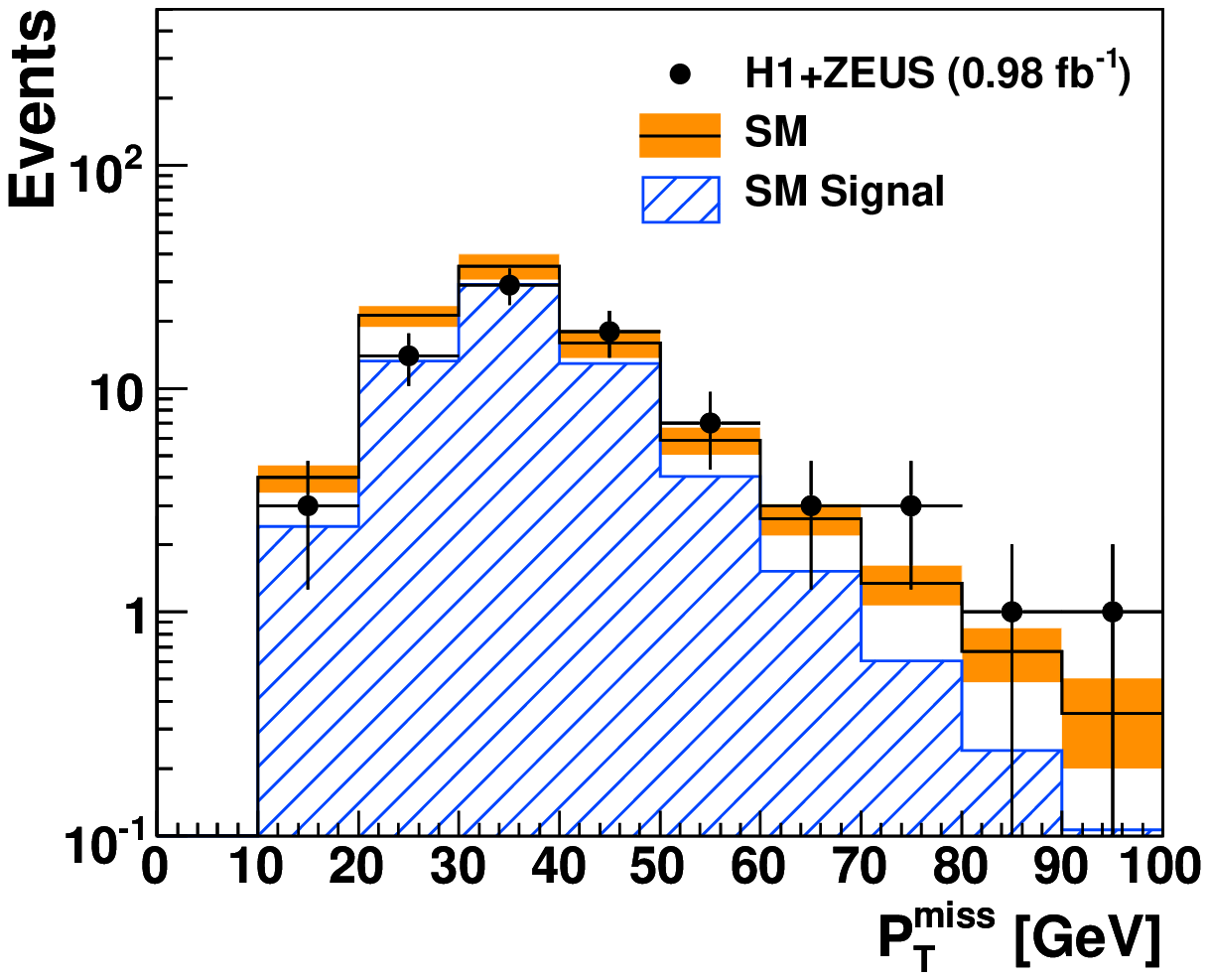}
    \includegraphics[width=0.475\textwidth]{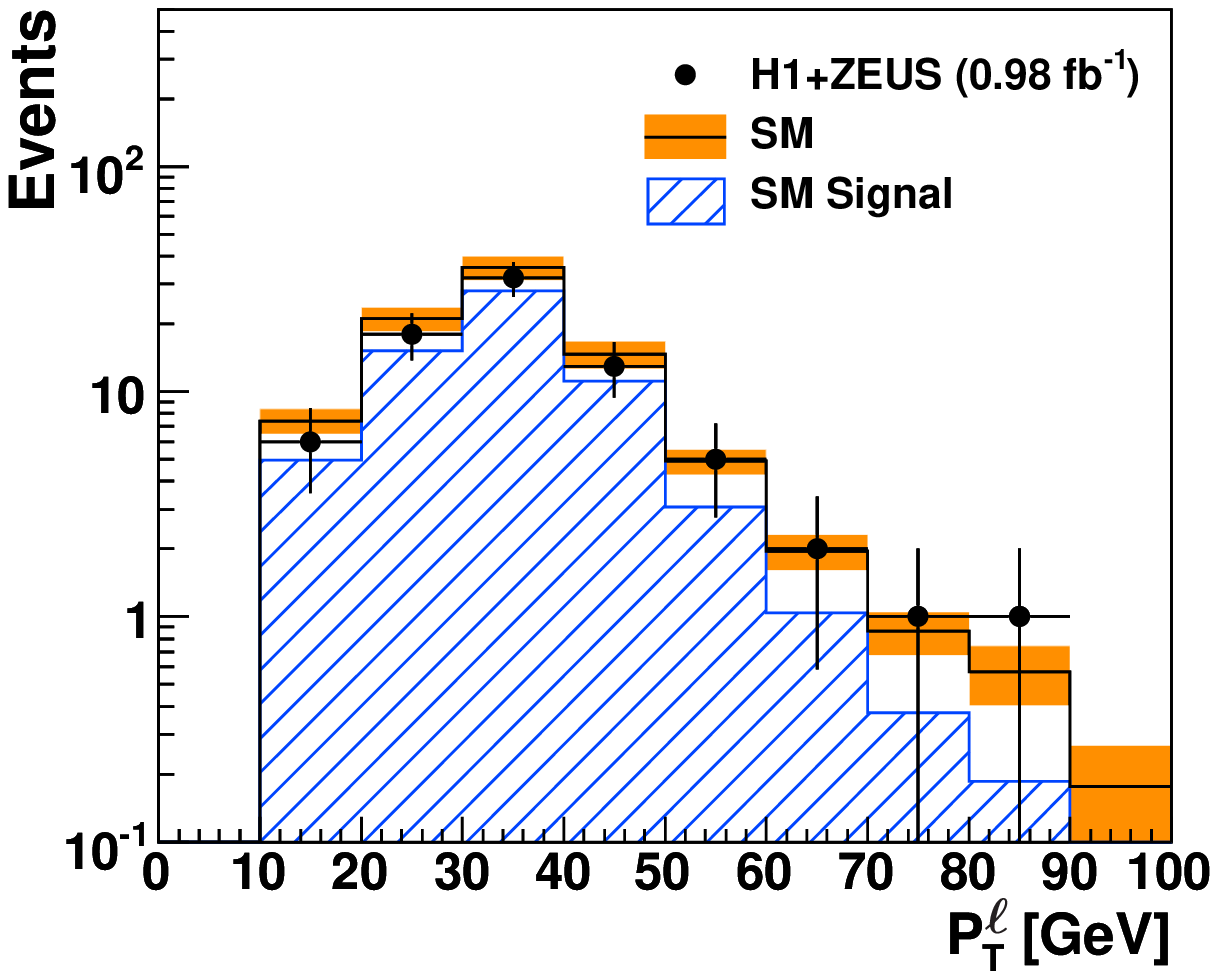}
  }
  \begin{picture} (0.,0.) 
    \setlength{\unitlength}{1.0cm}
    \put (1.9,19.6){\bf\normalsize  (a)} 
    \put (10.7,19.6){\bf\normalsize  (b)} 
    \put (1.9,12.9){\bf\normalsize  (c)} 
    \put (10.7,12.9){\bf\normalsize  (d)} 
    \put (1.9,6){\bf\normalsize  (e)} 
    \put (10.7,6){\bf\normalsize  (f)} 
  \end{picture} 
  \caption{Distributions of kinematic variables of events with an
    isolated electron or muon and missing transverse momentum in the
    full HERA $e^{\pm}p$ data. Shown are: the polar angle of the lepton
    $\theta_{\ell}$~(a), the difference in the azimuthal angle of the
    lepton and the hadronic systems $\Delta\phi_{\ell-X}$~(b), the
    lepton--neutrino transverse mass $M_{T}^{\ell\nu}$~(c), the hadronic
    transverse momentum $P_{T}^{X}$~(d), the missing transverse momentum
    $P_{T}^{\rm miss}$~(e) and the transverse momentum of the lepton
    $P_{T}^{\ell}$~(f). The data (points) are compared to the SM
    expectation (open histogram). The signal component of the SM
    expectation, dominated by single $W$ production, is shown as the
    striped histogram. The total uncertainty on the SM expectation is
    shown as the shaded band.}
  \label{fig:isolep-h1zfinalsample}
\end{figure*}

Figure~\ref{fig:isolep-h1zfinalsample} shows a variety of kinematic
distributions of the H1 and ZEUS combined analysis for the complete
$e^{\pm}p$ HERA I+II data, for the electron and muon channels together.
The shape and normalisation of the distributions are well described
within the uncertainties.
The distribution of events in $M_{T}^{l\nu}$ is compatible with the
Jacobian peak expected from $W$ production.
Similarly, the observed $P_{T}^{X}$ spectrum is compatible with that
expected from $W$ production, peaking at low values of hadronic
transverse momentum.

\begin{table*}
  \renewcommand{\arraystretch}{1.3}
  \caption{Summary of the combined H1 and ZEUS search for events with
    an isolated electron or muon and missing transverse momentum for the
    $e^{+}p$ data (top), $e^{-}p$ data (middle) and the full HERA data
    set (bottom). The results are shown for the full selected sample and
    for the subsample with hadronic transverse momentum
    $P_{T}^{X}>25$~GeV. The number of observed events is compared to the
    SM prediction. The SM signal (dominated by single $W$ production)
    and the total background contribution (mainly NC and CC DIS,
    together with lepton-pair production) are also shown. The quoted
    uncertainties contain statistical and systematic uncertainties
    added in quadrature.}
  \label{tab:isolep-h1zrates}
  \centerline{
  \begin{tabular*}{1.0\textwidth}{@{\extracolsep{\fill}} c c c c c c }
    \hline
    \multicolumn{6}{@{\extracolsep{\fill}} l}{\bf Search for Events with an
    Isolated Electron or Muon and Missing Transverse Momentum at HERA}\\
    \hline
    \multicolumn{6}{@{\extracolsep{\fill}} l}{\bf Combined H1 and ZEUS Analysis}\\
    \hline
    {\bf \boldmath $e^{+}p$ collisions (${\mathcal L} = 0.59$ fb$^{-1}$)} & & & & &\\
    %\hline
    Channel &  & Data & Total SM & SM signal & Other SM\\
    \hline
    Electron & Total   & $37$ & $38.6  \pm 4.7$ & $28.9 \pm 4.4$ &  $9.7 \pm 1.4$ \\
    & $P^X_T > 25$ GeV & $12$ &  $7.4  \pm 1.0$ &  $6.0 \pm 0.9$ & $1.5 \pm 0.3$ \\
    \hline
    Muon     & Total   & $16$ & $11.2 \pm 1.6$ &  $9.9 \pm 1.6$ &  $1.3 \pm 0.3$ \\
    & $P^X_T > 25$ GeV & $11$ &  $6.6 \pm 1.0$ &  $5.9 \pm 0.9$ &  $0.8 \pm 0.2$ \\
    \hline
    Combined & Total   & $53$ & $49.8 \pm 6.2$ & $38.8 \pm 5.9$ & $11.1 \pm 1.5$ \\
    & $P^X_T > 25$ GeV & $23$ & $14.0 \pm 1.9$ & $11.8 \pm 1.9$ &  $2.2 \pm 0.4$ \\
    \hline
    {\bf \boldmath $e^{-}p$ collisions (${\mathcal L} = 0.39$ fb$^{-1}$)} & & & & &\\
    %\hline
    Channel &  & Data & Total SM & SM signal & Other SM\\
    \hline
    Electron & Total   & $24$ & $30.6 \pm 3.6$ & $19.4 \pm 3.0$ & $11.2 \pm 1.9$ \\
    & $P^X_T > 25$ GeV &  ~~$4$ &  $5.6 \pm 0.8$ &  $4.0 \pm 0.6$ &  $1.6 \pm 0.4$ \\
    \hline
    Muon     & Total   &  ~~$4$ &  $7.4 \pm 1.1$ &  $6.6 \pm 1.0$ &  $0.9 \pm 0.3$ \\
    & $P^X_T > 25$ GeV &  ~~$2$ &  $4.3 \pm 0.7$ &  $3.9 \pm 0.6$ &  $0.4 \pm 0.2$ \\
    \hline
    Combined & Total   & $28$ & $38.0 \pm 3.4$ & $26.0 \pm 3.4$ & $12.0 \pm 2.0$ \\
    & $P^X_T > 25$ GeV &  ~~$6$ & $10.0 \pm 1.3$ &  $7.9 \pm 1.2$ &  $2.1 \pm 0.5$ \\
    \hline
    {\bf \boldmath $e^{\pm}p$ collisions (${\mathcal L} = 0.98$ fb$^{-1}$)} & & & & &\\
    %\hline
    Channel &  & Data & Total SM & SM signal & Other SM\\
    \hline
    Electron & Total   & $61$ & $69.2 \pm 8.2$ & $48.3 \pm 7.4$ & $20.9 \pm 3.2$ \\
    & $P^X_T > 25$ GeV & $16$ & $13.0 \pm 1.7$ & $10.0 \pm 1.6$ &  $3.1 \pm 0.7$ \\
    \hline
    Muon     & Total   & $20$ & $18.6 \pm 2.7$ & $16.4 \pm 2.6$ &  $2.2 \pm 0.5$ \\
    & $P^X_T > 25$ GeV & $13$ & $11.0 \pm 1.6$ &  $9.8 \pm 1.6$ &  $1.2 \pm 0.3$ \\
    \hline
    Combined & Total   & $81$ & $87.8 \pm 11.0$  & $64.7 \pm 9.9$ & $23.1 \pm 3.3$ \\
    & $P^X_T > 25$ GeV & $29$ & $24.0\pm 3.2$  & $19.7 \pm 3.1$ &  $4.3 \pm 0.8$ \\
    \hline
  \end{tabular*}
}
\end{table*}

A measurement of the visible cross section for the isolated lepton and
missing transverse momentum topology in $e^{\pm}p$ collisions is performed
by H1 using the electron and muon channels in the phase space
$5^{\circ}<\theta_{\ell}<140^{\circ}$, $P_{T}^{\ell}>10$~GeV,
$P_{T}^{\rm miss}>12$~GeV and $D_{\rm jet}>1.0$ at a centre of
mass energy\footnote{Assuming a linear dependence of the cross section
on the proton beam energy.} of $\sqrt{s}=317$~GeV.
The cross section is calculated using equation~\ref{eq:xsection},
where the EPVEC generator is used to calculate the acceptance,
$\mathcal{A}$, which is predicted to be about the same
for $e^{+}p$ and $e^{-}p$ collisions.
The total visible cross section for events with an isolated
lepton and missing transverse momentum is measured by H1 as:
\[
\sigma_{\ell+P_{T}^{\rm miss}} = 0.23 \pm 0.05~({\rm stat.}) \pm 0.04~({\rm sys.})~{\rm pb},
\]
where the first error is statistical and the second systematic, in
agreement with the SM NLO value of $0.25 \pm 0.04$~pb from EPVEC.

%%%

The single $W$ boson production cross section is measured by H1
and ZEUS, individually and in the combined analysis described above.
The branching ratio corresponding to the leptonic $W$ boson decay to
any final state with an electron or muon, including the contribution
from leptonic tau-decay, is also included in
the calculation.
This cross section is also calculated using equation~\ref{eq:xsection}, where
$\mathcal{A}$ is again calculated using EPVEC but is now defined
with respect to the full phase space and the contribution from $Z^{0}$
production illustrated in figure~\ref{fig:isolep-feynman}(d) is
considered as background.
The acceptances for the two experiments are found to be similar in
each $P_{T}^{X}$ bin and vary between $27\%$ and $37\%$ in the
electron channel and between $18\%$ and $38\%$ in the
muon channel~\cite{Aaron:2009ab}.

%%%

\begin{figure}
  \includegraphics[width=0.98\columnwidth]{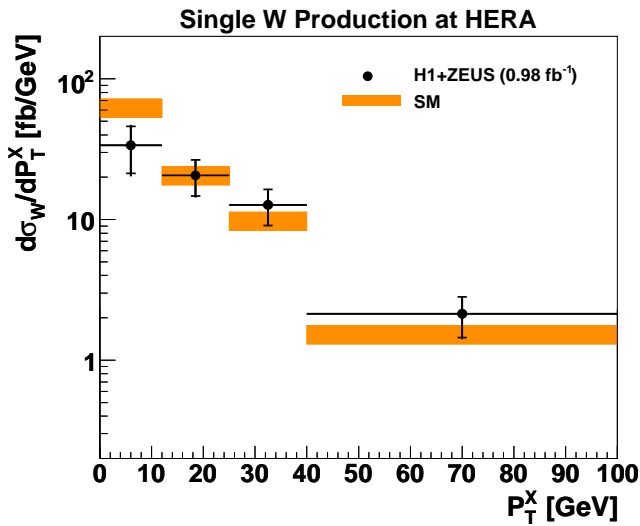}
  \caption{The single $W$ production cross section as a function of
  the hadronic transverse momentum, $P_{T}^{X}$, measured using the
  combined H1 and ZEUS data at a centre of mass energy of
  $\sqrt{s}=317$~GeV. The inner error bar represents the statistical
  error and the outer error bar indicates the statistical and
  systematic uncertainties added in quadrature. The shaded band
  represents the uncertainty on the SM prediction.}
\label{fig:isolep-h1zxs}
\end{figure}

The combined H1-ZEUS single $W$ production cross section,
evaluated using a weighted mean of the values measured by the two
collaborations in the common phase space, is measured as:
\[
\sigma_{W} = 1.06 \pm 0.16~{\rm (stat.)} \pm 0.07~{\rm (sys.)}~{\rm pb},
\]
where the first uncertainty is statistical and the second systematic.
The measurement agrees well with the NLO SM prediction of $1.26 \pm
0.19$~pb from EPVEC and the measurements by H1 and ZEUS in their
individual publications~\cite{Aaron:2009wp,Chekanov:2008gn}.
The single $W$ boson production cross section is also measured
differentially as a function of $P_{T}^{X}$, the results of which are
displayed in figure~\ref{fig:isolep-h1zxs}, and is also found to
be in agreement with the SM prediction.

%%%

Some additional investigations into $W$ bosons are performed by H1
using the analysis of their full data set~\cite{Aaron:2009wp}.
The production of single $W$ bosons at HERA is sensitive to anomalous
triple gauge boson couplings~\cite{Baur:1989gh}  via the process
illustrated in figure~\ref{fig:isolep-feynman}~(b), which can be
parametrised using two free coupling parameters, $\kappa$ and
$\lambda$~\cite{Hagiwara:1986vm}.
In the SM $\kappa=1$ and $\lambda=0$ at tree level and in the
following $\Delta\kappa \equiv \kappa-1$ is used, such that
any non--zero value for $\Delta\kappa$ or $\lambda$ represents a
deviation from the SM.
The hadronic transverse momentum spectrum of $W$ events
is expected to be sensitive to anomalous values of $\Delta\kappa$
and $\lambda$~\cite{Baur:1989gh}, and a likelihood
analysis on the measured $P_{T}^{X}$ distribution is performed using a Bayesian
approach and employing Poisson statistics~\cite{Aaron:2009wp}.
This is done for $\Delta\kappa$ and $\lambda$ separately, keeping the
other parameter fixed to its SM value.
The following limits are derived at $95\%$~CL:
\[
-0.7 < \Delta\kappa < 1.4,
\]
\[
-2.5 < \lambda < 2.5.
\]
in good agreement with the SM prediction and
since the value of $\Delta\kappa=-1$ is excluded at $95\%$~CL, the
results obtained explicitly demonstrate the presence of a magnetic
coupling of the photon to the $W$ boson, in addition to the coupling
to the electric charge of the $W$ boson.
The most stringent limits on these couplings were obtained by the LEP
experiments in single $\gamma$, single $W$ and $W$ pair
production~\cite{Schael:2004tq,Abdallah:2008sf,Achard:2004ji,Abbiendi:2003mk}. 

%%%

H1 also performs a measurement of the $W$ boson polarisation fractions
at HERA by examining the \cosths\ distribution in the decay
$W\rightarrow e/\mu+\nu$, where $\theta^{*}$ is defined as the angle
between the $W$ boson momentum in the lab frame and the charged decay
lepton in the $W$ boson rest frame.
The \cosths\ distributions for $W^{+}$ bosons are given by~\cite{Hagiwara:1986vm}:
\begin{eqnarray}
\frac{dN}{d\cos\theta^{*}} 
&\propto& \left( 1 - F_{-} - F_{0} \right) \cdot \frac{3}{8} \left( 1 + \cos\theta^{*}\right)^{2} \nonumber\\
&+&       F_{0} \cdot \frac{3}{4} \left( 1 - \cos^{2}\theta^{*}\right) \\
&+&       F_{-} \cdot \frac{3}{8} \left( 1 - \cos\theta^{*}\right)^{2}, \nonumber
\label{eqn:wpolmodel}
\end{eqnarray}
where the left handed $F_{-}$, longitudinal $F_{0}$ and right handed $F_{+}$ polarisation
fractions are constrained by the relation $F_{+} \equiv 1 - F_{-} - F_{0}$.
%
%%%
%
Beginning with the H1 isolated electron or muon and missing transverse
momentum event sample, the additional requirement of a reliable
charge measurement is added, so that the resulting charge
misidentification is well below $1\%$~\cite{Aaron:2009wp}.
The final event sample consists of $21$ electron events and $9$ muon
events and a SM prediction with a $W$ production purity of $76\%$.
To allow the combination of the different $W$ boson charges, the value
of \cosths\ is multiplied by the sign of the lepton charge $q_{\ell} =
\pm 1$ and the measured distribution, corrected for acceptance and
detector effects.

%%%

\begin{figure}
  \includegraphics[width=0.98\columnwidth]{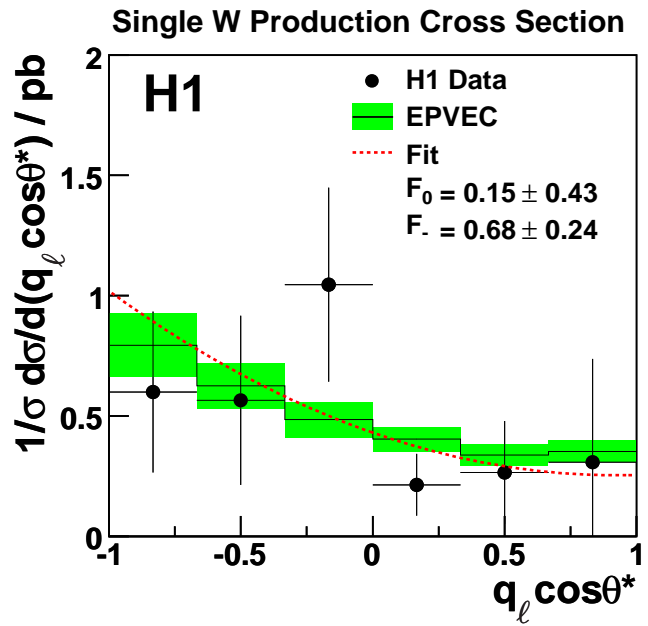}
  \caption{The H1 measured normalised differential cross section
    $1/\sigma$~$d\sigma/d\left(\qcosths\right)$ (points) as a function of
    $\qcosths$ for on--shell $W$ bosons. The EPVEC prediction is also
    shown (open histogram) with a $15\%$ theoretical uncertainty shown
    by the band. The result of the simultaneous fit of the $W$
    polarisation fractions is shown as the dashed histogram.}
  \label{fig:isolep-h1wpol}
\end{figure}

The normalised differential cross section as a function of \qcosths\
is shown in figure~\ref{fig:isolep-h1wpol}, compared to the SM prediction.
The cross section fit to the model defined in
equation~\ref{eqn:wpolmodel} is also shown.
The optimal values for the $W$ boson polarisation fractions $F_{-}$
and $F_{0}$ are simultaneously extracted from the fit as $0.68 \pm
0.24$ and $0.15 \pm 0.43$, respectively, in good agreement with the SM
values $0.62$ and $0.17$ as predicted by EPVEC.

\subsection{Events with isolated tau-leptons}
\label{sec:isolep:singletau}

The search for isolated tau leptons complements the analysis of the
electron and muon channels described in section~\ref{sec:isoelmu}.
This analysis has previously been performed by H1 and ZEUS
using their HERA~I data sets, where the tau is identified by its
hadronic decay~\cite{Aktas:2006fc,Chekanov:2003bf}.
The H1 collaboration have updated their search to include their full
HERA data set, corresponding to an integrated luminosity
of $474$~pb$^{-1}$~\cite{Aaron:2009wp}.

%%%

The selection of H1 events with isolated tau leptons and missing
transverse momentum is based on the HERA-I analysis~\cite{Aktas:2006fc}.
Only hadronic decays with one charged hadron (one-prong) are considered.
Tau decays to electrons and muons enter the electron and muon channels
described in section~\ref{sec:isoelmu}.

%%%

A tau identification algorithm selects narrow, low multiplicity jets typical
for hadronic tau decays, with transverse momentum $P_{T}^{\rm jet}>7$~GeV
in the central region $20^{\circ}<\theta_{\rm jet}<120^{\circ}$ of the detector.
Narrow jets are selected by requiring $R_{\rm jet}<0.12$
(see equation \ref{eq:taujet}).
At least one track measured in the central tracking detector with
transverse momentum $P_{T}^{\rm track}>5$ GeV is required to be
associated to the jet.
In the final selection, the tau jet is required to be isolated from
electrons, muons and other jets in the event by requiring a
minimum angular separation in $\eta-\phi$ space of $D>1.0$.

%%%

To ensure the presence of neutrinos in the event, large $P_{T}^{\rm
miss}> 12$~GeV and $P_{T}^{\rm calo}>12$~GeV, a significant azimuthal imbalance
$\delta_{\rm miss} =2E^{0}_{e}- \delta > 5$~GeV and low
$V_{\rm ap}/V_{\rm p} < 0.5$ are required.
The event has also to exhibit large inclusive hadronic transverse
momentum $P_{T}^{X}>12$~GeV.
Some degree of acoplanarity between the tau jet and the remaining
hadronic system $X'$ in the transverse plane
$\Delta\phi_{\tau-X'} < 170^{\circ}$ is required to suppress events
with back--to--back topologies, primarily NC events and photoproduction
events with jets.

%%%

\begin{table}
  \renewcommand{\arraystretch}{1.3}
  \caption{Summary of the H1 results of the search for events with tau
  leptons and missing transverse momentum for the $e^{+}p$ data, $e^{-}p$ data and
  the full H1 data set. The results are shown for the full selected sample
  and for the subsample at $P_{T}^{X}>25$~GeV. The number of observed
  events is compared to the SM prediction. The SM signal
  ($W\rightarrow\tau\nu_{\tau}$) and the background contributions are
  also shown. The quoted uncertainties contain statistical and systematic
  uncertainties added in quadrature.}
  \label{tab:isotau-h1rates}
  \begin{tabular*}{1.0\columnwidth}{@{\extracolsep{\fill}} c c c c c }
    \hline
    \multicolumn{5}{@{\extracolsep{\fill}} l}{\bf H1 Search for Events with an Isolated Tau-lepton}\\[-4pt]
    \multicolumn{5}{@{\extracolsep{\fill}} l}{\bf and Missing Transverse Momentum at HERA}\\
    \hline
    \multicolumn{5}{@{\extracolsep{\fill}} l}{\bf \boldmath $e^{+}p$ collisions (${\mathcal L} = 291$ pb$^{-1}$)}\\
    & Data & Total SM & SM signal & Other SM\\
    \hline
    Total  & ~~$9$ & $12.3  \pm 2.0$ & $1.66 \pm 0.25$ &  $10.6 \pm 1.8$ \\
    $P^X_T > 25$ GeV & ~~$0$ &  $0.82  \pm 0.12$ &  $0.38 \pm 0.06$ & $0.44 \pm 0.06$ \\
    \hline
    \multicolumn{5}{@{\extracolsep{\fill}} l}{\bf \boldmath $e^{-}p$ collisions (${\mathcal L} = 183$ pb$^{-1}$)}\\
    & Data & Total SM & SM signal & Other SM\\
    \hline
    Total   & ~~$9$ & $11.0 \pm 1.9$ & $1.00 \pm 0.15$ & $10.0 \pm 1.8$ \\
    $P^X_T > 25$ GeV &  ~~$1$ &  $0.68 \pm 0.11$ &  $0.21 \pm 0.03$ &  $0.47 \pm 0.07$ \\
    \hline
    \multicolumn{5}{@{\extracolsep{\fill}} l}{\bf \boldmath $e^{\pm}p$ collisions (${\mathcal L} = 474$ pb$^{-1}$)}\\
    & Data & Total SM & SM signal & Other SM\\
    \hline
    Total   & $18$ & $23.2 \pm 3.8$ & $2.66 \pm 0.40$ & $20.6 \pm 3.4$ \\
    $P^X_T > 25$ GeV & ~~$1$ & $1.50 \pm 0.21$ & $0.59 \pm 0.09$ &  $0.91 \pm 0.12$ \\
    \hline
  \end{tabular*}
\end{table}

\begin{figure*}
  \centerline{
    \includegraphics[width=0.98\columnwidth]{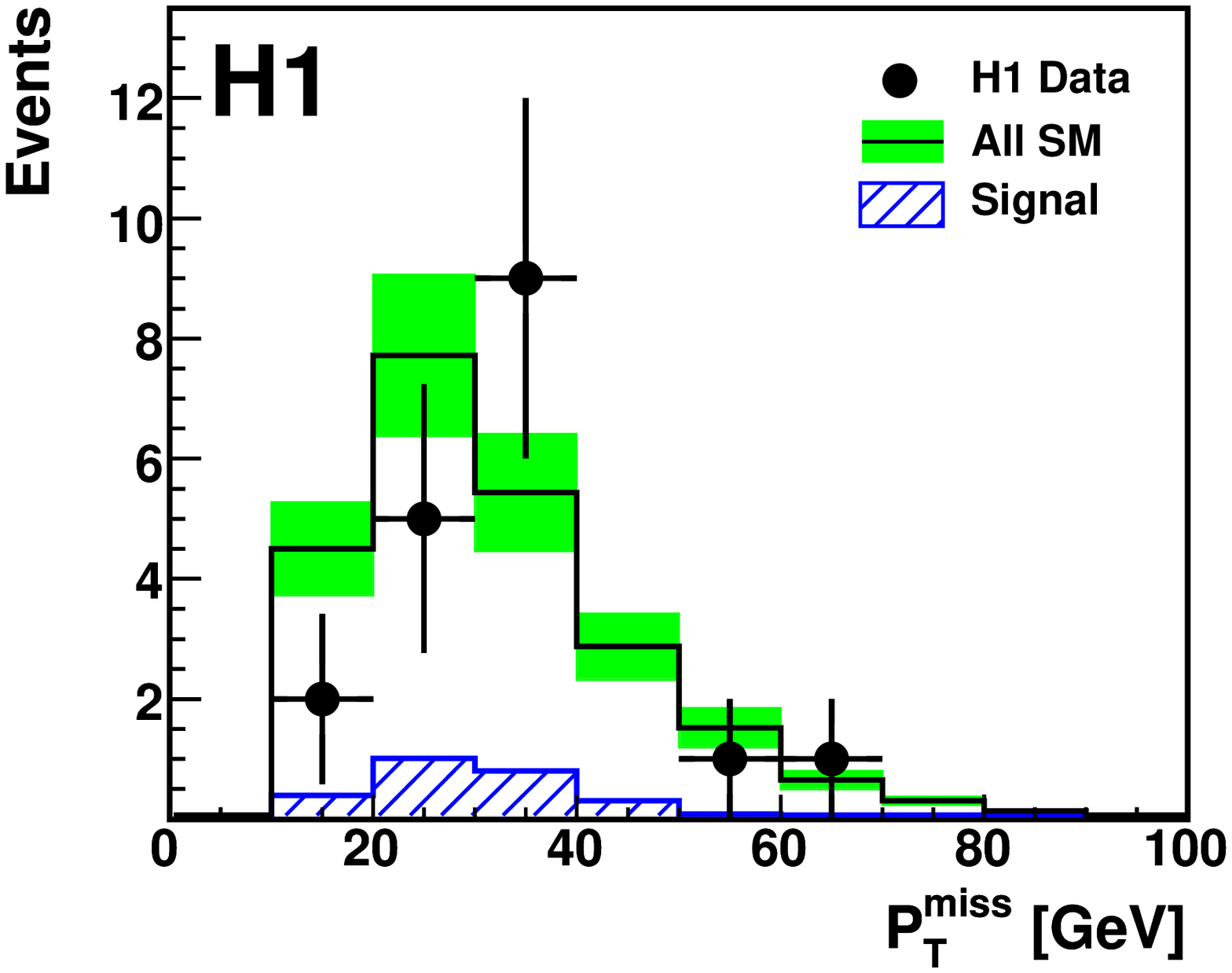}     
    \includegraphics[width=0.98\columnwidth]{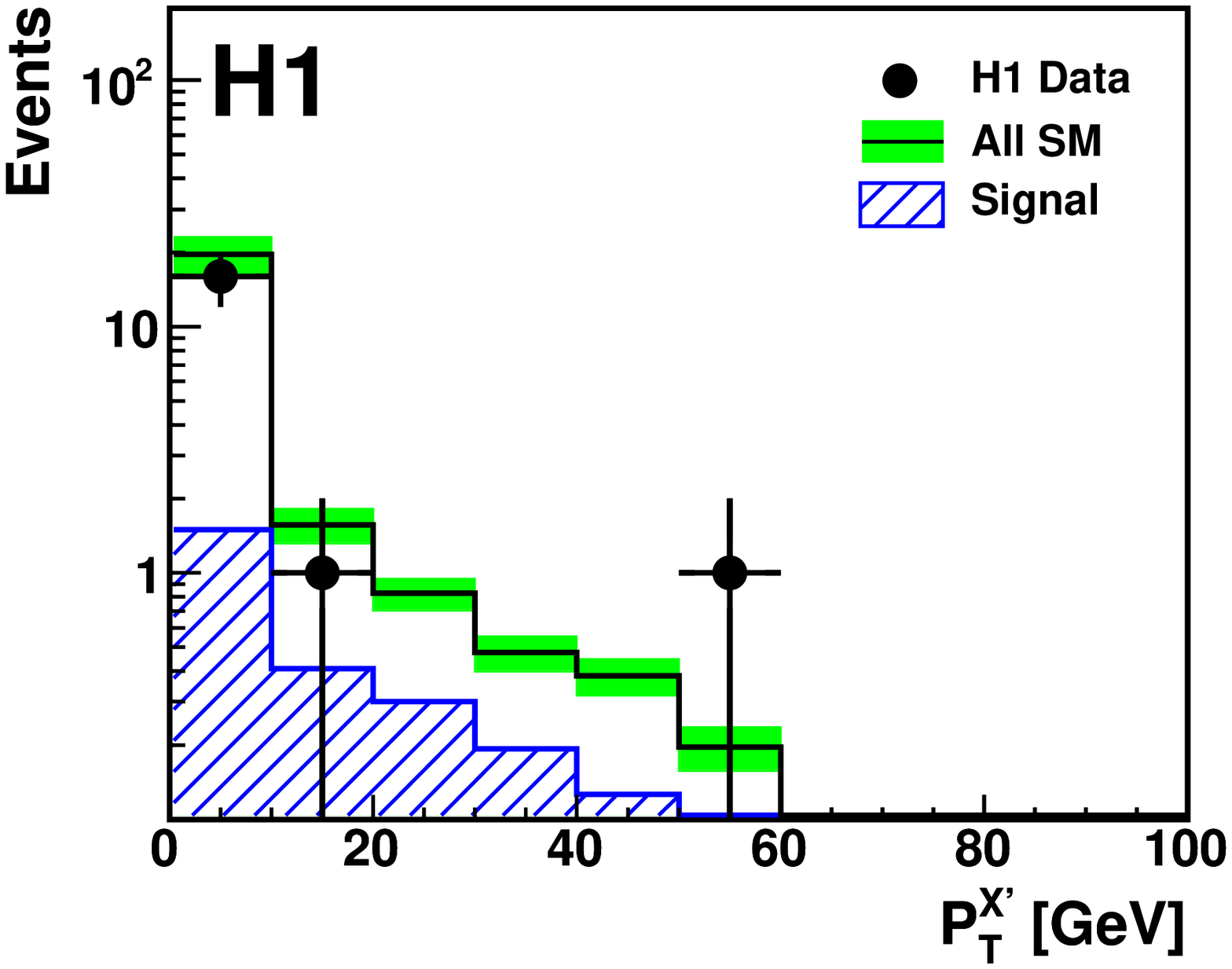}
  }
  \centerline{
    \includegraphics[width=0.98\columnwidth]{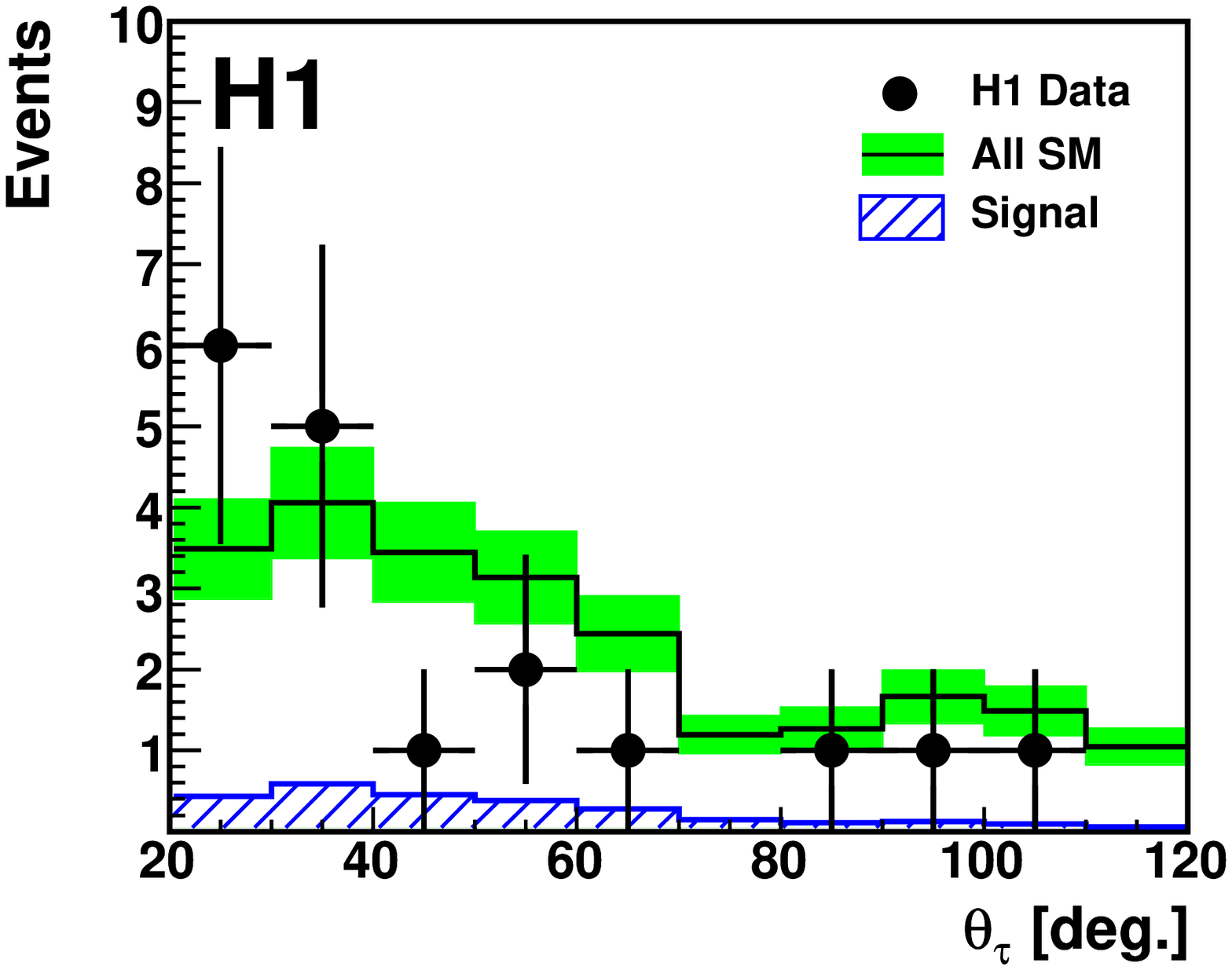}     
    \includegraphics[width=0.98\columnwidth]{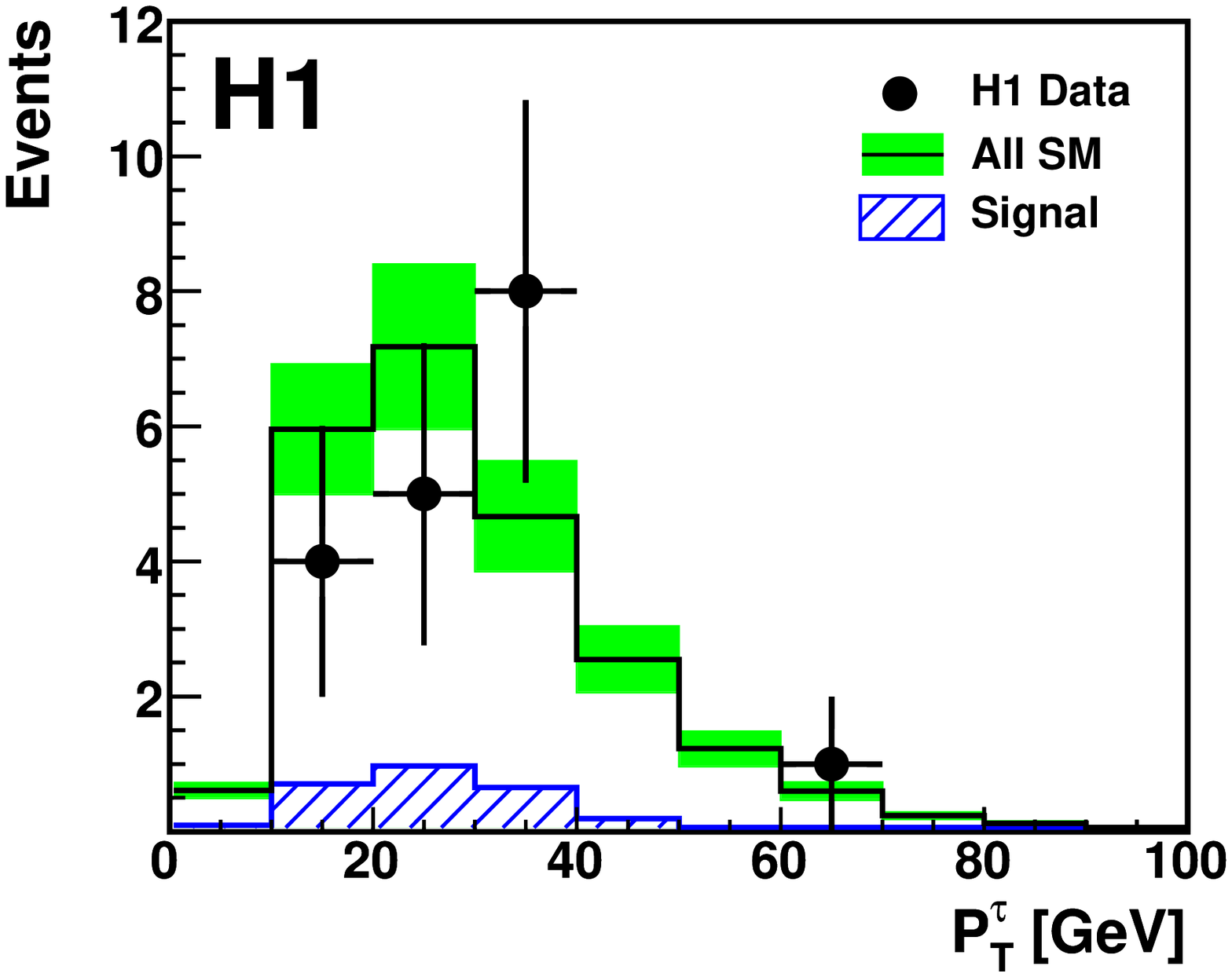}     
  }
  \begin{picture} (0.,0.)
    \setlength{\unitlength}{1.0cm}
    \put ( 6.5,10.5){\bf\normalsize (a)} 
    \put (15.8,10.5){\bf\normalsize (b)} 
    \put ( 6.5,3.7){\bf\normalsize (c)} 
    \put (15.8,3.7){\bf\normalsize (d)}  
  \end{picture} 
  \caption{Distributions in the tau channel for the H1 $e^{\pm}p$ data
    sample. Shown is the missing transverse momentum
    $P_T^{\mathrm{miss}}$~(a), the hadronic transverse momentum not
    including the tau-jet candidate $P_T^{X'}$~(b), the polar angle of
    the tau-jet candidate $\theta_{\tau}$~(c) and the tau-jet
    candidate transverse momentum $P_{T}^{\tau}$~(d). The data (points)
    are compared to the SM expectation (open histogram). The signal
    component of the SM expectation ($W\rightarrow\tau\nu_{\tau}$)
     is shown as the hatched histogram.
    The total uncertainty on the SM expectation is shown as the shaded
    band.}
  \label{fig:isotau-h1plots}
\end{figure*}

The results of the search in the tau channel are summarised in 
table~\ref{tab:isotau-h1rates}.
In the final event sample, $18$ events are selected, compared to a SM
expectation of $23.2 \pm 3.8$.
Unlike in the search for events with an isolated electron or muon and
missing transverse momentum, the SM expectation is dominated by CC DIS
background processes, and the signal contribution is only $11\%$.
Distributions of the events in the final sample are shown in
figure~\ref{fig:isotau-h1plots}.
Most of the events are observed at very low $P_{T}^{X'}$.
At $P_{T}^{X'}>25$ GeV one event is observed in the data, compared to
a SM expectation of $1.5\pm 0.2$. 
In this region the contribution of single $W$ boson production to the
SM expectation is about $38\%$.
The selected data event, shown in
figure~\ref{fig:isolep-eventdisplays}~(bottom), is observed in $e^{-}p$ collisions
and exhibits $P_{T}^{\tau}=14.3\pm1.2$~GeV, $P_{T}^{X'}=62\pm5$~GeV
and $P_{T}^{\rm miss}=68\pm6$~GeV.

%%%

In the ZEUS analysis of their HERA~I data, events are selected with higher
missing $P_T$ with respect to H1, $P_T^{\rm miss}> 20~{\rm GeV}$.
Tau candidates are reconstructed with $E_T^{\rm jet} > 5~{\rm GeV}$
in the region $-1.0 < \eta < 2.5$ and are distinguished from the QCD background
using a discriminant based on jet-shape observables~\cite{Chekanov:2003bf}.

After applying the discriminant cut, $3$ events are observed in the data,
compared to SM prediction of $0.40^{+0.12}_{-0.13}$ (where $43\%$ is due
to single-$W$ production).
In the region  $P_T^X > 25~{\rm GeV}$, $2$ events are observed
compared to an expectation of $0.20\pm 0.05$.
It can therefore be concluded that the data show a general good
agreement with the SM predictions and no hint of BSM phenomena is observed at HERA.
Recently, a review of single vector boson production at LHC at $\sqrt{s} = 7~{\rm TeV}$ has been 
published~\cite{Schott:2014sea}. There, all results have been found in agreement between the
two main experiments, ATLAS and CMS, and are furthermore consistent
with the presently available SM predictions.

\section{Search for single top production}
\label{sec:singletop}

The production of single top quarks at HERA is kinematically possible
due to the centre of mass energy, which is above the top mass threshold.
Within the SM, the dominant process for single top production is the
charged current reaction
$ep \rightarrow \nu t X$~\cite{Schuler:1987wj,Baur:1987ai,vanderBij:1990ju}, 
which has a tiny cross section of less than
$1$~fb~\cite{Stelzer:1997ns,Moretti:1997dz},
ruling out the observation of SM single top production at HERA.
However, flavour changing neutral current (FCNC) processes ($u \rightarrow t$ or $c \rightarrow t$, mediated
by a neutral vector boson, $\gamma$ or $Z$) could lead to a visible single top production cross section.
In several extensions of the SM the top quark is predicted to undergo FCNC
interactions~\cite{Atwood:1995ud,deDivitiis:1997sh,Peccei:1989kr,Fritzsch:1999rd} and the observation of top quarks
at HERA would thus be a clear indication of physics beyond the SM.

%%%

\begin{figure}
\begin{center}
\includegraphics[width=7cm]{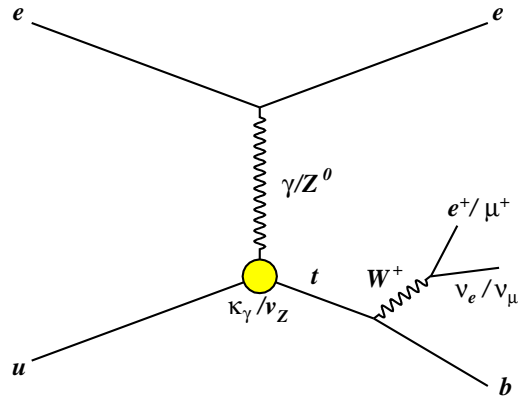}
\caption{Anomalous single-top production via flavour-changing neutral current transition with coupling $\kappa_{tU\gamma}$. }
\label{fig:sitopdiag}
\end{center}
\end{figure}

The diagram for anomalous single top production via FCNC is shown in
figure~\ref{fig:sitopdiag}, where the top quark coupling to a $U$-type quark via a photon ($Z^{0}$ boson) is
indicated as $\kappa_\gamma$ ($v_Z$).
As the top quark mass is comparable to the centre of mass energy at HERA, the initial
state quark originating from the proton needs to be at a significantly high value of $x$
for single top production.
As the charm quark density at high $x$ is low compared to the density of the
$u$ and $d$ valence quarks, the contribution to the cross section
given by charm quarks is neglected.
For the same reason, production of anti-top quarks is neglected, as this would
involve anti-quarks in the initial state. 

%%%

The anomalous single top production cross section can be parametrised in terms
describing the effect of the two FCNC couplings, $ A_\sigma$ and $ B_\sigma$, and of
their interference, $ C_\sigma$:
\begin{equation}
\sigma_{ep\rightarrow etX} = A_\sigma \kappa_\gamma^2 + B_\sigma v_Z^2+C_\sigma\kappa_\gamma v_Z.
\label{eq:sitop_sigma}
\end{equation}
Simulation of the anomalous single top signal is done using the package CompHEP~\cite{Boos:2009un},
which is also used to determine the parameters $ A_\sigma$, $ B_\sigma$ and $ C_\sigma$.
The interference parameter $C_\sigma$ has only a small effect, producing variations of the
cross section of less than $1\%$ in the full considered range of couplings and is therefore neglected. 

CompHEP is also used to determine the top decay widths in the different channels:
\begin{eqnarray}
\Gamma_{t \rightarrow u\gamma} & = & A_\Gamma \kappa_\gamma^2, \nonumber \\ 
\Gamma_{t \rightarrow u Z} & = & B_\Gamma v_Z^2, \\
\Gamma_{t \rightarrow qW} & = & C_\Gamma, \nonumber
\end{eqnarray}
where $A_\Gamma$ and $B_\Gamma$ are the partial widths of the top corresponding
to $u\gamma$ and $uZ$ unitary FCNC couplings and $C_\Gamma$ is the SM top width.

As any top quark immediately and exclusively decays into a $b$ quark and a $W$ boson,
the experimental signature of single top production at HERA is either
the leptonic, $\ell\nu$, or hadronic, $ q\bar{q}$, $W$ decay products
in combination with a high $P_T$ hadronic final state from the $b$-jet.
In the case of leptonic $W$ decay, this topology is of particular interest as it is the same
as the SM $W$ production events discussed in section~\ref{sec:isoelmu}, especially those
seen with a prominent hadronic final state.
In the case of hadronic $W$decay, the experimental signature of single top production is
the presence of three jets in an event, with a mass compatible with that of the top quark.
Both H1 and ZEUS have previously investigated FCNC using their HERA~I data
sets~\cite{Chekanov:2003yt,Aktas:2003yd} and have published searches for single top
events examining the HERA~II data sample~\cite{Aaron:2009vv,Abramowicz:2011tv}.

%%%

The main SM background to single top production in the $W$ leptonic decay
channel is from real $W$ production, which has a cross section of
about $1$~pb and is modelled using the EPVEC MC generator, which is
reweighted to NLO QCD.
In the $W$ hadronic decay channel, the main background contribution arises from
multi-jet production in photoproduction and NC and CC DIS, which are modelled
using PYTHIA, RAPGAP and DJANGOH.

%%%

Single top production is investigated by H1, employing both leptonic and hadronic
$W$ decays~\cite{Aaron:2009vv}.
In the leptonic channel, isolated electrons and muons with a transverse momentum
$P_T^{\ell} >10~{\rm GeV}$ in the polar angle range $5^\circ < \theta_l  < 140^\circ$ are
selected in events with a missing transverse momentum $P_T^{\rm miss} > 12~{\rm GeV}$,
where the selection is based on the H1 analysis described in section~\ref{sec:isoelmu}, where 
$39$ ($14$) electron (muon) events are observed in the data, compared to a SM prediction
of $43.1 \pm 6.0$ ($11.0 \pm 1.8$)~\cite{Aaron:2009wp}.

\begin{figure}
  \centerline{\includegraphics[width=1.0\columnwidth]{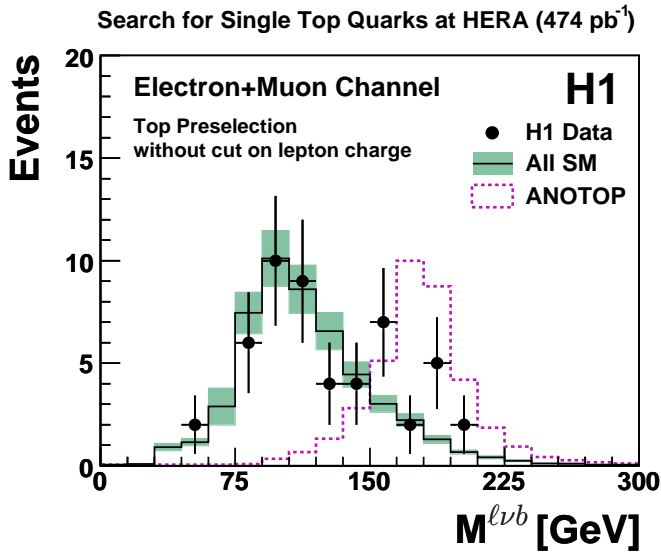}}
  \caption{The distribution of the reconstructed top mass $M^{l\nu b}$ in the H1 electron and muon
    channels, after neutrino reconstruction but before the cut on the lepton charge. 
    The data are shown as points, the total SM expectation as the open 
    histogram with systematic and statistical uncertainties added in
    quadrature (shaded band). The prediction from ANOTOP is also shown
    with arbitrary normalisation (dashed histogram).}
  \label{fig:sitop_h1mass}
\end{figure}

To estimate a potential top contribution to this sample, a top quark candidate is
reconstructed from its decay products (lepton $l$, neutrino $\nu$ and $b$ quark),
and the compatibility with single top quark production via FCNC was tested using a
multivariate discriminant method~\cite{Aaron:2009vv}.
The kinematic variables of the neutrino are reconstructed using the transverse and longitudinal
momentum balance of the event.
The four-momentum of the $b$-jet is taken as the four-momentum of the hadronic final state and
the mass of the top quark is reconstructed as the sum of the four-vectors of the isolated lepton,
the neutrino and the hadronic final state. 
The reconstructed top mass $M^{\ell \nu b}$ for the electron and muon channel combined is
shown in figure~\ref{fig:sitop_h1mass}, where the data are in overall agreement with the
SM prediction, and a slight excess of data events is observed in the
top quark mass range.
The prediciton from ANOTOP~\cite{Aktas:2003yd}, an anomolous top MC
MC generator, is also shown with arbitrary normalisation.
The results of the multivariate discriminant neural-net based analysis are cross checked
against a cut-based top selection, requiring the selected top events to have a $b$-jet
with transverse momentum $P_T^b > 30~{\rm GeV}$ and $M^{\ell \nu b} > 140~{\rm GeV}$.
In the final selection, five (four) events are observed in the electron (muon) channel,
compared to a SM expectation of $3.2 \pm 0.4$ ($2.1 \pm 0.3$ ).

%%%

To investigate the $W$ hadronic decay channel, events are selected by H1 containing at least three jets
in the pseudorapidity range $-0.5 < \eta_{\rm jet} < 2.5$, with $P_T^{\rm jet1} > 40~{\rm GeV}$,
$P_T^{\rm jet2} > 30~{\rm GeV}$ and $P_T^{\rm jet3} > 15~{\rm GeV}$, where the jets are ordered in
magnitude of their transverse momenta.
Two of the jets are required to have a mass compatible with that of the $W$ boson, within the
experimental resolution: $65 < M^{\rm jj} < 95~{\rm GeV}$.
The remaining jet is assumed to be that coming from the $b$ quark and is required to
have $P_T > 25~{\rm GeV}$.
After the preselection, $404$ events are selected, compared to a SM prediction of $388\pm 32$. 
Like in the analysis of the leptonic decay channels, a multivariate discriminant is used to
differentiate the signal from the background, and the analysis is cross checked with a
cut-based selection.%

The $b$-jet candidate is required have large transverse momentum
$P_T^b > 40~{\rm GeV}$ and the reconstructed top quark mass
must lie in the range $150 < M^{\rm jets} < 210 ~{\rm GeV}$.
The number of candidate events selected is $128$, compared to $123 \pm 13$ expected from SM processes.

%%%

A similar analysis is performed by ZEUS on the leptonic decay channels of the $W$ boson
using their HERA~II data sample~\cite{Abramowicz:2011tv}.
Events are selected with $P_T^{\rm miss} > 10~{\rm GeV}$ ($12~{\rm GeV}$ for electron events),
containing isolated electrons with $P_T^{e} > 10~{\rm GeV}$ in the angular region $17^\circ < \theta_{e} < 115^\circ$,
or isolated muons with $P_T^{\mu} > 8~{\rm GeV}$ in the angular region $11^\circ < \theta_{\mu} < 157^\circ$.

In the electron (muon) channel, a total of $245$ ($269$) events were observed,
compared to $253 \pm 6$ ($260 \pm 3$) expected from the SM. 
A cut on the transverse momentum of the hadronic final state, $P_T^X > 40~{\rm GeV}$,
is then applied to the data, resulting in a final sample containing $1$ event in the electron channel compared to
a SM prediction of $3.6\pm 0.6$, and $3$ events in the muon channel,
compared to a SM prediction of $3.0\pm 0.4$.

%%%

\begin{table}
  \renewcommand{\arraystretch}{1.3}
  \caption{Summary of the H1 and ZEUS searches for single top production in FCNC at HERA,
    where the observed number of events in each of the different $W$ decay channels
    is shown. The number of events predicted by the SM is also shown, where the
    quoted errors contain statistical and systematic uncertainties added in quadrature.}
  \label{tab:sitop_events}
  \begin{tabular*}{1.0\columnwidth}{@{\extracolsep{\fill}} c c c}
    \hline
    \multicolumn{3}{@{\extracolsep{\fill}} l}{\bf Search for Single Top Production in FCNC at HERA} \\
    \hline
    \multicolumn{3}{@{\extracolsep{\fill}} l}{\bf H1 Analysis ({\boldmath ${\mathcal L} = 474~{\rm pb}^{-1}$})}\\
    \hline                                        
    Decay channel & Data & Total SM\\
    \hline
    ~$W \rightarrow e\nu_{e}$ & ~~~$5$ & $3.2 \pm 0.4$\\
    ~~$W \rightarrow \mu\nu_{\mu}$ & ~~~$4$ & $2.1 \pm 0.3$\\
    $W \rightarrow qq$ & $128$ & $123 \pm 13$\\
    \hline
    \multicolumn{3}{@{\extracolsep{\fill}} l}{\bf ZEUS Analysis ({\boldmath ${\mathcal L} = 370~{\rm pb}^{-1}$})}\\
    \hline                                        
    Decay channel & Data & Total SM\\
    \hline
    ~$W \rightarrow e\nu_{e}$ & ~~~$1$ & $3.6 \pm 0.6$\\
    ~~$W \rightarrow \mu\nu_{\mu}$ & ~~~$3$ & $3.0 \pm 0.4$\\
    \hline
  \end{tabular*}
\end{table}

The number of selected events in each of the $W$ decay channels examined by H1 and ZEUS is summarised
in table~\ref{tab:sitop_events}, compared to the SM expectation.
As no significant deviation from the SM is observed, upper limits on the single top production cross section
are derived using the method of fractional event counting~\cite{Bock:2004xz}.
For all channels combined the H1 upper limit on the cross section for single top quark production $95\%$~CL is:
\[
\sigma(ep \rightarrow etX, \sqrt{s} = 319~{\rm GeV}) < 0.25~{\rm pb},
\]
where the limit is reported at $\sqrt{s}= 319$~GeV, taking into account~\cite{Aaron:2009vv} the ratio
of $0.70$ of the predicted cross sections at $\sqrt{s} = 301$~GeV and $319$~GeV~\cite{Belyaev:2001hf}.

%%%

The ZEUS analysis described above is based on the HERA~II data sample only, and the results are then combined
with those of the ZEUS publication on their HERA~I data~\cite{Chekanov:2003yt} which also includes the hadronic
decay channel of the $W$ boson.
The resulting ZEUS limit at $95\%$~CL on the single top production cross section is:
\[
\sigma(ep \rightarrow etX, \sqrt{s} = 315~{\rm GeV})  < 0.13~{\rm pb},
\]
reported at the average centre of mass energy $\sqrt{s} = 315$~GeV of
the complete $0.5$~fb$^{-1}$ ZEUS data sample.

%%%

\begin{figure*}
\begin{center}
\includegraphics[width=14cm]{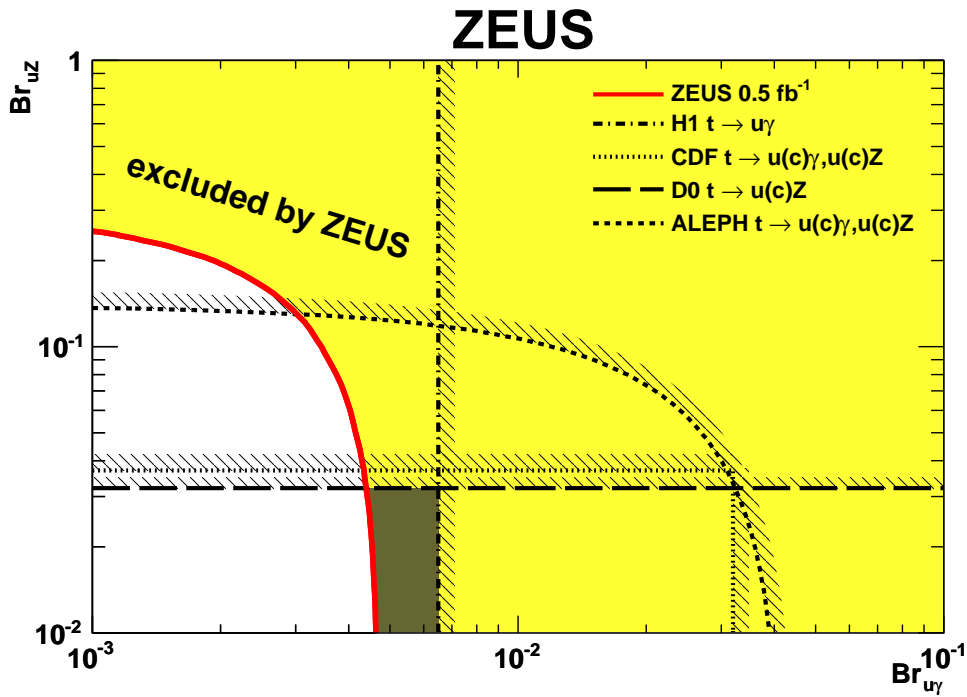}
\caption{Exclusion limits set by the H1 and ZEUS experiments on
  the branching ratios Br($t\rightarrow u\gamma$) and Br($t \rightarrow u Z$)
  from the search for single top production at HERA.
  Limits from other experiments are also shown.}
\label{fig:sitop_limits}
\end{center}
\end{figure*}

The limits on the cross sections are converted into $95\%$~CL limits on the anomalous 
FCNC coupling $k_\gamma$ using equation~\ref{eq:sitop_sigma}, which are found by H1 (ZEUS)
to be $\kappa_\gamma < 0.18$ ($\kappa_\gamma < 0.13$) for a
top mass of $175$~GeV.
These limits may in turn be transformed~\cite{Abramowicz:2011tv} into limits on the branching ratios
${\rm Br}(t \rightarrow u\gamma)$ and ${\rm Br}(t \rightarrow uZ)$ and H1 (ZEUS) sets a limit of
${\rm Br}(t \rightarrow u\gamma) < 0.64\%$ ($< 0.45\%$) for low values of ${\rm Br}(t \rightarrow uZ)$.
The limits on these branching ratios are displayed in figure~\ref{fig:sitop_limits}, compared to those
from other experiments ALEPH~\cite{Heister:2002xv} at LEP and CDF~\cite{Abe:1997fz,Aaltonen:2008ac}
and D{\O}~\cite{Abazov:2011qf} at the Tevatron.

Once again, the extended reach of the LHC data has extended these limits into a new domain,
where searches by the ATLAS~\cite{Aad:2012ij,Aad:2012gd} and CMS~\cite{Chatrchyan:2013nwa}
collaborations now limit these branching fractions to sub-permille
levels.

\section{{\boldmath $Z^0$} production at HERA}
\label{sec:z0}

Although the number of $W$ or $Z^0$ bosons produced at HERA is expected to be small,
their study at HERA provides the means to test the SM, as some anomalous couplings of
these bosons predict an increase in the production cross section.
A measurement of the $W$ production cross section at HERA was performed using events
with an isolated electron or muon and missing transverse momentum,
as discussed in section~\ref{sec:isoelmu}, where good agreement was observed
with the SM prediction.
An analysis of the production of $Z^{0}$ bosons performed by ZEUS is described in the following.

%%%

\begin{figure}
\centerline{\includegraphics[width=0.5\columnwidth, angle=270]{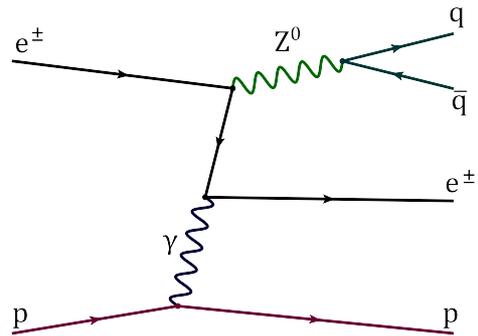}}
\caption{Example of a leading-order diagram of $Z^0$ boson production,
  $ep \rightarrow eZ^0p$ and subsequent hadronic decay into a quark $q$ and an antiquark $\bar{q}$.}
\label{fig:zeus_Z0_diagram}
\end{figure}

The full data sample collected with the ZEUS detector is used to study
the production of $Z^0$ bosons in the process $ep \rightarrow eZ^0X$.
Compared to the analyses described in section~\ref{sec:mlep}, which include $Z^0$
decays to lepton pairs, this analysis examines the hadronic decay mode of the $Z^0$, chosen
because of its large branching ratio and to exploit the excellent resolution of the ZEUS
hadronic calorimeter.
The analysis is restricted to elastic and quasi-elastic $Z^0$ production in order to suppress QCD
multi-jet background.
A diagram for $Z^0$ production at LO and the subsequent $Z^0$ hadronic decay is shown in
figure~\ref{fig:zeus_Z0_diagram}.
The decay products of the $Z^0$ boson form at least two hadronic jets with high transverse energies.
No energy deposit is found around the forward direction, in contrast to what would be expected in
inelastic collisions. 

The production of $Z^0$ bosons is simulated using the EPVEC program
to generate the events at the parton level, and PYTHIA to simulate
initial and final state parton showers.
As a reliable simulation for the SM background events, predominantly due to the
diffractive photoproduction of jets of high transverse momentum, was not available,
the background distributions are estimated from the data as described below. 

%%%

The event selection is performed requiring the presence of at least two jets in the final state
with $E_T > 25~{\rm GeV}$ and $|\eta| < 2$.
The two highest $E_{T}$ jets are required to be separated by at least $2$ radians in the
azimuthal plane, as the two leading jets from the $Z^0$ boson decays are expected to be nearly
back-to-back in the $x-y$ plane.   

%%%

To identify high $Q^2$ events, the electron produced in the DIS scattering 
in the $ep \rightarrow eZ^0X$ process is required to be
reconstructed in the final state.
The electron is required to have an energy $E_{e}^{\prime} > 5~{\rm
  GeV}$ and a track with momentum $p_{\rm track} > 3~{\rm GeV}$,
if found in the acceptance region of the tracking system.
Additional cuts are applied in order to suppress the background from low $Q^2$ and photoproduction events,
as well as background from beam-gas and cosmic interactions.
To select elastic and quasi-elastic processes, a cut on $\eta_{\rm max}<3.0$ is introduced, where $\eta_{\rm max}$
is defined as the pseudorapidity of the energy deposit in the calorimeter closest to the proton-beam direction
with energy greater than $400~{\rm MeV}$ as determined by calorimeter cells.
This cut also rejects signal events with energy deposits from the scattered electron in the calorimeter
around the forward beam pipe, causing an acceptance loss of about $30\%$.

%%%

\begin{figure}
  \centerline{\includegraphics[width=0.95\columnwidth]{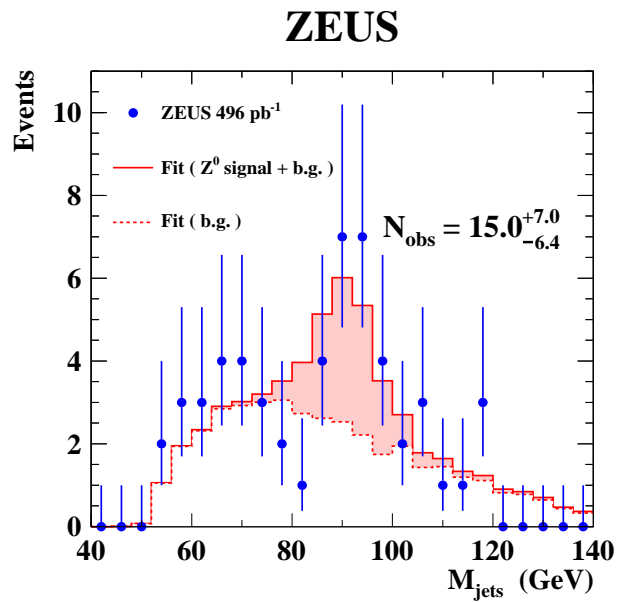}}
  \caption{The $M_{\rm jets}$ distribution and the fit result. The data are shown as points,
    and the fitting result of signal+background (background component) is shown as
    solid (dashed) line. The signal contribution is also indicated by the shaded area and
    amounts to a total number of $N_{\rm obs}$ events. The error bars represent the
    approximate $68\%$ CL intervals, calculated as $\pm \sqrt{n+0.25}+0.5$
    for a given entry $n$.}
  \label{fig:zeus_Z0_mjets}
\end{figure}

After all selection cuts are applied, $54$ events remain in the final data sample.
The total selection efficiency, estimated using the MC simulation, is
found to be $22\%$ for elastic and quasi-elastic processes and less
than $1\%$ for DIS and photoproduction events.
The number of signal and background events is estimated using  the
$M_{\rm jets}$ distribution as measured in the data.
To increase the statistics of the sample, the shape of the $M_{\rm jets}$ distribution outside the $Z^0$ mass
region is estimated from the inelastic data, obtained by removing the $\eta_{\rm max}$ cut, after verification
that this cut does not distort the distribution.
In this way, the $M_{\rm jets}$ distribution in the inelastic region is adopted as a background template in a fit
to the data in the elastic region, allowing the determination of the number of the signal events and therefore
the cross section.
The $M_{\rm jets}$ distribution of the selected events and the fit results are shown in
figure~\ref{fig:zeus_Z0_mjets}. 

The number of observed $Z^0$ events is measured to be $15.0^{+7.0}_{-6.4}~{\rm(stat.)}$, corresponding to a
statistical significance of the signal of $2.3\sigma$.
The cross section for the elastic and quasi-elastic production of $Z^0$ bosons,
$ep \rightarrow eZ^0X$, at $\sqrt{s} = 318~{\rm GeV}$, is measured as:
\[
\sigma_{Z^{0}} = 0.13~\pm~0.06~({\rm stat.})\pm~0.01~({\rm sys.})~{\rm pb,}
\]
in agreement with the SM prediction of $0.16~{\rm pb}$.

\section{A general search for new phenomena}
\label{sec:gs}

The searches for new physics reported so far in this review focus on the highest energy
regions accessible at HERA, as a large variety of possible extensions to the SM predict
new phenomena which may appear there.
Model-independent searches are performed without looking for signatures as
predicted by a particular model of BSM physics, but just looking for deviations between
the predicted and measured cross sections in regions in which the SM predictions are reliable.
Such model-independent analyses do not rely on any a priori definition of expected
signatures for exotic phenomena.
Therefore, they address the important question of whether unexpected phenomena may
occur through a new pattern, not predicted by the SM.

%%%

In this section, a more general, model-independent approach is described.
The analysis is performed by the H1 Collaboration using their complete
$e^{\pm}p$ data set~\cite{Aaron:2008aa},
corresponding to an integrated luminosity of $463~{\rm pb^{-1}}$, of
which $178~{\rm pb^{-1}}$ was recorded in $e^-p$ collisions and
$285~{\rm pb^{-1}}$ in $e^+p$ collisions. 
Following the analysis strategy of a previous H1 publication which used only the HERA~I
data set~\cite{Aktas:2004pz}, a search for differences between the observed number of
data events and the SM expectation in a large variety of different event topologies is performed.
The analysis includes all high $P_T$ final state topologies involving electrons ($e$),
muons ($\mu$), jets ($j$), photons ($\gamma$) or neutrinos ($\nu$) and searches for
deviations from the SM prediction in phase-space regions where the SM predictions
are reliable.
The identification of such particles follows that  described in
section~\ref{sec:pid}.

%%%

Calorimetric energy deposits and tracks are used to identify electron, photon
and muon candidates.
Electron and photon candidates are characterised by compact and isolated
electromagnetic showers in the LAr calorimeter with an associated
track from the inner tracking systems in the case of electrons or with
no associated track in the case of photons.
The identification of muon candidates is based on a track measured in the
inner tracking systems associated with signals in the muon detectors.
Calorimetric energy deposits and tracks not previously associated to
identified electron and muons are used to reconstruct jets using the
inclusive $k_{\rm T}$ algorithm.
If the event contains large $P_{\rm T}^{\rm miss}$ it is associated to an outgoing
neutrino candidate and the neutrino four-momentum is calculated assuming
transverse momentum conservation and using the relation in equation~\ref{eq:epz}. 
A detailed description of the further identification criteria applied to each type
of particle can be found in the H1 publication~\cite{Aaron:2008aa}.
Such criteria typically include strict isolation requirements and employ information from
multiple detector components to ensure a high efficiency and to reduce misidentification.

%%%

A precise estimate of all processes relevant at high transverse momentum in $ep$ interactions
is needed to ensure a reliable comparison to the SM.
Several MC generators, already introduced for the analyses described in the previous
sections, are therefore combined to simulate events in all classes.
The dominant SM processes at high transverse momenta are NC DIS, which
is simulated using RAPGAP and two-jet photoproduction, simulated using PYTHIA. 
Charged current DIS events are simulated using DJANGOH and contributions
from elastic and quasi-elastic QED Compton scattering are simulated with the
WABGEN generator.
Smaller contributions arising from the production of single $W$ bosons and
multi-lepton events are modelled using the EPVEC and GRAPE event generators,
respectively.

%%%

The common phase space for all the selected particles is defined as
$10^\circ < \theta < 140^\circ$ and $P_T > 20~{\rm GeV}$, except for
the neutrino, for which the phase space is defined as
$P_T^{\rm miss} > 20~{\rm GeV}$ and $\delta < 48~{\rm GeV}$.
All identified particles with $P_T > 20~{\rm GeV}$, including
the neutrino (where the $P_T$ is taken from its reconstructed
four-vector), are required to be isolated with respect to each other by
a minimum distance in pseudorapidity-azimuth of $R>1$.
All particles satisfying these requirements are referred to as {\it bodies}, and
events are classified into exclusive classes according to the number and the types
of bodies they contain, for example $e{-}j$, $\mu{-}\nu{-}j$, $j{-}j{-}j{-}j$ and so on.
All possible classes with at least two bodies are investigated, with the exception
of the $\mu\nu$ class, which originates mainly from poorly reconstructed muons
giving rise to missing transverse momentum.  

%%%

\begin{figure*}
\centerline{
    \includegraphics[width=2.0\columnwidth,angle=270]{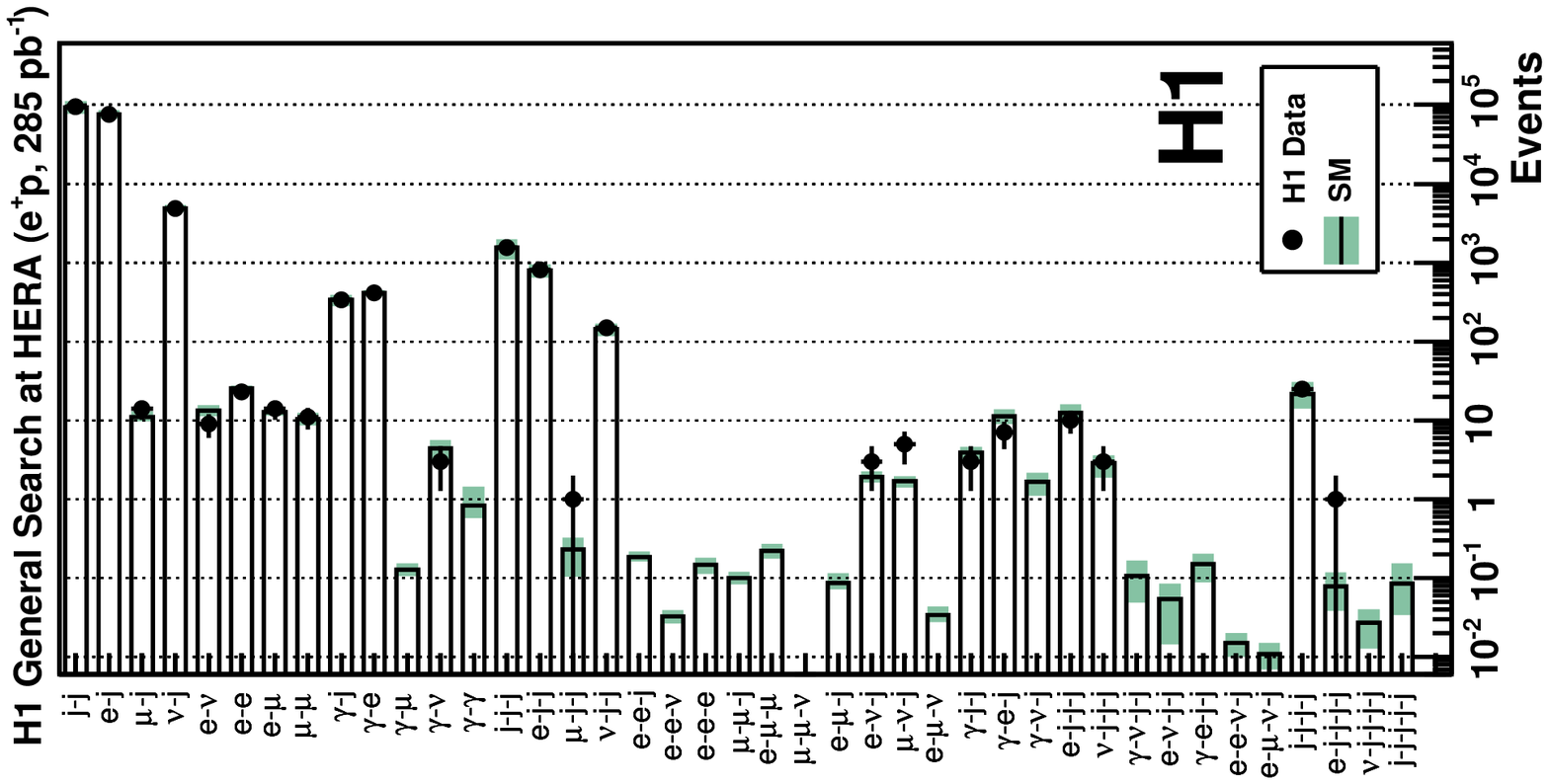}%
    \includegraphics[width=2.0\columnwidth,angle=270]{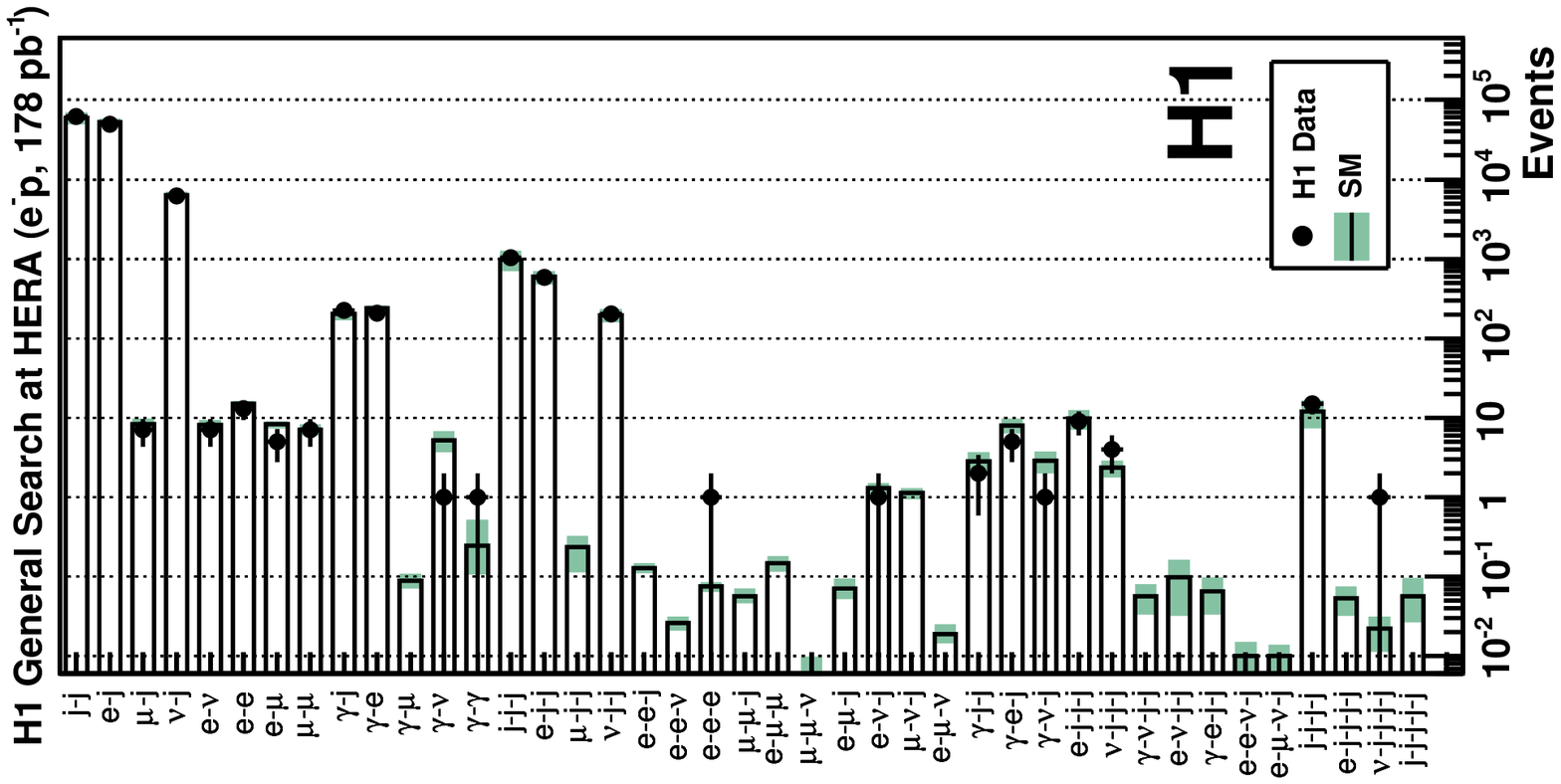}
    }
  \caption{The data and the SM expectation for all event classes in
    the H1 general search with observed data events or a SM
    expectation greater than $0.01$ events for $e^+p$ collisions (left)
    and $e^-p$ collisions (right). The error bands on the predictions include model
    uncertainties and experimental systematic uncertainties added in quadrature.}
  \label{fig:h1gs-yields}
\end{figure*}
\begin{table*}
  \renewcommand{\arraystretch}{1.3}
  \caption{Observed and predicted event yields for all event classes with observed
    data events or a SM expectation greater than $0.01$ for all $e^{\pm}p$ data. Each
    event class is labelled with the leading body listed first. The uncertainties on the predictions
    include model uncertainties and experimental systematic errors added in quadrature.
    The $\hat{P}$ values (see text) obtained in the scan of $\sum{P_T}$,  $M_{\rm all}$,
    $\cos\theta^*_{\rm lead}$ and $X_{\rm lead}$ distributions are also given.}
  \label{tab:h1gs-yields}
  \footnotesize{
    \begin{tabular*}{1.0\textwidth}{@{\extracolsep{\fill}} l c c l l l l}
      \hline
      \multicolumn{7}{@{\extracolsep{\fill}} l}{\bf H1 General Search for New Phenomena (\boldmath $e^{\pm}p$ collisions, ${\mathcal L} = 463$ pb$^{-1}$)}\\
      \hline
      Event class & Data & Total SM & $\hat{P}_{\sum{P_T}}$ & $\hat{P}_{M_{\rm all}}$ & $\hat{P}_{\cos\theta^*_{\rm lead}}$ & $\hat{P}_{X_{\rm lead}}$\rule[-6pt]{0pt}{19pt} \\
      \hline                                        
      $j-j$ & $156724$ & $153278 \pm 27400$ & $0.57$  & $0.33$  & $0.98$  &  \\ 
      $e-j$ & $125900$ & $127917 \pm 15490$ & $0.090$ & $0.99$  & $0.40$  &  \\ 
      $\mu-j$ & $21$ & $19.5 \pm 3.0$        & $0.30$  & $0.46$  & $0.024$ &\\ 
      $\nu-j$ & $11081$ & $11182 \pm 1165$   & $0.33$  & $0.31$  & $0.25$ & \\ 
      $e-\nu$ & $16$ & $21.5 \pm 3.5$        & $0.13$  & $0.084$  & $0.62$ &  \\ 
      $e-e$ & $36$ & $40.0 \pm 3.7$           & $0.35$  & $0.041$  & $0.52$ &  \\ 
      $e-\mu$ & $19$ & $21.0 \pm 2.1$        & $0.46$  & $0.83$   & $0.81$ &   \\ 
      $\mu-\mu$ & $18$ & $17.5 \pm 3.0$       & $0.31$  & $0.50$   & $0.88$ &    \\ 
      $\gamma-j$ & $563$ & $538 \pm 86$        & $0.31$  & $0.21$   & $0.77$ &     \\ 
      $\gamma-e$ & $619$ & $648 \pm 62$        & $0.93$  & $0.99$   & $0.10$ &    \\ 
        $\gamma-\mu$ & $0$ & ~~$0.22 \pm 0.04$            &  $1$ & $1$ & $1$ &    \\ 
        $\gamma-\nu$ & $4$ & ~~$9.6 \pm 2.8$      & $0.076$ & $0.33$   & $0.22$ &   \\ 
        $\gamma-\gamma$ & $1$ & ~~$1.1 \pm 0.6$             & $0.66$  & $0.35$ & $0.11$ &  \\ 
        \hline
        $j-j-j$ & $2581$ & $2520 \pm 725$      & $0.54$  & $0.65$ &  & $0.18$  \\ 
        $e-j-j$ & $1394$ & $1387 \pm 270$      & $0.0044$ & $0.70$ &  & $0.28$\\ 
        $\mu-j-j$ & $1$ & ~~$0.46 \pm 0.18$             & $0.12$  & $0.072$  &  & $0.99$\\ 
        $\nu-j-j$ & $355$ & $338 \pm 62$        & $0.80$  & $0.48$ & &$0.62$\\ 
        $e-e-j$ & $0$ & ~~$0.31 \pm 0.04$      &  $1$  &  $1$ &   & $1$\\ 
        $e-e-\nu$ & $0$ & ~~$0.06 \pm 0.01$             &  $1$  & $1$ &   &   $1$\\ 
        $e-e-e$ & $1$ & ~~$0.22 \pm 0.04$      & $0.15$  & $0.031$ &   & $0.14$ \\ 
        $\mu-\mu-j$ & $0$ & ~~$0.16 \pm 0.03$            &  $1$  & $1$ &  & $1$ \\ 
        $e-\mu-\mu$ & $0$ & ~~$0.37 \pm 0.07$            &  $1$  & $1$ & & $1$ \\ 
        $\mu-\mu-\nu$ & $0$ & ~~$0.010 \pm 0.005$         &  $1$  & $1$ &  & $1$ \\ 
        $e-\mu-j$ & $0$ & ~~$0.16 \pm 0.04$             &  $1$  & $1$ &  & $1$ \\ 
        $e-\nu-j$ & $4$ & ~~$3.2 \pm 0.5$       & $0.24$  & $0.57$ &   & $0.095$ \\ 
        $\mu-\nu-j$ & $5$ & ~~$2.8 \pm 0.5$      & $0.27$  & $0.30$ &   &   $0.35$ \\ 
        $e-\mu-\nu$ & $0$ & ~~$0.05 \pm 0.01$            &  $1$  & $1$  &   & $1$ \\ 
        $\gamma-j-j$ & $5$ & ~~$6.7 \pm 1.3$      & $0.41$  & $0.25$ &   &  $0.91$ \\ 
        $\gamma-e-j$ & $12$ & $19.4 \pm 4.0$      & $0.31$  & $0.28$ &   &  $0.53$ \\ 
        $\gamma-\nu-j$ & $1$ & ~~$4.5 \pm 1.5$             & $0.35$  & $0.62$ &   &  $0.47$ \\ 
        \hline
        $e-j-j-j$ & $19$ & ~~~$22 \pm 6.5$      & $0.84$  & $0.80$ &   & $0.14$\\ 
        $\nu-j-j-j$ & $7$ & ~~$5.2 \pm 1.4$      & $0.47$  & $0.39$ &  & $0.017$ \\ 
        $\gamma-\nu-j-j$ & $0$ & ~~$0.16 \pm 0.07$          &  $1$  &  $1$ &    &  $1$ \\ 
        $e-\nu-j-j$ & $0$ & ~~$0.15 \pm 0.09$            &  $1$  &  $1$ &    &  $1$ \\ 
        $\gamma-e-j-j$ & $0$ & ~~$0.22 \pm 0.07$           &  $1$  &  $1$ &    &  $1$ \\ 
        $e-e-\nu-j$ & $0$ & ~~$0.10 \pm 0.06$            &   $1$ &  $1$ &    &  $1$ \\ 
        $e-\mu-\nu-j$ & $0$ & ~~$0.08 \pm 0.05$           &  $1$  &  $1$ &    &  $1$ \\ 
        $j-j-j-j$ & $40$ & ~~$33 \pm 13$        &    &   &    & \\ 
        \hline
        $e-j-j-j-j$ & $1$ & ~~$0.13 \pm 0.06$            &   &   &    &  \\ 
        $\nu-j-j-j-j$ & $1$ & ~~$0.05 \pm 0.02$     &   &    &   & \\ 
        $j-j-j-j-j$ & $0$ & ~~$0.14 \pm 0.09$     &   &     &    & \\ 
        \hline
      \end{tabular*}
    }
\end{table*}

The event yields for all the event classes with observed events or with
SM expectations greater than 0.01 are given
in table~\ref{tab:h1gs-yields}. The class with the lowest SM expectation is the 
$\mu-\mu-\nu$, where the number of expected events given by the SM MC is $0.010 \pm 0.005$.
The yields are given for the full $e^\pm p$ data sample 
and a good agreement is observed between
the data and the SM prediction.
The same yields are shown in figure~\ref{fig:h1gs-yields},
separately for $e^+p$ and $e^-p$ collisions.
As expected, events with an electron (neutrino) and two or more jets are
dominated by NC (CC) DIS, while events with two or more jets and no
reconstructed leptons are dominated by photoproduction.
Event classes where an electron is observed together with a photon, with
or without an accompanying jet, arise from QED Compton processes.

%%%

Classes where more than one lepton is observed are mainly due to 
lepton pair production from $\gamma\gamma$ processes, as already
seen in the dedicated analyses described in section~\ref{sec:mlep}.
Compared to the H1 multi-lepton analysis~\cite{Aaron:2008jh}, the leptons in
the general search are identified in a wider polar angle region, down to
$10^\circ$ in the forward direction, and required to have higher transverse momenta.
All the multi-lepton events reconstructed in the dedicated analysis and falling into the
kinematic region of this analysis are also selected in the general search.
Similarly, events in which an electron or a muon are reconstructed together with a neutrino,
with or without a jet come mainly from single $W$ boson production, as already seen
in the searches for events with high $P_{T}$ leptons and missing transverse
momentum described in section~\ref{sec:isolep}.

%%%

\begin{figure*} 
  \centerline{
    \includegraphics[width=0.5\textwidth]{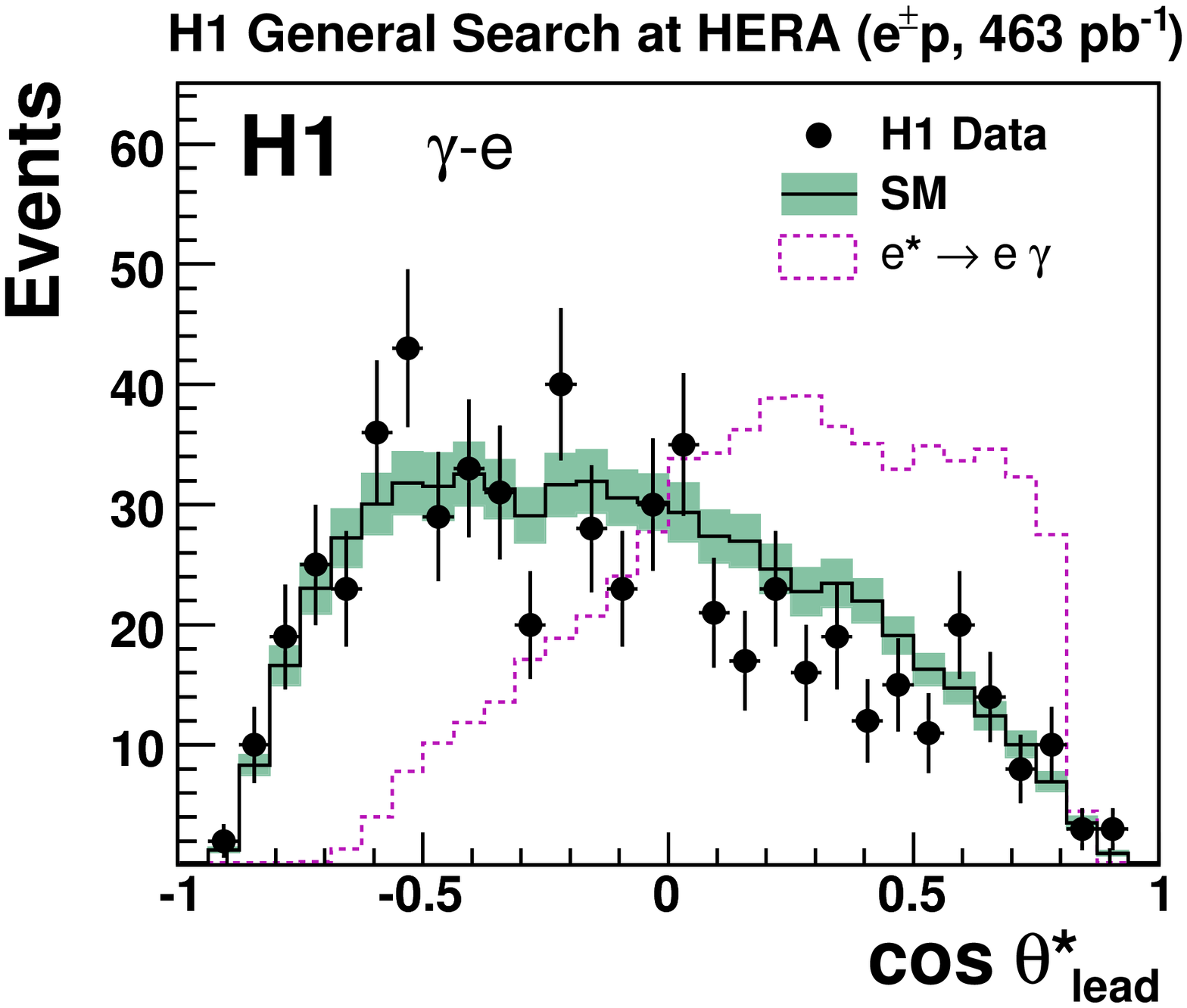}
    \includegraphics[width=0.5\textwidth]{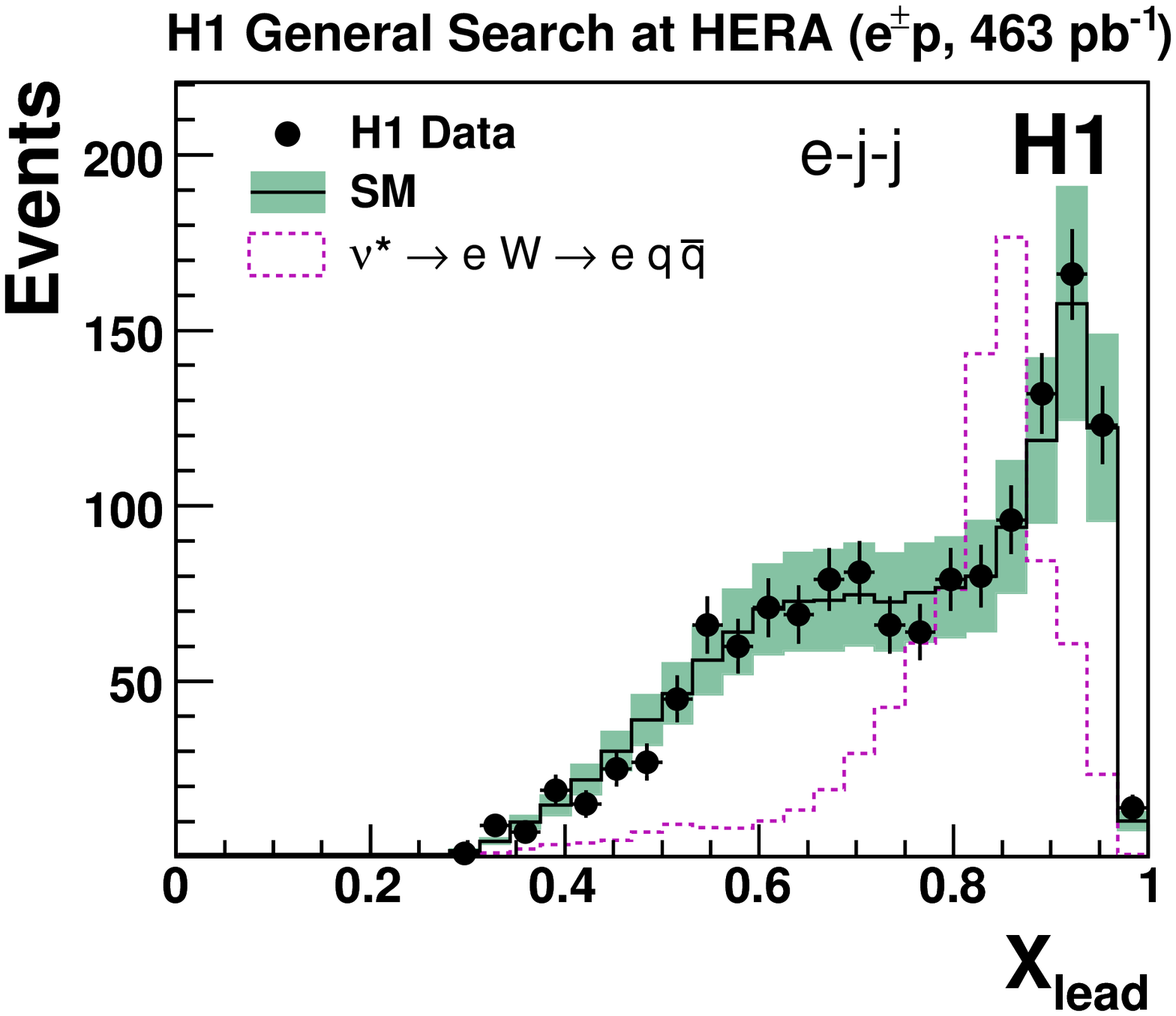}
  }
  \caption{The $\cos\theta^*_{\rm lead}$ distribution in the $\gamma{-}e$~event class (left)
    and the $X_{\rm lead}$ distribution in the $e{-}j{-}j$~event class
    (right) from the H1 general search. The points
    correspond to the observed data events and the histograms to the SM expectation.
    The error bands on the SM prediction include model uncertainties and experimental
    systematic uncertainties added in quadrature. The dashed line
    represents the distribution corresponding to an exotic resonance with a mass of
    $200$~GeV, with an arbitrary normalisation.}
 \label{fig:h1gs-topologies}  
\end{figure*}

In addition to the yields, the selected events are also analysed in terms of their topology.
The distributions of the transverse momenta of all bodies, $\sum{P_T}$, and the invariant
mass of all reconstructed bodies, $M_{\rm all}$, are analysed together with their angular
distributions and energy ratios, which are sensitive to spin and decay properties of hypothetical
high mass particles.
The variables employed are defined inspired by topological analyses of multi-jet
events~\cite{Geer:1995mp} and use the so-called \emph{main body}~\cite{Aaron:2008aa}
of the event.
Simplifying the picture, this main body is defined according to a priority list between
bodies of different types ($\gamma, e, \mu, \nu, j$) and using criteria based on its
transverse momentum, in the case where two bodies of the same kind are present in the event.
The variable $\cos \theta^*_{\rm lead}$ is then defined as the cosine of the polar angle
of the leading body relative to the incident proton in the centre of mass frame defined
by all bodies.
The variable $X_{\rm lead}$ is the energy fraction of the leading body and is defined for
systems with three or more bodies as:
\begin{equation}
X_{\rm lead} = \frac{2E^*_{\rm lead}}{\sum_i E_i^*},
\end{equation}
where the sum runs over all bodies and the energies are evaluated in
the centre of mass frame of all bodies.
The sensitivity of $\cos \theta^*_{\rm lead}$  and $X_{\rm lead}$ to new physics was tested
using MC samples of various exotics processes, such as leptoquarks, excited fermions or
anomalous top production.
The distribution of the variables $\sum{P_T}$, $M_{\rm all}$, 
$\cos \theta^*_{\rm lead}$ and $X_{\rm lead}$ from the data are compared to
SM predictions and, as also seen in the event yields in table~\ref{tab:h1gs-yields},
good agreement is found in all cases~\cite{Aaron:2008aa}. 
An example is given in figure~\ref{fig:h1gs-topologies}, where
a clear difference is observed between the
$\cos \theta^*_{\rm lead}$ ($X_{\rm lead}$) distribution of data and an
MC simulated exotic $e^*$ resonance~\cite{Aaron:2008cy}
($\nu^*$ resonance~\cite{Aaron:2008ae}) with a mass of $200$~GeV,
in the $\gamma e$ ($e{-}j{-}j$) event class. 
 
%%%

In order to quantify the observed difference between data and SM predictions
in terms of agreement or disagreement, a quantity $\hat{P}$ is introduced, according to 
a search algorithm described in the first H1 publication~\cite{Aktas:2004pz}.
The quantity $\hat{P}$ is defined such that the smaller the $\hat{P}$ value, 
the more significant the deviation between data and SM predictions. 
When $\hat{P} > 0.01$, the event class is considered to be in agreement with the SM.
The values of $\hat{P}$ for each of the event classes are also shown in
table~\ref{tab:h1gs-yields}.
For some of the classes, such as $j{-}j{-}j{-}j$, $e{-}j{-}j{-}j{-}j$ and $\nu{-}j{-}j{-}j{-}j$,
no reliable $\hat{P}$ values can be calculated due to the uncertainties of the SM predictions,
so they are not considered.
The fact that small values of $\hat{P}$ can be observed in some classes simply due
to the large number of analysed event classes is taken into account in the analysis.
The probability of a given value of $\hat{P}$ to be observed in the data is calculated by
performing a large number of MC experiments~\cite{Aaron:2008aa}. 

%%%

For the $\sum{P_T}$ and $M_{\rm all}$ distributions, in general good agreement is
observed between the $\hat{P}$ values observed in the data and those obtained
from the MC experiments.
The most significant deviation from SM is observed in $e^+p$ collisions in the
$e{-}e$ class, with $\hat{P}=0.0035$.
It is visible in figure~\ref{fig:h1gs-yields} but not in table~\ref{tab:h1gs-yields}, as
in the table the event yields are given for the full $e^\pm p$ data sample.
This corresponds to five data events being observed in the invariant mass region
$110 < M_{\rm all} < 120~{\rm GeV}$, where the SM expectation is $0.43\pm 0.04$.
This deviation has already been observed in the H1 analysis described in
section~\ref{sec:mlep}.
The global probability to find at least one class in the $e^+p$ data sample with
a smaller value of $\hat{P}$ is $12$\% as extracted from MC experiments.
It is however interesting to see that this very general analysis, performed with a
completely different method, finds as its most striking deviation from the SM an
event class already identified as interesting in a dedicated analysis,
focused on that particular topology. 
In the case of the $\cos \theta^*_{\rm lead}$ and $X_{\rm lead}$ distributions, no significant
deviation with respect to the SM is observed.
The full analysis is also performed at lower ($P_T>15~{\rm GeV}$) and higher ($P_T>40~{\rm GeV}$)
transverse momenta, and a good overall agreement between data and the
SM is also found in these regions. 

%%%

This general analysis of high transverse momentum, high-mass events, comprising different
kind of particles in the final states, demonstrates a very good understanding of the high
$P_T$ phenomena recorded at the HERA collider.
It also confirms a slight deviation already observed in a less general model-independent analysis,
the search for multi-lepton events, quantifying the probability of observing such a deviation as $12$\%.

\section{Searches for excited fermions}
\label{sec:fstar}

The existence of excited states of leptons and quarks is a natural
consequence of models of composite fermions~\cite{Harari:1982xy},
and their discovery would provide convincing evidence of a new scale
of matter.
The high energy electron-proton interactions at HERA provide a good
environment in which to search for excited states of first generation fermions.
Several results on searches for excited fermions have been
published by both H1 and ZEUS using partial HERA data
sets~\cite{Adloff:2002dy,Adloff:2001me,Adloff:2000gv,Chekanov:2001xk}.
In this review we focus on the H1
results~\cite{Aaron:2008cy,Aaron:2008ae,Aaron:2009iz} that use
the full data HERA sample, corresponding to an integrated luminosity
of $475$~pb$^{-1}$, which comprises $291$~pb$^{-1}$ of $e^{+}p$
collisions and $184$~pb$^{-1}$ of $e^{-}p$ collisions.
The ZEUS analyses are in agreement with that of H1 within their extracted larger 
limits.

%%%

\begin{figure*}
  \centerline{
    \includegraphics[width=0.60\columnwidth]{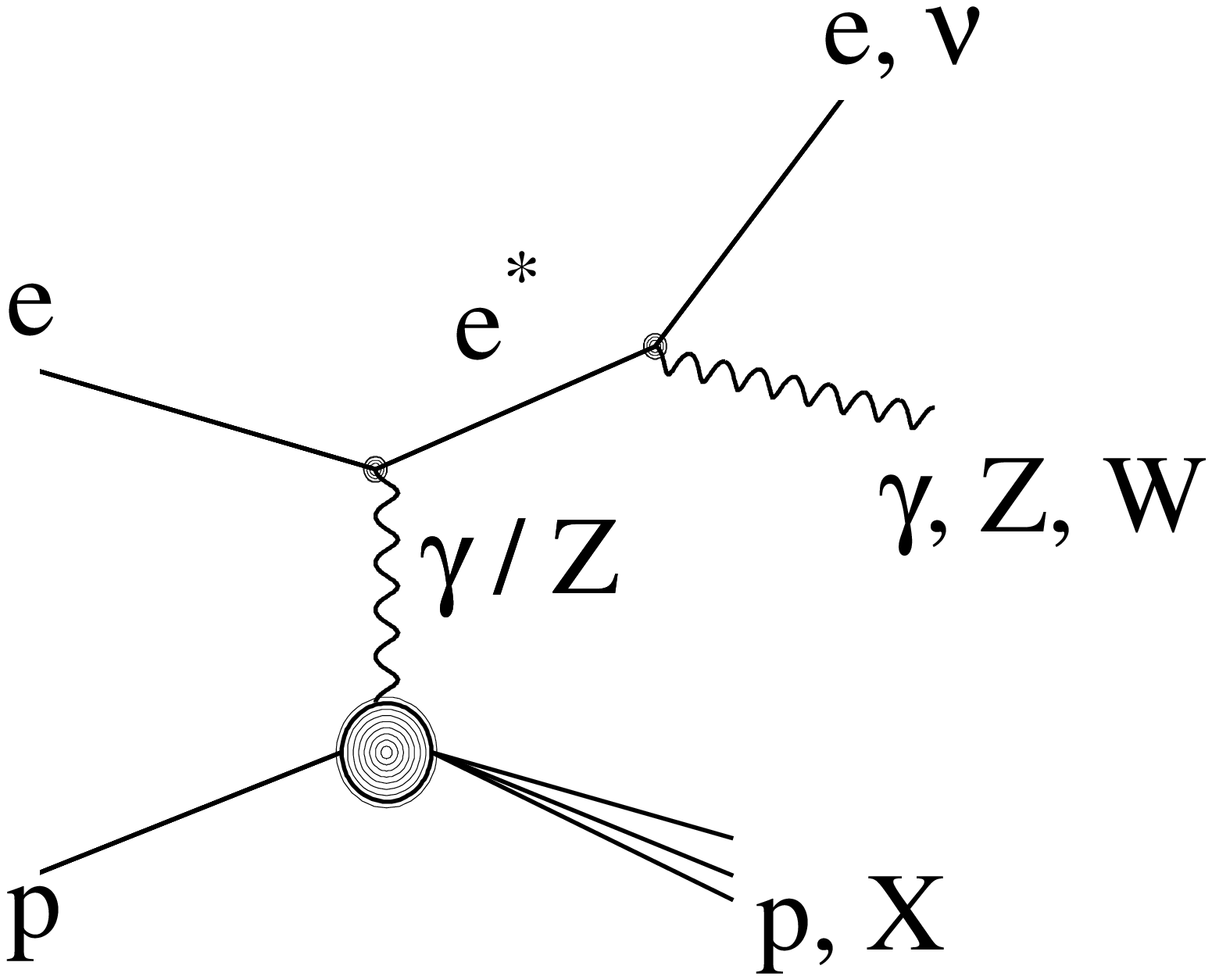}
    \hspace{0.5cm}
    \includegraphics[width=0.60\columnwidth]{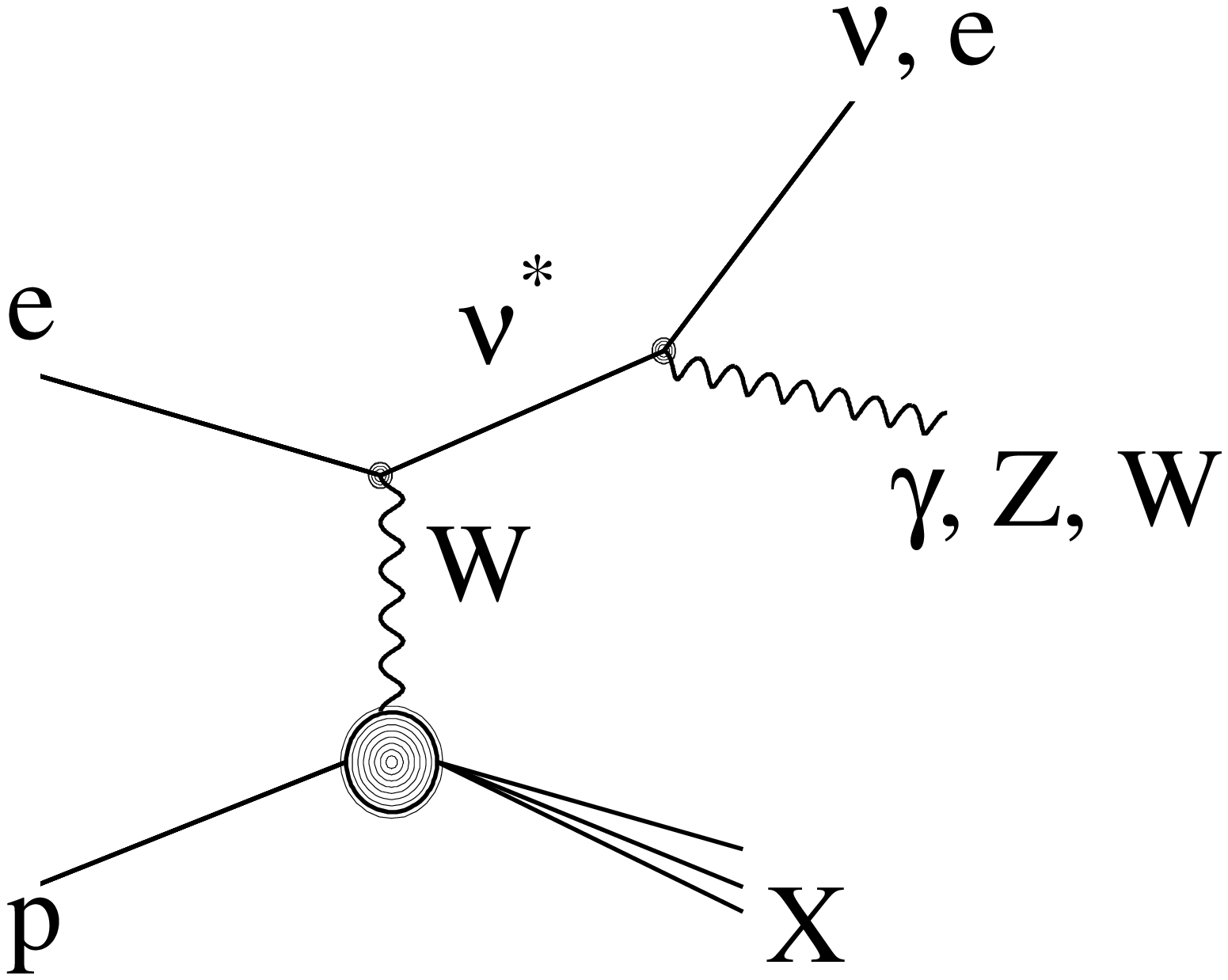}
    \hspace{0.5cm}
    \includegraphics[width=0.60\columnwidth]{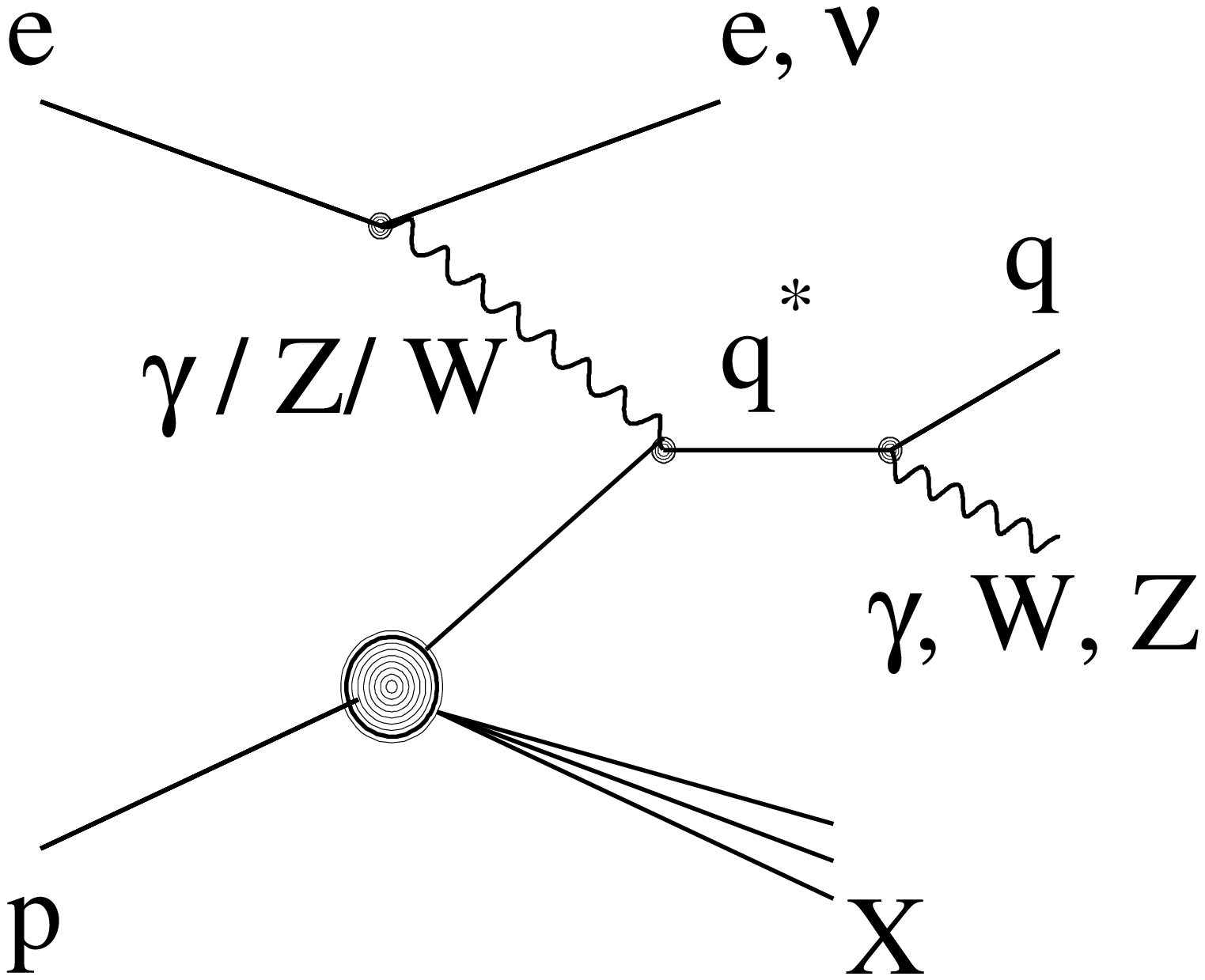}
  }
\vspace{1cm}
  \begin{picture} (0.,0.) 
    \setlength{\unitlength}{1.0cm}
    \put (2.0,0.5){\bf\normalsize  (a)} 
    \put (8.0,0.5){\bf\normalsize  (b)} 
    \put (14.0,0.5){\bf\normalsize  (c)} 
  \end{picture} 
  \caption{Diagrams showing the production of excited electrons (a), 
    excited neutrinos (b) and excited quarks (c) in $ep$ collisions at
    HERA, followed by decays into a SM fermion and a gauge boson.}
  \label{fig:fstar-feynman}
\end{figure*}

Leading order diagrams for gauge-mediated production and decay
of excited fermions at HERA are shown in figure~\ref{fig:fstar-feynman}.
The single production of an excited electron, $e^{*}$ (excited
neutrino, $\nu^{*}$), in $ep$ collisions may result from the
$t$-channel exchange of a photon or $Z^{0}$ boson ($W$ boson).
The single production of an excited quark, $q^{*}$, proceeds predominantly
via gauge boson exchange between the incoming electron and a quark
from the proton\footnote{The exchange of excited
  quarks in the $u$-channel is also possible and relevant to the
  analysis for high $q^*$ masses and low values of the compositeness
  scale, $\Lambda$.}.
Due to the helicity dependence of the weak interaction and the
valence quark composition and density distribution of the proton,
the $\nu^{*}$ production cross section is predicted to be much larger
for $e^{-}p$ collisions than for $e^{+}p$ and hence only the $e^{-}p$
data set is considered in that analysis.

%%%

%%%

The production and decay of such particles is primarily
studied\footnote{In addition to GM interactions, $e^*$
production and decay via contact interactions is also
investigated, where it is found to mediate less than $5\%$ of the
decays and is therefore neglected~\cite{Aaron:2008cy}.} by H1
in gauge-mediated (GM)
models~\cite{Hagiwara:1985wt,Boudjema:1992em,Baur:1989kv},
where the excited fermions are assumed to have spin $1/2$ and
isospin $1/2$ and the left-handed and right-handed components
form the weak iso-doublets $F_{L}^{*}$ and
$F_{R}^{*}$.
In such models, only the right-handed component of the excited
fermion $F_R^*$ is allowed to couple to light fermions, to prevent
the light leptons from radiatively acquiring a large anomalous
magnetic moment~\cite{Brodsky:1980zm,Renard:1982ij}.
The resulting effective
Lagrangian~\cite{Boudjema:1992em,Baur:1989kv} features a
compositeness scale $\Lambda$ with units of energy and three
form factors $f$, $f'$ and $f_s$ associated to the electroweak
and strong gauge groups.
%
%These form factors may be interpreted as parameters setting
%different scales $\Lambda_i = \Lambda/f_i$ for the different
%gauge groups, thus allowing the composite fermion to have
%arbitrary coupling strengths with the three gauge bosons.
%
For a given excited fermion mass, $M_{f^{*}}$, and assuming a
numerical relation between $f$, $f'$ and $f_{s}$, the excited fermion
branching ratios are fixed and the production cross section depends
only on $f/\Lambda$.

%%%

As neither the excited electron or excited neutrino are expected to have strong
interactions, these searches are insensitive to $f_s$ and the
assumption is made that the coupling parameters $f$ and $f'$ are
of comparable strength.
For excited leptons the usual complementary coupling assignments
$f = + f'$ and $f = - f'$ are considered, although in the case of
excited electrons $f = - f'$ means that the $e^{*}$ does not couple to
the photon resulting in a much smaller cross section, and as such
only $f = + f'$ is considered in that analysis.  
Only $\gamma$, $W$ and $Z^{0}$ decays of the $q^*$ are considered,
so the strong coupling parameter $f_{s} = 0$ and the assumption
is made that the coupling parameters $f$ and $f'$ are of comparable
strength, with the relation $f = f'$.

%%%

Excited fermion events are simulated using the cross section formulae
for GM interactions~\cite{Boudjema:1992em,Baur:1989kv}  in the
CompHEP program~\cite{Boos:2004kh,Pukhov:1999gg} for $\nu^{*}$ and
$q^{*}$ events, and using COMPOS~\cite{compos} for $e^{*}$ events.
The proton parton densities are taken from the
CTEQ5L~\cite{Pumplin:2002vw} parametrisation and 
are evaluated at the scale of the four-momentum transfer squared,
$\sqrt{Q^2}$, in the case of excited leptons and at
$\sqrt{\hat{s}}=\sqrt{sx}$ for excited quarks.
CompHEP includes the full transition matrix for the production and
decay modes, while COMPOS uses the narrow width approximation for the
production of $e^{*}$, also taking into account the natural width of
the subsequent $e^{*}$ decay~\cite{Aaron:2008cy}.

%%%

\begin{table}
  \renewcommand{\arraystretch}{1.3}
  \caption{Summary of the observed and predicted event yields for
    the various excited fermion decay channels in the H1 excited
    fermion analyses. The uncertainty
    on the SM prediction includes model and experimental systematic
    uncertainties added in quadrature. Typical selection efficiencies for
    excited fermion masses ranging from $120$ to $260$~GeV are also indicated.}
  \label{tab:fstar-rates}
  \begin{tabular*}{1.0\columnwidth}{@{\extracolsep{\fill}} l c c c}
    \hline
    \multicolumn{4}{@{\extracolsep{\fill}} l}{{\bf H1 Search for Excited Fermions at HERA}}\\
    \hline
    \multicolumn{4}{@{\extracolsep{\fill}} l}{\bf Excited electrons
    (\boldmath $e^{\pm}p$ collisions, ${\mathcal L} = 475$ pb$^{-1}$)}\\
    Channel & Data & Total SM & Signal Eff. [\%]\\
    \hline
    ${e}^{*} {\rightarrow} e \gamma$ (ela.) & $42$ & $48 \pm 4$ & $60$ -- $70$\\
    ${e}^{*} {\rightarrow} e \gamma$ (inel.) & $65$ & $65 \pm 8$ & $60$ -- $70$\\
    ${e}^{*} {\rightarrow} \nu W {\rightarrow} \nu q\bar{q}$ & $129$ & $133 \pm 32$ & $20$ -- $55$\\
    \hspace{-0.2cm}\begin{tabular}{l} ${e}^{*} {\rightarrow} \nu W {\rightarrow} \nu e \nu$ \\ ${e}^{*} {\rightarrow} e Z^{0} {\rightarrow} e \nu \nu $ \end{tabular} &  $4$ & $4.5 \pm 0.7$ & \begin{tabular}{l} $60$ \\ $35$ \end{tabular}\\
    ${e}^{*} {\rightarrow} e Z^{0} {\rightarrow} e q\bar{q}$ & $286$ & $277 \pm 62$ & $20$ -- $55$ \\
    ${e}^{*} {\rightarrow} e Z^{0} {\rightarrow} eee$ & $0$ & $0.72 \pm 0.06$ & $60$ \\
    ${e}^{*} {\rightarrow} e Z^{0} {\rightarrow} e\mu\mu$ & $0$ & $0.52 \pm 0.05$ & $40$ -- $15$ \\
    \hline
    \multicolumn{4}{@{\extracolsep{\fill}} l}{\bf Excited neutrinos (\boldmath $e^{-}p$
    collisions, ${\mathcal L} = 184$ pb$^{-1}$)}\\
    Channel & Data & Total SM & Signal Eff. [\%]\\
    \hline
    ${\nu}^{*} {\rightarrow} {\nu}{\gamma}$ & $7$ & $12.3~{\pm}~3.0$~~ & $50$ -- $55$\\
    ${\nu}^{*} {\rightarrow} {e} W {\rightarrow} eq\bar{q}$ & $220$ & $223~{\pm}~47$~~ &  $40$ -- $65$\\
    ${\nu}^{*} {\rightarrow} {e} W {\rightarrow} e\nu\mu$  & $0$ & $0.40~{\pm}~0.05$& $35$\\
    ${\nu}^{*} {\rightarrow} {e} W {\rightarrow} e \nu e$ & $0$ & $0.7~{\pm}~0.1$ & $45$ \\
    ${\nu}^{*} {\rightarrow} {\nu} Z^{0} {\rightarrow} \nu q\bar{q}$ & $89$ & $95~{\pm}~21$ &  $25$ -- $55$\\
    ${\nu}^{*} {\rightarrow} {\nu} Z^{0} {\rightarrow} \nu ee $ & $0$ & $0.19~{\pm}~0.05$ &  $45$\\
    \hline
    \multicolumn{4}{@{\extracolsep{\fill}} l}{\bf Excited quarks
    (\boldmath $e^{\pm}p$ collisions, ${\mathcal L} = 475$ pb$^{-1}$)}\\
    Channel & ~Data~ & Total SM & Signal Eff. [\%]\\
    \hline
    ${q}^{*} {\rightarrow}q \gamma$  & $44$ & $46 \pm \;8$~ & $35$ -- $45$\\
    ${q}^{*} {\rightarrow} q W/Z^{0} {\rightarrow} qq\bar{q}~~~$ & $341$ & $326 \pm 78$~~ &  ~~$5$ -- $55$\\
    ${q}^{*} {\rightarrow} q W {\rightarrow} q e \nu$ & $6$ & $6.0 \pm 0.8$ & $20$ -- $30$\\
    ${q}^{*} {\rightarrow} q W {\rightarrow} q \mu \nu$ & $5$ & $4.4 \pm 0.7$ & $20$ -- $40$\\
    ${q}^{*} {\rightarrow} q Z^{0} {\rightarrow} qee$ & $0$ & $0.44 \pm 0.08$ & $15$ -- $30$ \\
    ${q}^{*} {\rightarrow} q Z^{0} {\rightarrow} q\mu\mu$ & $0$ & $0.87 \pm 0.11$ & $15$ -- $30$\\
    \hline
  \end{tabular*}
\end{table}

%%%

\begin{figure*}
  \centerline{
    \includegraphics[width=0.475\textwidth]{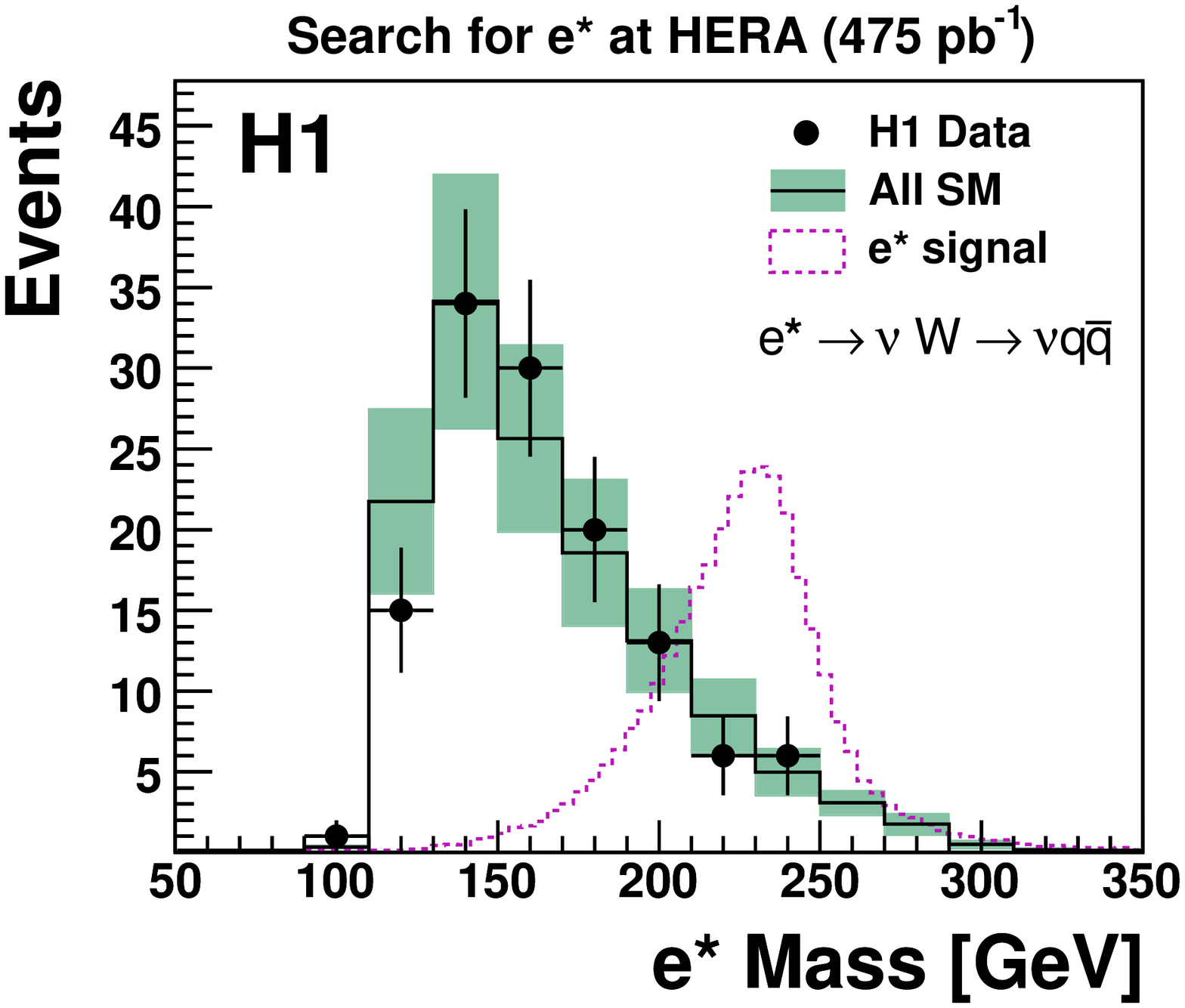}
    \includegraphics[width=0.475\textwidth]{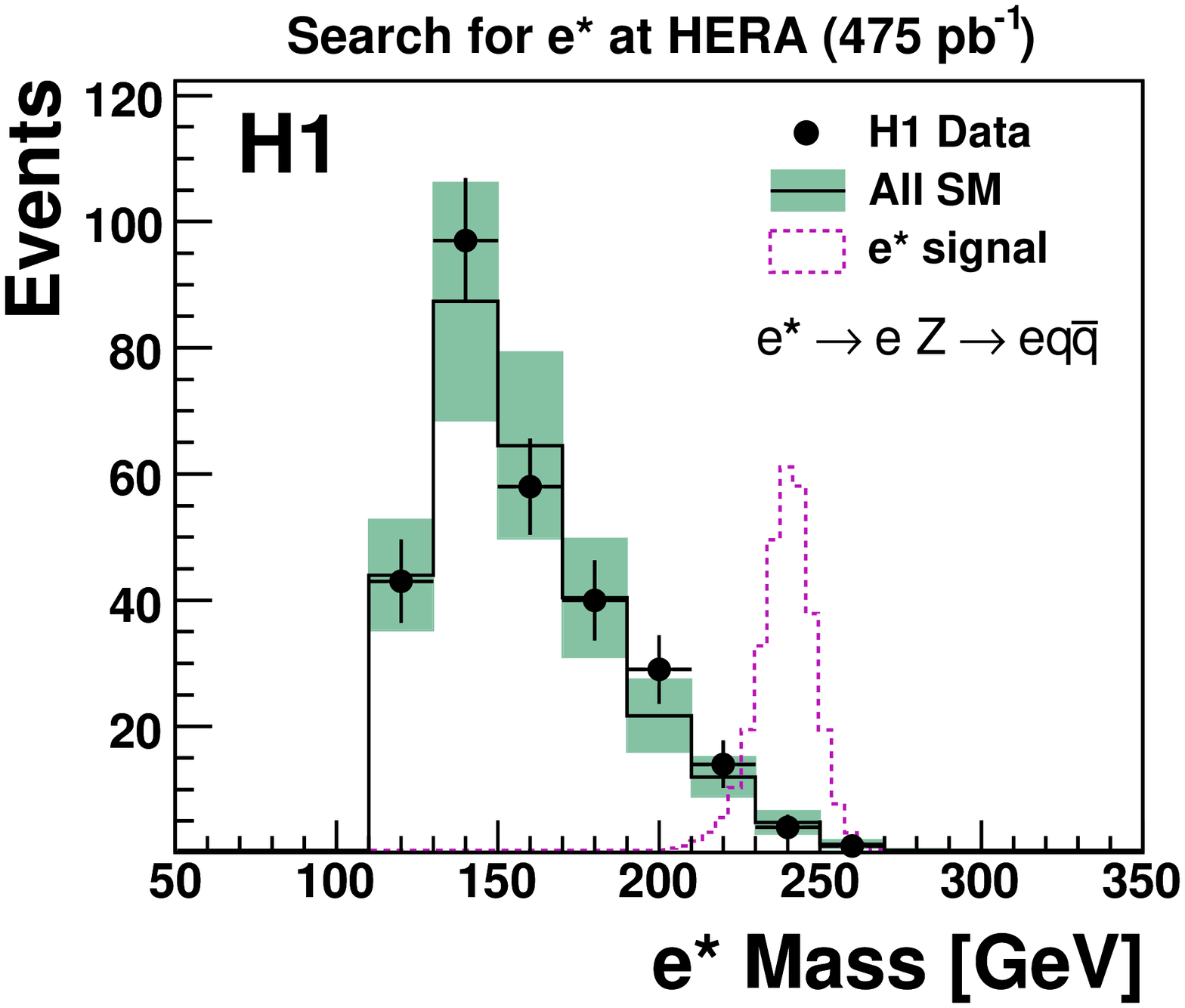}
  }
  \vspace{-0.15cm}
  \centerline{
    \includegraphics[width=0.475\textwidth]{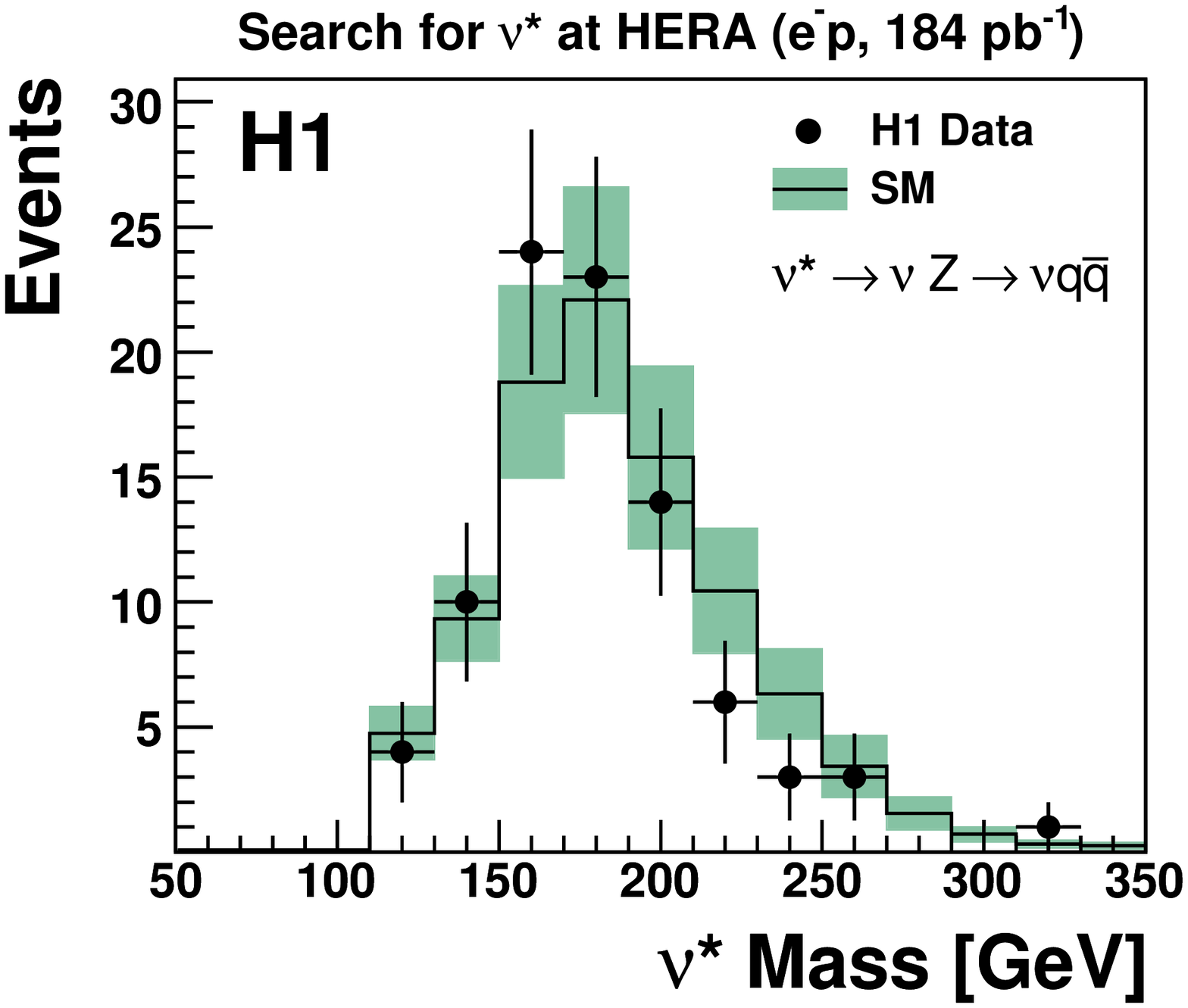}
    \includegraphics[width=0.475\textwidth]{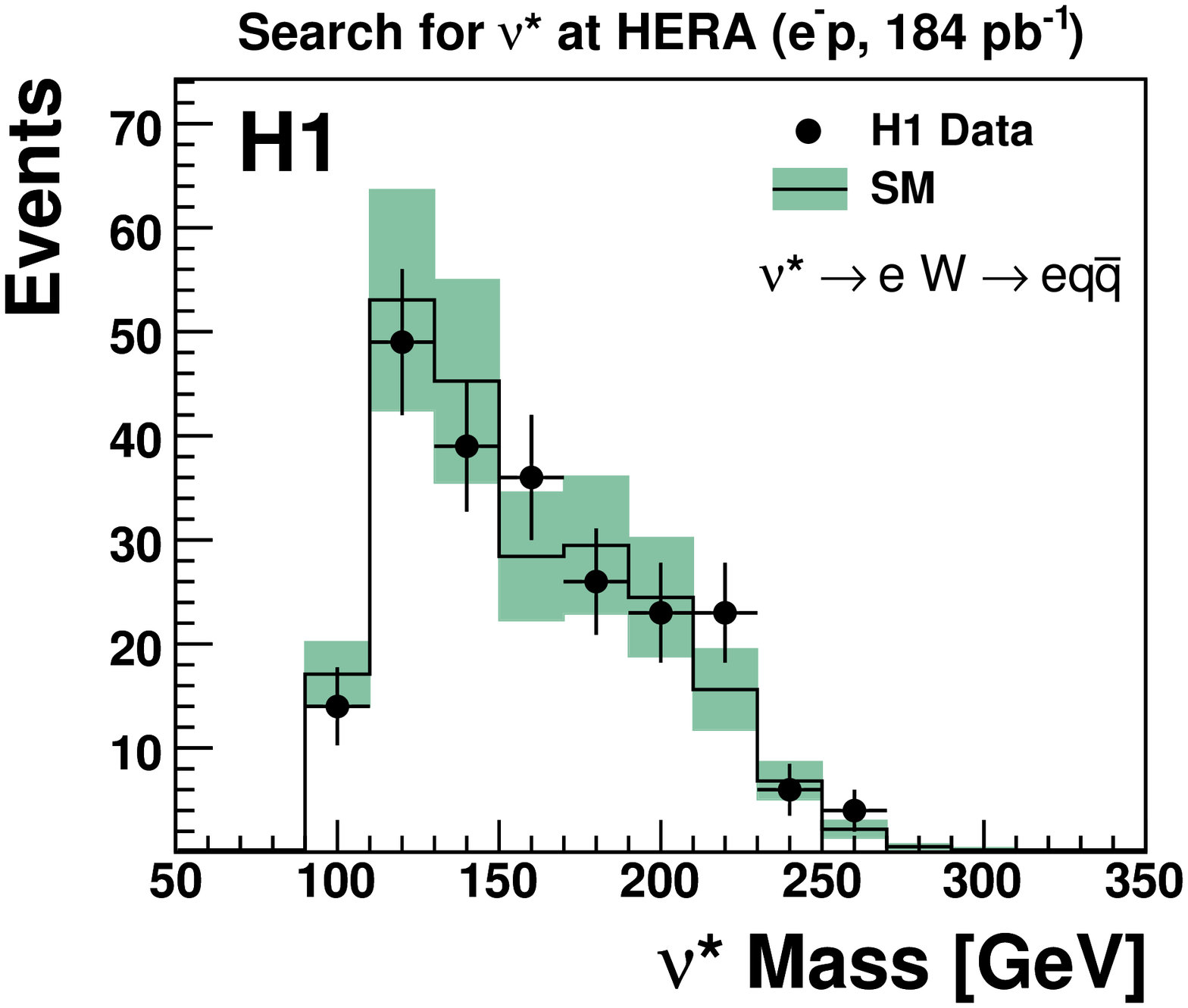}
  }
  \vspace{-0.15cm}
  \centerline{
    \includegraphics[width=0.475\textwidth]{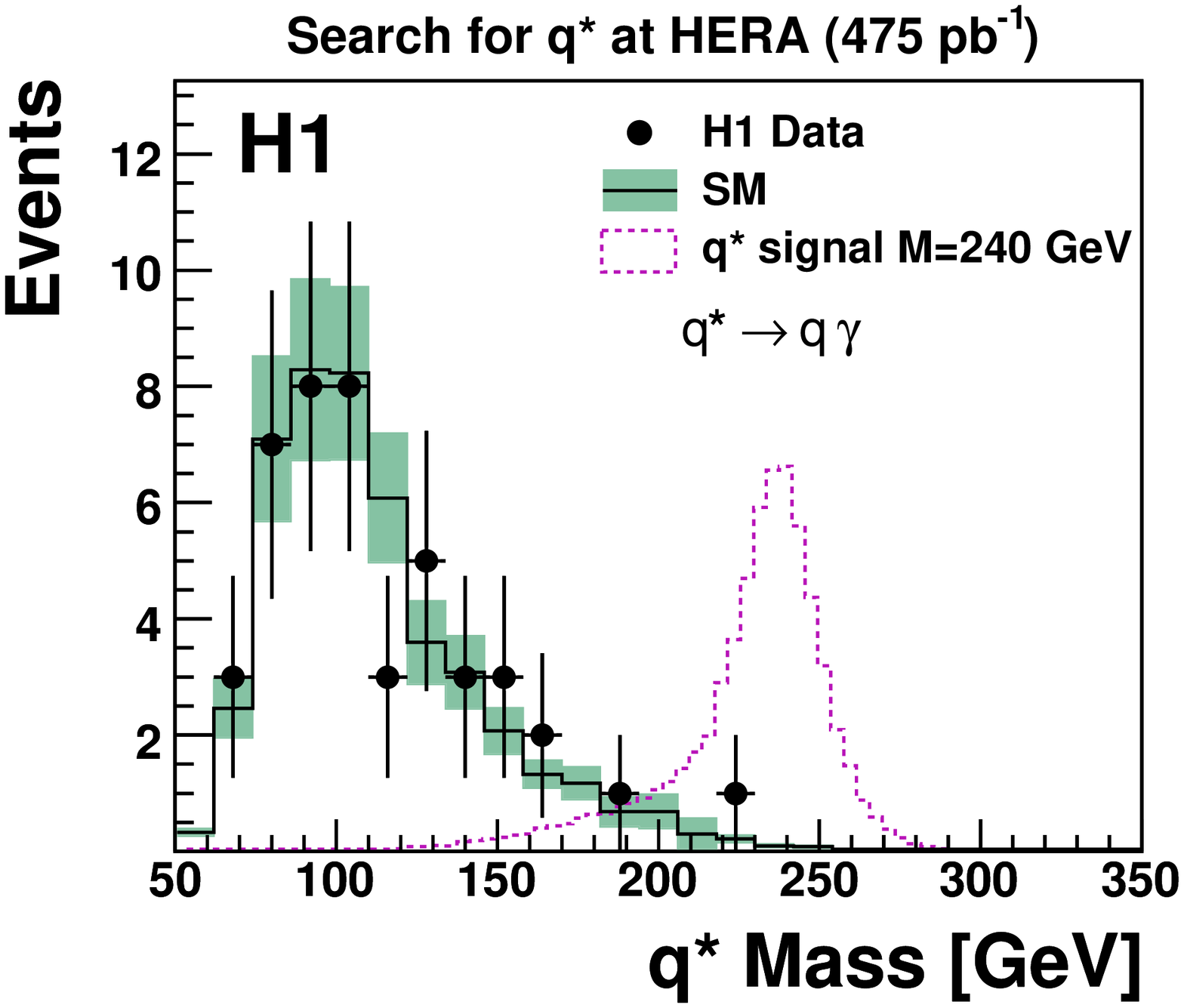}
    \includegraphics[width=0.475\textwidth]{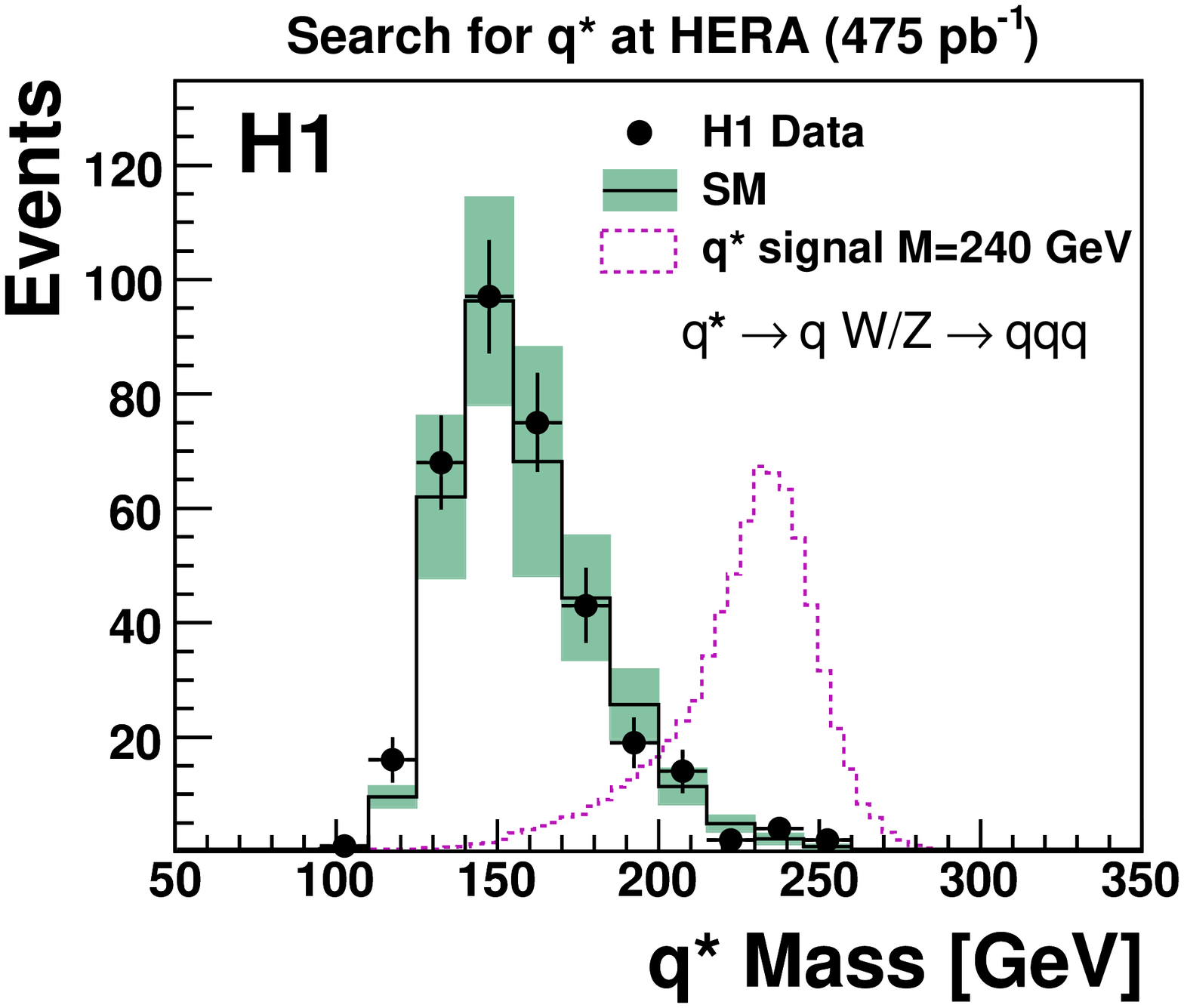}
  }
  \begin{picture} (0.,0.) 
    \setlength{\unitlength}{1.0cm}
    \put (7.2,17.5){\bf\normalsize  (a)} 
    \put (16,17.5){\bf\normalsize  (b)} 
    \put (7.2,10.5){\bf\normalsize  (c)} 
    \put (16,10.5){\bf\normalsize  (d)} 
    \put (7.2,3){\bf\normalsize  (e)} 
    \put (16,3){\bf\normalsize  (f)} 
  \end{picture} 
  \caption{Invariant mass distribution of the excited fermion candidates in the
   ${e}^{*} {\rightarrow} \nu W {\rightarrow} \nu q \bar{q}$ (a),
   ${e}^{*} {\rightarrow} e Z^{0} {\rightarrow} e q \bar{q}$ (b),
   ${\nu}^{*} {\rightarrow} {\nu} Z^{0} {\rightarrow} \nu q\bar{q}$ (c),
   ${\nu}^{*} {\rightarrow} {e} W {\rightarrow} e q\bar{q}$ (d),
   ${q}^{*} {\rightarrow} {q}{\gamma}$ (e) and 
   ${q}^{*} {\rightarrow} {q}{W/Z^{0}} {\rightarrow} qq\bar{q}$ (f) search
   channels. The points correspond to the observed data events and the
   histograms to the SM expectation after the final selections. The
   error bands on the SM prediction include model uncertainties and
   experimental systematic uncertainties added in quadrature. The dashed line
   in the $e^{*}$ and $q^{*}$ figures represents the reconstructed
   mass distribution of excited fermion events with $M_{f^*} = 240$~GeV
   with an arbitrary normalisation.}
  \label{fig:fstar-masses}
\end{figure*} 

%%%

As can be seen from figure~\ref{fig:fstar-feynman}, a large
number of final states are possible due to the decay of the $W$
or $Z^{0}$ produced in the excited fermion decay.
Each of these final states, which are listed in
table~\ref{tab:fstar-rates}, are analysed separately using
dedicated event selections as detailed in the H1
publications~\cite{Aaron:2008ae,Aaron:2008cy,Aaron:2009iz}.
These selections are based on the identification of high $P_{T}$
objects in the forward and central regions of the detector
such as electrons, muons, photons and jets, as well as missing
transverse momentum, similarly to what is done in the general
search described in section~\ref{sec:gs}.
The event selections are optimised to maximise the signal
efficiency in each channel.
The analysis of the excited electron decay ${e}^{*} \rightarrow
e\gamma$ is separated into inelastic and elastic parts using the
total hadronic energy in the event~\cite{Aaron:2008cy}.
When multiple objects of the same type are found in the event,
the $P_{T}$ threshold of the second and third objects are generally
lower.
Note that ${e}^{*} {\rightarrow} \nu W {\rightarrow} \nu e \nu$ and
${e}^{*} {\rightarrow} \nu Z^{0} {\rightarrow} e\nu\nu$ decays produce
identical final states and they are therefore examined together.

%%%

The main source of SM background in each channel depends on the
final state.
NC DIS is modelled using RAPGAP and CC DIS is modelled using DJANGOH.
Compton events are modelled using WABGEN and the GRAPE (EPVEC) program
is used to model the lepton pair ($W$ production) contribution.
Cuts to reduce SM background are applied similarly to what is done in
the analysis of events with isolated electrons or muons and missing 
transverse momentum (see section~\ref{sec:isoelmu}), using
quantities such as the total $E - P_{z}$ of the event (equation
\ref{eq:epz}) and $\xi_e = E_e \cos^2(\theta_e/2)$, which, in NC DIS,
is equivalent to the transferred four-momentum squared, $Q^2$.

%%%

The event yields observed in all decay channels are summarised
in table~\ref{tab:fstar-rates} and are in agreement
with the corresponding SM expectations.
%
% estar
QED Compton is the main SM background in the
$e^{*} {\rightarrow} e\gamma$ search, whereas
NC (CC) DIS dominates the SM prediction in the
$e^{*} {\rightarrow} eZ^{0} {\rightarrow}  e{q}{\bar{q}}$
($e^{*} {\rightarrow} \nu W {\rightarrow} \nu{q}{\bar{q}}$) channel.
In the search for excited neutrinos, the SM predictions
are dominated by NC DIS in the
$\nu^{*} {\rightarrow} eW {\rightarrow}  e{q}{\bar{q}}$ search and by
CC DIS for in the $\nu^{*} {\rightarrow} \nu\gamma$
and $\nu^{*} {\rightarrow} \nu Z^{0} {\rightarrow} \nu{q}{\bar{q}}$ searches. 
The SM contribution in the $q^{*} {\rightarrow} q\gamma$
and $q^{*} {\rightarrow} qW/Z^{0} {\rightarrow}  q{q}{\bar{q}}$ channels is mainly
due to NC DIS, whereas $W$ production is the main
contribution in the $q^{*} {\rightarrow} qW {\rightarrow} qe\nu$ and
$q^{*} {\rightarrow} qW {\rightarrow} q\mu\nu$ channels. 
%
%%%
%
The invariant mass distributions of some of the most populous
excited fermion decay channels are displayed in
figure~\ref{fig:fstar-masses}, where the data distributions are seen to be
in good agreement with the SM expectation.

%%%

Since no evidence of excited fermions is observed, limits are derived
at the $95\%$~CL on the scale $f/\Lambda$ as a function of the excited
fermion mass $M_{f^{*}}$.
For each excited fermion type, the decay channels are combined and the
limits are obtained from the mass spectra using a modified
frequentist approach, which takes statistical and systematic
uncertainties into account~\cite{Junk:1999kv}.
For excited electrons a pure gauge-mediated interaction is assumed and for the
standard scenario where $f = + f'$ and $f/\Lambda=1/M_{e^*}$, excited
electrons with a mass lower than $272$~GeV are excluded at
$95\%$~CL.
With the assumption $f/\Lambda = 1/M_{\nu^*}$ excited neutrinos with
masses up to $213$~GeV ($196$~GeV)
are excluded for $f = - f'$ ($f = + f'$).
Finally, assuming $f = +f'$, no strong interactions $f_{s}=0$ and
$f/\Lambda=1/M_{q^*}$, excited quarks with a mass lower than
$259$~GeV are excluded at $95\%$~CL.

%%%

The H1 limits as a function of excited fermion mass are shown in
figure~\ref{fig:fstar-limits}, where they are compared to limits from
LEP (excited electrons:~\cite{Abbiendi:2002wf,Abdallah:2004rc},
excited neutrinos:~\cite{Achard:2003hd}, excited
quarks:~{\cite{Abreu:1998jw}) and the Tevatron
(excited electrons:~\cite{Acosta:2004ri}).
Analyses from the LHC by ATLAS and CMS using their
$\sqrt{s} = 8$~TeV data have recently pushed these limits into a
new mass regime, where, under the assumption $f = f' = 1$ and
$\Lambda = M_{f^{*}}$, excited electrons (quarks) are ruled out by
ATLAS at the $95\%$~CL for masses lower than
$2.2$~TeV~\cite{Aad:2013jja} ($3.5$~TeV~\cite{Aad:2013cva}).
However, the H1 limits on excited neutrinos currently remain the
most stringent at the time of writing.

%%%

\begin{figure}
  \begin {center}
    \includegraphics[width=0.95\columnwidth]{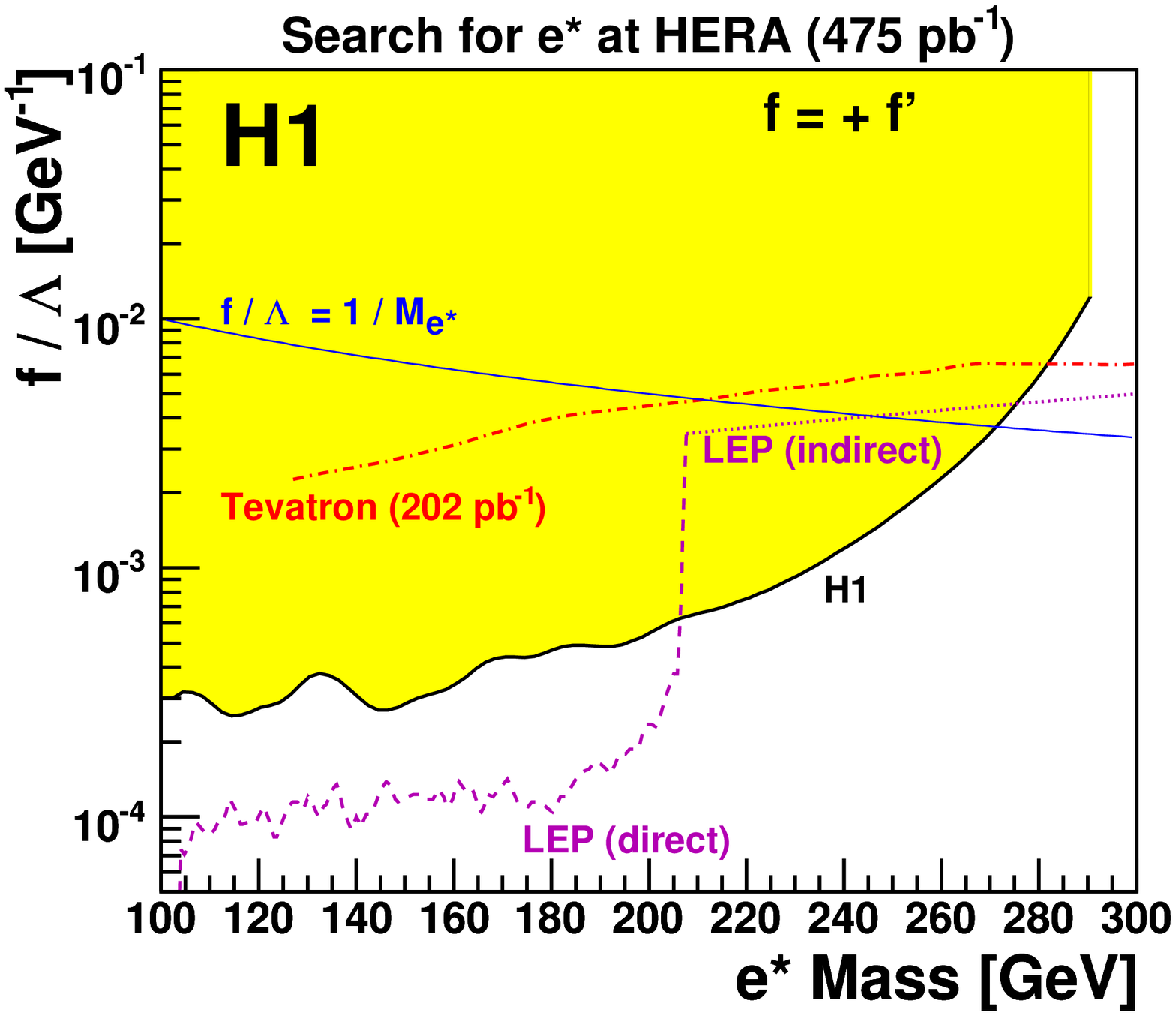}
    \includegraphics[width=0.95\columnwidth]{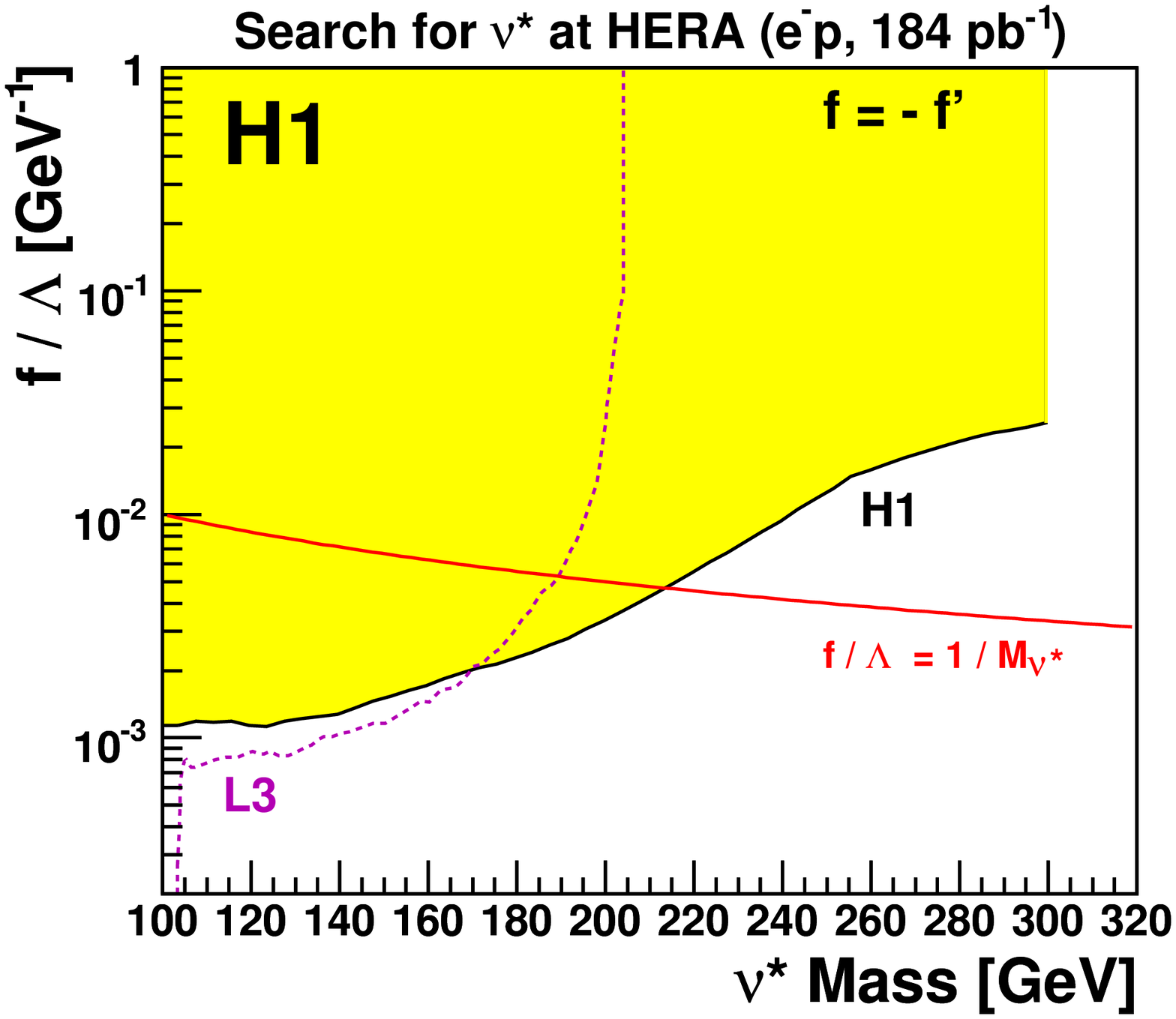}
    \includegraphics[width=0.95\columnwidth]{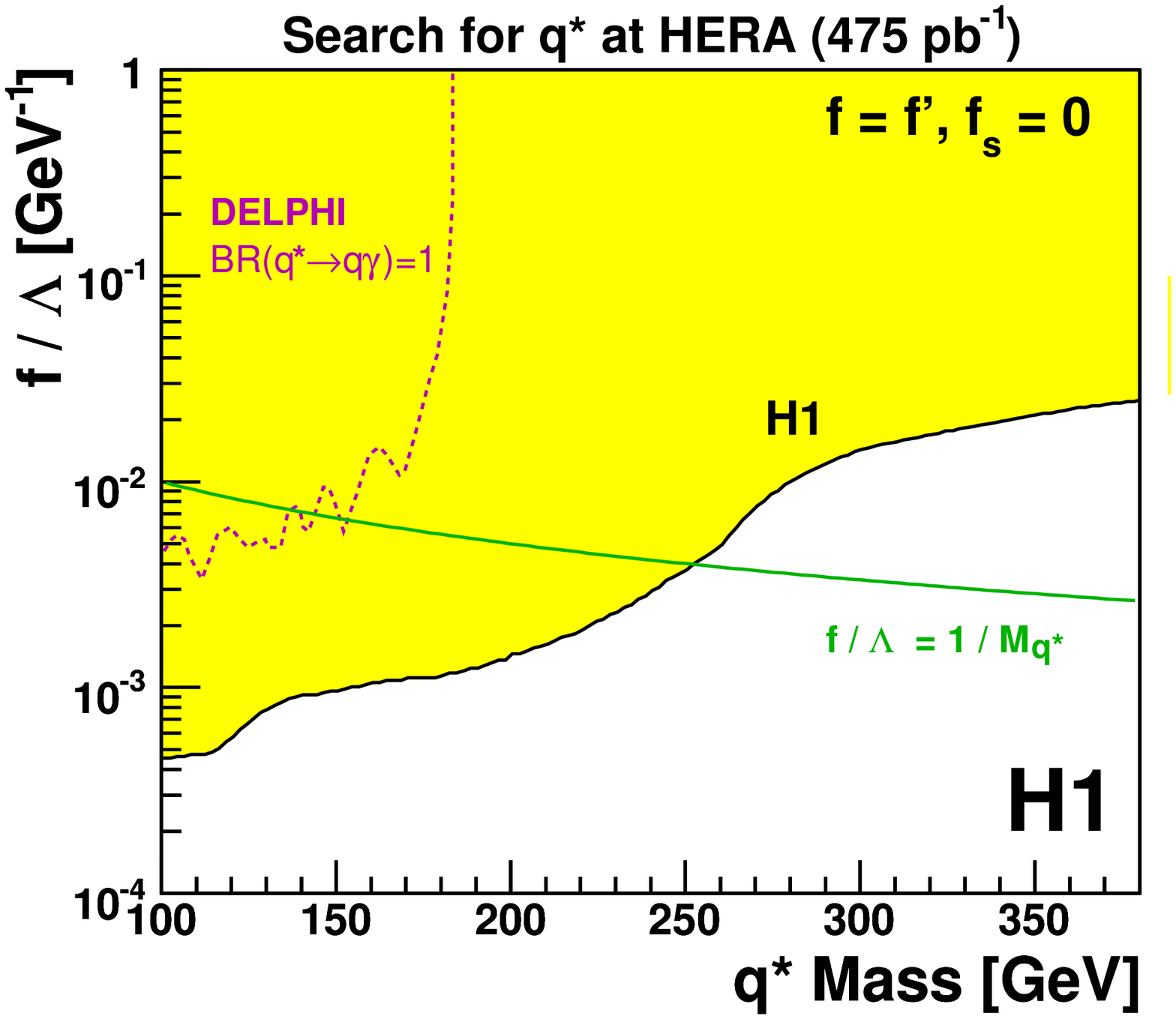}
    \caption{Exclusion limits at $95\%$~CL on the ratio $f/\Lambda$
      as a function of $M_{f^{*}}$, for excited electrons (top),
      excited neutrinos (middle) and excited quarks (bottom) with the
      assumptions given in the text and on the plot.
      Values of $f/\Lambda$ above the curves are excluded.
      Limits from the LEP and Tevatron colliders are also indicated.}  
    \label{fig:fstar-limits}
  \end{center}
\end{figure}

\section{Searches for supersymmetry}
\label{sec:susy}

Supersymmetric (SUSY) extensions of the SM~\cite{Wess:1974tw} introduce new elementary
particles which are the superpartners of SM particles but
differ in spin by half a unit~\cite{Nilles:1983ge,Haber:1984rc}. 
The hypothesis of the existence of supersymmetric particles has had a huge impact 
on both theory and experiments. A large variety of new states and a rich phenomenology 
of the Higgs sector was predicted, which came into focus during the running of the LEP and 
HERA experiments, and is now thoroughly investigated at the LHC.
The masses of the new particles, called {\it sparticles}, are related
to the symmetry breaking mechanism. 
A new quantum number $R_p=(-1)^{3B+L+2S}$ is defined, denoted $R$-parity, 
where $B$ is the baryon number, $L$ the lepton number and $S$ the spin of a particle. 
For particles $R_p=1$ and for their supersymmetric partners $R_p=-1$.

%%%

Most collider searches focus on SUSY models that conserve
$R$-parity, only allowing the production of sparticles in pairs.
However, the most general supersymmetric theory that is renormalisable
and gauge invariant with respect to the SM gauge group does not
impose $R$-parity conservation~\cite{Weinberg:1981wj,Sakai:1981pk},
allowing couplings between two SM fermions and a squark ($\tilde{q}$) or a 
slepton ($\tilde{l}$).
This makes the resonant, single production of sparticles via $R$-parity
violating (\rpv) couplings possible, and the $ep$ collisions at
HERA provide an ideal environment to search for such new particles.
Searches have been performed at HERA for three different
\rpv SUSY signatures with squarks, bosonic stop decays and light
gravitinos, which are described in the remainder of this
section.

\subsection{Search for squarks in $\boldmath R$-parity violating supersymmetry}
\label{sec:squarks}

In SUSY models with $R$-parity violation squarks can couple to
electrons and quarks via the \rpv Yukawa couplings $\lambda'_{1jk}$, and
then subsequently decay to a number of different final state
topologies which may be investigated at HERA~\cite{Butterworth:1992tc}.
Searches for such final states have been performed by both
H1~\cite{Aktas:2004ij} and ZEUS~\cite{Chekanov:2006aa} using their
HERA~I data sets.
In this review we focus on the most recent H1
analysis~\cite{Aaron:2010ez}, which employs the full HERA data set,
corresponding to an integrated luminosity $183$~pb$^{-1}$ for $e^{-}p$
collisions and $255$~pb$^{-1}$ for $e^{+}p$ collisions.

%%%

Feynman diagrams of the resonant production and decay of
squarks at HERA via \rpv couplings $\lambda'_{1jk}$ are shown in
figure~\ref{fig:squark-directdecays}.
For simplicity, it is assumed here that one of the $\lambda'_{1jk}$ couplings
dominates over all the other trilinear \rpv couplings.
At the high values of Bjorken-$x$ which are required to produce squarks of
significant mass, the valence quarks are the dominant contribution to
the proton PDFs.
Therefore, $e^-p$ scattering gives sensitivity to the couplings
$\lambda'_{11k}$ $(k=1,2,3)$ which dominate the production of
$\tilde{d}_R$-type squarks (i.e. the superpartners $\tilde{d}_R$,
$\tilde{s}_R$ and $\tilde{b}_R$ of down-type quarks).
Conversely, $e^+p$ scattering provides sensitivity to the
couplings $\lambda'_{1j1}$ $(j=1,2,3)$, which dominate the production
of $\tilde{u}_L$-type squarks (i.e. the superpartners $\tilde{u}_L$,
$\tilde{c}_L$ and $\tilde{t}_L$ of up-type quarks).
Due to the larger $u$-quark density in the proton at large $x$ with
respect to the $d$-quark density, larger production cross sections are
expected in $e^-p$ interactions for identical couplings and squark masses.

%%%

All sparticles are unstable in \rpv SUSY and squarks can decay
directly via the Yukawa coupling $\lambda'_{ijk}$ into SM fermions.
The $\tilde{d}^k_R$-type ($k=1,2,3$) squarks can decay via the
coupling $\lambda'_{11k}$ either into $e^{-}u$ or $\nu_{e}d$,
while the $\tilde{u}^j_L$-type ($j=1,2,3$) squarks decay via the
coupling $\lambda'_{1j1}$ into $e^{+}d$ only, as illustrated in
figure~\ref{fig:squark-directdecays}.
Squarks may also decay via $R_{p}$ conserving gauge couplings with
subsequent \rpv decay into SM particles via the Yukawa coupling
$\lambda'_{1jk}$.
In this case, the $\tilde{u}_L$-type squarks can undergo a gauge decay into states
involving a neutralino $\chi^0_i$ ($i=1,2,3,4$), a chargino $\chi^{+}_i$ ($i=1,2$) 
or a gluino $\tilde{g}$.
However, $\tilde{d}_R$-type squarks mainly decay to $\chi^0_i$ or $\tilde{g}$ 
and decays into charginos are suppressed~\cite{Butterworth:1992tc}.
The resulting final states observed in the detector from such cascade
decays may contain multiple leptons and jets as well as missing
transverse momentum.
Such a cascade decay is illustrated in figure~\ref{fig:squark-examplegaugedecay},
where a $\tilde{d}^k_R$-type squark decays via a neutralino $\chi^0_1$
to a selectron-electron pair, and finally the selectron decays to SM
fermions.

%%%

Squark signal events are simulated using cross sections obtained in
the narrow width approximation from the leading order amplitudes in
leptoquark production~\cite{Buchmuller:1986zs}, adjusted to NLO QCD
using multiplicative correction factors~\cite{Plehn:1997az}.
The parton densities are evaluated at the hard scale $M_{\tilde{q}}^2$.
A dedicated MC simulation is performed for each of the signal topologies:
for the direct lepton-quark decay channels $eq$ and $\nu q$ shown in
figure~\ref{fig:squark-directdecays}, events are generated using
LEGO~\cite{LEGO}, whereas for the gauge decays of squarks such as the one
in figure~\ref{fig:squark-examplegaugedecay}, events are generated using
SUSYGEN3~\cite{Ghodbane:1999va}.
To allow for a model independent interpretation of the results,
the squark decay processes are simulated for a wide range of sparticle
masses.
Further details on the signal event simulation can be found in the H1
publication~\cite{Aaron:2010ez}.
\begin{figure} 
  \begin{center}
    \includegraphics[width=0.475\columnwidth]{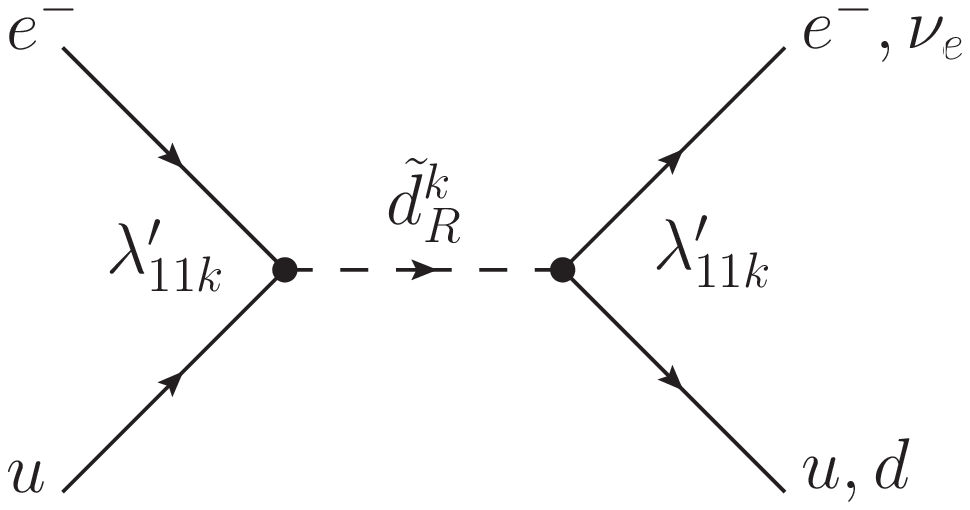}
    \hspace{0.1cm}
    \includegraphics[width=0.475\columnwidth]{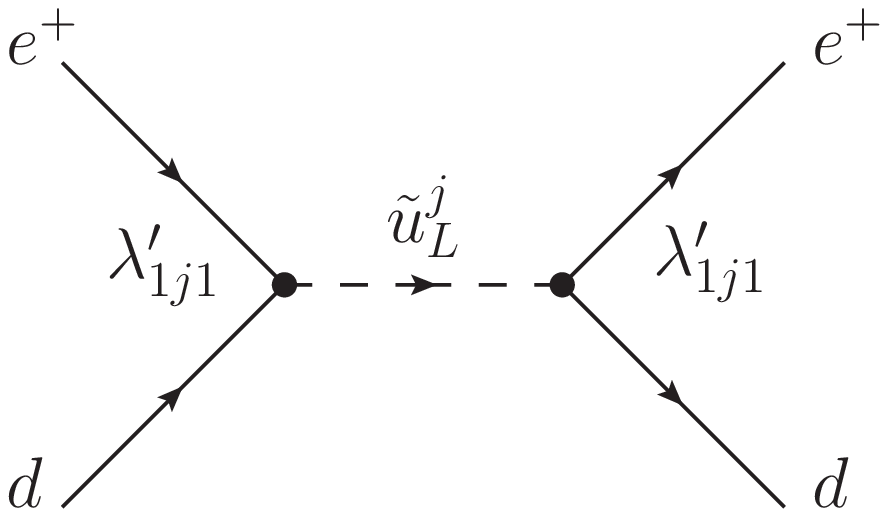}
  \caption{Feynman diagrams for the single resonant s-channel production of right-handed
    down-type squarks in $e^-p$ collisions (left) and left-handed up-type squarks in $e^+p$ collisions (right)
    with subsequent decays into SM particles via Yukawa couplings $\lambda'_{11k}$ or 
    $\lambda'_{1j1}$, respectively. The right-handed down-type squarks 
    can decay either into $e^{-}u$ or $\nu_{e}d$, while the left-handed up-type 
    squarks decay into $e^{+}d$ only.}
  \label{fig:squark-directdecays}
\end{center}
\end{figure}
\begin{figure}
  \begin{center}
   \includegraphics[width=0.8\columnwidth]{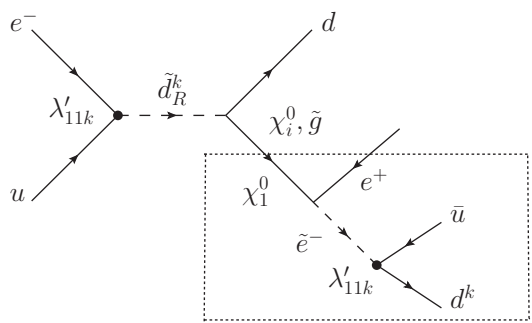}
   \caption{Feynman diagram for squark production and subsequent
     cascade decay via gauginos, shown here for the case of right-handed down-type
     squarks with subsequent \rpv slepton decay into SM fermions via Yukawa couplings $\lambda'_{11k}$.
     Decays of down-type squarks to charginos are suppressed, and the
     $\tilde{d}_R$ decays to either $\chi^0_i$ (where $i=1,2,3,4$) or
    $\tilde{g}$.}
  \label{fig:squark-examplegaugedecay}
  \end{center}
\end{figure} 

%%%

\begin{table*}
  \renewcommand{\arraystretch}{1.3}
  \caption{Summary of the observed and predicted event yields for the
    different decay channels considered in H1 search for squarks in
    \rpv SUSY, in the $e^-p$ and $e^+p$ data. The total uncertainties on the
    SM prediction is determined by
    adding the effects of all model and experimental systematic
    uncertainties in quadrature. The range of signal efficiencies is
    also given for each channel, for squark masses ranging from
    $100$~GeV to $290$~GeV and gaugino masses ranging from
    $30$~GeV up to the squark mass.}
  \label{tab:squark-rates}
  \begin{tabular*}{1.0\textwidth}{@{\extracolsep{\fill}} lccccc}  
    \hline
    \multicolumn{6}{@{\extracolsep{\fill}} l}{{\bf \boldmath H1 Search for Squarks in $R_{p}$ Violating Supersymmetry}}\\
    \hline
    & \multicolumn{2}{@{\extracolsep{\fill}} l}{\bf \boldmath $e^{-}p$
      collisions, ${\mathcal L} = 183$ pb$^{-1}$} & \multicolumn{2}{@{\extracolsep{\fill}} l}{\bf \boldmath $e^{+}p$
      collisions, ${\mathcal L} = 255$ pb$^{-1}$} & \\
    Channel & Data & Total SM & Data & Total SM & Signal Eff. [\%]\\
    \hline
    $eq$           & $3121$ & $3215 \pm 336$    & $2946$ & $ 2899 \pm 302$ & $30$ -- $40$\\
    $\nu q$      & $2858$ & $2983 \pm 358$    &     --     &  ~--                      & $50$ -- $60$\\
    \hline
    $eMJ$ (RC)   & $147$  & $158.3 \pm 23.9$  & $140$   & $146.0 \pm 21.4$ & $10$ -- $40$\\
    $eMJ$ (WC)  & $0$      & ~$1.3 \pm 0.3 $   & $1$       & ~$0.6 \pm  0.4 $  & ~$5$ -- $20$\\
    $ee MJ$       & $0$      & ~$1.5 \pm  0.5 $    & $2$       & ~$1.7 \pm 0.5 $  & ~$5$ -- $35$\\
    $e\mu MJ$ & $0$     & ~$0.03 \pm 0.02 $  & $0$    & ~$0.03 \pm 0.03 $ & ~$5$ -- $15$\\
    $e \nu MJ$  & $3$     & ~$5.6 \pm 1.2 $   & $5$       & ~$8.2 \pm 2.0 $ & ~$5$ -- $40$\\
    \hline
    $\nu MJ$     & $204$ & $235.5 \pm 63.3 $  & $113$ & $134.0 \pm 33.8 $ & ~$5$ -- $15$\\
    $\nu\mu MJ$  & $0$ & ~$0.04 \pm 0.02 $ & $0$ & ~$0.06 \pm 0.03 $ & ~$5$ -- $20$\\
    \hline
  \end{tabular*}
\end{table*}
\begin{figure*}
  \centerline{
    \includegraphics[width=0.33\textwidth]{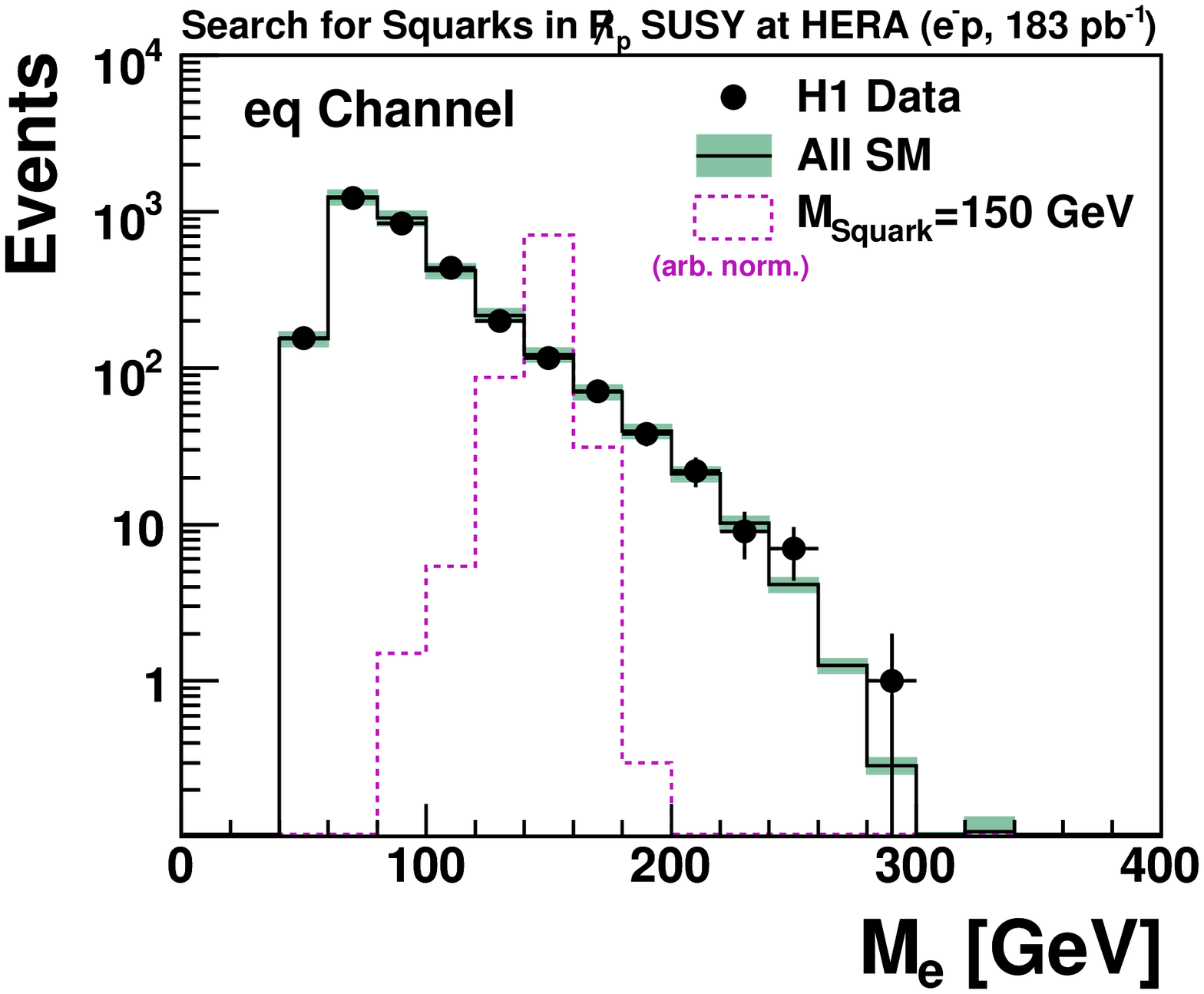}
    \includegraphics[width=0.33\textwidth]{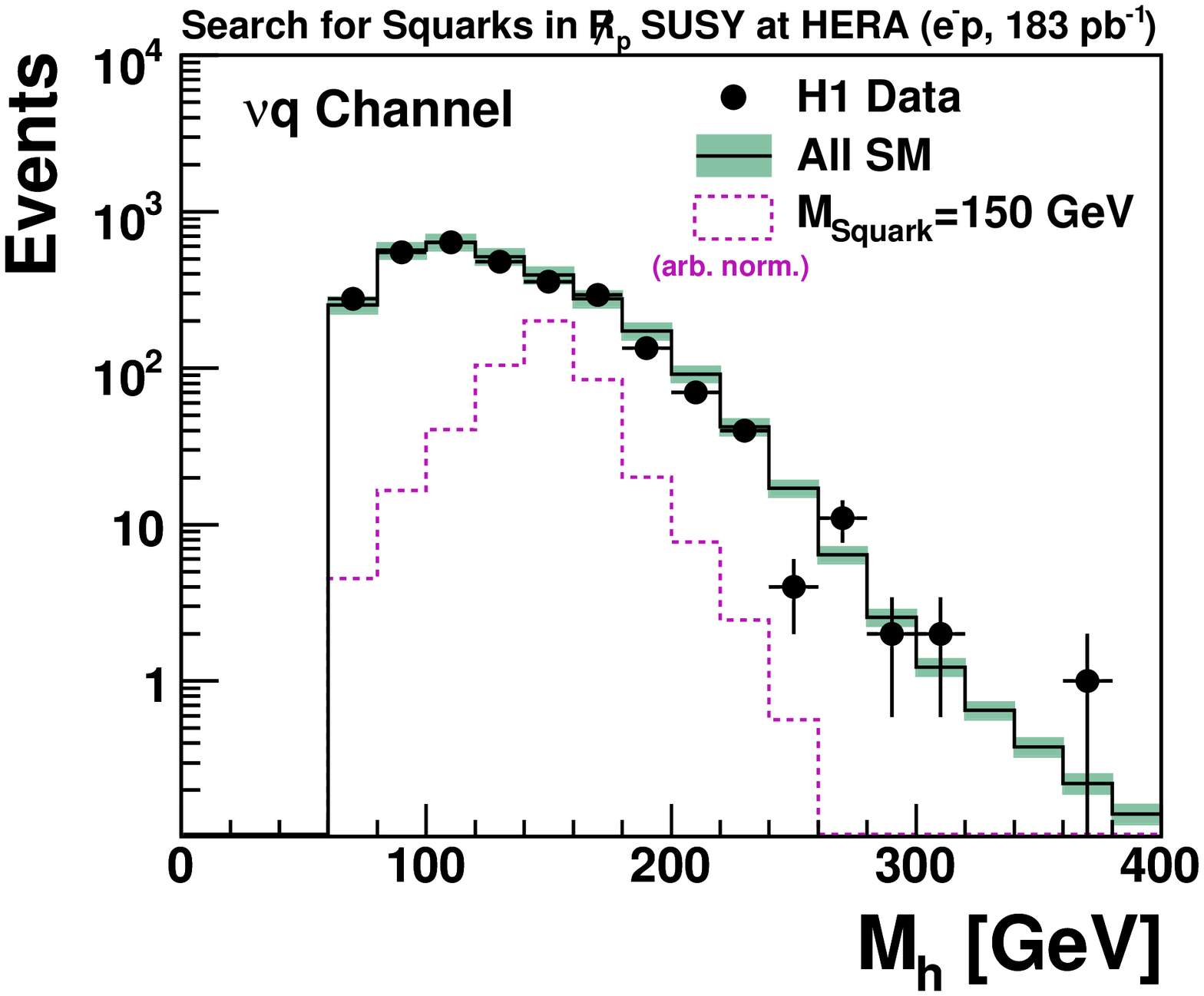}
    \includegraphics[width=0.33\textwidth]{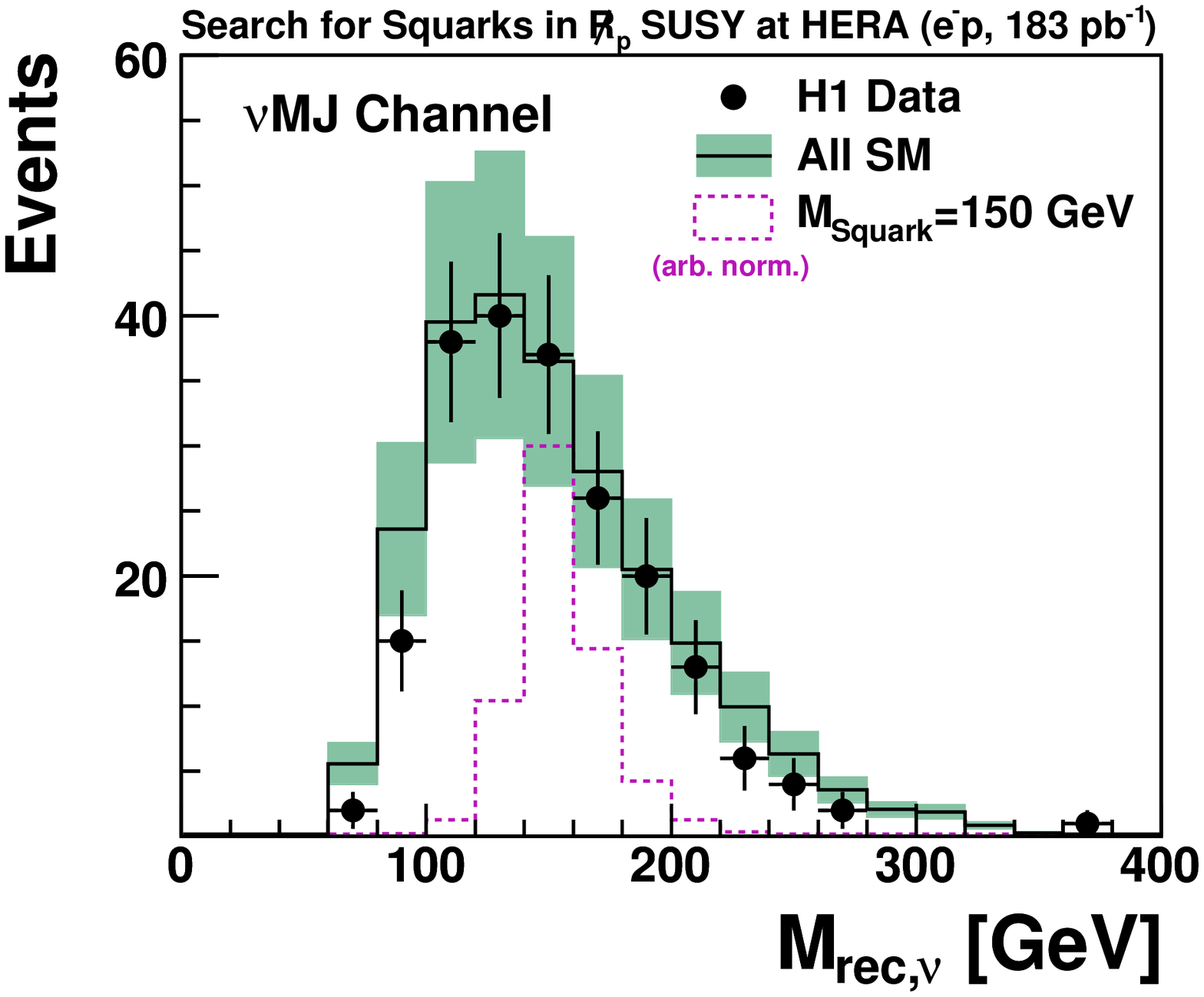}
  }
  \centerline{
    \includegraphics[width=0.33\textwidth]{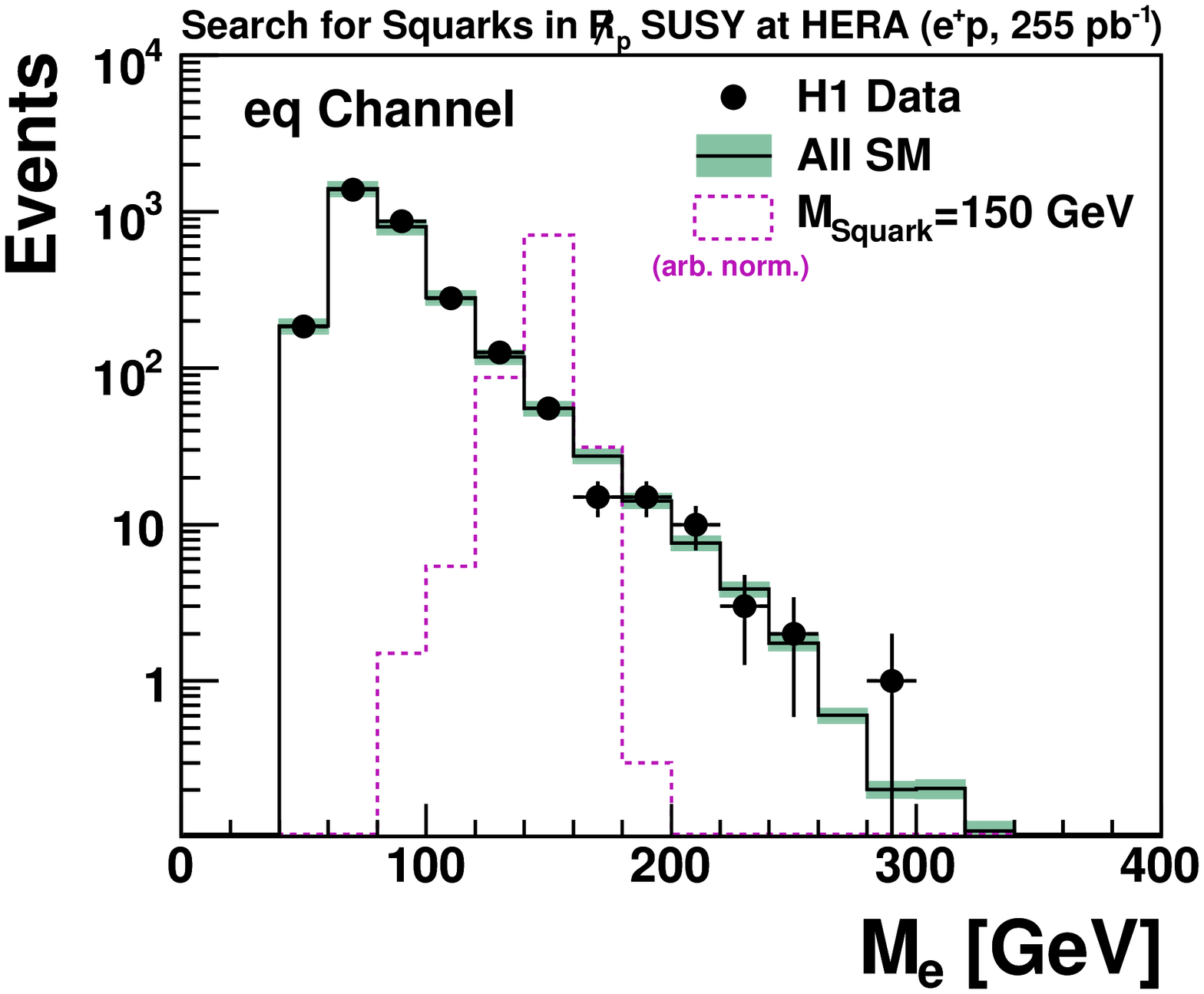}
    \includegraphics[width=0.33\textwidth]{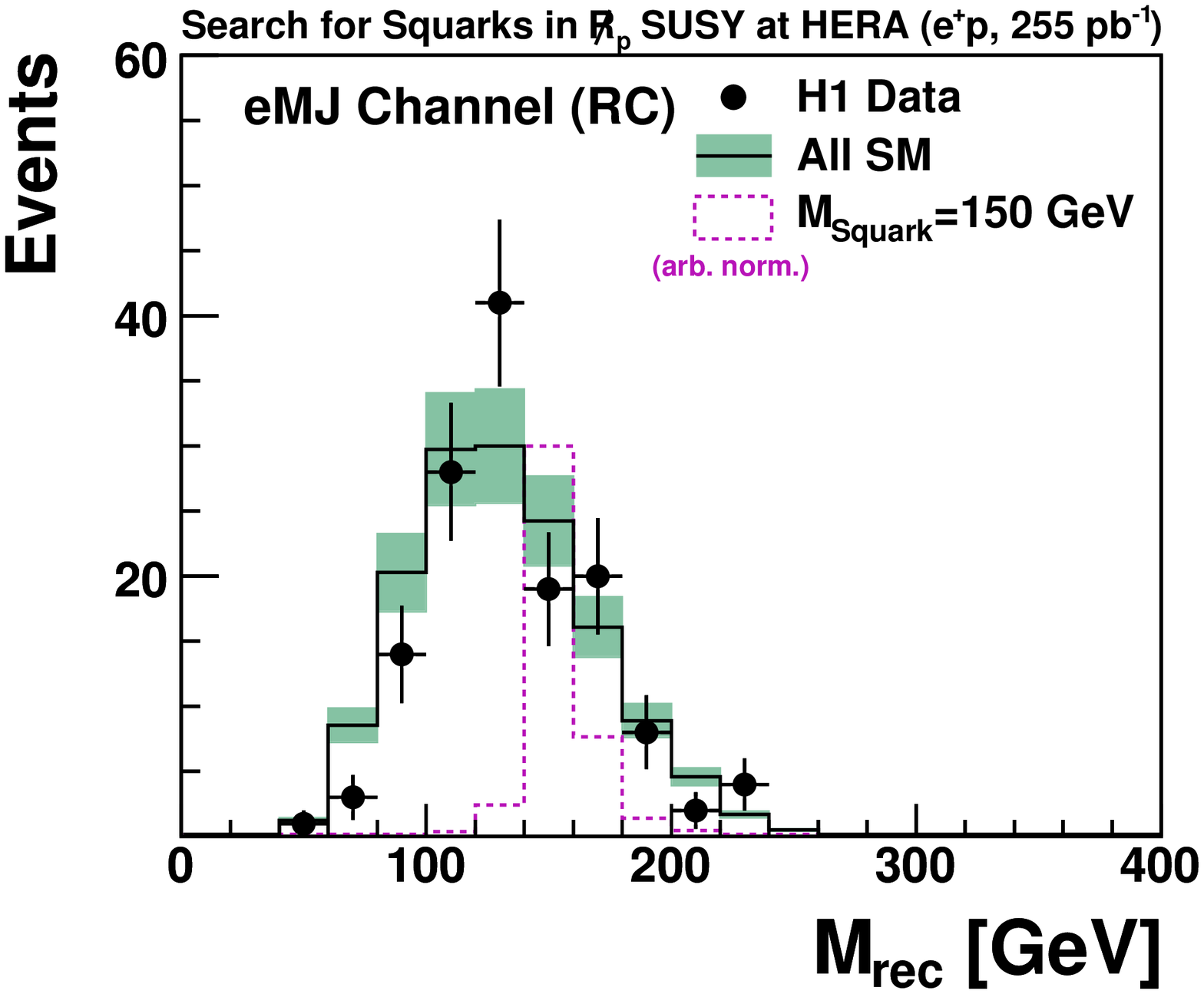}
    \includegraphics[width=0.33\textwidth]{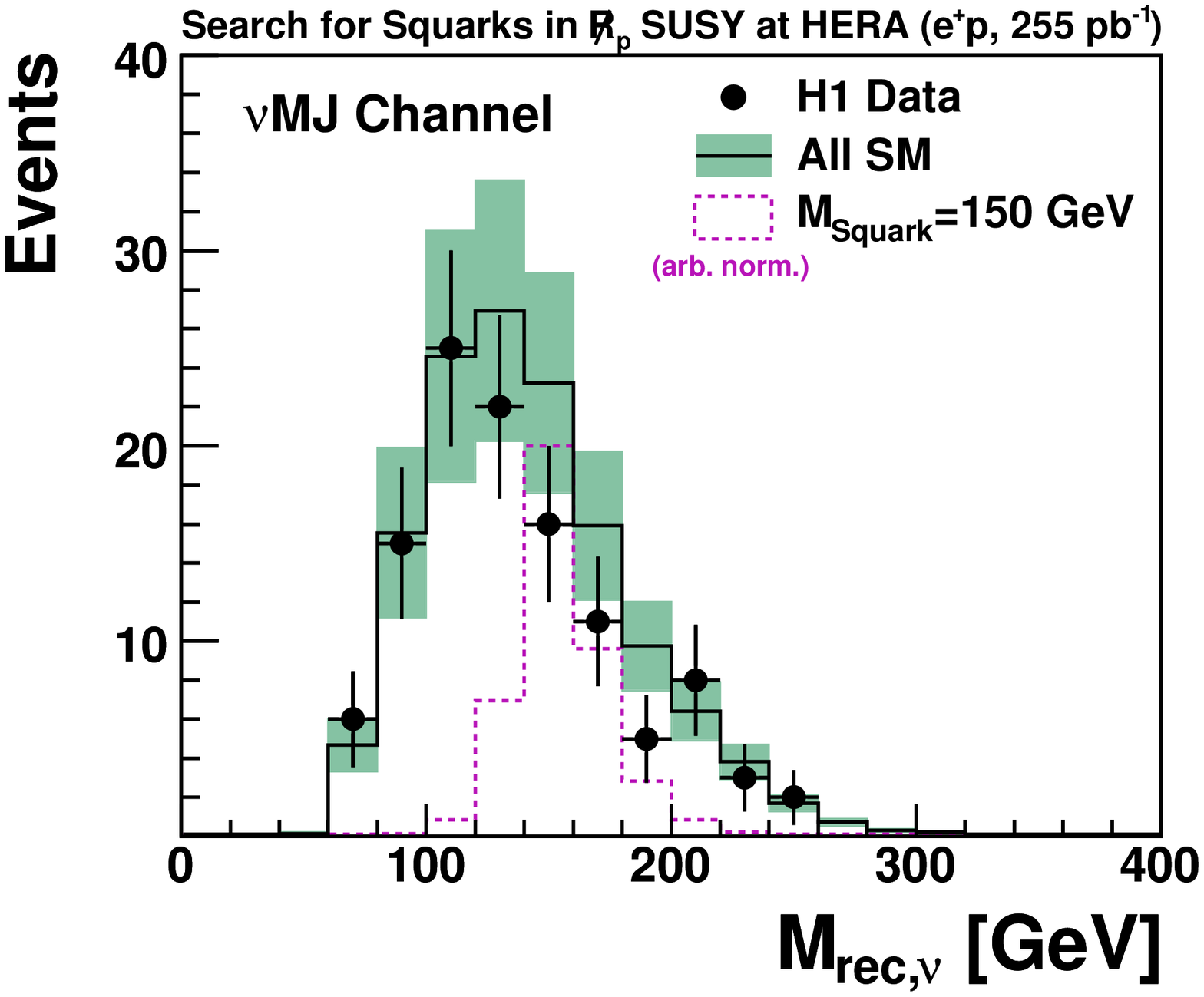}
  }
  \caption{Selected reconstructed invariant mass distributions from the H1 search
    for squarks in the $e^{-}p$ data (top row) and $e^{+}p$ data
    (bottom row). The data (points) are compared to SM MC predictions. 
    The error band represents all model and experimental systematic
    uncertainties on the SM prediction (solid histogram) added in
    quadrature. The dashed histogram indicates the signal from a squark with 
    $M_{\tilde q} = 150\,\text{GeV}$ with arbitrary normalisation.}
  \label{fig:squark-masses}
\end{figure*}

As explained above, the squark decays in this analysis can
produce a large variety of final states, which are
classified~\cite{Aktas:2004ij} into event topologies, or channels,
to be examined.
This classification relies on the number of isolated electrons,
muons and hadronic jets in the final state, as well as on the presence
of missing transverse momentum, indicating undetected neutrinos.
A list of channels investigated in this analysis can be found in
table~\ref{tab:squark-rates}.
The channels labelled $eq$ and $\nu q$ are the squark decay modes
that proceed directly via \rpv couplings resulting in event topologies
with an isolated electron or neutrino and a single jet. 
The remaining channels result from the gauge decays of the squark and
are characterised by final states with more than one jet,
(``multijet'', $M\!J$) with additional leptons.
The $eM\!J$ and $\nu M\!J$ channels involve one or two gauginos
in the decay cascade.
In the $eM\!J$ channel, an electron or positron can be found in the
final state and a distinction may be made with respect to the
incident beam lepton charge and therefore, two discrete channels
with the ``right'' (same) sign lepton charge $eM\!J$(RC) and
``wrong'' (opposite) sign lepton charge $eM\!J$(WC) are formed.
Channels with an electron or neutrino and a further charged lepton
are denoted $ee M\!J$, $e\mu M\!J$ and $e \nu M\!J$, $\nu \mu M\!J$,
respectively, and necessarily involve two gauginos.

%%%

Several SM processes may mimic the final states produced by squark
decays and therefore a standard selection of MC generators
is employed to compare the observed data events to the prediction
from the SM.
NC DIS events are simulated using RAPGAP and both direct and resolved
photoproduction of jets, as well as prompt photon production, are
simulated using PYTHIA.
Inclusive CC DIS events are simulated using DJANGOH.
The leading order MC prediction of processes with two or more high
transverse momentum jets in NC DIS, CC DIS and photoproduction is
scaled by a factor of $1.2$ to account for the incomplete description of 
higher orders in the MC generators~\cite{Adloff:2002au,Aktas:2004pz}. 
Additional, smaller contributions from single $W$ boson production and
lepton pair production are modelled using EPVEC and GRAPE, respectively.

%%%

Each of the final states listed in table~\ref{tab:squark-rates}
is analysed separately using a dedicated event
selection~\cite{Aaron:2010ez}.
These selections are based on the identification of high $P_{T}$
electrons, muons, jets, as well as missing transverse momentum.
%a
The final state of a squark decaying into an electron and a high $P_T$
jet is identical to that from a NC DIS event at high $x$ and $Q^2$, and
therefore the event selection in the $eq$ channel closely resembles the
selection described in section~\ref{sec:sm}: An isolated electron is
required in the event with $P_{T}^{e}>16$~GeV in the region
$5^{\circ}<\theta_e<145^{\circ}$, and the event must be in the
kinematic phase space $Q^2_e>2500$~GeV$^2$, $y_e<0.9$,
$40$~GeV $< \delta< 70$~GeV and ${P}_T^{\rm{miss}}<15$~GeV.
The squark mass is reconstructed as $M_e= \sqrt{x_{e} s}$ and a
mass dependent $y_e$ cut is added to separate the NC background from
the signal~\cite{Aaron:2010ez}.

%%%

Squarks decaying into a neutrino and a high $P_T$ jet lead
to the same signature as CC DIS events with high missing transverse
momentum, and so the event selection in the $\nu q$ channel is
accordingly based on the analysis of such events.
Events with a neutrino are selected by requiring ${P}_T^{\rm{miss}}>30$~GeV
and $\delta < 50$~GeV} and the phase space is restricted to
$Q^2_h>2500$~GeV$^2$ and $y_h<0.9$.
Similarly to the $eq$ channel, a $y_h$ cut dependent on the
reconstructed mass $M_h = \sqrt{x_h s}$ is applied.
The $\nu q$ channel is not relevant for $e^+p$ data since the
$\tilde{u}_L$-type squarks produced in $e^+p$ do not undergo this decay.

%%%

Squarks decaying via neutralinos or charginos are expected to have
a higher multiplicity of jets and leptons in the final state and their
signatures correspond to final states detectable in higher order
NC DIS processes.
Squark decays with single or multiple neutrinos produced via
neutralino or chargino decays can result in final states similar to
that of higher order CC DIS processes.
The remaining channels can therefore be divided into two groups, for
which common preselections are employed: on the one side
electron-multijet and electron-lepton-multijet final states and on the
other side secondly neutrino-multijet and neutrino-muon-multijet final states.
Further cuts are then applied depending on the number and flavour of
the leptons in the event, as well as the charge, to separate the $eM\!J(RC)$
and $eM\!J(WC)$ channels.
The event selections are optimised to maximise
the signal efficiency in each channel and are described
in the publication~\cite{Aaron:2010ez}.
For each selected event a squark mass
$M_{\rm{rec}}= \sqrt{4E^0_e (\sum{E_i}-E^0_e)}$ is calculated, where
the sum includes the energies of the reconstructed electrons,
muons and jets with $P_T^{\rm{jet}}>5$~GeV in the event, in
addition to the neutrino in the case of the $\nu MJ$ and $\nu \mu MJ$
channel.

%%%

The number of events observed in the data in each channel is shown
in table~\ref{tab:squark-rates} compared to the SM prediction, where
a good agreement is observed in all channels.
Of the gauge decays, only the $eM\!J(RC)$  and $\nu M\!J$ channels
have significant event yields.
The invariant mass distributions of some of the most populous
squark decay channels are displayed in figure~\ref{fig:squark-masses},
where the data are in good agreement with the SM expectation.
As no significant deviation from the SM is observed, all analysis channels are 
combined to set constraints on various supersymmetric models as
described in the following.
Mass dependent exclusion limits are obtained~\cite{Aaron:2010ez}
on the production of squarks parameterised by the strength of the
\rpv couplings $\lambda'_{1j1}$ and $\lambda'_{11k}$.

%%%

\begin{figure*} 
  \centerline{
    \includegraphics[width=0.5\textwidth]{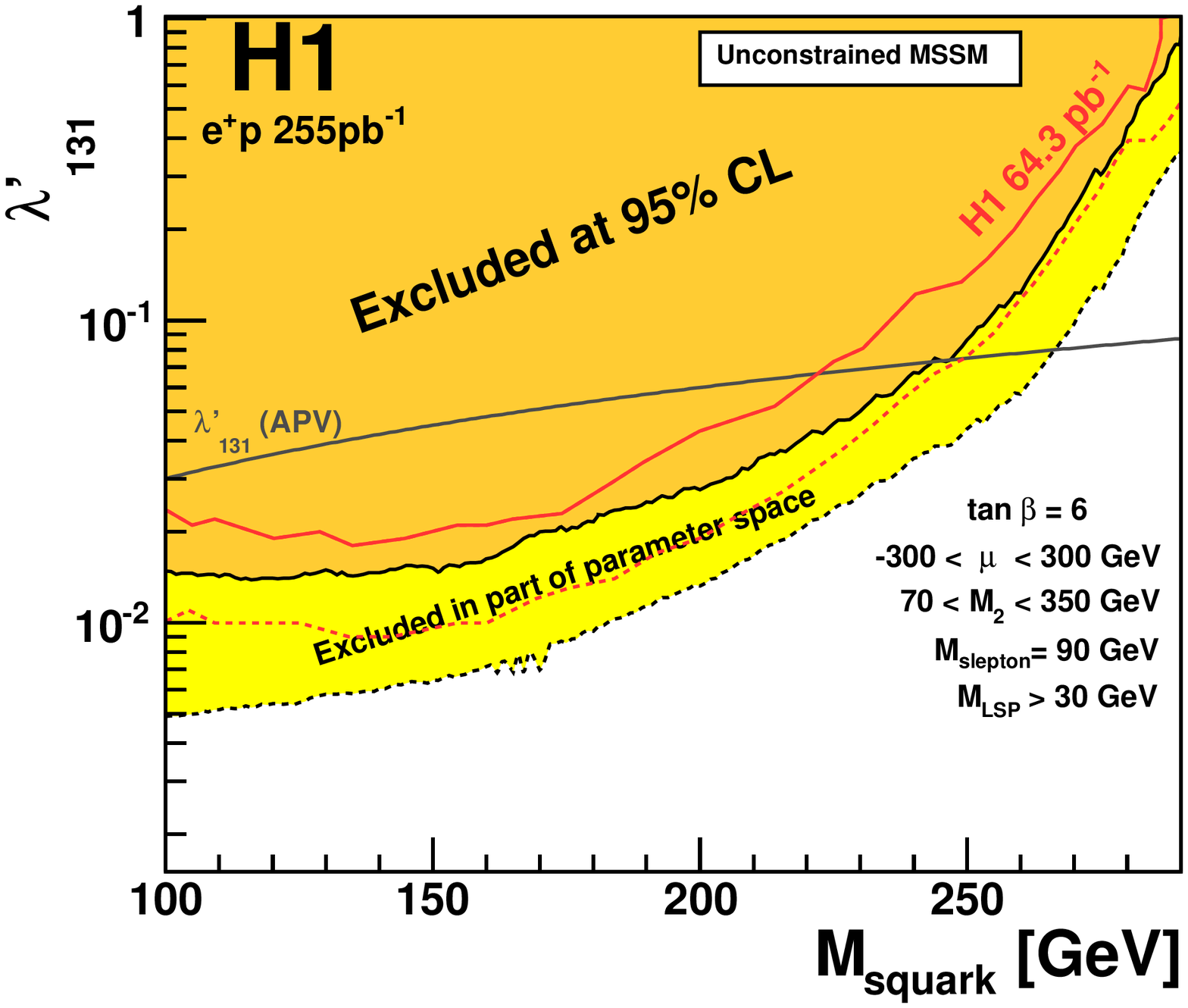}
    \includegraphics[width=0.5\textwidth]{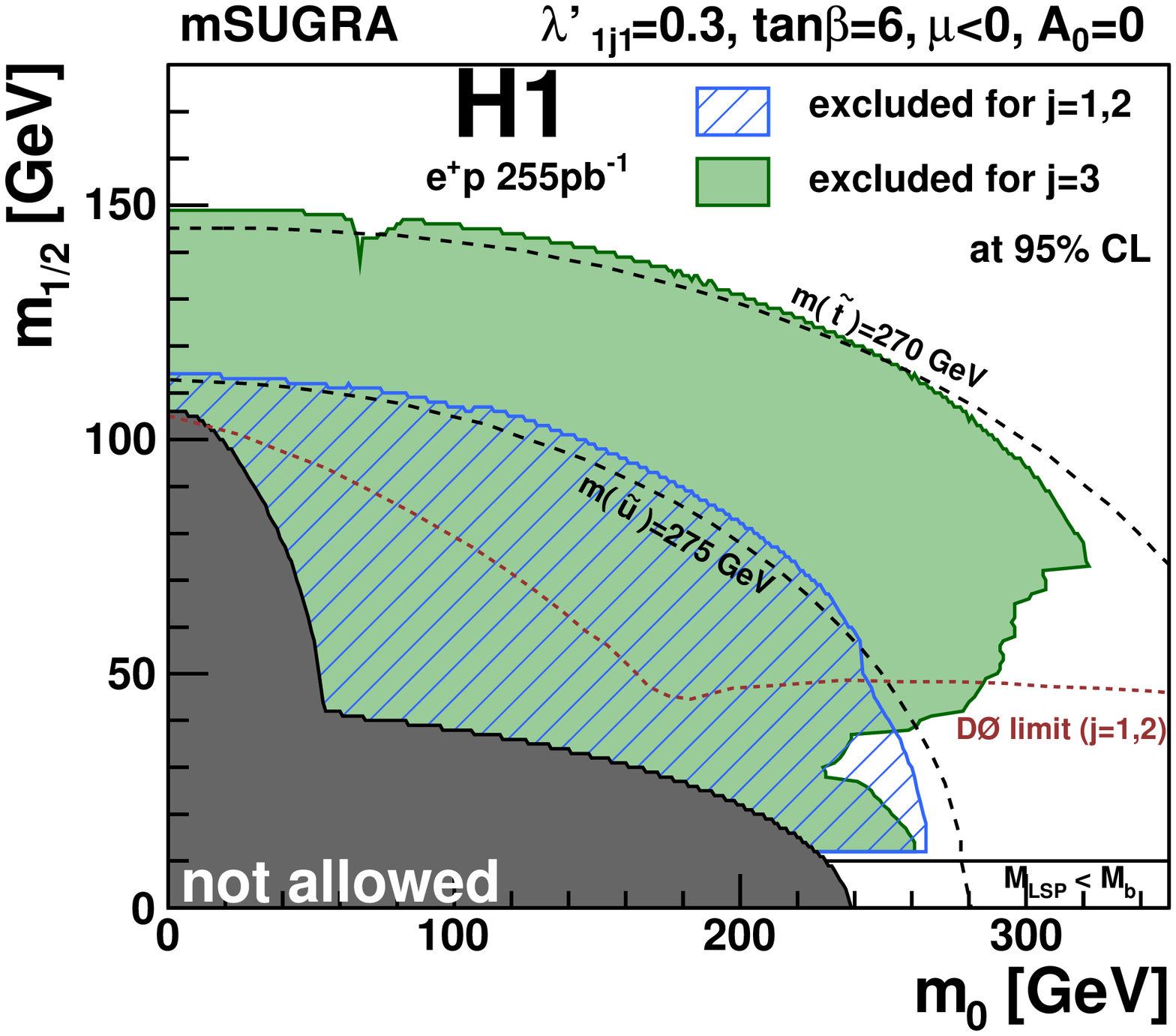}
  }
  \caption{Left: Exclusion limits at the $95\%$ CL on $\lambda'_{131}$
    as a function of the squark mass from a scan of the MSSM parameter
    space as indicated in the figure. The dark filled region indicates
    values of the coupling $\lambda'_{131}$ excluded in all
    investigated scenarios whereas the light filled region is excluded
   only in part of the parameter space. An indirect limit from atomic parity violation 
   (APV) is also shown, as well as the the limit from the previous H1 analysis.
   Right: Exclusion limits at the $95\%$ CL in the $m_0,m_{1/2}$ plane in
   the mSUGRA parameter space indicated in the figure for $j=1,2$
   (striped region) and $j=3$ (light filled region). Curves of constant squark
   mass are illustrated for $m(\tilde{u})=275$~GeV and $m(\tilde{t})=270$~GeV.
   A constraint obtained by the D\O\ experiment at the
   Tevatron is also indicated. The dark filled region
   labelled as ``not allowed'' indicates where no radiative
   electroweak symmetry breaking solution is possible or where
   the LSP is a sfermion.}
 \label{fig:squark-limits}  
\end{figure*}
 
%%%

For the interpretation of the results a version of the Minimal Supersymmetric 
Standard Model (MSSM) is considered where the masses of the
neutralinos, charginos and gluinos are determined via the usual SUSY parameters:
the Higgs mass term $\mu$, which mixes the Higgs super-fields;
the SUSY soft-breaking mass parameter $M_2$; and the ratio of the
vacuum expectation values of the two neutral scalar Higgs fields
$\tan \beta$~\cite{Nilles:1983ge,Haber:1984rc}.
The parameters are defined at the electroweak scale and the lightest
supersymmetric particle (LSP) is the neutralino $\chi^0_1$.
Slepton masses $M_{\tilde{l}}$ are fixed at $90$~GeV.
Exclusion limits are calculated for two scenarios with $\tan \beta =
2$: a photino-like neutralino ($\mu = -200$~GeV, $M_2 =
80$~GeV) and a zino-like neutralino ($\mu = 200$~GeV, $M_2 =
150$~GeV).
A combination of both scenarios is also achieved by performing a full parameter
scan, where the parameters $M_2$ and $\mu$ are varied in the range
$70$~GeV~$< M_2 < 350$~GeV and $-300$~GeV~$ < \mu < 300$~GeV
for $\tan \beta = 6$.
As an example, exclusion limits at $95\%$ CL from the parameter
scan for the $\lambda'_{131}$ coupling as a function of squark mass
using the full H1 $e^{+}p$ data are shown in figure~\ref{fig:squark-limits}
(left), compared to the previous H1 limit~\cite{Aktas:2004ij} and an
indirect limit from atomic parity violation~\cite{Langacker:1990jf}.
In the parameter space considered in the analysis, Yukawa couplings of
electromagnetic strength, $\lambda'_{1j1}$ or
$\lambda'_{11k}=\sqrt{4\pi\alpha_{\rm em}}= 0.3$, are excluded up to
masses of $275$~GeV at $95\%$~CL for up-type squarks and up to
masses of $290$~GeV for down-type squarks.

%%%

Constraints are also obtained on the Minimal Supergravity
Model (mSUGRA)~\cite{Freedman:1976xh,Drees:1991ab,Baer:1992dc,Kane:1993td}, which
is a model that assumes gauge coupling unification and radiative
electroweak symmetry breaking with the choice of 5 parameters: the
common mass of scalar sparticles $m_0$; the common mass of fermionic
sparticles $m_{1/2}$; the common trilinear coupling $A_0$; the sign of
the Higgs mixing parameter $\mu$ and $\tan \beta$ as defined above.
The masses of squarks, sleptons and gauginos as well as the branching
ratios in the analysis channels are determined by the parameter set
for given values of the couplings $\lambda'_{11k}$ and $\lambda'_{1j1}$.
$A_0$ enters only marginally in the interpretation and is set to zero. 
The parameter $\mu$ is taken with negative sign.
Figure~\ref{fig:squark-limits} (right) shows example exclusion
limits at $95\%$ in the $m_0,m_{1/2}$ plane assuming $\lambda'_{1j1} =
0.3$ and $\tan \beta = 6$, obtained using the full H1 $e^{+}p$ data.
The excluded region typically covers masses of $m(\tilde{u})=275$~GeV
and $m(\tilde{t})=270$~GeV, as indicated in the figures.
A constraint from the D\O\ experiment~\cite{Abbott:1999nh} at the Tevatron
using di-electron events is also indicated, where the region excluded by
H1 is considerably larger.

\subsection{Search for bosonic stop decays in $\boldmath R$-parity violating supersymmetry}
\label{sec:bosonicstop}

\begin{figure}
  \begin{center}
    \includegraphics[clip,width=1.0\columnwidth]{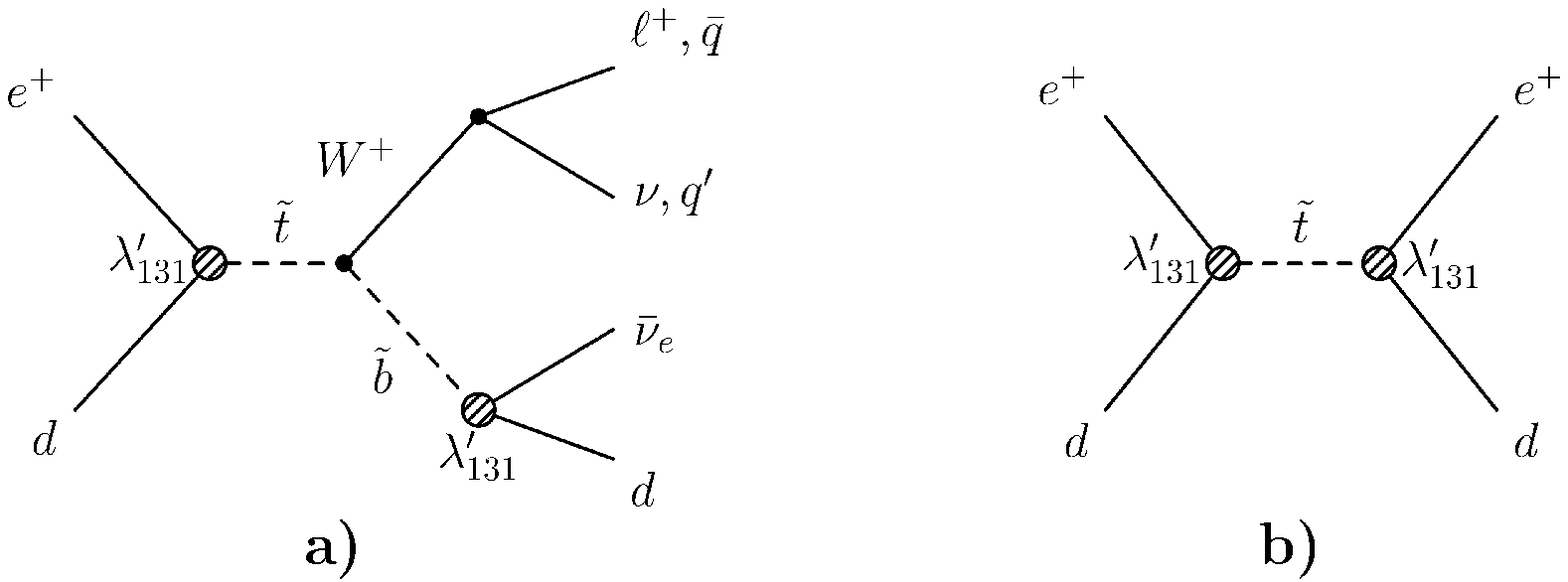}
  \end{center}
  \caption{Lowest order $s$-channel diagram for \rpv stop production
    at HERA followed by (left) the bosonic decay of the stop and (right) the
    \rpv decay of the stop.}
  \label{fig:stop-feynman}
\end{figure}

In most SUSY models the third generation squarks, namely stop ($\tilde
t$) and sbottom ($\tilde b$), are the lightest squarks.
If the sbottom mass is smaller than the stop mass,
$M_{\tilde b} < M_{\tilde t}$, a stop quark resonantly produced in
$eq$-fusion at HERA via the \rpv coupling $\lambda'_{131}$
may then decay bosonically, providing a SUSY scenario and
final states complementary to the \rpv squark production described
in section~\ref{sec:squarks}.
In this scenario, the only possible decay modes are
$\tilde{t} \rightarrow \tilde{b} W$
with $W \rightarrow f\bar{f}'$ and \rpv sbottom decay into SM particles,
$\tilde b \rightarrow \bar\nu_{e} d$.
In addition, the \rpv decay of the stop into SM fermions,
$\tilde t \rightarrow e^+d$, a more general version of which is
described in the previous section, also contributes.
Feynman diagrams of these two processes are presented in
figure~\ref{fig:stop-feynman}.
The diagram where the $W$ decays leptonically is particularly
interesting, as it results in a final state similar to that in the analysis
of events with isolated leptons and missing transverse momentum
presented in section~\ref{sec:isolep}, which has provided hints of
physics beyond the SM.

The analysis presented below uses data collected with the H1 detector in $e^{+}p$
scattering during the HERA~I period, corresponding to an integrated luminosity
of $106$~pb$^{-1}$ \cite{Aktas:2004tm}.
Simulation of SUSY signal events is done using
SUSYGEN3~\cite{Ghodbane:1999va}, which relies on the LO amplitudes
for $ed\rightarrow \tilde b W$ \cite{Kon:1997bz}.
The parton densities are taken from the CTEQ5L parameterisation
and evaluated at the scale of the stop mass, $M_{\tilde t}$.
The various SM background contributions are estimated using the same
generators as described in section~\ref{sec:squarks}.

%%%

The bosonic stop decay leads to three different final state
topologies, as illustrated in figure~\ref{fig:stop-feynman}~(left).
If the $W$ decays into leptons, the signature is a jet, a lepton
(electron or muon)\footnote {The $W$ decay into $\nu_{\tau}\tau$,
  where $\tau\rightarrow hadrons +\nu$, is not
  investigated.} and missing transverse momentum ($je
P_{T}^{\rm {miss}}$ channel and $j\mu P_{T}^{\rm {miss}}$ channels).
The selection criteria in these channels closely resembles those
used in the H1 search for events with isolated leptons and missing
transverse momentum based on the HERA~I data~\cite{Andreev:2003pm},
with the additional requirement of a jet with $P_T^{\rm jet}>10$~GeV within
the angular range $7^\circ < \theta_{\rm jet} < 140^\circ$.
If the $W$ decays into hadrons the event signature is the presence of
three jets and missing transverse momentum ($jjj P_{T}^{\rm {miss}}$ channel).
Events with three jets with $P_T^{\rm jet 1}>20$~GeV, $P_{T}^{\rm jet 2}>15$~GeV 
and $P_{T}^{\rm jet 3}>10$ GeV are selected, each with polar
angle $7^\circ<\theta_{\rm jet}<140^\circ$.  A total missing
transverse momentum $P_{T}^{\rm {miss}}>25$ GeV is also required.
In the analysis of both the leptonic and hadronic $W$ decay channels
a cut on the inelasticity $y$ is employed to separate the SM
background from the stop signal~\cite{Aktas:2004tm}.
%
%%%
%
For stop and sbottom masses $M_{\tilde{t}}\approx M_{\tilde{b}}+M_W$,
the \rpv decay $\tilde t \rightarrow ed$, as illustrated in
figure~\ref{fig:stop-feynman}~(right), becomes dominant.
These events are selected using criteria similar, though not
identical, to that presented in section~\ref{sec:squarks}.

%%%

\begin{table}
  \renewcommand{\arraystretch}{1.3}
  \caption{Summary of the observed and predicted event yields for
    the various stop decay channels in the H1 analysis. The uncertainty on
    the SM predictions includes model and experimental systematic
    uncertainties added in quadrature. For the $je P_{T}^{\rm {miss}}$ and
    $j\mu P_{T}^{\rm {miss}}$ channels the $W$ production component of
    the SM is given in the last column.}
  \label{tab:stop-rates}
  \begin{tabular*}{1.0\columnwidth}{@{\extracolsep{\fill}} l c c c}
    \hline
    \multicolumn{4}{@{\extracolsep{\fill}} l}{{\bf H1 Search for Bosonic Stop Decays}}\\[-4pt]
    \multicolumn{4}{@{\extracolsep{\fill}} l}{{\bf \boldmath in $R_{p}$ Violating Supersymmetry}}\\
    \hline
    \multicolumn{4}{@{\extracolsep{\fill}} l}{\bf \boldmath $e^{\pm}p$ collisions, ${\mathcal L} = 106$ pb$^{-1}$}\\
    Channel & Data & Total SM & $W$ production\\
    \hline
    $je P_{T}^{\rm {miss}}$ & 3 & ~$3.84 \pm 0.92$ & $2.55 \pm 0.41$\\
    $j\mu P_{T}^{\rm {miss}}$ & 8 & ~$2.69 \pm 0.47$ & $1.93 \pm 0.31$\\
    \hline
    $jjj  P_{T}^{\rm {miss}}$ & $5$ & ~$6.24 \pm 1.74$ & -- \\
    $ed$ & $1100$ & $1120 \pm 131$ & -- \\
    \hline
  \end{tabular*}
\end{table}

%%%

The number of events observed in the data in each channel is shown
in table~\ref{tab:stop-rates}, compared to the SM prediction.
The dominant contribution to the SM expectation in the $je P_{T}^{\rm miss}$
and $j \mu P_{T}^{\rm miss}$ channels arises from the
production of real $W$ bosons, which is also given in table~\ref{tab:stop-rates}.
The main SM contribution in the $jjj P_{T}^{\rm miss}$ ($ed$) channel
is due to CC (NC) DIS events.
A good agreement is observed in all channels except in
the $j\mu P_{T}^{\rm miss}$ channel, where $8$ events are observed in
the data compared to a SM expectation of $2.69 \pm 0.47$.

%%%

\begin{figure*} 
  \centerline{
    \includegraphics[height=0.33\textheight]{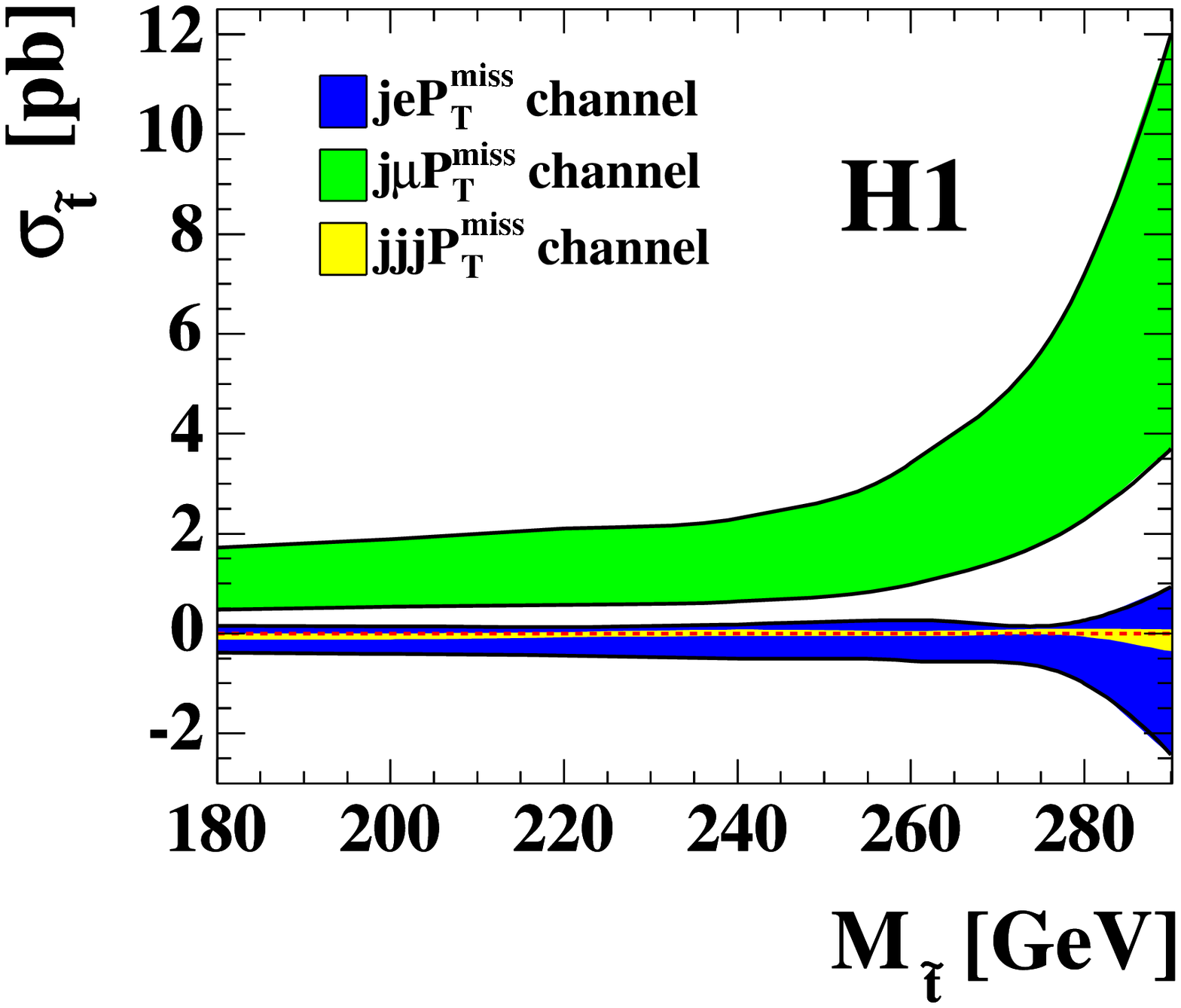}
    \includegraphics[height=0.33\textheight]{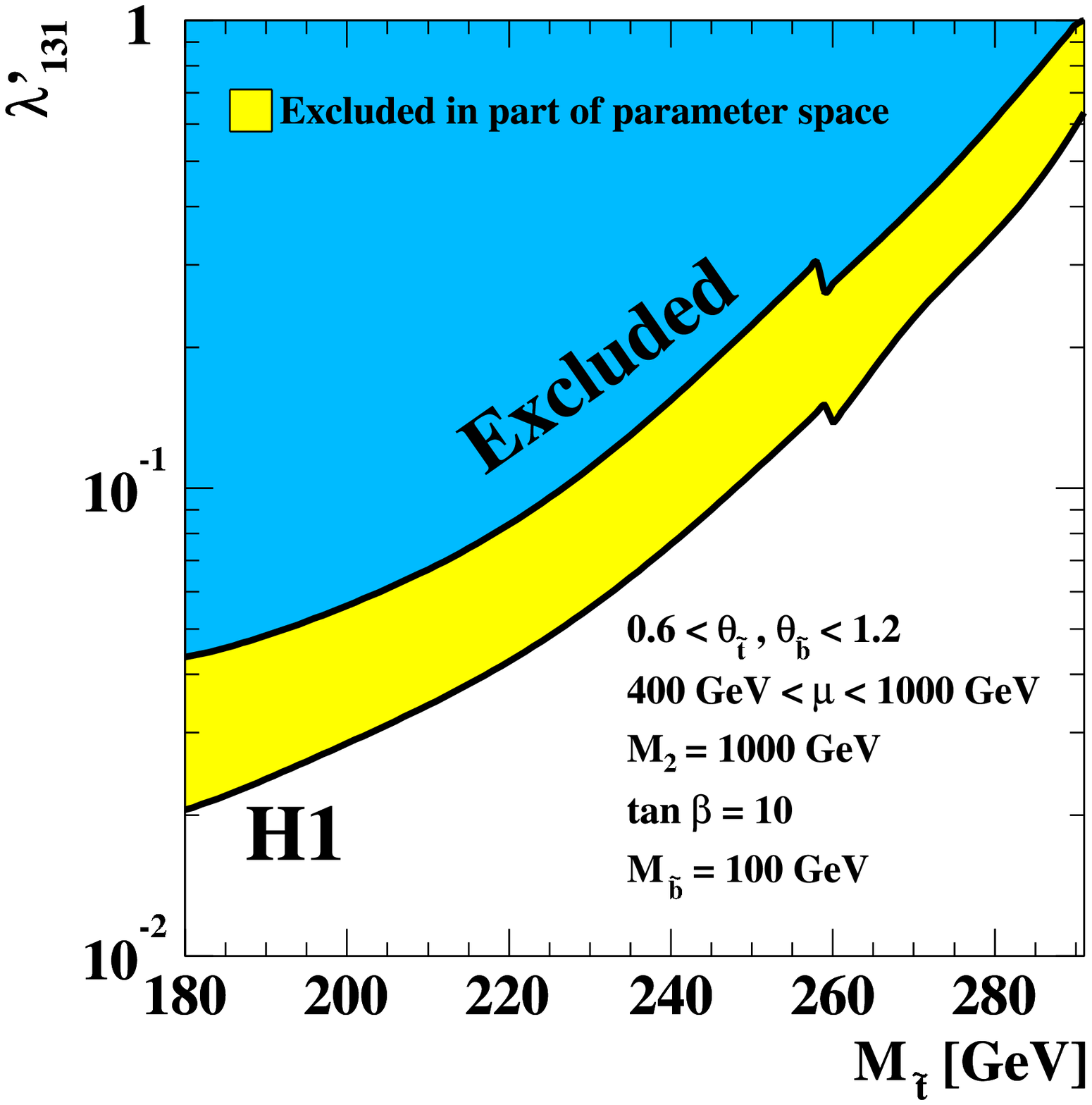}
  }
  \caption{Left: Bands representing the allowed stop cross section
    regions $\sigma_{\tilde{t}}\pm \Delta\sigma_{\tilde{t}}$ as a
    function of the stop mass as obtained from the analysis of each
    bosonic stop decay channel. Right: Exclusion limits at the $95\%$
    CL on the \rpv coupling $\lambda'_{131}$ as a function of the stop
    mass for $M_{\tilde b}=100$~GeV. The limits are derived from a
    scan of the MSSM parameter space described in the text. The two
    full curves indicate the regions excluded in all (dark) or part
    (light) of the parameter space investigated.}
 \label{fig:stop-limits}  
\end{figure*}

Assuming the presence of a stop of mass $M_{\tilde{t}}$ decaying
bosonically, the observed event yields are used to determine the
allowed range for a stop production cross section
$\sigma_{\tilde t}$, and to examine the compatibility of the different
decay modes.
The calculation takes into account the signal efficiency, the $\tilde t$ and $W$ branching
ratios $BR_{\tilde t \rightarrow \tilde b W}\cdot  BR_{W\rightarrow f
  \bar{f}'}$ and the relative integrated luminosities of the HERA~I data
sets taken at different centre of mass energies~\cite{Aktas:2004tm}.
The bands in figure~\ref{fig:stop-limits}~(left) represent the allowed
cross section regions  $\sigma_{\tilde t}\pm \Delta\sigma_{\tilde t}$
for all bosonic decay channels.
It can be seen that the stop interpretation of the excess seen in the
$j\mu P_{T}^{\rm  miss}$ channel is not supported by the other decay modes. 

%%%

As no stop signal is observed, exclusion limits on the stop production
cross section are derived at the $95\%$ CL in the framework of the 
MSSM~\cite{Nilles:1983ge,Haber:1984rc} using a modified frequentist
approach based on likelihood ratios \cite{Junk:1999kv}.
A scan of the SUSY parameter space is performed, to systematically
investigate the dependence of the sensitivity on the MSSM parameters.
The SUSY soft-breaking mass parameter $M_2$ is set to $1000$~GeV
and the Higgs mass term is restricted to $400 < \mu < 1000$~GeV,
which ensures that the gaugino masses are large.
The mixing angles $\theta_{\tilde t}$ and $\theta_{\tilde b}$ are
allowed to vary between $0.6$ rad and $1.2$ rad.
An example of the limit projected in the
$(M_{\tilde t},\lambda'_{131})$ plane is shown in
figure~\ref{fig:stop-limits}~(right), for $\tan \beta = 10$ and
$M_{\tilde b}=100$~GeV.
For $M_{\tilde t}=200$~GeV, couplings $\lambda'_{131} \gsim 0.03$ are
ruled out and for $M_{\tilde t}=275$~GeV the allowed domain
is $\lambda'_{131} \lsim 0.3$.

\subsection{Search for light gravitinos in events with photons and missing transverse momentum}
\label{sec:gravitinos}

\begin{figure}
  \begin{center}
    \includegraphics[width=1.0\columnwidth]{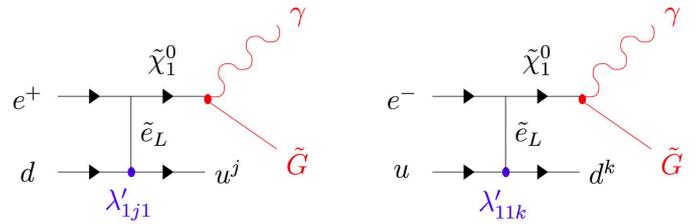}
    \caption{Diagrams for neutralino production via \rpv selectron
      exchange in $e^+p$ (left) and $e^-p$ (right) scattering, with
      subsequent neutralino decay into a gravitino and a photon.}
    \label{fig:gravitino-feynman}
  \end{center}
\end{figure}

In Gauge Mediated Supersymmetry Breaking (GMSB) SUSY models, new
``messenger'' fields are introduced which couple to the source of
supersymmetry breaking.
The breaking is then transmitted to the SM fields and their
superpartners by gauge interactions~\cite{Giudice:1998bp}. 
The gravitino, $\tilde{G}$, is the lightest supersymmetric particle
(LSP) and can be as light as $10^{-3}$~eV.
The next-to-lightest supersymmetric particle (NLSP) is generally
either the lightest neutralino $\neu$ or a slepton $\tilde{\ell}$,
which decays to the stable gravitino via $\neu \rightarrow
\gamma \tilde{G}$ or  $\tilde{\ell} \rightarrow \ell \tilde{G}$. 
At HERA, the presence of the \rpv couplings $\lambda'_{1j1}$ and
$\lambda'_{11k}$ could lead to neutralino production in $e^+p$
and $e^-p$ collisions, respectively, via $t$-channel selectron
exchange, as illustrated in figure~\ref{fig:gravitino-feynman}.
The hard scattering process at large Bjorken-$x$ is dominated by
the valence quarks in the proton, and therefore if the initial state
lepton is a positron (electron) the scatter involves a down (up) quark from the 
proton, as shown in the left (right) part of figure~\ref{fig:gravitino-feynman}.
For a given \rpv\ coupling, the $\neu$ production cross section
for an initial state electron is roughly a factor of two larger than that
for an initial positron, reflecting the different parton
densities for valence up and down quarks in the proton.

%%%

A search for \rpv resonant single neutralino production
via $t$-channel selectron exchange, $e^\pm q \rightarrow \neu q'$ 
is performed by H1 using their HERA~I data set taken at $\sqrt{s} =
319$~GeV, corresponding to an integrated luminosity of $64.3$~pb$^{-1}$ for
$e^+p$ collisions and $13.5$~pb$^{-1}$ for $e^-p$
collisions~\cite{Aktas:2004cc}.
It is assumed that the $\neu$ is the NLSP and that the decay $\neu
\rightarrow \gamma \tilde{G}$ occurs with an unobservably small
lifetime and dominates over \rpv neutralino decays.
It is also assumed that one of the couplings $\lambda'_{1j1}$ ($j=1,2$)
or $\lambda'_{11k}$ ($k=1,2,3$) dominates\footnote{The
coupling $\lambda'_{131}$ is not studied in this analysis because the production of
a top quark together with a neutralino is suppressed due to the high top
quark mass.}.
The process considered in this analysis is independent
of the squark sector, and so is a complementary approach to those
presented in sections~\ref{sec:squarks} and~\ref{sec:bosonicstop}.

%%%

%%%

The GMSB model~\cite{Dimopoulos:1996vz} examined in the analysis is
characterised by six new parameters in addition to those of the SM:
the parameter $\Lambda$, which sets the overall mass scale for the SUSY
particles; the mass of the messenger particles $M$;
the number $f$ sets of messenger particles, $N$; the intrinsic SUSY
breaking scale $\sqrt{F}$, which also determines the $\grav$
mass according to $m_{\tilde{G}} \simeq 2.5 \cdot F/(100\ {\rm TeV})^2
\, {\rm eV}$; the ratio of the Higgs vacuum expectation values
$\tan\beta$; and the sign of the Higgs sector mixing
parameter $\mu$. 
The signal topology is simulated using the SUSYGEN3
generator~\cite{Ghodbane:1999va}.%
The parton densities are evaluated at the scale of the Mandelstam
variable $-t$.

%%%

The final state resulting from the process
$ e^\pm q \rightarrow \neu q' \rightarrow \gamma \tilde{G} q'$
contains a photon, a jet originating from the scattered quark and
missing transverse momentum due to the escaping gravitino.
The SM background almost exclusively arises from radiative
CC DIS, which features a jet, a photon and a neutrino in the final
state and is modelled using DJANGOH.
Smaller contributions from NC DIS, photoproduction and $W$ production
are estimated using the DJANGOH, PYTHIA and EPVEC generators, respectively.

%%%

Events are selected with large missing transverse momentum
determined from the calorimetric energy deposits,
$P_{T}^{\rm calo} > 25$~GeV.
The events are also required to contain at least one hadronic jet
in the range $10^{\circ} < \theta_{\rm jet} < 145^{\circ}$ and an
identified photon in the LAr, both with transverse momenta
greater than $5$~GeV.
Photons are identified using a shower shape analysis of
energy deposits in the LAr calorimeter and for
$\theta_{\gamma} > 20^{\circ}$ an electromagnetic cluster is
only accepted as a photon candidate if it is not associated
with a charged track in the central tracking system.
In addition, the photon must be isolated from any reconstructed
jet with $P_{T}^{\rm jet} > 5$~GeV.
At this point in the selection, $12$ events are observed in the data 
compared to a SM expectation of $11.5\pm 1.5$, predominantly
from radiative CC DIS.

%%%

Based on a study of this preselection, additional cuts are then
applied to reduce the SM background, increasing the minimum
photon transverse momentum $P_{T}^{\gamma}$ to $15$~GeV and
requiring that the sum of the $E - P_{z}$ in the event is larger
than $15$~GeV~\cite{Aktas:2004cc}.
In the final selection no candidate event is found in the $e^+p$ data,
compared to a SM prediction of $1.8\pm0.2$.
In the $e^-p$ data sample, $1$ candidate event is observed compared
to a SM prediction of $ 1.1 \pm 0.2$.
The SM expectation arises predominantly from CC~DIS
with small contributions from NC~DIS and the production of $W$ and
$Z$ bosons where the final state electron is misidentified as a photon.

%%%

\begin{figure}
  \begin{center}
    \includegraphics[clip,width=1.0\columnwidth]{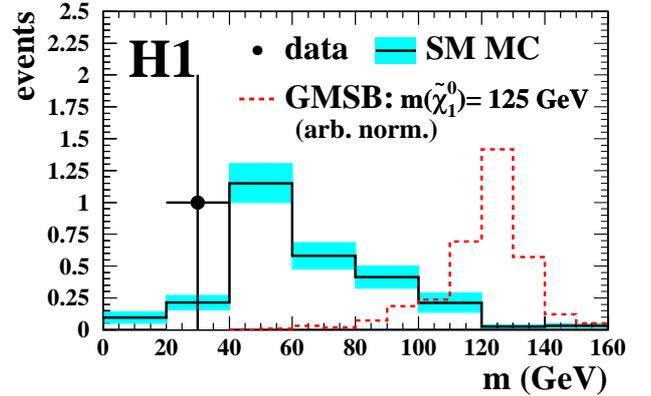}
    \caption{Distribution of the invariant mass of the photon candidate
      and the reconstructed missing particle in the H1 search for light
      gravitinos in events with photons. The data (points) are compared with the SM
      prediction (solid histogram). The signal expected for a
      neutralino with a mass of $125$~GeV is shown 
      with arbitrary normalisation (dashed histogram).}
    \label{fig:gravitinos-mass}
  \end{center}
\end{figure}

Assuming that the massless gravitino is the only non-interacting
particle in the event, the gravitino kinematics are reconstructed by
exploiting the conservation of transverse momentum and the
constraint $(E-p_z) + (E_{\tilde{G}} - p_{z, \tilde{G}}) = 2E_e$.
The four-vector of this particle is then added to that of the 
photon to reconstruct the invariant mass $m$ of the decaying
neutralino.
The data and the SM expectation for this distribution are shown in
figure~\ref{fig:gravitinos-mass}.
From the simulation of the SUSY signal, also shown in figure~\ref{fig:gravitinos-mass},
the mass resolution is determined to be around $10$~GeV.
The candidate event has a reconstructed invariant neutralino mass
of $36\pm 4$~GeV. 

%%%

\begin{figure*} 
  \centerline{
    \includegraphics[width=0.5\textwidth]{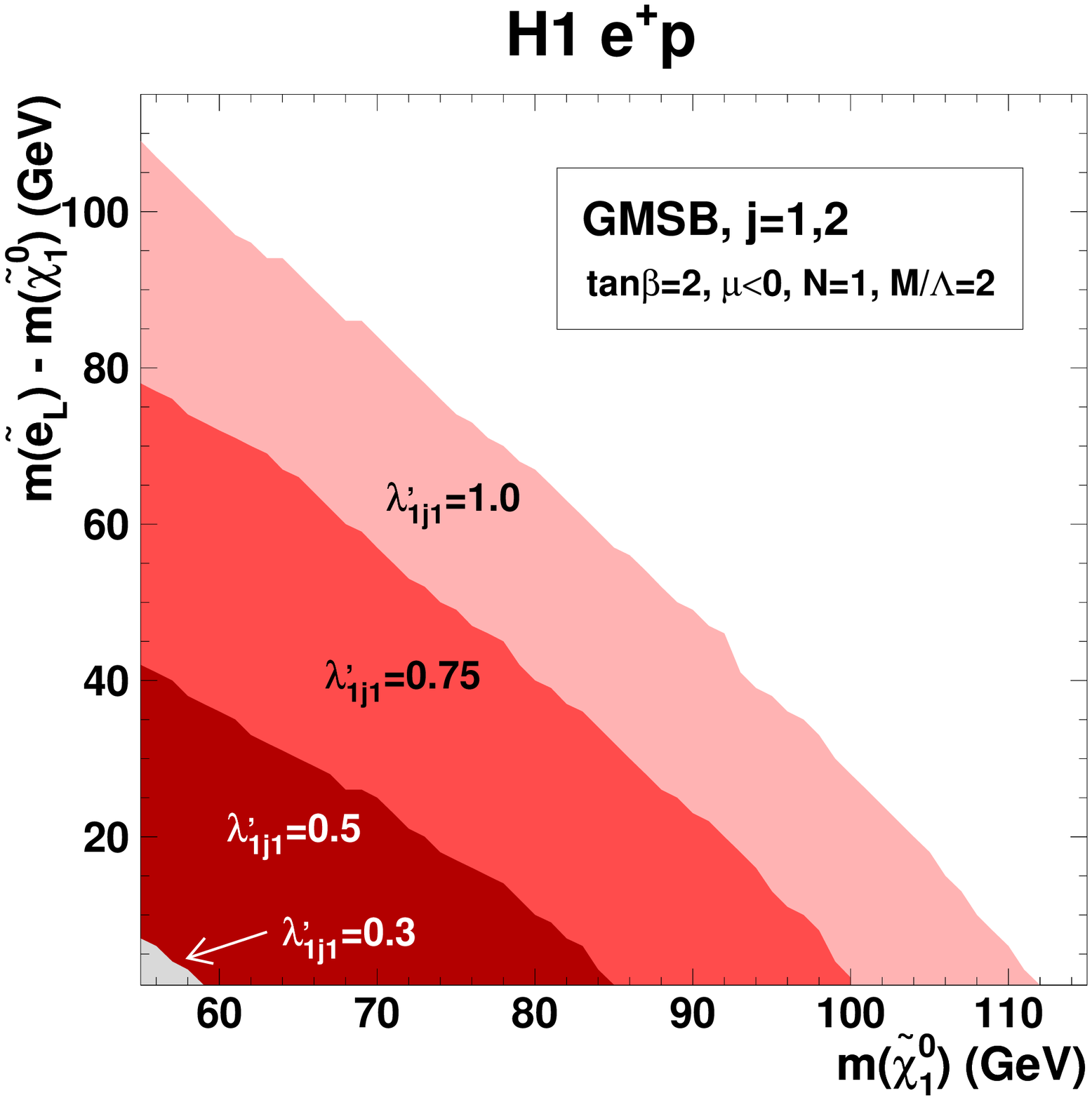}
    \includegraphics[width=0.5\textwidth]{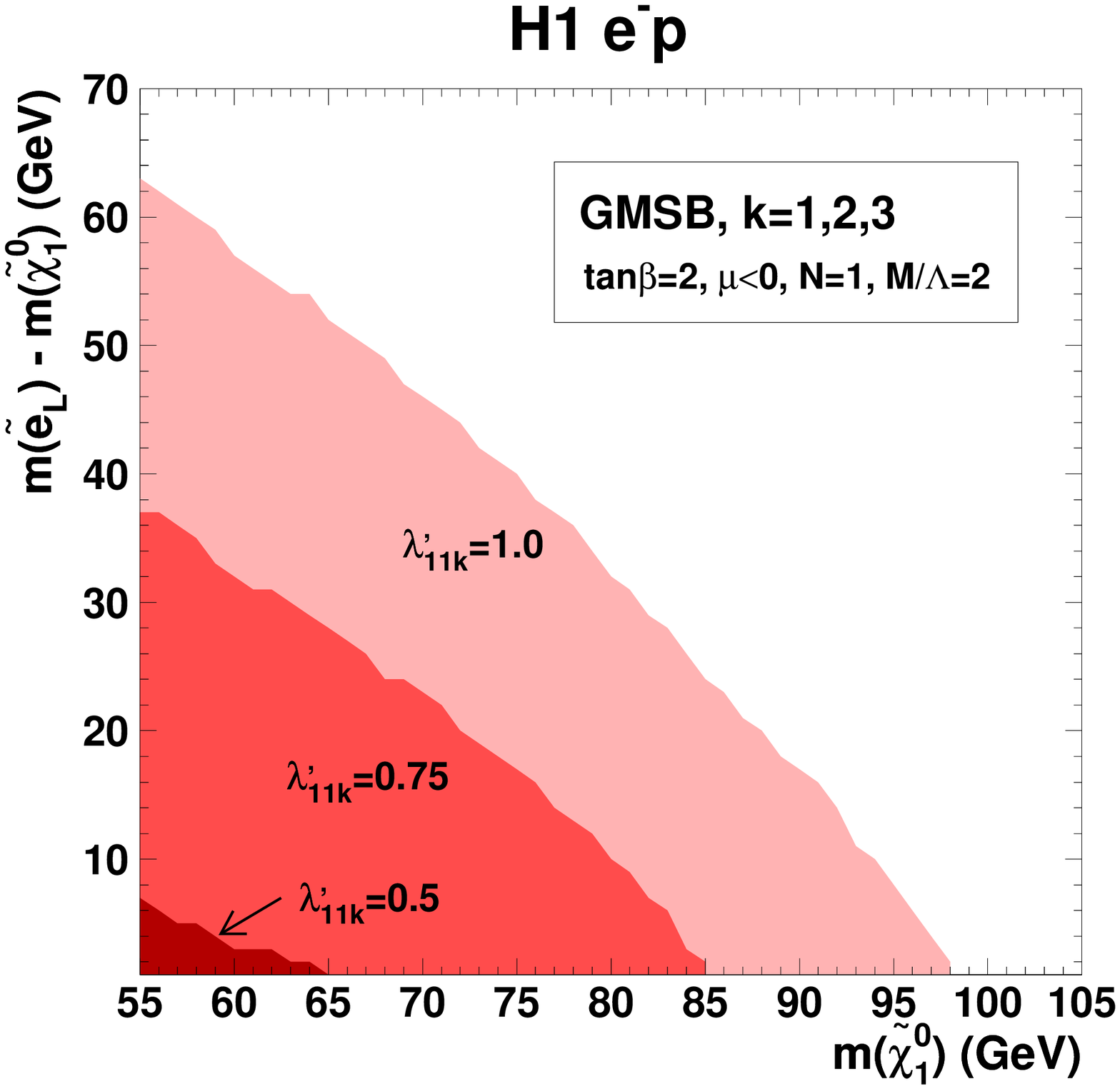}
  }
  \caption{Exclusion limits at the $95\%$ CL in the $m$($\neu$), $m$($\tilde{e}_L$)--$m$($\neu$)
    plane from the H1 search for light gravitinos in events with
    photons, for various values of the coupling
    $\lambda'_{1j1}$ ($j=1,2$) (left) and $\lambda'_{11k}$ ($k=1,2,3$)
    (right) in the GMSB SUSY parameter space indicated on the plot.}
  \label{fig:gravitino-limits}
\end{figure*} 

%%%

As no significant deviation from the SM is observed, the results are
used to derive constraints on GMSB models for different values of the
\rpv coupling $\lambda'_{1jk}$ for fixed values of the SUSY
parameters $\tan \beta$, $N$ and (sign)$\mu$.
As an example, figure~\ref{fig:gravitino-limits} displays excluded
regions in the $m$($\neu$), $m$($\tilde{e}_L$)--$m$($\neu$) plane
derived from the H1 $e^{+}p$ (left) and $e^{-}p$ (right) data for
various values of $\lambda'_{1jk}$ in the parameter space indicated
in the figure.
It can be seen that for small mass differences between the neutralino
and selectron,
neutralino masses up to $112$~GeV are ruled out at $95\%$ CL for \rpv
coupling $\lambda'_{1jk} =1$.
Furthermore, for neutralino masses close to $55$~GeV, $\lambda'_{1j1}$
Yukawa couplings of electromagnetic strength are excluded.
These are the only constraints from HERA on SUSY models independent
of the squark sector.

\section{A direct search for stable magnetic monopoles}
\label{sec:monopoles}

\begin{figure*}
\begin{center}
  \includegraphics[clip,width=0.8\textwidth]{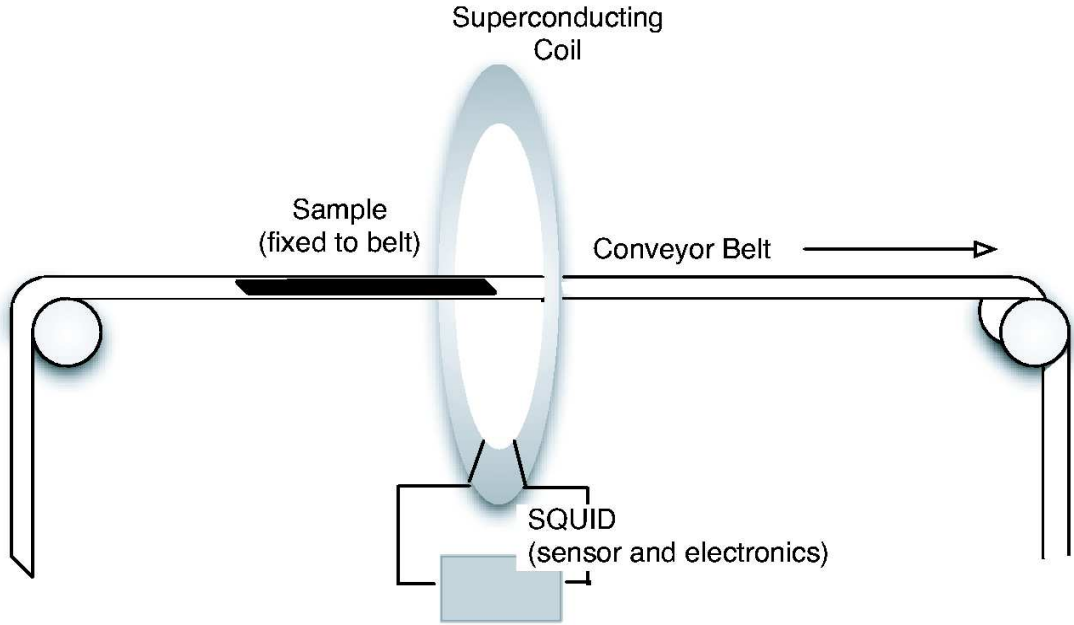}
  \caption{Method for the analysis of the H1 beampipe strips in the direct search for magnetic monopoles. The conveyor belt travelled in steps of typically 5 cm until the sample traversed completely the superconducting coil. At each step the conveyor belt stopped for $1$ sec before the current in the superconducting coil (magnetometer current) was read to avoid the effects of eddy currents. The time for each step was typically $3$ secs.}
  \label{fig:h1mono_1}
\end{center}
\end{figure*}

The existence of magnetic monopoles is one of the open issues in modern physics and their discovery would allow a better explanation of some well established aspects of nature.
The quantisation of the electric charge was explained by Dirac~\cite{Dirac:1931kp} by postulating the existence of particles with a magnetic charge, which shall be a multiple of the Dirac charge, $g_D$, given by:
\begin{equation}
\frac{g_De}{\hbar c} = \frac{1}{2} \Rightarrow \frac{g_D}{e} = \frac{1}{2\alpha_{\rm e}} \approx 68.5,
\end{equation}
where $e$ is the elementary electric charge and $\alpha_{\rm e}$ is the fine structure constant.
With the presence of a magnetic monopole, considering the duality of Maxwell's equations the
(very large) magnetic coupling can be expressed as:
\begin{equation}
\alpha_{\rm m} = \frac{g_D^2}{\hbar c} = \frac{1}{4\alpha_{\rm e}}.
\end{equation}
The large value of this coupling constant prevents the use of perturbative field theory for a reliable
calculation of the expected rates of processes involving magnetic monopoles.
It also implies that the energy released by ionisation by a magnetic monopole is much larger
than for minimum ionising electrically charged particles, as a magnetic monopole will effectively behave,
in terms of ionisation energy loss, as a highly charged stable particle, with a charge $\approx 68.5$. 

%%%

In addition to the Dirac argument, magnetic monopoles arise naturally in field theories unifying
the fundamental forces, like string theory~\cite{Sorkin:1983ns,Gross:1983hb,Bergshoeff:1998ef} and
Supersymmetric Grand Unified theories~\cite{Polyakov:1974ek,'tHooft:1974qc,Minakata:1984an,Lopez:1996gy}.
Although most of these theories tend to predict heavy monopoles, with masses higher than $10^{15}$GeV,
there are some Grand Unified scenarios~\cite{Rizzo:1981su,Preskill:1984gd,Weinberg:1983bf} in which
mass values of the order of $10^{4}$ GeV are allowed.
Other approaches~\cite{Kirkman:1981ck,Yang:1998sh,Cho:2002bk,Bais:1979gv,Banks:1988rj} also exist,
in which a light monopole is allowed, and postulates on values of the classical radius of a monopole lead
to estimates of a monopole mass of the order of tens of GeV~\cite{Particle_Astropysics}.   

%%%

One of the techniques that can be used in the direct search for
magnetic monopoles is the search for the induction of a
persistent current within a superconducting loop~\cite{Alvarez_science}.
This approach is used by the H1 Collaboration in a direct search for
magnetic monopoles~\cite{Aktas:2004qd}, exploiting the idea that 
heavily ionising magnetic monopoles may stop in the beam pipe
surrounding the interaction region of the H1 detector.
Such monopoles, if stable, would then remain permanently trapped in the beam pipe, as the binding
energy between the monopoles and the aluminium of the beam pipe is expected to be large, of the order
of hundreds of keV~\cite{Gamberg:1998xf}.
The magnetic field of the trapped monopoles would induce a persistent current on a superconducting
coil, after their complete passage through the coil.
In contrast, a passage through the coil of material with no trapped monopole would induce no current,
as the permanent dipole moment in the material would cancel in the passage of the material through the coil. 

%%%

The aluminium beam pipe used in the search was in place in the H1 delector 1995-1997,
during which time it was exposed to a luminosity of $62$~pb$^{-1}$.
The beam pipe had a diameter of $9$~cm and thickness $1.7$~mm in the range $-0.3 < z < 0.5$~m
and a diameter of $11$~cm and thickness $2$~mm in the range $0.5 < z < 2.0$~m.
The analysis was performed by cutting the H1 beam pipe in long thin strips and then passing them through a superconducting coil coupled to a Superconducting Quantum Mechanical Interference Device (SQUID), as shown in figure~\ref{fig:h1mono_1}. 
This full length of the beampipe, covering $-0.3 < z < +2.0~{\rm m}$, was cut into 45 longitudinal strips
each of length on average of $573$~mm ($\sim 2~{\rm mm}$ was lost in each cut).
The central section ($-0.3 < z < 0.3~{\rm m}$) was cut into $15$ long strips of width $\sim 18~{\rm mm}$,
two of which were further divided into $32$ short segments varying in length from $1$ to $10$~cm.
The downstream section ($0.3 < z < 2.0~{\rm m}$) was divided into $3$ longitudinal sections, each of which
was cut into $10$ long strips of width $\sim 32~{\rm mm}$. 

%%%

All the samples were passed along the axis of the 2G Enterprises type 760 magnetometer~\cite{SQUID}
hosted at the Southampton Oceanography Centre, in the United Kingdom.
After each sample was passed through the coil, the current in the superconducting loop was measured,
and the current induced by the sample estimated as the difference between the currents measured before
and after its passage through the coil.
This procedure was repeated many times for each sample in order to estimate the reproducibility
of the results. 

%%%

The sensitivity of the SQUID magnetometer to a magnetic monopole was assessed using a long,
thin solenoid, as the magnetic field at the end of a long solenoid is similar to that produced by a
magnetic monopole.
A long solenoid can be considered as composed by two magnetic "pseudopoles",
each of strength $g = N \cdot I \cdot S/g_D$, where $N$ is the number of turns per unit length,
$I$ the current and $S$ the section area of the solenoid.
The current and radius of the solenoid can be chosen to mimic the pole strength.
The calibration was performed by passing a solenoid with different current through the
magnetometer, measuring the induced current and subtracting the current induced at
zero solenoid current.
The current in the magnetometer following the passage of one end of the solenoid was found
to increase linearly with the current in the solenoid. 

%%%

\begin{figure}
  \begin{center}
    \includegraphics[width=1.0\columnwidth]{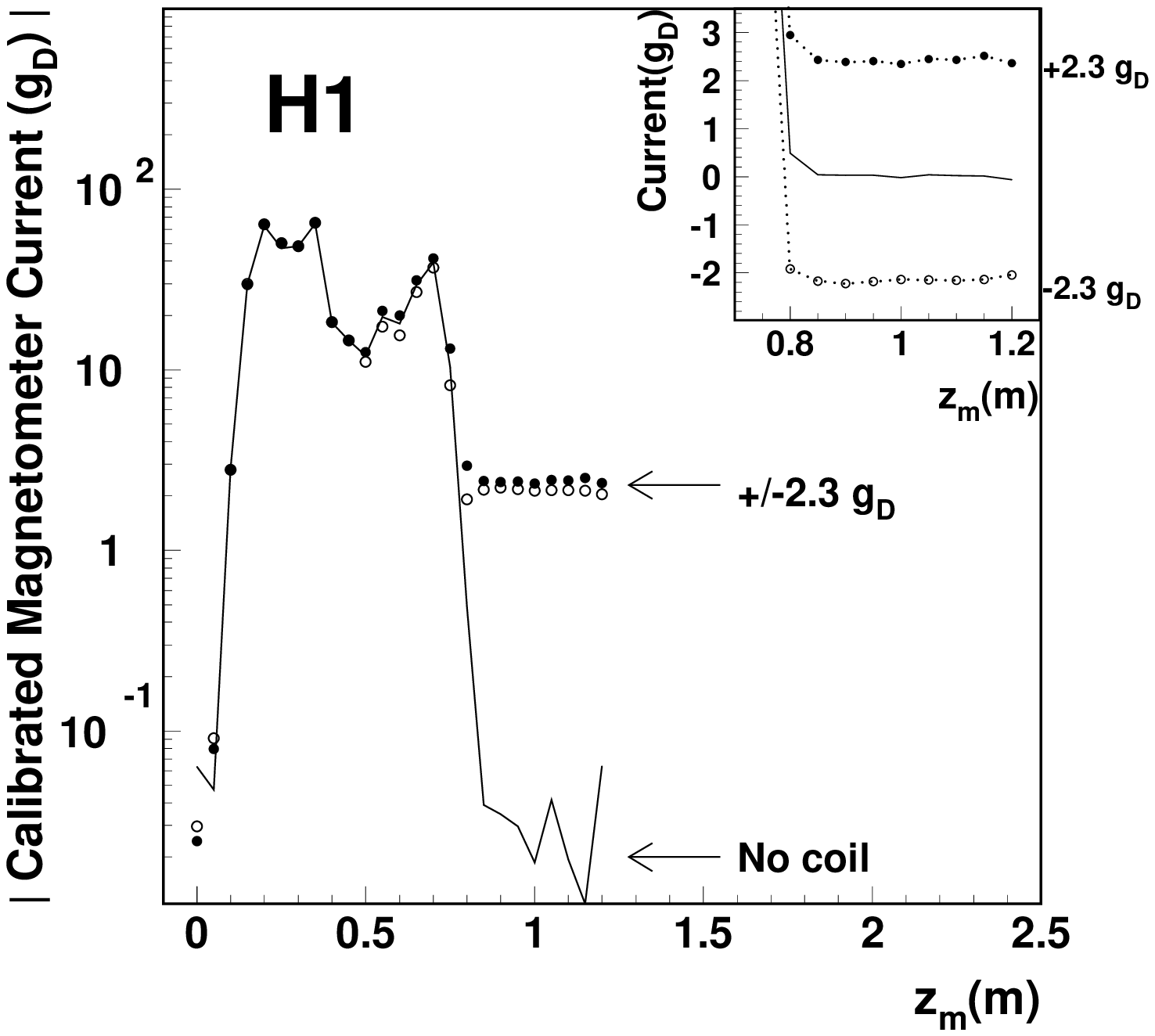}%
    \caption{The absolute value of the calibrated magnetometer current 
      versus step position ($z_m$) for a strip from the central beam pipe region
      ($-0.3 < z < 0.3~{\rm m}$). The solid line shows the measurements with the long strip alone.
      The closed (open) points show the measurements with the long strip together with the calibration
      solenoid excited to simulate a pole of strength $+2.3g_D$ ($-2.3g_D$). The inset shows the
      signed measurements of the calibrated magnetometer currents versus the step position for
      $z_m > 0.8$ m on a linear scale. The expected persistent currents for monopoles of
      strength $\pm 2.3 g_D$ are shown by the arrow on the logarithmic plot and by the numbers
      in the margin on the inset linear plot.}
\label{fig:h1mono_3}
\end{center}
\end{figure}

To simulate the magnetometer behaviour at the passage of a magnetic monopole, the long solenoid
was attached to a beam pipe section and the two passed jointly through the coil, allowing the passage
of only one end of the solenoid through the magnetometer.
The results are shown in figure~\ref{fig:h1mono_3}, where the passage of the beam pipe
strip alone is compared to the passage of the same strip  attached to a solenoid with a "pseudopole"
strength of $2.3g_D$ and $-2.3g_D$.
A large structure is visible in the centre of the figure, and this is due to the magnetic field of
the permanent magnetic dipole moments in the aluminium.
A persistent current is induced when the pseudopoles are present; when they are absent, the
large permanent dipole moment of the aluminium does not prevent the current to go back to zero.
In the inset of the figure, the current measured as the strips leaves the magnetometer coil is shown.
The values are equal and opposite, and equal to that measured with the calibration solenoid alone.
This demonstrates the sensitivity of the apparatus to a magnetic monopole trapped in the beam pipe strip. 

%%%

The sample first analysed corresponds to the $15$ strips cut from the region
$-0.3 < z < 0.3$~m, out of which two were cut in 32 short samples.
No persistent current was observed in the solenoid, and in the few cases in which
it was observed, it disappeared in a subsequent measurement.
The non-zero readout was therefore attributed to random jumps in the baseline of
the magnetometer electronics.
The conclusion was therefore that no magnetic monopole has been trapped in the
central region of the beam pipe.

%%%

\begin{figure}
\begin{center}
    \includegraphics[width=1.0\columnwidth]{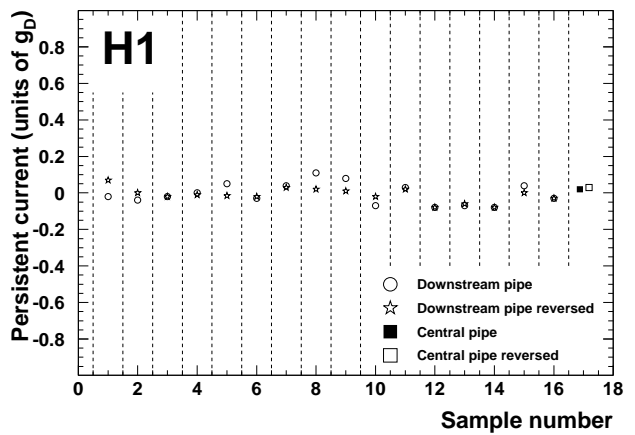}%
  \caption{The persistent currents in the long beam pipe strips in units of $g_D$, after their
    passage through the magnetometer versus sample number, measured after the samples had
    been demagnetised (see text). Samples $1$-$16$ consist of several long strips (usually two or three)
    from the downstream beam pipe bundled together.
    Sample $17$ consists of the thirteen long strips of the central beam pipe bundled together.}
  \label{fig:h1mono_6}
\end{center}
\end{figure}

For the strips of the downstream beam pipe, some problems were encountered in the analysis.
The permanent dipole moment of the downstream strips was found to be much larger than that
of the strips of the central beam pipe, and gave readings in the magnetometer more than three
orders of magnitude larger than those expected from a magnetic monopole.
Moreover, fluctuations in the baseline level of the magnetometer up to $0.7g_D$ were induced.
To mitigate these effects, the strips were demagnetised in a low frequency magnetic field of
initial intensity of $0.1$~T.
This procedure was not expected to dislodge any potential magnetic monopoles trapped in the
beam pipe, due to their large binding energy.
Also in this case, no significant persistent current is observed and it is therefore concluded
that no monopole of strength greater than $0.1~g_D$ had been trapped in the beam pipe strips,
which made up $93\%$ of the total beam pipe, the rest being lost in the cutting procedure. 
Figure~\ref{fig:h1mono_6} shows a summary of the measurements performed on the whole sample.

%%%

\begin{figure*}
\centerline{
    \includegraphics[width=1.0\columnwidth]{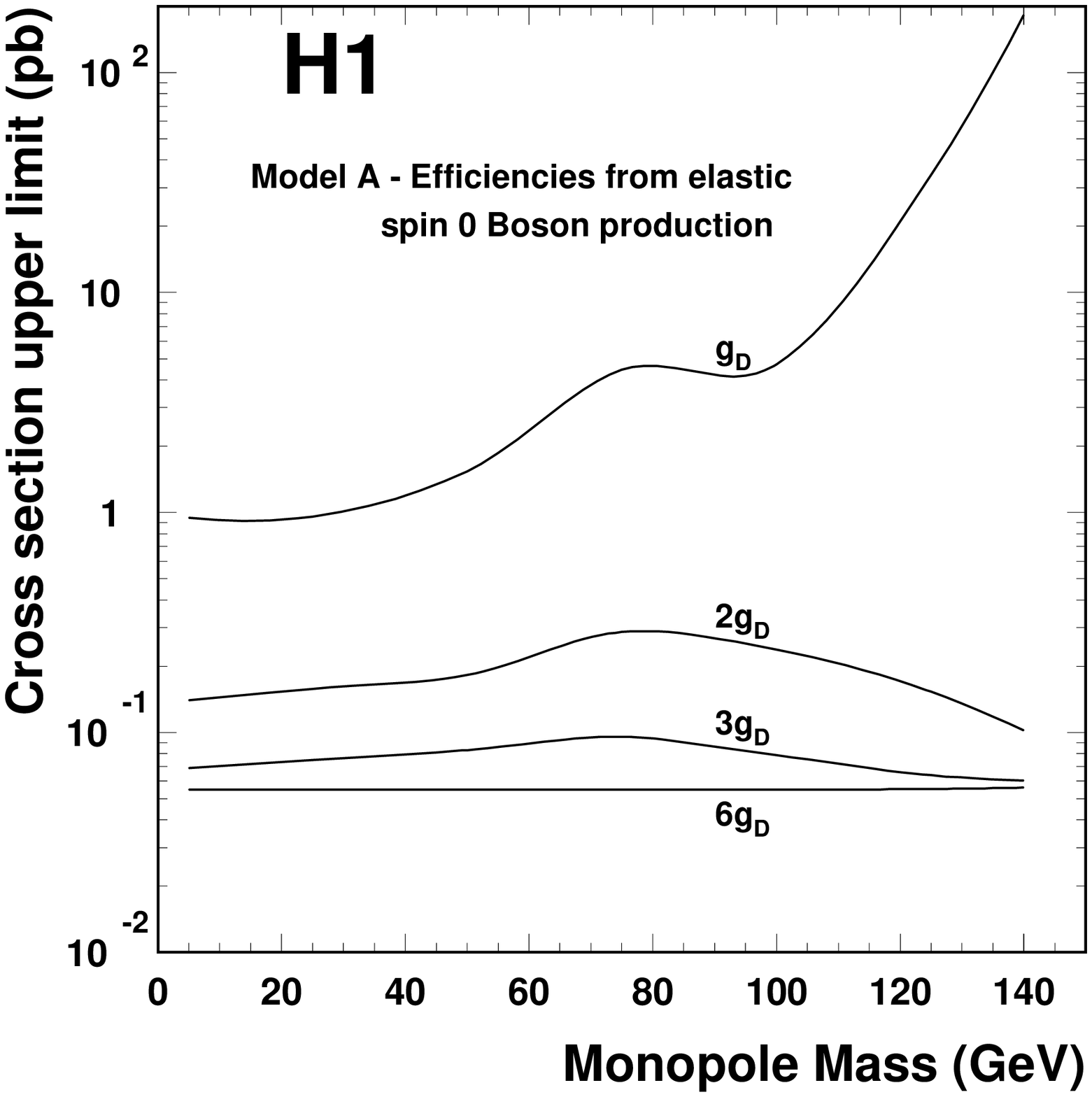}%
    \includegraphics[width=1.0\columnwidth]{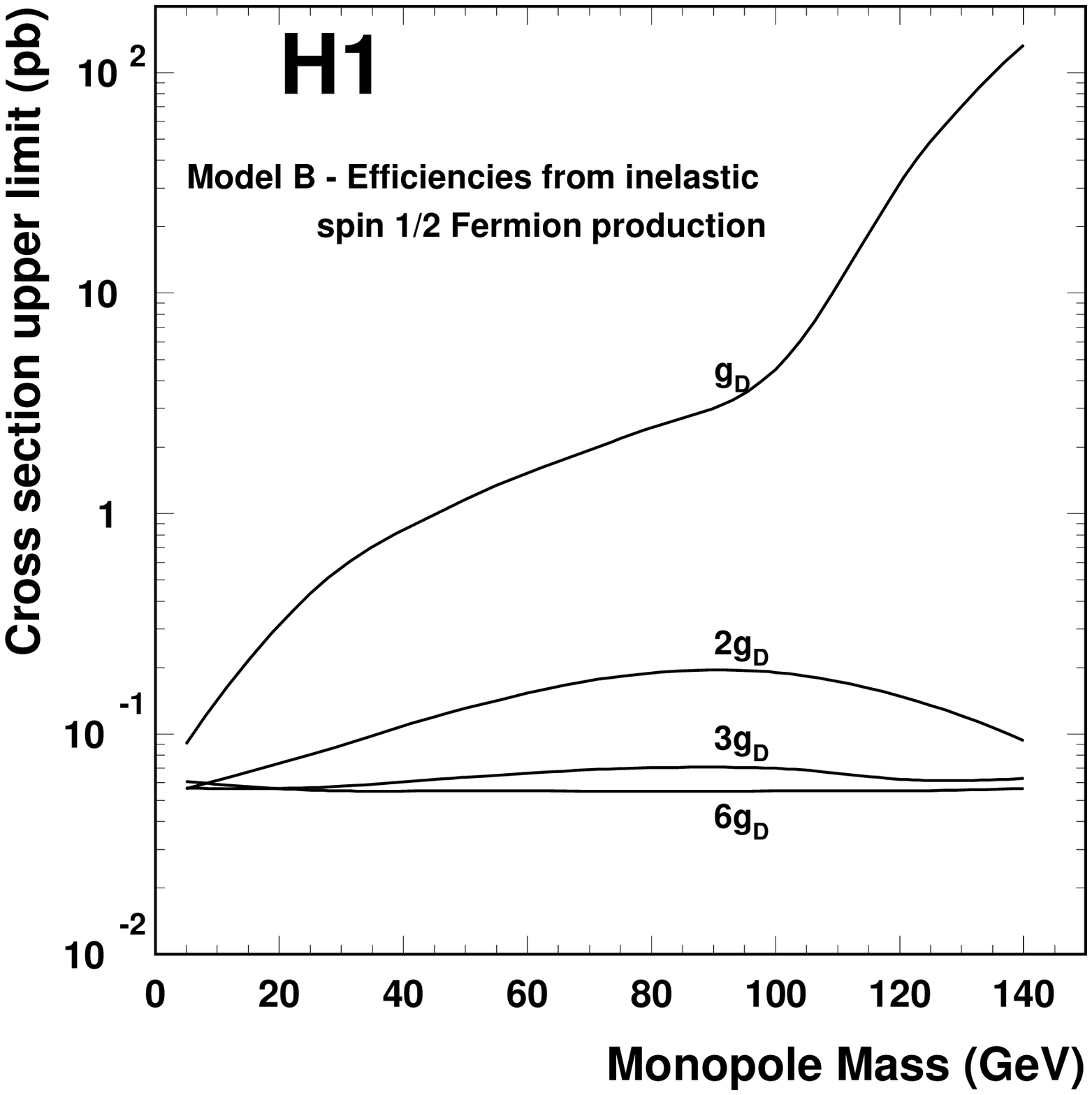}%
    }
\caption{Upper limits on the cross section for the production of a monopole-antimonopole pair,
  determined within the context of model A (left) and model B (right), for monopole-antimonopole
  pair production in $e^+p$ collisions as a function of monopole mass for monopoles of
  strength $g_D$, $2g_D$, $3g_D$ and $6g_D$ or more.}
  \label{fig:h1mono_1011}
\end{figure*}    

As no magnetic monopole was observed, upper limits on the production cross section
are determined.
The cross section extraction needs the evaluation of the detector acceptance, therefore a
model for the production is needed and two differnet models are considered.
In each model, a monopole-antimonopole ($M \bar{M}$) pair was assumed to be produced in a
photon-photon interaction.
In the first model (model A), elastic production of a spin-$0$ monopole pair in the process
$e^+p \rightarrow e^+ M \bar{M} p$, through the interaction of a photon from the electron
and a photon from the proton, was assumed.
In the second model (model B), spin-1/2 monopoles were produced in the inelastic
process $e^+p \rightarrow e^+ M \bar{M} X$, through photon-photon fusion with a
photon from the electron and one from a quark in the proton.
Events for model A were generated using the program CompHEP~\cite{Boos:2009un},
while a dedicated program was used for model B~\cite{Aktas:2004qd}. 
The generated final state particles were tracked in the H1 interaction region of the beam pipe,
and if the range of the monopole in aluminium was lower than the thickness of the beam pipe,
the monopole was assumed to stop there. 
The cross section upper limit was derived considering that no observation translates
into a $95\%$ CL upper limit of $3$ monopoles pair events produced.
The upper limits on the cross sections for model A and B, and for monopoles of different
charges are shown in figure~\ref{fig:h1mono_1011}.

%%%

The direct search for magnetic monopoles has been performed with different techniques
in various fields of physics.
A number of papers on magnetic monopole searches have been published,
in cosmic rays~\cite{Ambrosio:2002qq,Balestra:2008ps,Hogan:2008sx,Detrixhe:2010xi,Abbasi:2012eda,Aartsen:2014awd},
in matter~\cite{Ross:1973it,Kovalik:1986zz,Jeon:1995rf,Bendtz:2013tj}
and at colliders~\cite{Fairbairn:2006gg,Aubert:1982zi,Musset:1983ii,Abbott:1998mw,Abulencia:2005hb,Abbiendi:2007ab,Aad:2012qi,DeRoeck:2012wua}.
At colliders, although a universal production mechanism can be postulated, the comparison of
the cross section upper limit obtained in the H1 analysis with those of other colliders in different
types of collisions ($e^+e^-$, $p\bar{p}$, $pp$) is difficult.
Recently, the ATLAS collaboration has published~\cite{Aad:2015kta} a search for magnetic monopoles and stable particles with high electric charge, setting a model-independent upper limit on the production cross section of $0.5~{\rm fb}$ for signal particles with magnetic charge $0.5 g_D \le |g| \le 2.0 g_D$ and electric charge in the range between $20$ and $60$ times the elementary charge, with masses between $200$ and $2500$~GeV.
This result is valid in well-defined fiducial regions of high and uniform event selection efficiency.
The H1 result described above remains the only one obtained in $e^+p$ collisions at high energy. 

\section{Summary and outlook}
\label{sec:summary}

The HERA $ep$ collider at DESY is a unique machine, which has brought a
unrivalled insight into the structure of the proton via the precision
analysis of deep inelastic scattering over many orders of magnitude
in Bjorken $x$ and negative four-momentum transfer squared $Q^{2}$.
The data harvest collected by the H1 and ZEUS experiments has also
provided the opportunity to search for rare processes and physics
beyond the Standard Model.
Combining the data of both experiments, resulting in a HERA
dataset with an integrated luminosity of $1$~fb$^{-1}$, has resulted
in an increased sensitivity to such processes and allowed a
thorough search of the high $P_{T}$ kinematic region.

%%%

Cross sections of the rare production of $W$ and $Z^{0}$ bosons are
measured, as well as the production of high $P_{T}$ lepton
pairs via two photon exchange.
These analyses have also provided tantalising glimpses at
physics beyond the Standard Model, and although no significant signal
excess is observed in the complete dataset, a number of interesting
events remain in high $P_{T}$ regions of the H1 analyses of their $e^{+}p$ data.
The single and pair production of tau leptons is also observed at HERA,
utilising both leptonic and hadronic decays of the tau.

%%%

The initial $ep$ state at HERA provides a complimentary environment to
searches using the $e^{+}e^{-}$ collisions at  LEP and the $p\bar{p}$
collisions at the Tevatron.
In particular, HERA is an ideal place to search for the single, resonant
production of hypothetical particles such as leptoquarks, both within
the kinematic limit given by the available centre of mass energy, and
beyond using contact interaction models.
Model dependent searches, such as for excited fermions and
supersymmetry are also performed, as well as a general search
at high $P_{T}$, which confirms the results found in dedicated
analyses.

%%%

No significant deviation from the Standard Model is observed in
any of the search analyses performed by H1 and ZEUS and, where
appropriate, mass dependent model exclusion limits are derived. 
Whereas the limits from HERA are competitive and complementary to
the analyses from colliders of the same generation, the advent of the
LHC has meant that many have been superseded at the time of writing.
Nevertheless, a few exceptions remain, in particular the H1 limits on
excited neutrino production using the $e^{-}p$ data remain the best
currently available, and the pair production leptoquark limits from
the LHC are so far only applicable to scalar models, whereas vector
leptoquark limits are set by both H1 and ZEUS.

%%%

Future searches in $ep$ physics may be possible if the LHeC collider
is to be built, where the program will closely follow the analyses
described in this review, albeit at the TeV energy scale.
At this energy, the addition of the Higgs sector opens up a whole new
physics program to explore.
For example, the study of the $WW \rightarrow H \rightarrow b\bar{b}$
process, which is extremely difficult at the LHC but essentially clean 
at the LHeC, allows the $WWH$ coupling to be probed and the properties
of the still relatively new Higgs boson to be compared to those
predicted by the Standard Model.


\begin{thebibliography}{100}

\bibitem{Glashow:1961tr}
S.L. Glashow.
\newblock {Partial symmetries of weak interactions}.
\newblock {\em Nucl.Phys.}, 22:579, 1961.

\bibitem{Weinberg:1967tq}
S.~Weinberg.
\newblock {A model of leptons}.
\newblock {\em Phys. Rev. Lett.}, 19:1264, 1967.

\bibitem{Salam:1968}
A.~Salam.
\newblock Weak and electromagnetic interactions.
\newblock In N.~Svartholm, editor, {\em Elementary particle theory}, page 367.
  Almquist \& Wiksell.

\bibitem{higgsATLAS}
G.~Aad et~al.
\newblock {Observation of a new particle in the search for the Standard Model
  Higgs boson with the ATLAS detector at the LHC}.
\newblock {\em Phys. Lett.}, B716:1, 2012.

\bibitem{higgsCMS}
S.~Chatrchyan et~al.
\newblock {Observation of a new boson at a mass of {$125$} {GeV} with the CMS
  experiment at the LHC}.
\newblock {\em Phys. Lett.}, B716:30, 2012.

\bibitem{Englert:1964et}
F.~Englert and R.~Brout.
\newblock {Broken symmetry and the mass of gauge vector mesons}.
\newblock {\em Phys. Rev. Lett.}, 13:321, 1964.

\bibitem{Higgs:1964pj}
P.W. Higgs.
\newblock {Broken symmetries and the masses of gauge bosons}.
\newblock {\em Phys. Rev. Lett.}, 13:508, 1964.

\bibitem{Guralnik:1964eu}
G.S. Guralnik, C.R. Hagen, and T.W.B. Kibble.
\newblock {Global conservation laws and massless particles}.
\newblock {\em Phys. Rev. Lett.}, 13:585, 1964.

\bibitem{Collins:1989gx}
J.C. Collins, D.E. Soper, and G.F. Sterman.
\newblock {Factorization of hard processes in QCD}.
\newblock {\em Adv. Ser. Direct. High Energy Phys.}, 5:1, 1988.

\bibitem{Gribov:1972ri}
V.N. Gribov and L.N. Lipatov.
\newblock Deep inelastic $ep$ scattering in perturbation theory.
\newblock {\em Sov. J. Nucl. Phys.}, 15:438, 1972.

\bibitem{Altarelli:1977zs}
G.~Altarelli and G.~Parisi.
\newblock {Asymptotic freedom in parton language}.
\newblock {\em Nucl. Phys.}, B126:298, 1977.

\bibitem{Dokshitzer:1977sg}
Y.L. Dokshitzer.
\newblock {Calculation of the structure functions for deep inelastic scattering
  and $e^{+}e^{-}$ annihilation by perturbation theory in Quantum
  Chromodynamics}.
\newblock {\em Sov. Phys. JETP}, 46:641, 1977.

\bibitem{heraacc}
{HERA: A proposal for a large electron proton beam facility at DESY}.
\newblock DESY-HERA-81-10, 1981.

\bibitem{AbelleiraFernandez:2012cc}
J.L. Abelleira~Fernandez et~al.
\newblock {A large hadron electron collider at CERN: Report on the physics and
  design concepts for machine and detector}.
\newblock {\em J. Phys.}, G39:075001, 2012.

\bibitem{Abt:1996hi}
I.~Abt et~al.
\newblock {The H1 detector at HERA}.
\newblock {\em Nucl. Instrum. Meth.}, A386:310, 1997.

\bibitem{Abt:1996xv}
I.~Abt et~al.
\newblock {The tracking, calorimeter and muon detectors of the H1 experiment at
  HERA}.
\newblock {\em Nucl. Instrum. Meth.}, A386:348, 1997.

\bibitem{ZEUSdetector}
ZEUS Coll.
\newblock {U. Holm (ed.), The ZEUS Detector. Status report (unpublished)}.
\newblock DESY (1993), available on {\tt
  http://www-zeus.desy.de/bluebook/bluebook.html}.

\bibitem{Ackerstaff:1998av}
K.~Ackerstaff et~al.
\newblock {The HERMES spectrometer}.
\newblock {\em Nucl. Instrum. Meth.}, A417:230, 1998.

\bibitem{Hartouni:1995cf}
E.P. Hartouni, M.~Kreisler, G.~Van~Apeldoorn, H.~van~der Graaf, W.~Ruckstuhl,
  et~al.
\newblock {HERA-B: An experiment to study CP violation in the B system using an
  internal target at the HERA proton ring. Design report}.
\newblock DESY-PRC-95-01, 1995.

\bibitem{HERAB:2000aa}
{HERA-B: Report on status and prospects October 2000. Executive summary}.
\newblock DESY-PRC-00-04, 2000.

\bibitem{heraupgrade}
{HERA} luminosity upgrade report for funding agencies.
\newblock DESY-PRC-98-03, 1998.

\bibitem{suszyski}
L.~Suszycki.
\newblock Luminosity monitoring, photon tagging and {QED} tests, in \emph{Proc.
  of the HERA Workshop 1987}, {edited by R.D. Peccei (DESY, Hamburg, 1988),
  p.505}.

\bibitem{levonian}
S.~Levonian.
\newblock Luminosity measurements with {H1} luminosity monitor, in \emph{Proc.
  of the Harz Seminar}, {edited by F. Willeke (Bad Lauterberg, 1992,
  DESY-HERA-92-07), p.247}.

\bibitem{Ahmed:1995cf}
T.~Ahmed et~al.
\newblock {Experimental study of hard photon radiation processes at HERA}.
\newblock {\em Z. Phys.}, C66:529, 1995.

\bibitem{lumi1}
J.~Andruszk\'ow et~al.
\newblock First measurement of {HERA} luminosity by {ZEUS} lumi monitor.
\newblock DESY-92-066, 1993.

\bibitem{zfp:c63:391}
M.~Derrick et~al.
\newblock Measurement of total and partial photon proton cross-sections at
  $180$ {GeV} center-of-mass energy.
\newblock {\em Z. Phys.}, C63:391, 1994.

\bibitem{app:b32:2025}
J.~Andruszk\'ow et~al.
\newblock Luminosity measurement in the {ZEUS} experiment.
\newblock {\em Acta Phys. Pol.}, B32:2025, 1994.

\bibitem{Helbich:2005qf}
M.~Helbich et~al.
\newblock The spectrometer system for measuring {ZEUS} luminosity at {HERA}.
\newblock {\em Nucl. Instrum. Meth.}, A565:572, 2006.

\bibitem{Aaron:2012kn}
F.D. Aaron et~al.
\newblock {Determination of the integrated luminosity at {HERA} using elastic
  {QED} Compton events}.
\newblock {\em Eur. Phys. J.}, C72:2163, 2012.
\newblock [Erratum: \emph{Eur. Phys. J.}, C74:2733, 2012].

\bibitem{sokolov-ternov1}
A.A. Sokolov and I.M. Ternov.
\newblock On polarization and spin effects in the theory of synchrotron
  radiation.
\newblock {\em Sov. Phys. Dokl.}, 8:1203, 1964.

\bibitem{sokolov-ternov2}
V.N. Baier and V.A. Khoze.
\newblock Determination of the transverse polarization of high-energy
  electrons.
\newblock {\em Sov. J. Nucl. Phys.}, B9:238, 1969.

\bibitem{tpol}
D.P.~Barber et~al.
\newblock The {HERA} polarimeter and the first observation of electron spin
  polarization at {HERA}.
\newblock {\em Nucl. Instrum. Meth.}, A329:79, 1993.

\bibitem{lpol}
M.~Beckmann et~al.
\newblock The longitudinal polarimeter at {HERA}.
\newblock {\em Nucl. Instrum. Meth.}, A479:334, 2002.

\bibitem{Sobloher:2012rc}
B.~Sobloher, R.~Fabbri, T.~Behnke, J.~Olsson, D.~Pitzl, et~al.
\newblock {Polarisation at HERA: Reanalysis of the HERA II polarimeter data}.
\newblock DESY-11-259, 2012.

\bibitem{Lippmann:2011bb}
C.~Lippmann.
\newblock {Particle identification}.
\newblock {\em Nucl. Instrum. Meth.}, A666:148, 2012.

\bibitem{Kogler:2011zz}
R.~Kogler.
\newblock {Measurement of jet production in deep-inelastic $ep$ scattering at
  {HERA}}.
\newblock 2011.

\bibitem{Pitzl:2000wz}
D.~Pitzl, O.~Behnke, M.~Biddulph, K.~Bosiger, R.~Eichler, et~al.
\newblock {The H1 silicon vertex detector}.
\newblock {\em Nucl. Instrum. Meth.}, A454:334, 2000.

\bibitem{Eick:1996gv}
W.~Eick, H.~Henschel, H.H. Kaufmann, M.~Klein, P.~Kostka, et~al.
\newblock {Development of the H1 backward silicon strip detector}.
\newblock {\em Nucl. Instrum. Meth.}, A386:81, 1997.

\bibitem{Andrieu:1993kh}
B.~Andrieu et~al.
\newblock {The H1 liquid argon calorimeter system}.
\newblock {\em Nucl. Instrum. Meth.}, A336:460, 1993.

\bibitem{compensate}
H.~Wellisch et~al.
\newblock {Hadronic calibration of the H1 LAr calorimeter using software
  weighting techniques}.
\newblock H1 Internal Note, H1-IN-346, 1994.

\bibitem{Andrieu:1993tz}
B.~Andrieu et~al.
\newblock {Results from pion calibration runs for the H1 liquid argon
  calorimeter and comparisons with simulations}.
\newblock {\em Nucl. Instrum. Meth.}, A336:499, 1993.

\bibitem{Andrieu:1994yn}
B.~Andrieu et~al.
\newblock {Beam tests and calibration of the H1 liquid argon calorimeter with
  electrons}.
\newblock {\em Nucl. Instrum. Meth.}, A350:57, 1994.

\bibitem{Appuhn:1996na}
R.D. Appuhn et~al.
\newblock {The H1 lead / scintillating fiber calorimeter}.
\newblock {\em Nucl. Instrum. Meth.}, A386:397, 1997.

\bibitem{Nicholls:1995di}
T.~Nicholls et~al.
\newblock {Performance of an electromagnetic lead / scintillating fiber
  calorimeter for the H1 detector}.
\newblock {\em Nucl. Instrum. Meth.}, A374:149, 1996.

\bibitem{Appuhn:1996xh}
R.D. Appuhn et~al.
\newblock {H1 backward upgrade with a SPACAL calorimeter: The hadronic
  section}.
\newblock DESY-96-013, 1996.

\bibitem{nim:a279:290}
N.~Harnew et~al.
\newblock Vertex triggering using time difference measurements in the {ZEUS}
  central tracking detector.
\newblock {\em Nucl. Instrum. Meth.}, A279:290, 1989.

\bibitem{npps:b32:181}
B.~Foster et~al.
\newblock The performance of the {ZEUS} central tracking detector z-by-timing
  electronics in a transputer based data acquisition system.
\newblock {\em Nucl. Phys. Proc. Suppl.}, B32:181, 1993.

\bibitem{nim:a338:254}
B.~Foster et~al.
\newblock The design and construction of the {ZEUS} central tracking detector.
\newblock {\em Nucl. Instrum. Meth.}, A338:254, 1994.

\bibitem{nim:a581:656}
A.~Polini et~al.
\newblock The design and performance of the {ZEUS} micro vertex detector.
\newblock {\em Nucl. Instrum. Meth.}, A581:656, 2007.

\bibitem{nim:a535:191}
S.~Fourletov.
\newblock Straw tube tracking detector {(STT)} for {ZEUS}.
\newblock {\em Nucl. Instrum. Meth.}, A535:191, 2004.

\bibitem{nim:a309:77}
M.~Derrick et~al.
\newblock Design and construction of the {ZEUS} barrel calorimeter.
\newblock {\em Nucl. Instrum. Meth.}, A309:77, 1991.

\bibitem{nim:a309:101}
A.~Andresen et~al.
\newblock Construction and beam test of the {ZEUS} forward and rear
  calorimeter.
\newblock {\em Nucl. Instrum. Meth.}, A309:101, 1991.

\bibitem{nim:a321:356}
A.~Caldwell et~al.
\newblock Design and implementation of a high precision readout system for the
  {ZEUS} calorimeter.
\newblock {\em Nucl. Instrum. Meth.}, A321:356, 1992.

\bibitem{nim:a336:23}
A.~Bernstein et~al.
\newblock Beam tests of the {ZEUS} barrel calorimeter.
\newblock {\em Nucl. Instrum. Meth.}, A336:23, 1993.

\bibitem{Bamberger:1996hi}
A.~Bamberger et~al.
\newblock The presampler for the forward and rear calorimeter in the {ZEUS}
  detector.
\newblock {\em Nucl. Instrum. Meth.}, A382:419, 1996.

\bibitem{zeus-pres2}
S.~Magill and S.~Chekanov.
\newblock Jet energy corrections with the {ZEUS} barrel preshower detector, in
  \emph{Proc. of the IX International Conference on Calorimetry}, {edited by B.
  Aubert (Frascati Physics Series, Annecy, France, 2000), vol.XXI, p.625}.

\bibitem{Dwurazny:1988fj}
A.~Dwurazny et~al.
\newblock Experimental study of electron-hadron separation in calorimeters
  using silicon diodes.
\newblock {\em Nucl. Instrum. Meth.}, A277:176, 1989.

\bibitem{nim:a333:342}
G.~Abbiendi et~al.
\newblock The {ZEUS} barrel and rear muon detector.
\newblock {\em Nucl. Instrum. Meth.}, A333:342, 1993.

\bibitem{nim:a313:126}
H.~Abramowicz et~al.
\newblock Intercalibration of the {ZEUS} high resolution and backing
  calorimeters.
\newblock {\em Nucl. Instrum. Meth.}, A313:126, 1992.

\bibitem{Aaron:2012qi}
F.D. Aaron et~al.
\newblock {Inclusive deep inelastic scattering at high {$Q^2$} with
  longitudinally polarised lepton beams at {HERA}}.
\newblock {\em JHEP}, 1209:061, 2012.

\bibitem{Salam:2009jx}
G.P. Salam.
\newblock {Towards jetography}.
\newblock {\em Eur. Phys. J.}, C67:637, 2010.

\bibitem{Catani:1993hr}
S.~Catani, Y.L. Dokshitzer, M.H. Seymour, and B.R. Webber.
\newblock Longitudinally invariant $k_{t}$ clustering algorithms for
  hadron-hadron collisions.
\newblock {\em Nucl. Phys.}, B406:187, 1993.

\bibitem{Andreev:2014wwa}
V.~Andreev et~al.
\newblock {Measurement of multijet production in $ep$ collisions at high
  {$Q^2$} and determination of the strong coupling $\alpha _s$}.
\newblock {\em Eur. Phys. J.}, C75(2):65, 2015.

\bibitem{Cacciari:2008gp}
M.~Cacciari, G.P. Salam, and G.~Soyez.
\newblock {The anti-k(t) jet clustering algorithm}.
\newblock {\em JHEP}, 0804:063, 2008.

\bibitem{Derrick:1996hn}
M.~Derrick et~al.
\newblock Measurement of the {$F_{2}$} structure function in deep inelastic
  $e^{+}p$ scattering using 1994 data from the {ZEUS} detector at {HERA}.
\newblock {\em Z. Phys.}, C72:399, 1996.

\bibitem{yjb}
F.~Jacquet and A.~Blondel.
\newblock Detection of the charged current event - method {II}, in \emph{Proc.
  of the Study of an $ep$ Facility for Europe}, {edited by U. Amaldi (Hamburg,
  1979, DESY 79/48) p.391}.

\bibitem{Callan:1969uq}
C.G. Callan, Jr. and D.J. Gross.
\newblock High-energy electroproduction and the constitution of the electric
  current.
\newblock {\em Phys. Rev. Lett.}, 22:156, 1969.

\bibitem{Bassler:1994uq}
U.~Bassler and G.~Bernardi.
\newblock On the kinematic reconstruction of deep inelastic scattering at
  {HERA}: The sigma method.
\newblock {\em Nucl. Instrum. Meth.}, A361:197, 1995.

\bibitem{Bassler:1997tv}
U.~Bassler and G.~Bernardi.
\newblock {Structure function measurements and kinematic reconstruction at
  HERA}.
\newblock {\em Nucl. Instrum. Meth.}, A426:583, 1999.

\bibitem{Heinemann:1999ry}
B.~Heinemann.
\newblock {Measurement of charged current and neutral current cross-sections in
  positron proton collisions at $\sqrt{s} \simeq 300$ {GeV}}.
\newblock 1999.

\bibitem{standa}
J.~Engelen S.~Bentvelsen and P.~Kooijman.
\newblock Reconstruction of {($x$, $Q^{2}$)} and extration of structure
  functions in neutral current scattering, in \emph{Proc. of the Workshop
  ``Physics at HERA"}, {edited by W. Buchm{\"u}ller and G. Ingelman (DESY,
  Hamburg, 1991), vol.1, p.23}.

\bibitem{hoegerda}
K.~Hoeger.
\newblock Measurement of $x$, $y$, {$Q^{2}$} in neutral current events, in
  \emph{Proc. of the Workshop ``Physics at HERA"}, {edited by W. Buchm{\"u}ller
  and G. Ingelman (DESY, Hamburg, 1991), vol.1, p.43}.

\bibitem{Collins:1988ig}
{J.C. Collins, D.E. Soper, and G.F. Sterman}.
\newblock Soft gluons and factorization.
\newblock {\em Nucl. Phys.}, B308:833, 1988.

\bibitem{Corcella:2000bw}
I.G. Corcella, G.~Knowles, G.~Marchesini, S.~Moretti, K.~Odagiri, et~al.
\newblock {HERWIG 6: An event generator for hadron emission reactions with
  interfering gluons (including supersymmetric processes)}.
\newblock {\em JHEP}, 01:010, 2001.

\bibitem{Andersson:1983ia}
B.~Andersson, G.~Gustafson, G.~Ingelman, and T.~Sj{\"o}strand.
\newblock {Parton fragmentation and string dynamics}.
\newblock {\em Phys. Rept.}, 97:31, 1983.

\bibitem{Sjostrand:2006za}
T.~Sj{\"o}strand, S.~Mrenna, and P.Z. Skands.
\newblock {PYTHIA 6.4 physics and manual}.
\newblock {\em JHEP}, 05:026, 2006.

\bibitem{jetset}
{T. Sj{\"o}strand and M. Bengtsson}.
\newblock The {Lund Monte Carlo} for jet fragmentation and $e^{+}e^{-}$
  physics. {Jetset} version 6.3: An update.
\newblock {\em Comp. Phys. Comm.}, 43:367, 1987.

\bibitem{geant}
{R. Brun}.
\newblock {GEANT 3.13}.
\newblock {CERN DD/EE 84-1, 1984}.

\bibitem{Gribov:1972rt}
V.N. Gribov and L.N. Lipatov.
\newblock $e^{+}e^{-}$ pair annihilation and deep inelastic $ep$ scattering in
  perturbation theory.
\newblock {\em Sov. J. Nucl. Phys.}, 15:675, 1972.

\bibitem{Lipatov:1974qm}
L.N. Lipatov.
\newblock The parton model and perturbation theory.
\newblock {\em Sov. J. Nucl. Phys.}, 20:94, 1975.

\bibitem{Abramowicz:2015mha}
H.~Abramowicz et~al.
\newblock {Combination of measurements of inclusive deep inelastic ${e^{\pm
  }p}$ scattering cross sections and QCD analysis of HERA data}.
\newblock {\em Eur. Phys. J.}, C75(12):580, 2015.

\bibitem{Aaron:2009bp}
F.D. Aaron et~al.
\newblock Measurement of the inclusive $ep$ scattering cross section at low
  {$Q^{2}$} and $x$ at {HERA}.
\newblock {\em Eur. Phys. J.}, C63:625, 2009.

\bibitem{Glazov:2005rn}
A.~Glazov.
\newblock Averaging of {DIS} cross section data, in \emph{Proc. of the 13th
  International Workshop on Deep Inelastic Scattering (DIS 2005)}, {edited by
  W.H. Smith, S.R. Dasu (Madison, Wisconsin (USA), 2005), vol.792, p.237}.

\bibitem{Breitweg:2000yn}
J.~Breitweg et~al.
\newblock Measurement of the proton structure function {$F_{2}$} at very low
  {$Q^{2}$} at {HERA}.
\newblock {\em Phys. Lett.}, B487:53, 2000.

\bibitem{Breitweg:1998dz}
J.~Breitweg et~al.
\newblock {ZEUS} results on the measurement and phenomenology of {$F_{2}$} at
  low $x$ and low {$Q^{2}$}.
\newblock {\em Eur. Phys. J.}, C7:609, 1999.

\bibitem{Adloff:1997mf}
C.~Adloff et~al.
\newblock Measurement of inclusive jet cross-sections in photoproduction at
  {HERA}.
\newblock {\em Nucl. Phys.}, B497:3, 1997.

\bibitem{django}
G.A. Schuler and H.~Spiesberger.
\newblock {DJANGO}: The interface for the event generators {HERACLES} and
  {LEPTO}, in \emph{Proc. of the Workshop ``Physics at HERA"}, {edited by W.
  Buchm{\"u}ller and G. Ingelman (DESY, Hamburg, 1991), vol.3, p.1419}.

\bibitem{heracles}
{A. Kwiatkowski, H. Spiesberger, and H.-J. and M{\"o}hring}.
\newblock Heracles: An event generator for $ep$ interactions at {HERA} energies
  including radiative processes: Version 1.0.
\newblock {\em Comp. Phys. Comm.}, 69:155, 1992.

\bibitem{lepto}
G.~Ingelman.
\newblock {LEPTO} version 6.1: The {Lund Monte Carlo} for deep inelastic
  lepton-nucleon scattering, in \emph{Proc. of the Workshop ``Physics at
  HERA"}, {edited by W. Buchm{\"u}ller and G. Ingelman (DESY, Hamburg, 1991),
  vol.3, p.1366}.

\bibitem{ariadne}
{L. L{\"o}nnblad}.
\newblock {ARIADNE} version 4: A program for simulation of {QCD} cascades
  implementing the color dipole model.
\newblock {\em Comp. Phys. Comm.}, 71:15, 1992.

\bibitem{Benvenuti:1989rh}
A.C. Benvenuti et~al.
\newblock A high statistics measurement of the proton structure functions
  {$F_{2}$} ($x$, {$Q^{2}$}) and $r$ from deep inelastic muon scattering at
  high {$Q^{2}$}.
\newblock {\em Phys. Lett.}, B223:485, 1989.

\bibitem{Arneodo:1996qe}
M.~Arneodo et~al.
\newblock Measurement of the proton and deuteron structure functions,
  {$F_{2}^{p}$} and {$F_{2}^{d}$}, and of the ratio {sigma-L / sigma-T}.
\newblock {\em Nucl. Phys.}, B483:3, 1997.

\bibitem{Altarelli:1978tq}
G.~Altarelli and G.~Martinelli.
\newblock {Transverse momentum of jets in electroproduction from quantum
  chromodynamics}.
\newblock {\em Phys. Lett.}, B76:89, 1978.

\bibitem{Aubert:1982ts}
J.J. Aubert et~al.
\newblock {Measurement of $r = \sigma_l / \sigma_t$ in deep inelastic
  muon-proton scattering}.
\newblock {\em Phys. Lett.}, B121:87, 1983.

\bibitem{Whitlow:1990gk}
L.W. Whitlow, S.~Rock, A.~Bodek, E.M. Riordan, and S.~Dasu.
\newblock {A precise extraction of {R = sigma-L / sigma-T} from a global
  analysis of the SLAC deep inelastic $ep$ and $ed$ scattering cross-sections}.
\newblock {\em Phys. Lett.}, B250:193, 1990.

\bibitem{Andreev:2013vha}
V.~Andreev et~al.
\newblock Measurement of inclusive $ep$ cross sections at high {$Q^{2}$} at
  {$\sqrt{s} = 225$} and {$252$ GeV} and of the longitudinal proton structure
  function {$F_{L}$} at {HERA}.
\newblock {\em Eur. Phys. J.}, C74(4):2814, 2014.

\bibitem{Chekanov:2009na}
S.~Chekanov et~al.
\newblock Measurement of the longitudinal proton structure function at {HERA}.
\newblock {\em Phys. Lett.}, B682:8, 2009.

\bibitem{Ahmed:1994fa}
T.~Ahmed et~al.
\newblock {First measurement of the charged current cross-section at HERA}.
\newblock {\em Phys. Lett.}, B324:241, 1994.

\bibitem{Derrick:1995us}
M.~Derrick et~al.
\newblock {Measurement of charged and neutral current $e^{-}p$ deep inelastic
  scattering cross-sections at high {$Q^{2}$}}.
\newblock {\em Phys. Rev. Lett.}, 75:1006, 1995.

\bibitem{Abe:2000cv}
T.~Abe.
\newblock {GRAPE dilepton (version 1.1): A generator for dilepton production in
  $ep$ collisions}.
\newblock {\em Comput. Phys. Commun.}, 136:126, 2001.

\bibitem{Baur:1991pp}
U.~Baur, J.A.M. Vermaseren, and D.~Zeppenfeld.
\newblock Electroweak vector boson production in high-energy $ep$ collisions.
\newblock {\em Nucl. Phys.}, B375:3, 1992.

\bibitem{Adloff:1999ah}
C.~Adloff et~al.
\newblock {Measurement of neutral and charged current cross-sections in
  positron proton collisions at large momentum transfer}.
\newblock {\em Eur. Phys. J.}, C13:609, 2000.

\bibitem{Aaron:2009aa}
F.D. Aaron et~al.
\newblock Combined measurement and {QCD} analysis of the inclusive $e^{\pm}p$
  scattering cross sections at {HERA}.
\newblock {\em JHEP}, 01:109, 2010.

\bibitem{Fanchiotti:1992tu}
S.~Fanchiotti, B.A. Kniehl, and A.~Sirlin.
\newblock Incorporation of {QCD} effects in basic corrections of the
  electroweak theory.
\newblock {\em Phys. Rev.}, D48:307, 1993.

\bibitem{Giele:1998gw}
W.T. Giele and S.~Keller.
\newblock Implications of hadron collider observables on parton distribution
  function uncertainties.
\newblock {\em Phys. Rev.}, D58:094023, 1998.

\bibitem{Abramowicz:1900rp}
H.~Abramowicz et~al.
\newblock {Combination and QCD analysis of charm production cross section
  measurements in deep-inelastic $ep$ scattering at HERA}.
\newblock {\em Eur. Phys. J.}, C73(2):2311, 2013.

\bibitem{dis_book}
R.~Devenish and A.~Cooper-Sarkar.
\newblock {\em Deep Inelastic Scattering}.
\newblock 2003.
\newblock Oxford University Press.

\bibitem{Radescu:2013mka}
V.~Radescu.
\newblock Combination and {QCD} analysis of the {HERA} inclusive cross
  sections, in \emph{Proc. of the 35th International Conference on High Energy
  Physics}, {PoS(ICHEP 2010)168}.

\bibitem{CooperSarkar:2011aa}
A.M. Cooper-Sarkar.
\newblock {PDF} fits at {HERA}, in \emph{Proc. of the 21st International
  Europhysics Conference on High Energy Physics}, {PoS(EPS-HEP 2011)320}.

\bibitem{Collaboration:2010xc}
H.~Abramowicz et~al.
\newblock Measurement of {high-$Q^{2}$} charged current deep inelastic
  scattering cross sections with a longitudinally polarised positron beam at
  {HERA}.
\newblock {\em Eur. Phys. J.}, C70:945, 2010.

\bibitem{Eichten:1983hw}
E.~Eichten, K.D. Lane, and M.E. Peskin.
\newblock {New tests for quark and lepton substructure}.
\newblock {\em Phys. Rev. Lett.}, 50:811, 1983.

\bibitem{Ruckl:1983hz}
R.~Ruckl.
\newblock {Effects of compositeness in deep inelastic scattering}.
\newblock {\em Phys. Lett.}, B129:363, 1983.

\bibitem{Ruckl:1983ag}
R.~Ruckl.
\newblock {Probing lepton and quark substructure in polarized $e^\mp N$
  scattering}.
\newblock {\em Nucl. Phys.}, B234:91, 1984.

\bibitem{haberl}
{P. Haberl, F. Schrempp and H.U. Martyn}.
\newblock Contact interactions and new heavy bosons at {HERA} - {A model
  independent analysis}, in \emph{Proc. of the Workshop ``Physics at HERA"},
  {edited by W. Buchm{\"u}ller and G. Ingelman (DESY, Hamburg, 1991), vol.2,
  p.1133}.

\bibitem{Buchmuller:1986zs}
W.~Buchm{\"u}ller, R.~Ruckl, and D.~Wyler.
\newblock {Leptoquarks in lepton-quark collisions}.
\newblock {\em Phys. Lett.}, B191:442, 1987.
\newblock [Erratum: \emph{Phys. Lett.}, B448:320, 1999].

\bibitem{schrempp}
F.~Schrempp.
\newblock Leptoquarks and leptogluons - {Theoretical perspectives}, in
  \emph{Proc. of the Workshop ``Physics at HERA"}, {edited by W. Buchm{\"u}ller
  and G. Ingelman (DESY, Hamburg, 1991), vol.2, p.1034}.

\bibitem{ArkaniHamed:1998rs}
N.~Arkani-Hamed, S.~Dimopoulos, and G.R. Dvali.
\newblock {The hierarchy problem and new dimensions at a millimeter}.
\newblock {\em Phys. Lett.}, B429:263, 1998.

\bibitem{Antoniadis:1998ig}
I.~Antoniadis, N.~Arkani-Hamed, S.~Dimopoulos, and G.R. Dvali.
\newblock {New dimensions at a millimeter to a Fermi and superstrings at a
  {TeV}}.
\newblock {\em Phys. Lett.}, B436:257, 1998.

\bibitem{ArkaniHamed:1998nn}
N.~Arkani-Hamed, S.~Dimopoulos, and G.R. Dvali.
\newblock {Phenomenology, astrophysics and cosmology of theories with
  submillimeter dimensions and {TeV} scale quantum gravity}.
\newblock {\em Phys. Rev.}, D59:086004, 1999.

\bibitem{Chekanov:2003pw}
S.~Chekanov et~al.
\newblock {Search for contact interactions, large extra dimensions and finite
  quark radius in $ep$ collisions at HERA}.
\newblock {\em Phys. Lett.}, B591:23, 2004.

\bibitem{Adloff:2000dp}
C.~Adloff et~al.
\newblock {Search for compositeness, leptoquarks and large extra dimensions in
  $eq$ contact interactions at HERA}.
\newblock {\em Phys. Lett.}, B479:358, 2000.

\bibitem{Giudice:1998ck}
G.F. Giudice, R.~Rattazzi, and J.D. Wells.
\newblock {Quantum gravity and extra dimensions at high-energy colliders}.
\newblock {\em Nucl. Phys.}, B544:3, 1999.

\bibitem{Cheung:1999qh}
K.~Cheung.
\newblock {Global lepton quark neutral current constraint on low scale gravity
  model}.
\newblock {\em Phys. Lett.}, B460:383, 1999.

\bibitem{Breitweg:1999ssa}
J.~Breitweg et~al.
\newblock {Search for contact interactions in deep inelastic $e^{+} p \to e^{+}
  X$ scattering at {HERA}}.
\newblock {\em Eur. Phys. J.}, C14:239, 2000.

\bibitem{Adloff:2003jm}
C.~Adloff et~al.
\newblock {Search for new physics in $e^\pm q$ contact interactions at HERA}.
\newblock {\em Phys. Lett.}, B568:35, 2003.

\bibitem{Aaron:2011mv}
F.D. Aaron et~al.
\newblock {Search for contact interactions in {$e^{\pm}p$} collisions at
  {HERA}}.
\newblock {\em Phys. Lett.}, B705:52, 2011.

\bibitem{Pumplin:2002vw}
J.~Pumplin, D.R. Stump, J.~Huston, H.L. Lai, P.M. Nadolsky, and W.K. Tung.
\newblock {New generation of parton distributions with uncertainties from
  global {QCD} analysis}.
\newblock {\em JHEP}, 07:012, 2002.

\bibitem{Aad:2014wca}
G.~Aad et~al.
\newblock {Search for contact interactions and large extra dimensions in the
  dilepton channel using proton-proton collisions at $\sqrt{s}$ = 8 TeV with
  the ATLAS detector}.
\newblock {\em Eur. Phys. J.}, C74(12):3134, 2014.

\bibitem{Pati:1974yy}
J.C. Pati and A.~Salam.
\newblock {Lepton number as the fourth color}.
\newblock {\em Phys. Rev.}, D10:275, 1974.
\newblock [Erratum: \emph{Phys. Rev.}, D11:703, 1975].

\bibitem{Georgi:1974sy}
H.~Georgi and S.L. Glashow.
\newblock {Unity of all elementary particle forces}.
\newblock {\em Phys. Rev. Lett.}, 32:438, 1974.

\bibitem{Langacker:1980js}
P.~Langacker.
\newblock {Grand unified theories and proton decay}.
\newblock {\em Phys. Rept.}, 72:185, 1981.

\bibitem{Schrempp:1984nj}
B.~Schrempp and F.~Schrempp.
\newblock {Light leptoquarks}.
\newblock {\em Phys. Lett.}, B153:101, 1985.

\bibitem{Wudka:1985ef}
J.~Wudka.
\newblock {Composite leptoquarks}.
\newblock {\em Phys. Lett.}, B167:337, 1986.

\bibitem{Dimopoulos:1979es}
S.~Dimopoulos and L.~Susskind.
\newblock {Mass without scalars}.
\newblock {\em Nucl. Phys.}, B155:237, 1979.

\bibitem{Dimopoulos:1979sp}
S.~Dimopoulos.
\newblock {Technicolored signatures}.
\newblock {\em Nucl. Phys.}, B168:69, 1980.

\bibitem{Farhi:1979zx}
E.~Farhi and L.~Susskind.
\newblock {A technicolored G.U.T.}
\newblock {\em Phys. Rev.}, D20:3404, 1979.

\bibitem{Farhi:1980xs}
E.~Farhi and L.~Susskind.
\newblock {Technicolor}.
\newblock {\em Phys. Rept.}, 74:277, 1981.

\bibitem{Nilles:1983ge}
H.-P. Nilles.
\newblock {Supersymmetry, supergravity and particle physics}.
\newblock {\em Phys. Rept.}, 110:1, 1984.

\bibitem{Haber:1984rc}
H.E. Haber and G.L. Kane.
\newblock {The search for supersymmetry: Probing physics beyond the Standard
  Model}.
\newblock {\em Phys. Rept.}, 117:75, 1985.

\bibitem{Aktas:2007ji}
A.~Aktas et~al.
\newblock {Search for lepton flavour violation in $ep$ collisions at HERA}.
\newblock {\em Eur. Phys. J.}, C52:833, 2007.

\bibitem{Adloff:1999tp}
C.~Adloff et~al.
\newblock {A Search for leptoquark bosons and lepton flavor violation in $e^{+}
  p$ collisions at HERA}.
\newblock {\em Eur. Phys. J.}, C11:447, 1999.
\newblock [Erratum: \emph{Eur. Phys. J.}, C14:553, 2000].

\bibitem{Aktas:2005pr}
A.~Aktas et~al.
\newblock {Search for leptoquark bosons in $ep$ collisions at HERA}.
\newblock {\em Phys. Lett.}, B629:9, 2005.

\bibitem{Breitweg:2000sa}
J.~Breitweg et~al.
\newblock {Search for resonances decaying to $e^{+}$-jet in $e^{+} p$
  interactions at HERA}.
\newblock {\em Eur. Phys. J.}, C16:253, 2000.

\bibitem{Breitweg:2000ks}
J.~Breitweg et~al.
\newblock {Search for resonance decays to an anti-neutrino plus jet in $e^{+}p$
  scattering at {DESY HERA}}.
\newblock {\em Phys. Rev.}, D63:052002, 2001.

\bibitem{Chekanov:2003af}
S.~Chekanov et~al.
\newblock {A search for resonance decays to lepton $+$ jet at HERA and limits
  on leptoquarks}.
\newblock {\em Phys. Rev.}, D68:052004, 2003.

\bibitem{Adloff:1997fg}
C.~Adloff et~al.
\newblock {Observation of events at very high {$Q^{2}$} in $ep$ collisions at
  {HERA}}.
\newblock {\em Z. Phys.}, C74:191, 1997.

\bibitem{Breitweg:1997ff}
J.~Breitweg et~al.
\newblock {Comparison of {ZEUS} data with standard model predictions for $e^{+}
  p \to e^{+} X$ scattering at high x and {$Q^{2}$}}.
\newblock {\em Z. Phys.}, C74:207, 1997.

\bibitem{Aaron:2011qaa}
F.D. Aaron et~al.
\newblock {Search for first generation leptoquarks in $ep$ collisions at
  {HERA}}.
\newblock {\em Phys. Lett.}, B704:388, 2011.

\bibitem{Abramowicz:2012tg}
H.~Abramowicz et~al.
\newblock {Search for first-generation leptoquarks at HERA}.
\newblock {\em Phys. Rev.}, D86:012005, 2012.

\bibitem{Lai:1999wy}
H.L. Lai, J.~Huston, S.~Kuhlmann, J.~Morfin, F.I. Olness, J.F. Owens,
  J.~Pumplin, and W.K. Tung.
\newblock {Global QCD analysis of parton structure of the nucleon: CTEQ5 parton
  distributions}.
\newblock {\em Eur. Phys. J.}, C12:375, 2000.

\bibitem{Bock:2004xz}
P.~Bock.
\newblock Computation of confidence levels for exclusion or discovery of a
  signal with the method of fractional event counting.
\newblock {\em JHEP}, 01:080, 2007.

\bibitem{Abbiendi:1998ea}
G.~Abbiendi et~al.
\newblock {Tests of the Standard Model and constraints on new physics from
  measurements of fermion pair production at {$183$} {GeV} at {LEP}}.
\newblock {\em Eur. Phys. J.}, C6:1, 1999.

\bibitem{Acciarri:2000uh}
M.~Acciarri et~al.
\newblock {Search for manifestations of new physics in fermion pair production
  at LEP}.
\newblock {\em Phys. Lett.}, B489:81, 2000.

\bibitem{Abazov:2009ab}
V.M. Abazov et~al.
\newblock {Search for pair production of first-generation leptoquarks in
  {$p\bar{p}$} collisions at {$\sqrt{s}=1.96$} {TeV}}.
\newblock {\em Phys. Lett.}, B681:224, 2009.

\bibitem{Abazov:2011qj}
V.M. Abazov et~al.
\newblock {Search for first generation leptoquark pair production in the
  electron + missing energy + jets final state}.
\newblock {\em Phys. Rev.}, D84:071104, 2011.

\bibitem{Aad:2011ch}
G.~Aad et~al.
\newblock {Search for first generation scalar leptoquarks in $pp$ collisions at
  $\sqrt{s}=7$ TeV with the ATLAS detector}.
\newblock {\em Phys. Lett.}, B709:158, 2012.
\newblock [Erratum: \emph{Phys. Lett.}, B711:442, 2012].

\bibitem{Chatrchyan:2012vza}
S.~Chatrchyan et~al.
\newblock {Search for pair production of first- and second-generation scalar
  leptoquarks in $pp$ collisions at $\sqrt{s}= 7$ TeV}.
\newblock {\em Phys. Rev.}, D86:052013, 2012.

\bibitem{Aad:2015caa}
G.~Aad et~al.
\newblock {Searches for scalar leptoquarks in $pp$ collisions at $\sqrt{s}= 8$
  TeV with the ATLAS detector}.
\newblock {\em Eur. Phys. J.}, C76(1):5, 2016.

\bibitem{Barbieri:2011ci}
R.~Barbieri, G.~Isidori, J.~Jones-Perez, P.~Lodone, and D.M. Straub.
\newblock {$U(2)$ and minimal flavour violation in supersymmetry}.
\newblock {\em Eur. Phys. J.}, C71:1725, 2011.

\bibitem{Chekanov:2005au}
S.~Chekanov et~al.
\newblock {Search for lepton-flavor violation at HERA}.
\newblock {\em Eur. Phys. J.}, C44:463, 2005.

\bibitem{Aaron:2011zz}
F.D. Aaron et~al.
\newblock {Search for lepton flavour violation at {HERA}}.
\newblock {\em Phys. Lett.}, B701:20, 2011.

\bibitem{Aktas:2006fc}
A.~Aktas et~al.
\newblock {Tau lepton production in $ep$ collisions at HERA}.
\newblock {\em Eur. Phys. J.}, C48:699, 2006.

\bibitem{Jung:1993gf}
H.~Jung.
\newblock {Hard diffractive scattering in high-energy $ep$ collisions and the
  Monte Carlo generator {RAPGAP}}.
\newblock {\em Comput. Phys. Commun.}, 86:147, 1995.

\bibitem{Barate:2003sz}
R.~Barate et~al.
\newblock {Search for the Standard Model Higgs boson at LEP}.
\newblock {\em Phys. Lett.}, B565:61, 2003.

\bibitem{Khachatryan:2015bsa}
V.~Khachatryan et~al.
\newblock {Search for third-generation scalar leptoquarks in the t$\tau$
  channel in proton-proton collisions at $\sqrt{s}= 8$ TeV}.
\newblock {\em JHEP}, 07:042, 2015.

\bibitem{Khachatryan:2014ura}
V.~Khachatryan et~al.
\newblock {Search for pair production of third-generation scalar leptoquarks
  and top squarks in proton-proton collisions at $\sqrt(s) = 8$ TeV}.
\newblock {\em Phys. Lett.}, B739:229, 2014.

\bibitem{Vermaseren:1982cz}
J.A.M. Vermaseren.
\newblock {Two photon processes at very high-energies}.
\newblock {\em Nucl. Phys.}, B229:347, 1983.

\bibitem{Aktas:2003jg}
A.~Aktas et~al.
\newblock {Multi-electron production at high transverse momenta in $ep$
  collisions at HERA}.
\newblock {\em Eur. Phys. J.}, C31:17, 2003.

\bibitem{Aktas:2003sz}
A.~Aktas et~al.
\newblock {Muon pair production in $ep$ collisions at HERA}.
\newblock {\em Phys. Lett.}, B583:28, 2004.

\bibitem{Aaron:2008jh}
F.D. Aaron et~al.
\newblock {Multi-lepton production at high transverse momenta in $ep$
  collisions at {HERA}}.
\newblock {\em Phys. Lett.}, B668:268, 2008.

\bibitem{Chekanov:2009cv}
S.~Chekanov et~al.
\newblock {Multi-lepton production at high transverse momentum at {HERA}}.
\newblock {\em Phys. Lett.}, B680:13, 2009.

\bibitem{Berger:1998kp}
C.~Berger and P.~Kandel.
\newblock A new generator for wide angle bremsstrahlung, in \emph{Proc. of the
  Workshop ``Monte Carlo Generators for HERA Physics"}, {edited by A.T. Doyle
  et al. (DESY, Hamburg, 1999), p.596}.

\bibitem{Aaron:2009ad}
F.D. Aaron et~al.
\newblock {Multi-leptons with high transverse momentum at {HERA}}.
\newblock {\em JHEP}, 10:013, 2009.

\bibitem{Levy:2010aa}
A.~Levy et~al.
\newblock {Study of tau-pair production at HERA}.
\newblock {\em JHEP}, 02:117, 2011.

\bibitem{Carli:2002jp}
T.~Carli and B.~Koblitz.
\newblock A multivariate discrimination technique based on range searching.
\newblock {\em Nucl. Instrum. Meth.}, A501:576, 2003.

\bibitem{Accomando:1993ar}
E.~Accomando and S.~Petrarca.
\newblock {Searching a doubly charged Higgs boson at {HERA}}.
\newblock {\em Phys. Lett.}, B323:212, 1994.

\bibitem{Aktas:2006nu}
A.~Aktas et~al.
\newblock {Search for doubly-charged Higgs boson production at HERA}.
\newblock {\em Phys. Lett.}, B638:432, 2006.

\bibitem{Junk:1999kv}
T.~Junk.
\newblock {Confidence level computation for combining searches with small
  statistics}.
\newblock {\em Nucl. Instrum. Meth.}, A434:435, 1999.

\bibitem{Abbiendi:2003pr}
G~Abbiendi et~al.
\newblock {Search for the single production of doubly charged Higgs bosons and
  constraints on their couplings from Bhabha scattering}.
\newblock {\em Phys. Lett.}, B577:93, 2003.

\bibitem{Abbiendi:2001cr}
G.~Abbiendi et~al.
\newblock {Search for doubly charged Higgs bosons with the OPAL detector at
  LEP}.
\newblock {\em Phys. Lett.}, B526:221, 2002.

\bibitem{Abdallah:2002qj}
J.~Abdallah et~al.
\newblock {Search for doubly charged Higgs bosons at LEP-2}.
\newblock {\em Phys. Lett.}, B552:127, 2003.

\bibitem{Achard:2003mv}
P.~Achard et~al.
\newblock {Search for doubly charged Higgs bosons at LEP}.
\newblock {\em Phys. Lett.}, B576:18, 2003.

\bibitem{Aaltonen:2011rta}
T.~Aaltonen et~al.
\newblock {Search for new physics in high $p_T$ like-sign dilepton events at
  CDF~II}.
\newblock {\em Phys. Rev. Lett.}, 107:181801, 2011.

\bibitem{Aad:2014hja}
G.~Aad et~al.
\newblock {Search for new phenomena in events with three or more charged
  leptons in $pp$ collisions at $\sqrt{s}=8$ TeV with the ATLAS detector}.
\newblock {\em JHEP}, 08:138, 2015.

\bibitem{ATLAS:2014kca}
G.~Aad et~al.
\newblock {Search for anomalous production of prompt same-sign lepton pairs and
  pair-produced doubly charged Higgs bosons with $ \sqrt{s}=8 $ TeV $pp$
  collisions using the ATLAS detector}.
\newblock {\em JHEP}, 03:041, 2015.

\bibitem{firstH1event}
T.~Ahmed et~al.
\newblock {Observation of an $e^{+}p \rightarrow \mu + X$ event with high
  transverse momenta at HERA}.
\newblock DESY-94-248, 1994.

\bibitem{Diener:2003df}
K.-P.O. Diener, C.~Schwanenberger, and M.~Spira.
\newblock {Photoproduction of $W$ bosons at HERA: Reweighting method for
  implementing QCD corrections in Monte Carlo programs}.
\newblock 2003.

\bibitem{Diener:2002if}
K.-P.~O. Diener, C.~Schwanenberger, and M.~Spira.
\newblock {Photoproduction of $W$ bosons at HERA: QCD corrections}.
\newblock {\em Eur. Phys. J.}, C25:405, 2002.

\bibitem{Nason:1999xs}
P.~Nason, R.~Ruckl, and M.~Spira.
\newblock {A note on $W$ production at HERA}.
\newblock {\em J. Phys.}, G25:1434, 1999.

\bibitem{Spira:1999ja}
M.~Spira.
\newblock {$W$ boson production at NLO}, in \emph{Proc. of the Workshop ``Monte
  Carlo Generators for HERA Physics"}, {edited by A.T. Doyle et al. (DESY,
  Hamburg, 1999), p.623}.

\bibitem{Adloff:1998aw}
C.~Adloff et~al.
\newblock {Observation of events with an isolated high-energy lepton and
  missing transverse momentum at HERA}.
\newblock {\em Eur. Phys. J.}, C5:575, 1998.

\bibitem{Andreev:2003pm}
V.~Andreev et~al.
\newblock {Isolated electrons and muons in events with missing transverse
  momentum at HERA}.
\newblock {\em Phys. Lett.}, B561:241, 2003.

\bibitem{Breitweg:1999ie}
J.~Breitweg et~al.
\newblock {$W$ production and the search for events with an isolated
  high-energy lepton and missing transverse momentum at HERA}.
\newblock {\em Phys. Lett.}, B471:411, 2000.

\bibitem{Chekanov:2003yt}
S.~Chekanov et~al.
\newblock {Search for single top production in $ep$ collisions at HERA}.
\newblock {\em Phys. Lett.}, B559:153, 2003.
\newblock [Addendum: DESY-03-188, 2003].

\bibitem{Aaron:2009wp}
F.D. Aaron et~al.
\newblock {Events with isolated leptons and missing transverse momentum and
  measurement of $W$ production at HERA}.
\newblock {\em Eur. Phys. J.}, C64:251, 2009.

\bibitem{Chekanov:2008gn}
S.~Chekanov et~al.
\newblock {Search for events with an isolated lepton and missing transverse
  momentum and a measurement of $W$ production at HERA}.
\newblock {\em Phys. Lett.}, B672:106, 2009.

\bibitem{Diaconu:2006qs}
C.~Diaconu.
\newblock {Events with isolated leptons and missing transverse momentum at
  HERA}, in \emph{Proc. of the 33rd International Conference on High Energy
  Physics (ICHEP 2006)}, {edited by A. Sissakian et al., C060726, p.1151}.

\bibitem{Aaron:2009ab}
F.D. Aaron et~al.
\newblock {Events with an isolated lepton and missing transverse momentum and
  measurement of $W$ production at HERA}.
\newblock {\em JHEP}, 03:035, 2010.

\bibitem{Baur:1989gh}
U.~Baur and D.~Zeppenfeld.
\newblock {Measuring the $WW\gamma$ vertex in single $W$ production at $ep$
  colliders}.
\newblock {\em Nucl. Phys.}, B325:253, 1989.

\bibitem{Hagiwara:1986vm}
K.~Hagiwara, R.D. Peccei, D.~Zeppenfeld, and K.~Hikasa.
\newblock {Probing the weak boson sector in $e^{+}e^{-} \rightarrow
  W^{+}W^{-}$}.
\newblock {\em Nucl. Phys.}, B282:253, 1987.

\bibitem{Schael:2004tq}
S.~Schael et~al.
\newblock {Improved measurement of the triple gauge-boson couplings $\gamma WW$
  and $ZWW$ in $e^{+}e^{-}$ collisions}.
\newblock {\em Phys. Lett.}, B614:7, 2005.

\bibitem{Abdallah:2008sf}
J.~Abdallah et~al.
\newblock {Study of $W$ boson polarisations and triple gauge boson couplings in
  the reaction $e^{+}e^{-} \rightarrow W^{+}W^{-}$ at LEP 2}.
\newblock {\em Eur. Phys. J.}, C54:345, 2008.

\bibitem{Achard:2004ji}
P.~Achard et~al.
\newblock {Measurement of triple gauge boson couplings of the $W$ boson at
  LEP}.
\newblock {\em Phys. Lett.}, B586:151, 2004.

\bibitem{Abbiendi:2003mk}
G.~Abbiendi et~al.
\newblock {Measurement of charged current triple gauge boson couplings using
  $W$ pairs at LEP}.
\newblock {\em Eur. Phys. J.}, C33:463, 2004.

\bibitem{Chekanov:2003bf}
S.~Chekanov et~al.
\newblock {Isolated tau leptons in events with large missing transverse
  momentum at HERA}.
\newblock {\em Phys. Lett.}, B583:41, 2004.

\bibitem{Schott:2014sea}
M.~Schott and M.~Dunford.
\newblock {Review of single vector boson production in $pp$ collisions at
  $\sqrt{s} = 7$ TeV}.
\newblock {\em Eur. Phys. J.}, C74:2916, 2014.

\bibitem{Schuler:1987wj}
G.A. Schuler.
\newblock Heavy flavor production at {HERA}.
\newblock {\em Nucl. Phys.}, B299:21, 1988.

\bibitem{Baur:1987ai}
U.~Baur and J.J. van~der Bij.
\newblock Top quark production at {HERA}.
\newblock {\em Nucl. Phys.}, B304:451, 1988.

\bibitem{vanderBij:1990ju}
J.J. van~der Bij and G.J. van Oldenborgh.
\newblock {QCD} radiative corrections to charged current heavy quark
  production.
\newblock {\em Z. Phys.}, C51:477, 1991.

\bibitem{Stelzer:1997ns}
T.~Stelzer, Z.~Sullivan, and S.~Willenbrock.
\newblock {Single top quark production via $W$-gluon fusion at next-to-leading
  order}.
\newblock {\em Phys. Rev.}, D56:5919, 1997.

\bibitem{Moretti:1997dz}
S.~Moretti and K.~Odagiri.
\newblock {Single top production at future $ep$ colliders}.
\newblock {\em Phys. Rev.}, D57:3040, 1998.

\bibitem{Atwood:1995ud}
D.~Atwood, L.~Reina, and A.~Soni.
\newblock {Probing flavor changing top-charm-scalar interactions in
  $e^{+}e^{-}$ collisions}.
\newblock {\em Phys. Rev.}, D53:1199, 1996.

\bibitem{deDivitiis:1997sh}
G.M. de~Divitiis, R.~Petronzio, and L.~Silvestrini.
\newblock {Flavor changing top decays in supersymmetric extensions of the
  Standard Model}.
\newblock {\em Nucl. Phys.}, B504:45, 1997.

\bibitem{Peccei:1989kr}
R.D. Peccei and X.~Zhang.
\newblock Dynamical symmetry breaking and universality breakdown.
\newblock {\em Nucl. Phys.}, B337:269, 1990.

\bibitem{Fritzsch:1999rd}
H.~Fritzsch and D.~Holtmannspotter.
\newblock {The production of single $t$ quarks at LEP and HERA}.
\newblock {\em Phys. Lett.}, B457:186, 1999.

\bibitem{Boos:2009un}
E.~Boos et~al.
\newblock {CompHEP} 4.5 status report, in \emph{Proc. of the 12th International
  Workshop on Advanced Computing and Analysis Techniques in Physics Research},
  {PoS(ACAT2008)008}.

\bibitem{Aktas:2003yd}
A.~Aktas et~al.
\newblock {Search for single top quark production in $ep$ collisions at HERA}.
\newblock {\em Eur. Phys. J.}, C33:9, 2004.

\bibitem{Aaron:2009vv}
F.D. Aaron et~al.
\newblock {Search for single top quark production at HERA}.
\newblock {\em Phys. Lett.}, B678:450, 2009.

\bibitem{Abramowicz:2011tv}
H.~Abramowicz et~al.
\newblock {Search for single-top production in $ep$ collisions at HERA}.
\newblock {\em Phys. Lett.}, B708:27, 2012.

\bibitem{Belyaev:2001hf}
A.~Belyaev and N.~Kidonakis.
\newblock {QCD corrections to FCNC single top production at HERA}.
\newblock {\em Phys. Rev.}, D65:037501, 2002.

\bibitem{Heister:2002xv}
A.~Heister et~al.
\newblock {Search for single top production in $e^{+}e^{-}$ collisions at
  $\sqrt{s}$ up to $209$ GeV}.
\newblock {\em Phys. Lett.}, B543:173, 2002.

\bibitem{Abe:1997fz}
F.~Abe et~al.
\newblock {Search for flavor-changing neutral current decays of the top quark
  in $p\bar{p}$ collisions at $\sqrt{s} = 1.8$ TeV}.
\newblock {\em Phys. Rev. Lett.}, 80:2525, 1998.

\bibitem{Aaltonen:2008ac}
T.~Aaltonen et~al.
\newblock {Search for the flavor changing neutral current decay $t \to Zq$ in
  $p\bar{p}$ collisions at $\sqrt{s} = 1.96$ TeV}.
\newblock {\em Phys. Rev. Lett.}, 101:192002, 2008.

\bibitem{Abazov:2011qf}
V.M. Abazov et~al.
\newblock {Search for flavor changing neutral currents in decays of top
  quarks}.
\newblock {\em Phys. Lett.}, B701:313, 2011.

\bibitem{Aad:2012ij}
G.~Aad et~al.
\newblock {A search for flavour changing neutral currents in top-quark decays
  in $pp$ collision data collected with the ATLAS detector at $\sqrt{s}=7$
  TeV}.
\newblock {\em JHEP}, 09:139, 2012.

\bibitem{Aad:2012gd}
G.~Aad et~al.
\newblock {Search for FCNC single top-quark production at $\sqrt{s}=7$ TeV with
  the ATLAS detector}.
\newblock {\em Phys. Lett.}, B712:351, 2012.

\bibitem{Chatrchyan:2013nwa}
S.~Chatrchyan et~al.
\newblock {Search for flavor-changing neutral currents in top-quark decays $t
  \to Zq$ in $pp$ collisions at $\sqrt{s}=8$ TeV}.
\newblock {\em Phys. Rev. Lett.}, 112(17):171802, 2014.

\bibitem{Aaron:2008aa}
F.D. Aaron et~al.
\newblock {A general search for new phenomena at HERA}.
\newblock {\em Phys. Lett.}, B674:257, 2009.

\bibitem{Aktas:2004pz}
A.~Aktas et~al.
\newblock {A general search for new phenomena in $ep$ scattering at HERA}.
\newblock {\em Phys. Lett.}, B602:14, 2004.

\bibitem{Geer:1995mp}
S.~Geer and T.~Asakawa.
\newblock {The analysis of multi-jet events produced at high-energy hadron
  colliders}.
\newblock {\em Phys. Rev.}, D53:4793, 1996.

\bibitem{Aaron:2008cy}
F.D. Aaron et~al.
\newblock {Search for excited electrons in $ep$ collisions at HERA}.
\newblock {\em Phys. Lett.}, B666:131, 2008.

\bibitem{Aaron:2008ae}
F.D. Aaron et~al.
\newblock {A search for excited neutrinos in $e^{-}p$ collisions at HERA}.
\newblock {\em Phys. Lett.}, B663:382, 2008.

\bibitem{Harari:1982xy}
H.~Harari.
\newblock {Composite models for quarks and leptons}.
\newblock {\em Phys. Rept.}, 104:159, 1984.

\bibitem{Adloff:2002dy}
C.~Adloff et~al.
\newblock {Search for excited electrons at HERA}.
\newblock {\em Phys. Lett.}, B548:35, 2002.

\bibitem{Adloff:2001me}
C.~Adloff et~al.
\newblock {Search for excited neutrinos at HERA}.
\newblock {\em Phys. Lett.}, B525:9, 2002.

\bibitem{Adloff:2000gv}
C.~Adloff et~al.
\newblock {A search for excited fermions at HERA}.
\newblock {\em Eur. Phys. J.}, C17:567, 2000.

\bibitem{Chekanov:2001xk}
S.~Chekanov et~al.
\newblock {Searches for excited fermions in $e p$ collisions at HERA}.
\newblock {\em Phys. Lett.}, B549:32, 2002.

\bibitem{Aaron:2009iz}
F.D. Aaron et~al.
\newblock {Search for excited quarks in $ep$ collisions at HERA}.
\newblock {\em Phys. Lett.}, B678:335, 2009.

\bibitem{Hagiwara:1985wt}
K.~Hagiwara, D.~Zeppenfeld, and S.~Komamiya.
\newblock {Excited lepton production at LEP and HERA}.
\newblock {\em Z. Phys.}, C29:115, 1985.

\bibitem{Boudjema:1992em}
F.~Boudjema, A.~Djouadi, and J.L. Kneur.
\newblock {Excited fermions at $e^{+}e^{-}$ and $eP$ colliders}.
\newblock {\em Z. Phys.}, C57:425, 1993.

\bibitem{Baur:1989kv}
U.~Baur, M.~Spira, and P.M. Zerwas.
\newblock {Excited quark and lepton production at hadron colliders}.
\newblock {\em Phys. Rev.}, D42:815, 1990.

\bibitem{Brodsky:1980zm}
S.J. Brodsky and S.D. Drell.
\newblock {The anomalous magnetic moment and limits on fermion substructure}.
\newblock {\em Phys. Rev.}, D22:2236, 1980.

\bibitem{Renard:1982ij}
F.M. Renard.
\newblock {Limits on masses and couplings of excited electrons and muons}.
\newblock {\em Phys. Lett.}, B116:264, 1982.

\bibitem{Boos:2004kh}
E.~Boos, V.~Bunichev, M.~Dubinin, L.~Dudko, V.~Ilyin, A.~Kryukov, V.~Edneral,
  V.~Savrin, A.~Semenov, and A.~Sherstnev.
\newblock {CompHEP 4.4: Automatic computations from Lagrangians to events}.
\newblock {\em Nucl. Instrum. Meth.}, A534:250, 2004.

\bibitem{Pukhov:1999gg}
A.~Pukhov, E.~Boos, M.~Dubinin, V.~Edneral, V.~Ilyin, D.~Kovalenko, A.~Kryukov,
  V.~Savrin, S.~Shichanin, and A.~Semenov.
\newblock {CompHEP: A package for evaluation of Feynman diagrams and
  integration over multiparticle phase space}.
\newblock NP-MSU-98-41-542, 1999.

\bibitem{compos}
T.~K{\"o}hler.
\newblock {Exotic proesses at HERA: The event generator COMPOS}, in \emph{Proc.
  of the Workshop ``Physics at HERA"}, {edited by W. Buchm{\"u}ller and G.
  Ingelman (DESY, Hamburg, 1991), vol.3, p.1526}.

\bibitem{Abbiendi:2002wf}
G.~Abbiendi et~al.
\newblock {Search for charged excited leptons in $e^+e^-$ collisions at
  $\sqrt{s} = 183-209$ GeV}.
\newblock {\em Phys. Lett.}, B544:57, 2002.

\bibitem{Abdallah:2004rc}
J.~Abdallah et~al.
\newblock {Determination of the $e^+e^- \rightarrow \gamma \gamma (\gamma)$
  cross-section at LEP 2}.
\newblock {\em Eur. Phys. J.}, C37:405, 2004.

\bibitem{Achard:2003hd}
P.~Achard et~al.
\newblock {Search for excited leptons at LEP}.
\newblock {\em Phys. Lett.}, B568:23, 2003.

\bibitem{Abreu:1998jw}
P.~Abreu et~al.
\newblock {Search for composite and exotic fermions at LEP-2}.
\newblock {\em Eur. Phys. J.}, C8:41, 1999.

\bibitem{Acosta:2004ri}
D.~Acosta et~al.
\newblock {Search for excited and exotic electrons in the $e \gamma$ decay
  channel in $p \bar{p}$ collisions at $\sqrt{s} = 1.96$ TeV}.
\newblock {\em Phys. Rev. Lett.}, 94:101802, 2005.

\bibitem{Aad:2013jja}
G.~Aad et~al.
\newblock {Search for excited electrons and muons in $\sqrt{s}$=8 TeV
  proton-proton collisions with the ATLAS detector}.
\newblock {\em New J. Phys.}, 15:093011, 2013.

\bibitem{Aad:2013cva}
G.~Aad et~al.
\newblock {Search for new phenomena in photon+jet events collected in
  proton--proton collisions at sqrt(s) = 8 TeV with the ATLAS detector}.
\newblock {\em Phys. Lett.}, B728:562, 2014.

\bibitem{Wess:1974tw}
J.~Wess and B.~Zumino.
\newblock {Supergauge transformations in four-dimensions}.
\newblock {\em Nucl. Phys.}, B70:39, 1974.

\bibitem{Weinberg:1981wj}
S.~Weinberg.
\newblock {Supersymmetry at ordinary energies. 1. Masses and conservation
  laws}.
\newblock {\em Phys. Rev.}, D26:287, 1982.

\bibitem{Sakai:1981pk}
N.~Sakai and T.~Yanagida.
\newblock {Proton decay in a class of supersymmetric grand unified models}.
\newblock {\em Nucl. Phys.}, B197:533, 1982.

\bibitem{Butterworth:1992tc}
J.~Butterworth and H.K. Dreiner.
\newblock {$R$-parity violation at {HERA}}.
\newblock {\em Nucl. Phys.}, B397:3, 1993.

\bibitem{Aktas:2004ij}
A.~Aktas et~al.
\newblock {Search for squark production in $R$-parity violating supersymmetry
  at HERA}.
\newblock {\em Eur. Phys. J.}, C36:425, 2004.

\bibitem{Chekanov:2006aa}
S.~Chekanov et~al.
\newblock {Search for stop production in R-parity-violating supersymmetry at
  HERA}.
\newblock {\em Eur. Phys. J.}, C50:269, 2007.

\bibitem{Aaron:2010ez}
F.D. Aaron et~al.
\newblock {Search for squarks in $R$-parity violating supersymmetry in $ep$
  collisions at HERA}.
\newblock {\em Eur. Phys. J.}, C71:1572, 2011.

\bibitem{Plehn:1997az}
T.~Plehn, H.~Spiesberger, M.~Spira, and P.M. Zerwas.
\newblock {Formation and decay of scalar leptoquarks / squarks in $ep$
  collisions}.
\newblock {\em Z. Phys.}, C74:611, 1997.

\bibitem{LEGO}
K.~Rosenbauer.
\newblock {Suche nach Leptoquarks und Leptogluonen im H1-Experiment bei HERA}.
\newblock 1995.

\bibitem{Ghodbane:1999va}
N.~Ghodbane.
\newblock {SUSYGEN3: An event generator for linear colliders}, in \emph{Proc.
  of the 4th International Workshop on Linear Colliders (LCWS 99)},
  {LYCEN-9992, p.745}.

\bibitem{Adloff:2002au}
C.~Adloff et~al.
\newblock {Measurement of dijet cross-sections in photoproduction at HERA}.
\newblock {\em Eur. Phys. J.}, C25:13, 2002.

\bibitem{Langacker:1990jf}
P.~Langacker.
\newblock {Parity violation in muonic atoms and cesium}.
\newblock {\em Phys. Lett.}, B256:277, 1991.

\bibitem{Freedman:1976xh}
D.Z. Freedman, P.~van Nieuwenhuizen, and S.~Ferrara.
\newblock {Progress toward a theory of supergravity}.
\newblock {\em Phys. Rev.}, D13:3214, 1976.

\bibitem{Drees:1991ab}
M.~Drees and M.M. Nojiri.
\newblock {Radiative symmetry breaking in minimal $N=1$ supergravity with large
  Yukawa couplings}.
\newblock {\em Nucl. Phys.}, B369:54, 1992.

\bibitem{Baer:1992dc}
H.~Baer and X.~Tata.
\newblock {Probing charginos and neutralinos beyond the reach of LEP at the
  Tevatron collider}.
\newblock {\em Phys. Rev.}, D47:2739, 1993.

\bibitem{Kane:1993td}
G.L. Kane, C.F. Kolda, L.~Roszkowski, and J.D. Wells.
\newblock {Study of constrained minimal supersymmetry}.
\newblock {\em Phys. Rev.}, D49:6173, 1994.

\bibitem{Abbott:1999nh}
B.~Abbott et~al.
\newblock {Search for $R-$parity violating supersymmetry in the dielectron
  channel}.
\newblock {\em Phys. Rev. Lett.}, 83:4476, 1999.

\bibitem{Aktas:2004tm}
A.~Aktas et~al.
\newblock {Search for bosonic stop decays in $R$-parity violating supersymmetry
  in $e^{+}p$ collisions at HERA}.
\newblock {\em Phys. Lett.}, B599:159, 2004.

\bibitem{Kon:1997bz}
T.~Kon, T.~Matsushita, and T.~Kobayashi.
\newblock {Possible excess in charged current events with high {$Q^{2}$} at
  {HERA} from stop and sbottom production}.
\newblock {\em Mod. Phys. Lett.}, A12:3143, 1997.

\bibitem{Giudice:1998bp}
G.F. Giudice and R.~Rattazzi.
\newblock {Theories with gauge mediated supersymmetry breaking}.
\newblock {\em Phys. Rept.}, 322:419, 1999.

\bibitem{Aktas:2004cc}
A.~Aktas et~al.
\newblock {Search for light gravitinos in events with photons and missing
  transverse momentum at HERA}.
\newblock {\em Phys. Lett.}, B616:31, 2005.

\bibitem{Dimopoulos:1996vz}
S.~Dimopoulos, M.~Dine, S.~Raby, and S.D. Thomas.
\newblock {Experimental signatures of low-energy gauge mediated supersymmetry
  breaking}.
\newblock {\em Phys. Rev. Lett.}, 76:3494, 1996.

\bibitem{Dirac:1931kp}
P.A.M. Dirac.
\newblock {Quantized singularities in the electromagnetic field}.
\newblock {\em Proc. Roy. Soc. Lond.}, A133:60, 1931.

\bibitem{Sorkin:1983ns}
R.d. Sorkin.
\newblock {Kaluza-Klein monopole}.
\newblock {\em Phys. Rev. Lett.}, 51:87, 1983.

\bibitem{Gross:1983hb}
D.J. Gross and M.J. Perry.
\newblock {Magnetic monopoles in Kaluza-Klein theories}.
\newblock {\em Nucl. Phys.}, B226:29, 1983.

\bibitem{Bergshoeff:1998ef}
E.~Bergshoeff, E.~Eyras, and Y.~Lozano.
\newblock {The massive Kaluza-Klein monopole}.
\newblock {\em Phys. Lett.}, B430:77, 1998.

\bibitem{Polyakov:1974ek}
A.M. Polyakov.
\newblock {Particle spectrum in the quantum field theory}.
\newblock {\em JETP Lett.}, 20:194, 1974.

\bibitem{'tHooft:1974qc}
G.~'t~Hooft.
\newblock {Magnetic monopoles in unified gauge theories}.
\newblock {\em Nucl. Phys.}, B79:276, 1974.

\bibitem{Minakata:1984an}
H.~Minakata.
\newblock {Magnetic monopoles in supersymmetric grand unification}.
\newblock {\em Phys. Lett.}, B155:352, 1985.

\bibitem{Lopez:1996gy}
J.L. Lopez.
\newblock {Supersymmetry: From the Fermi scale to the Planck scale}.
\newblock {\em Rept. Prog. Phys.}, 59:819, 1996.

\bibitem{Rizzo:1981su}
T.G. Rizzo and G.~Senjanovic.
\newblock {Can there be low intermediate mass scales in grand unified
  theories?}
\newblock {\em Phys. Rev. Lett.}, 46:1315, 1981.

\bibitem{Preskill:1984gd}
J.~Preskill.
\newblock {Magnetic monopoles}.
\newblock {\em Ann. Rev. Nucl. Part. Sci.}, 34:461, 1984.

\bibitem{Weinberg:1983bf}
E.J. Weinberg, D.~London, and J.L. Rosner.
\newblock {Magnetic monopoles with $Z(n$) charges}.
\newblock {\em Nucl. Phys.}, B236:90, 1984.

\bibitem{Kirkman:1981ck}
T.W. Kirkman and C.K. Zachos.
\newblock {Asymptotic analysis of the monopole structure}.
\newblock {\em Phys. Rev.}, D24:999, 1981.

\bibitem{Yang:1998sh}
Y.S. Yang.
\newblock {Dually charged particle-like solutions in the Weinberg-Salam
  theory}.
\newblock {\em Proc. Roy. Soc. Lond.}, A454:155, 1998.

\bibitem{Cho:2002bk}
Y.M. Cho.
\newblock {Analytic electroweak dyon}.
\newblock 2002.

\bibitem{Bais:1979gv}
F.A. Bais and W.~Troost.
\newblock {Zero modes and bound states of the supersymmetric monopole}.
\newblock {\em Nucl. Phys.}, B178:125, 1981.

\bibitem{Banks:1988rj}
T.~Banks, M.~Dine, H.~Dykstra, and W.~Fischler.
\newblock {Magnetic monopole solutions of string theory}.
\newblock {\em Phys. Lett.}, B212:45, 1988.

\bibitem{Particle_Astropysics}
H.V. Klapdor-Kleingrothaus and K.~Zuber.
\newblock {\em Patricle Astrophysics}.
\newblock 1999.
\newblock The Institute of Physics, UK.

\bibitem{Alvarez_science}
L.W. Alvarez et~al.
\newblock Search for magnetic monopoles in the lunar sample.
\newblock {\em Science}, 167:701, 1970.

\bibitem{Aktas:2004qd}
A.~Aktas et~al.
\newblock {A direct search for stable magnetic monopoles produced in
  positron-proton collisions at HERA}.
\newblock {\em Eur. Phys. J.}, C41:133, 2005.

\bibitem{Gamberg:1998xf}
L.P. Gamberg, G.R. Kalbfleisch, and K.A. Milton.
\newblock {Difficulties with photonic searches for magnetic monopoles}.
\newblock 1998.

\bibitem{SQUID}
2G Enterprises, {\tt http://www.2genterprises.com}.

\bibitem{Ambrosio:2002qq}
M.~Ambrosio et~al.
\newblock {Final results of magnetic monopole searches with the MACRO
  experiment}.
\newblock {\em Eur. Phys. J.}, C25:511, 2002.

\bibitem{Balestra:2008ps}
S.~Balestra et~al.
\newblock {Magnetic monopole search at high altitude with the SLIM experiment}.
\newblock {\em Eur. Phys. J.}, C55:57, 2008.

\bibitem{Hogan:2008sx}
D.P. Hogan, D.Z. Besson, J.P. Ralston, I.~Kravchenko, and D.~Seckel.
\newblock {Relativistic magnetic monopole flux constraints from RICE}.
\newblock {\em Phys. Rev.}, D78:075031, 2008.

\bibitem{Detrixhe:2010xi}
M.~Detrixhe et~al.
\newblock {Ultra-relativistic magnetic monopole search with the ANITA-II
  balloon-borne radio interferometer}.
\newblock {\em Phys. Rev.}, D83:023513, 2011.

\bibitem{Abbasi:2012eda}
R.~Abbasi et~al.
\newblock {Search for relativistic magnetic monopoles with IceCube}.
\newblock {\em Phys. Rev.}, D87(2):022001, 2013.

\bibitem{Aartsen:2014awd}
M.G. Aartsen et~al.
\newblock {Search for non-relativistic magnetic monopoles with IceCube}.
\newblock {\em Eur. Phys. J.}, C74(7):2938, 2014.

\bibitem{Ross:1973it}
R.R. Ross, P.H. Eberhard, L.W. Alvarez, and R.D. Watt.
\newblock {Search for magnetic monopoles in lunar material using an
  electromagnetic detector}.
\newblock {\em Phys. Rev.}, D8:698, 1973.

\bibitem{Kovalik:1986zz}
J.M. Kovalik and J.L. Kirschvink.
\newblock {New superconducting quantum interface device based constraints on
  the abundance of magnetic monopoles trapped in matter: An investigation of
  deeply buried rocks}.
\newblock {\em Phys. Rev.}, A33:1183, 1986.

\bibitem{Jeon:1995rf}
H.~Jeon and M.J. Longo.
\newblock {Search for magnetic monopoles trapped in matter}.
\newblock {\em Phys. Rev. Lett.}, 75:1443, 1995.
\newblock [Erratum: \emph{Phys. Rev. Lett.}, 76:159, 1996].

\bibitem{Bendtz:2013tj}
K.~Bendtz, D.~Milstead, H.-P. H{\"a}chler, A.M. Hirt, P.~Mermod, P.~Michael,
  T.~Sloan, C.~Tegner, and S.B. Thorarinsson.
\newblock {Search for magnetic monopoles in polar volcanic rocks}.
\newblock {\em Phys. Rev. Lett.}, 110(12):121803, 2013.

\bibitem{Fairbairn:2006gg}
M.~Fairbairn, A.C. Kraan, D.A. Milstead, T.~Sj{\"o}strand, Peter~Z. Skands, and
  T.~Sloan.
\newblock {Stable massive particles at colliders}.
\newblock {\em Phys. Rept.}, 438:1, 2007.

\bibitem{Aubert:1982zi}
B.~Aubert, P.~Musset, M.~Price, and J.P. Vialle.
\newblock {Search for magnetic monopoles in proton-antiproton interactions at
  $540$ GeV cm energy}.
\newblock {\em Phys. Lett.}, B120:465, 1983.

\bibitem{Musset:1983ii}
P.~Musset, M.~Price, and E.~Lohrmann.
\newblock {Search for magnetic monopoles in electron-positron collisions at
  $34$ GeV cm energy}.
\newblock {\em Phys. Lett.}, B128:333, 1983.

\bibitem{Abbott:1998mw}
B.~Abbott et~al.
\newblock {A search for heavy point-like Dirac monopoles}.
\newblock {\em Phys. Rev. Lett.}, 81:524, 1998.

\bibitem{Abulencia:2005hb}
A.~Abulencia et~al.
\newblock {Direct search for Dirac magnetic monopoles in $p\bar{p}$ collisions
  at $\sqrt{s} = 1.96$ TeV}.
\newblock {\em Phys. Rev. Lett.}, 96:201801, 2006.

\bibitem{Abbiendi:2007ab}
G.~Abbiendi et~al.
\newblock {Search for Dirac magnetic monopoles in $e^{+}e^{-}$ collisions with
  the OPAL detector at LEP2}.
\newblock {\em Phys. Lett.}, B663:37, 2008.

\bibitem{Aad:2012qi}
G.~Aad et~al.
\newblock {Search for magnetic monopoles in $\sqrt{s}=7$ TeV $pp$ collisions
  with the ATLAS detector}.
\newblock {\em Phys. Rev. Lett.}, 109:261803, 2012.

\bibitem{DeRoeck:2012wua}
A.~De~Roeck, H.P. H{\"a}chler, A.M. Hirt, M.-D. Joergensen, A.~Katre,
  P.~Mermod, D.~Milstead, and T.~Sloan.
\newblock {Development of a magnetometer-based search strategy for stopped
  monopoles at the Large Hadron Collider}.
\newblock {\em Eur. Phys. J.}, C72:2212, 2012.

\bibitem{Aad:2015kta}
G.~Aad et~al.
\newblock {Search for magnetic monopoles and stable particles with high
  electric charges in 8 TeV $pp$ collisions with the ATLAS detector}.
\newblock {\em Phys. Rev.}, D93(5):052009, 2016.

\end{thebibliography}
\end{document}